\documentclass[11pt,a4paper,oneside]{book}

\usepackage[top=25mm,bottom=25mm,left=32.5mm,right=32.5mm,includehead,includefoot]{geometry}  %%% equally sized margins
\usepackage{fancyhdr,setspace}
\usepackage[utf8]{inputenc}

\usepackage{amsfonts,amsmath,amssymb}
\usepackage{hyperref}
\usepackage{comment}
\usepackage[font=footnotesize,labelfont=bf,margin=1cm]{caption}

\usepackage{chngcntr}
\counterwithout{footnote}{chapter}

\numberwithin{equation}{chapter}
\allowdisplaybreaks[4]

\usepackage{graphicx}
\usepackage{chngcntr}
\counterwithout{figure}{chapter}
\counterwithout{table}{chapter}

\usepackage{cite}
\usepackage{multirow}					
\usepackage{multicol}
\newcommand{\dd}{\mathrm{d}}	
\newcommand{\qandq}{\qquad\mathrm{and}\qquad}
\newcommand{\qorq}{\qquad\mathrm{or}\qquad}
\newcommand{\diag}{\mathrm{diag}}

\newcommand{\beq}{\begin{equation}}
\newcommand{\eeq}{\end{equation}}
\newcommand{\be}{\begin{equation}}
\newcommand{\ee}{\end{equation}}

\newcommand{\cf}{{\it cf.\ }}
\newcommand{\ie}{{\it i.e.\ }}
\newcommand{\eg}{{\it e.g.\ }}
\newcommand{\etal}{{\it et\;al.\ }}

\newcommand{\Odd}{O(D,D)}	
\newcommand{\Slf}{SL(5)}	
\newcommand{\Es}{E_7}	
\newcommand{\Slt}{SL(2)\times\mathbb{R}^+}	
\newcommand{\G}{\Slt}
\newcommand{\Edd}{E_{D}}

\newcommand{\sfC}{\mathsf{C}}
\newcommand{\sfD}{\mathsf{D}}
\newcommand{\sfE}{\mathsf{E}}

\newcommand{\HH}{\mathcal{H}}
\newcommand{\cH}{\mathcal{H}}					
\newcommand{\MM}{\mathcal{M}}
\newcommand{\gM}{\mathcal{M}}
\newcommand{\LL}{\mathcal{L}}
\newcommand{\EE}{\mathcal{E}}
\newcommand{\RR}{\mathcal{R}}
\newcommand{\AAA}{\mathcal{A}}
\newcommand{\FF}{\mathcal{F}}
\newcommand{\DD}{\mathcal{D}}

\newcommand{\II}{\mathbb{I}}
\newcommand{\JJ}{\mathbb{J}}
\newcommand{\KKK}{\mathbb{K}}
\newcommand{\XX}{\mathbb{X}}

\newcommand{\g}{\mathfrak{g}}

\newcommand{\hR}{\hat{R}}
\newcommand{\hg}{\hat{g}}

\newcommand{\hx}{\hat{x}}
\newcommand{\hp}{\hat{p}}

\newcommand{\ha}{\hat{a}}
\newcommand{\hta}{\hat{\tilde{a}}}
\newcommand{\hmu}{{\hat{\mu}}}
\newcommand{\hnu}{{\hat{\nu}}}
\newcommand{\hrho}{{\hat{\rho}}}
\newcommand{\hphi}{{\hat{\phi}}}
\newcommand{\htphi}{\hat{\tilde{\phi}}}

\newcommand{\hlambda}{\hat{\lambda}}

\newcommand{\ba}{{\bar{a}}}
\newcommand{\bb}{{\bar{b}}}
\newcommand{\bi}{{\bar{i}}}
\newcommand{\bj}{{\bar{j}}}
\newcommand{\bk}{{\bar{k}}}
\newcommand{\bl}{{\bar{l}}}
\newcommand{\bm}{{\bar{m}}}
\newcommand{\bn}{{\bar{n}}}
\newcommand{\bp}{{\bar{p}}}
\newcommand{\bq}{{\bar{q}}}
\newcommand{\bt}{{\bar{t}}}
\newcommand{\bz}{{\bar{z}}}
\newcommand{\bw}{\bar{w}}
\newcommand{\btheta}{{\bar{\theta}}}
\newcommand{\btau}{{\bar{\tau}}}
\newcommand{\bg}{\bar{g}}
\newcommand{\bmu}{{\bar{\mu}}}

\newcommand{\bF}{\bar{F}}

\newcommand{\tpartial}{\tilde{\partial}}
\newcommand{\ttt}{\tilde{t}}
\newcommand{\tz}{{\tilde{z}}}
\newcommand{\tx}{\tilde{x}}
\newcommand{\ty}{\tilde{y}}
\newcommand{\ta}{\tilde{a}}
\newcommand{\tg}{\tilde{g}}
\newcommand{\ttheta}{{\tilde{\theta}}}
\newcommand{\tphi}{{\tilde{\phi}}}

\newcommand{\tu}{\tilde{u}}
\newcommand{\tv}{\tilde{v}}
\newcommand{\tC}{\tilde{C}}
\newcommand{\tF}{\tilde{F}}
\newcommand{\tMM}{\tilde{\MM}}
\newcommand{\tQ}{\tilde{Q}}

\newcommand{\dm}{{\dot{m}}}
\newcommand{\dn}{{\dot{n}}}
\newcommand{\dpp}{{\dot{p}}}
\newcommand{\dq}{{\dot{q}}}
\newcommand{\dg}{\dot{g}}	

\newcommand{\ua}{{\underline{a}}}
\newcommand{\ub}{{\underline{b}}}

\newcommand{\PP}[4]{P_{#1#2}^{\phantom{#1#2}#3#4}}

\newcommand{\Aa}{A}
\newcommand{\Ab}{B}
\newcommand{\Ac}{C}
\newcommand{\Ad}{D}
\newcommand{\Ae}{E}
\newcommand{\Af}{F}

\newcommand{\bA}{\bar{C}}
\newcommand{\bB}{\bar{C}}
\newcommand{\bC}{\bar{C}}

\newcommand{\Fa}{\mathcal{F}}
\newcommand{\Fb}{\mathcal{H}}
\newcommand{\Fc}{\mathcal{J}}
\newcommand{\Fd}{\mathcal{K}}
\newcommand{\Fe}{\mathcal{L}}

\newcommand{\TAw}{\TA(1/7)}
\newcommand{\TBw}{\TB(2/7)}
\newcommand{\TCw}{\TC(3/7)}
\newcommand{\TDw}{\TD(4/7)}
\newcommand{\TEw}{\TE(5/7)}
\newcommand{\TFw}{\TF(6/7)}
\newcommand{\TSw}{\TS(1)}
\newcommand{\TA}{\mathbb{A}}
\newcommand{\TB}{\mathbb{B}}
\newcommand{\TC}{\mathbb{C}}
\newcommand{\TD}{\mathbb{D}}
\newcommand{\TE}{\mathbb{E}}
\newcommand{\TF}{\mathbb{F}}
\newcommand{\TS}{\mathbb{S}}

\newcommand{\1}{{\mu_1}}
\newcommand{\2}{{\mu_2}}
\newcommand{\3}{{\mu_3}}
\newcommand{\4}{{\mu_4}}
\newcommand{\5}{{\mu_5}}
\newcommand{\6}{{\mu_6}}
\newcommand{\7}{{\mu_7}}
\newcommand{\8}{{\mu_8}}

\newcommand{\mt}{{\mu_{10}}}

\newcommand{\gL}{\LL}

\newcommand{\D}{\mathcal{D}}
\newcommand{\hpartial}{\hat{\partial}}
\newcommand{\hd}{\hpartial}

\newcommand{\p}{\bullet}
\newcommand{\pl}{\!\!\p\!}

\newcommand{\im}{\mathrm{Im\, }}

\begin{document}

\newcommand{\HRule}{\rule{\linewidth}{0.5mm}}

\begin{titlepage}

\begin{centering}

\begin{figure}[h]
\centering
\includegraphics[scale=0.8]{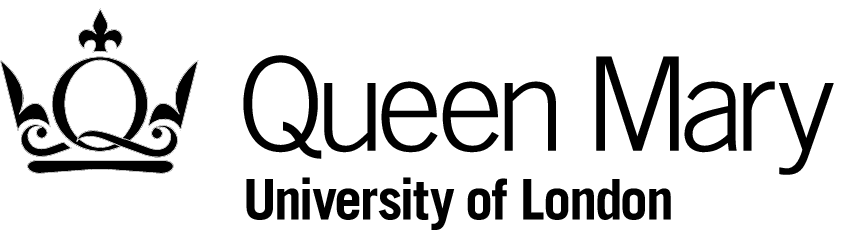}
\end{figure}\ \\

{\LARGE
\textsc{Thesis submitted for the Degree of \\[0.1cm] Doctor of Philosophy}
}\\[1.1cm]

\HRule\\[0.5cm]
{\huge
\bfseries{DUALITY COVARIANT SOLUTIONS \\[0.5cm] IN EXTENDED FIELD THEORIES}
}\\[0.5cm]
\HRule\\[1.5cm]

{\huge
\textsc{Felix J. Rudolph}
}\\[2cm]

{\Large
Supervisor \\[0.3cm]
\textsc{Prof. David S. Berman}\\[1.5cm]
September 2016}\\[0.5cm]

{\large
\textsf{
Centre for Research in String Theory \\[0.1cm]
School of Physics and Astronomy \\[0.1cm]
Queen Mary University of London}}

\end{centering}

\end{titlepage}

\thispagestyle{empty}
\topskip0pt

\vspace*{7cm}
\begin{center}
\textit{ This thesis is dedicated to my parents \\ for their unconditional support and love.}
\end{center}
\vspace*{\fill}

\newpage
\thispagestyle{empty}
\topskip0pt
\vspace*{7cm}
\begin{center}
\begin{minipage}[][][c]{0.8\textwidth}
\textit{
\hspace{-5pt}What a piece of work is a man! \\
How noble in reason, %\\
how infinite in faculty! \\
In form and moving how express and admirable! \\
In action how like an angel! %\\
In apprehension how like a god! \\
The beauty of the world! %\\
The paragon of animals! \\
And yet to me, what is this quintessence of dust?} \\
\begin{flushright}
--- Hamlet, Act II, Scene 2
\end{flushright}
\end{minipage}
\end{center}

\vspace*{\fill}

\chapter*{Abstract}

Double field theory and exceptional field theory are formulations of supergravity that make certain dualities manifest symmetries of the action. To achieve this, the geometry is extended by including dual coordinates corresponding to winding modes of the fundamental objects. This geometrically unifies the spacetime metric and the gauge fields (and their local symmetries) in a generalized geometry. Solutions to these extended field theories take the simple form of waves and monopoles in the extended space. From a supergravity point of view they appear as 1/2 BPS objects such as the string, the membrane and the fivebrane in ordinary spacetime. In this thesis double field theory and exceptional field theory are introduced, solutions to their equations of motion are constructed and their properties are analyzed. Further it is established how isometries in the extended space give rise to duality relations between the supergravity solutions. Extensions to these core ideas include studying Goldstone modes, probing singularities at the core of solutions and localizing them in winding space. The relation of exceptional field theory to F-theory is also covered providing an action for the latter and incorporating the duality between M-theory and F-theory.
\chapter*{Acknowledgements}

First and foremost, I wish to thank my supervisor, Professor David Berman, to whom I will always be grateful for his support, guidance and encouragement throughout my time as a Ph.D. student. His wisdom, intuition and experience, but most of all his enthusiasm for the subject, were the driving force behind my research. He is a tremendous mentor and has become a true friend.

I also want to thank my other collaborators, Joel Berkeley, Chris Blair and Emanuel Malek, for our excellent cooperation, and the other academics in the Centre for Research in String Theory for many interesting and stimulating conversations. My years of research were supported by an STFC studentship.

To my fellow Ph.D. students I want to extend the warmest of gratitudes for making my time at Queen Mary such an enjoyable experience. They have been a great source of inspirational discussion and helpful advice. In particular I want to thank James McGrane, Brenda Penante, Edward Hughes, Zac Kenton, Paolo Mattioli, Martyna Kostacinska, Emanuele Moscato, Joseph Hayling and Rodolfo Panerai. I greatly value our companionship and I am sure the friendships we have built over the years will last a lifetime.  

Finally I wish to thank my family and friends for their support and belief in me. Especially my parents, to whom this thesis is dedicated, deserve all my thanks. Without their backing this thesis would not have been possible.

\chapter*{Declaration}

I, Felix Jakob Rudolph, confirm that the research included within this thesis is my own work or that where it has been carried out in collaboration with, or supported by others, that this is duly acknowledged below and my contribution indicated. Previously published material is also acknowledged below. \\

\noindent I attest that I have exercised reasonable care to ensure that the work is original, and does not to the best of my knowledge break any UK law, infringe any third party’s copyright or other Intellectual Property Right, or contain any confidential material. \\

\noindent I accept that the College has the right to use plagiarism detection software to check the electronic version of the thesis. \\

\noindent I confirm that this thesis has not been previously submitted for the award of a degree by this or any other university. \\

\noindent The copyright of this thesis rests with the author and no quotation from it or information derived from it may be published without the prior written consent of the author. \\

\noindent Signature: 

\begin{figure}[h]
%\hspace{1cm}
\includegraphics[scale=0.5]{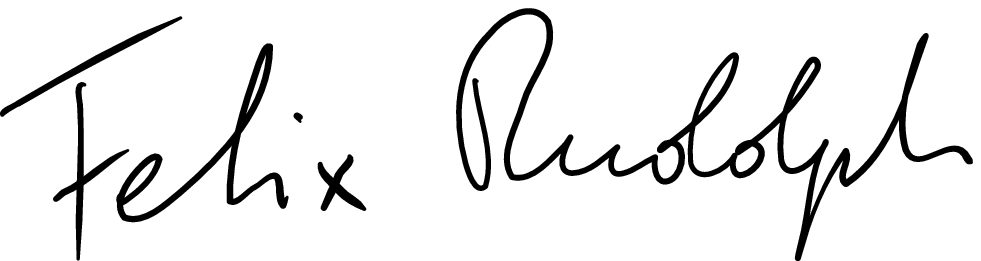}
\end{figure}

\noindent Date: $1^{\mathrm{st}}$ July 2016 \\
 
\noindent  Details of collaboration and publications: \\

\noindent This thesis describes research carried out with my supervisor David S. Berman which was published in \cite{Berkeley:2014nza,Berman:2014jsa,Berman:2014hna,Berman:2015rcc}. It also contains some unpublished material. We collaborated with Joel Berkeley on \cite{Berkeley:2014nza} and Chris Blair and Emanual Malek on \cite{Berman:2015rcc}. Where other sources have been used, they are cited in the bibliography.

\tableofcontents
\chapter{Introduction}

In many disciplines of science symmetries provide a guiding principle in the quest for classifying and unifying different concepts and ideas. Even more so in theoretical physics, symmetries are a powerful tool to discover the underpinning rules of the universe. A very rich and fruitful example of the employment of symmetries to enhance our knowledge of the physical world is superstring theory.

Since its formative years, string theory has grown into a giant edifice firmly rooted in its mathematical foundations with branches reaching into multiple diverse directions scaling new heights of insights into and understanding of the fundamental laws of nature.  At the heart of string theory are three intimately connected pillars: the sigma model, supergravity and dualities. The string sigma model describes the dynamics of strings on general backgrounds. Supergravity provides the low energy effective description of the theory, its solutions giving string theory vacua. A web of string theory dualities then relates seemingly different supergravity backgrounds. 

Whereas a \emph{symmetry} is is a property of an individual object or theory expressing an invariance under certain transformations, a \emph{duality} allows for an identification of different theories or frameworks. This can be very useful if a given problem is very hard to solve in one formulation, but might be tackled much more easily in the dual description. An interesting question now is how to promote a duality to a symmetry of a grander theory such that the distinct yet dual frameworks simply become different aspects of a single formulation which provides some kind of unification. The analysis of dualities of string theory and how to make them manifest symmetries is one of the key topics of this thesis.

\bigskip

Due to their extended nature, strings experience geometry with its extra dimensions very differently than point particles. This is exemplified by one of the dualities, namely T-duality. Some of the dimensions of string theory might be compact, then a string can not only move along such a direction in space (giving it quantized momentum) but can also wind around it. Associated to this mode of the string is a quantum number called \emph{winding}. The presence of these winding modes of the string which see spacetime differently than momentum modes leads to a much more varied and complex structure of spacetime, the objects populating it and the interactions between them. 

T-duality states that a string propagating in a background geometry with a circle of a given radius is equivalent to a string moving in a background with inverse radius if one exchanges the winding and momentum quantum numbers. This can be illustrated by considering the mass-shell condition for a closed string in a geometry with a single compact dimension of radius $R$. Then we have
\begin{equation}
	m^2 = \frac{2}{\alpha'}\left( N + \tilde{N} - 2 \right) + \frac{p^2}{R^2} + \frac{w^2R^2}{\alpha'^2}
\label{eq:masspectrum}
\end{equation}
and
\begin{equation}
	N - \tilde{N} = p\cdot w \, .
\label{eq:levelmatching}
\end{equation}
Here $N$ and $\tilde{N}$ are left- and right-moving oscillator modes of the string, $p$ is the momentum and $w$ the winding quantum number. As usually in string theory, the scale is set by $\alpha'=\frac{1}{2\pi T}$ which is inversely proportional to the string tension $T$. The first equation shows the contributions of the oscillator modes, the zero-point energy and the momentum and winding modes of the string to the mass spectrum. The second equation is called the level-matching condition and relates left- and right-moving modes.

These two equations are clearly invariant under the exchange of winding and momentum and the radius with its inverse
\begin{equation}
 	p\leftrightarrow w \qandq R \leftrightarrow \frac{\alpha'}{R} \, .
\end{equation} 
This is a manifestation of T-duality; it exemplifies how strings differ from point particles when probing the geometry of spacetime. More generally, a duality -- the presence of a hidden symmetry or relation between theories -- once found, immediately provokes a set of questions. The very presence of the duality seems to  imply a lack of understanding of the theory; one hopes to discover the reason for the hidden symmetry and perhaps construct a theory in which this duality is a manifest symmetry. The question is therefore if there is a more appropriate geometrical language to formulate string theory which takes T-duality into account. We want to consider a more unified description of spacetime where momentum and winding modes are treated on an equal footing. In such a formalism T-duality would be manifest symmetry.

From a sigma model point of view, there have been a several approaches of a duality symmetric string in a spacetime with double the number of dimension \cite{Duff:1989tf,Tseytlin:1990nb,Tseytlin:1990va,Maharana:1992my,Cremmer:1997ct,Hull:2004in,Hull:2006va}. By extending the target space, geometries related by T-duality can be included in a symmetric way. The description of the geometry of a doubled target space including an action and equations of motion for the background fields is provided by \emph{Double Field Theory} (DFT). It was proposed by Hull and Zwiebach \cite{Hull:2009mi} and was constructed utilizing string field theory. It is related to earlier work by Siegel \cite{Siegel:1993xq,Siegel:1993th} and Tseytlin \cite{Tseytlin:1990nb,Tseytlin:1990va}. The extended tangent space of the theory resembles the development of generalized geometry by Hitchin \cite{Hitchin:2004ut} and Gualtieri \cite{Gualtieri:2003dx}.

The basic idea of double field theory is to not only double the dimensions to treat T-dual modes uniformly but also to encode the dynamical field content in a \emph{generalized metric} on the doubled space. The generalized metric provides a geometric unification of the usual spacetime metric with the Kalb-Ramond two-form field of the NSNS-sector into a single object. This is a very elegant formulation in the spirit of worldsheet theory where both fields arise from the level two mode of the closed string as opposed to the traditional supergravity picture where the metric forms a background on which the two-form field lives. Now the action of DFT can be expressed in terms of the generalized metric together with a shifted form of the string dilaton in a T-duality manifest form.

Let us now briefly return to the mass spectrum of the closed string given in \eqref{eq:masspectrum}. The contribution of the winding mode is proportional to the string tension via $\alpha'$ while the momentum mode contribution is independent of it. We can thus look at different limits of the radius $R$. For $R^2\gg\alpha'$ the momentum modes are light and should dominate the low energy physics while for $R^2\ll\alpha'$ the winding modes become light. The oscillator modes also depend on the string tension, but are independent of $R$. Therefore there are different regimes we can consider.

For $R\gg\sqrt{\alpha'}$ the momentum modes are lighter than both winding and oscillator modes. In this scenario supergravity provides a consistent low energy effective theory. In the opposite case where $R\ll\sqrt{\alpha'}$ the winding modes are the lightest and again supergravity is a valid description, but this time in the T-dual picture. If $R\approx\sqrt{\alpha'}$, \ie around the string scale, all modes -- momentum, oscillator, winding -- have to be kept. At other scales there is a hierarchy with a supergravity action for the lightest modes. The conclusion of this is that there is no regime where one can only consider momentum and winding modes but cast way the oscillator modes.

Therefore double field theory is not a low energy effective theory but rather a truncation of string theory. Nevertheless DFT has proven itself as a consistent framework without pathologies by passing the usual quantum tests such as modular invariance \cite{Berman:2007vi} and vanishing of the beta function \cite{Berman:2007xn,Berman:2007yf}.

\bigskip

The non-perturbative version of string theory is M-theory whose low energy effective action is eleven-dimensional supergravity. It is related to the ten-dimensional string theories via various compactification limits and a web of dualities. Besides T-duality there is also S-duality (strong/weak or electric/magnetic duality) which together are part of U-duality. The fundamental objects of M-theory are the membrane and the fivebrane. They give rise to the string, fivebrane and D-branes in string theory under dimensional reduction.

The M-theory dualities arise when the eleven-dimensional theory is compactified on $D$-dimensional tori. The relevant symmetry groups $G=E_D$ of these ``hidden symmetries'' \cite{Cremmer:1997ct, Cremmer:1998px} in the low energy limit -- $(11-D)$-dimensional supergravity -- are listen in Table \ref{tab:GH}. These are the continuous version of the discrete U-duality groups of M-theory. In order to obtain a better understanding of these dualities and their influence on the full uncompactified eleven-dimensional supergravity, the aim is to make them manifest symmetries along the lines of T-duality in double field theory.

By introducing extra coordinates which now correspond to wrapping modes of the membrane and fivebrane, a generalized metric on the extended space can be constructed which geometrically unifies the spacetime metric with the three- and six-form fields of M-theory the branes couple to. This \emph{exceptional extended geometry} has been constructed for several U-duality groups but was originally restricted to truncations of the eleven-dimensional theory \cite{Hull:2007zu, Pacheco:2008ps, Hillmann:2009ci, Berman:2010is, Coimbra:2011ky, Coimbra:2012af} where the ``external'' metric (for the $11-D$ dimensions) was taken to be flat and off-diagonal terms (the ``gravi-photon'') were set to zero. Furthermore, coordinate dependence was restricted to the internal extended coordinates. The term ``exceptional'' is taken from the presence of the exceptional groups $E_D$.

The full, non-truncated \emph{Exceptional Field Theory} (EFT) allows for a dependence on all coordinates, external, internal and extended. This allows for eleven-dimensional supergravity to be embedded into a theory that is fully covariant under the exceptional groups $E_D$ \cite{Hohm:2013pua}. These exceptional field theories have been constructed for the groups in Table \ref{tab:GH} in \cite{Hohm:2013vpa,Hohm:2013uia,Hohm:2014fxa,Abzalov:2015ega,Musaev:2015ces,Hohm:2015xna,Berman:2015rcc}.

\bigskip

Like the spacetime metric in general relativity, the generalized metrics of DFT and EFT belong to a symmetric space G/H. In relativity we have the global symmetry group $G=GL(D)$ with its maximal compact subgroup $H=SO(D)$ in the Euclidean case or $H=SO(1,D-1)$ in the Lorentzian case where local $H$-transformations are a symmetry of the metric. In DFT we have $G=\Odd$ and $H=O(D)\times O(D)$ or $H=O(1,D-1)\times O(1,D-1)$. For the various EFTs we have the exceptional groups $G=E_D$, their subgroups $H$ are also listed in Table \ref{tab:GH}. Additional local symmetries of the theories are the diffeomorphisms from general coordinate transformations and the gauge transformations of the $p$-form fields. These local symmetries do not commute, they form an algebroid\footnote{This algebroid is not of the Lie-type but a generalization of a Courant algebroid whose bracket is the so called C-bracket which reduces to the usual Courant bracket if certain necessary constraints are imposed.}. The diffeomorphisms and gauge transformations can be combined into a single local symmetry of the generalized geometry which is generated infinitesimally by a \emph{generalized Lie derivative}.

Besides the global and local symmetry groups G and H, Table \ref{tab:GH} lists the two representations\footnote{These are the relevant representations of gauged or maximal supergravities in the given spacetime dimensions, see for example \cite{deWit:2008ta}.} $R_1$ and $R_2$ of $G$. The first one is the representation of the generalized coordinates with dimension of the extended space. The second one is the representation of a constraint that is required by all extended theories. The necessity of this can be seen by considering the following.

\begin{table}[h]
\centering
\begin{tabular}{|c|c|c|c|c|}
\hline
$D$ & G & H & $R_1$ & $R_2$ \\ \hline
2 & $SL(2)\times\mathbb{R}^+$ & $SO(2)$ & $(\mathbf{2},\mathbf{1})$ & $(\mathbf{1},\mathbf{1})$ \\
3 & $SL(3)\times SL(2)$ & $SO(3)\times SO(2)$ & $(\mathbf{3},\mathbf{2})$ & $(\bar{\mathbf{3}},\mathbf{1})$ \\
4 & $SL(5)$ & $SO(5)$ & $\mathbf{10}$ & $\bar{\mathbf{5}}$ \\
5 & $SO(5,5)$ & $SO(5)\times SO(5)$ & $\mathbf{16}$ & $\mathbf{10}$ \\
6 & $E_6$ & $USp(8)$ & $\mathbf{27}$ & $\bar{\mathbf{27}}$ \\
7 & $E_7$ & $SU(8)$ & $\mathbf{56}$ & $\mathbf{133}$ \\
8 & $E_8$ & $SO(16)$ & $\mathbf{248}$ & $\mathbf{1}\oplus\mathbf{3875}$ \\ \hline
\end{tabular}
\caption{The duality groups G and their maximal compact subgroup H that appear in the reduction of eleven-dimensional supergravity to $11-D$ dimensions. The two columns $R_1$ and $R_2$ give the representation of the extended coordinates and the section condition respectively.}
\label{tab:GH}
\end{table}

The formulations of DFT and EFT are defined on an extended spacetime. Since string theories are notoriously tied to a critical dimension, the dynamics have to be restricted to end up with the correct physical degrees of freedom and a consistent theory. The constraint which achieves this is called the \emph{physical section condition}. It specifies locally a subspace of the extended space which forms the physical spacetime in a given duality frame. In DFT this amounts to choosing a maximal isotropic $D$-dimensional subspace of the $2D$-dimensional doubled space. In the EFTs the section condition is more complicated and projects onto a $D$-dimensional subspace of the much larger exceptionally extended space.

The section condition has its physical origin in the level-matching condition of the closed string given in \eqref{eq:levelmatching} with the oscillator modes set to zero. It can be stated in terms of the $\Odd$ invariant structure $\eta^{MN}$ for DFT  or a projection on the representation $R_2$ for the EFTs (here $M,N$ are a generalized coordinate indices in the representation $R_1$)
\begin{equation}
	\eta^{MN}\partial_M\otimes\partial_N = 0 \qorq 
	\partial_M\otimes\partial_N\Big|_{R_2} = 0
\end{equation}
which will be explained in more detail in the main text where we will also discuss a subtle difference between the weak and strong form of this constraint and how it might be relaxed. Here the notation indicates that the derivatives act on any field or gauge parameter and also products thereof. 

One way of satisfying the section condition in DFT is to restrict all coordinate dependences of fields and parameters to the usual spacetime coordinates, \ie they have no dependence on the extra winding coordinates introduced. In this case the theory reduces to the formalism of generalized geometry where the tangent space is extended but the underlying space itself is not. This ultimately means there is no extra physical content in the theory beyond supergravity, but it provides a very elegant and potentially powerful reformulation of the theory which makes the duality a manifest symmetry. 

\bigskip

Once the duality manifest theories have been established, one can look for solutions to their equations of motions. The observation that the setup of DFT and EFT is very similar to Kaluza-Klein theories is very useful in the search for and construction of solutions. The form of the generalized metric in terms of the spacetime metric and the form field(s) is the same as for a KK-ansatz to reduce the extra dimensions. This shared structure with KK-theory will be a guiding principle when considering solutions to the extended theories. For example, a massless, uncharged state in the full (say five-dimensional) KK-theory acquires a mass and charge when viewed from the reduced point of view (in four dimensions). It is the momentum along the KK-direction which becomes a fundamental charge when writing down the KK-ansatz to split the five dimensions into $4+1$. This can be easily seen by considering the mass-shell condition for the massless state in five dimensions
\begin{equation}
	\hp_\hmu \hp^\hmu = 0 \, .
\end{equation}
Now the five-dimensional momentum is split into $\hp^\hmu = (p^\mu,p^z)$ where $z$ is the direction of the KK-reduction (taken to be compact and an isometry of the full theory). Then the above relation can be rearranged as
\begin{equation}
	p_\mu p^\mu + p_zp^z = 0 \qorq p_\mu p^\mu = - p_zp^z = -m^2
\end{equation}
and we see that from a four-dimensional point of view we have a state with mass given by the momentum along the KK-direction $z$. Note that in this example the off-diagonal term in the KK-ansatz for the metric was taken to vanish. If this KK-vector is non-zero, the state in the reduced picture acquires a charge with respect to this $U(1)$ gauge field. Therefore, in Kaluza-Klein theories the origin of electric charge is from momentum in the KK-direction. The quantization of momentum (due to the KK-direction being a circle) then results in the quantization of electric charge. The origin of magnetic charges comes from twisting the KK-circle to produce a non-trivial circle bundle with non-vanishing first Chern class. The first Chern class is the magnetic charge. The construction of such a non-trivial solution for traditional Kaluza-Klein theory was first given in \cite{Sorkin:1983ns,Gross:1983hb} and is now called the KK-monopole.

In analogy, we want to consider solutions in the dual directions of the extended field theories and interpret them from an ordinary supergravity point of view. This has been done for the wave solution in DFT and in the truncated $\Slf$ EFT \cite{Berkeley:2014nza} which gives the fundamental string and the membrane respectively; and for the monopole solution in DFT and in the truncated $\Es$ EFT \cite{Berman:2014jsa} which gives the NS5 and M5 fivebranes respectively. By going to the full non-truncated $\Es$ EFT and considering a self-dual solution with both wave-like and monopole-like aspects \cite{Berman:2014hna}, a single solution can be constructed which gives all 1/2 BPS branes of ten- and eleven-dimensional supergravity, including bound states (an executive summary of this can be found in \cite{Rudolph:2015pgz}).

Besides the analogy with solutions in Kaluza-Klein theory, the work on null states in the dual space has been influenced by the fact that the non-linear realization construction central to the $E_{11}$ programme (see below) has its origins in the theory of pions as Goldstone modes of the spontaneously broken chiral Lagrangian. This led to the idea that the duality invariant theory may contain massless Goldstone modes from spontaneously breaking the duality symmetry. Analyzing the Goldstone modes of the DFT wave solution gives the duality symmetric string of Tseytlin \cite{Tseytlin:1990nb,Tseytlin:1990va} in a doubled space. Whether the null states identified are the relevant Goldstone modes though is an open question.

Other aspects of DFT, EFT and their solutions that have been studied in \cite{Berkeley:2014nza,Berman:2014jsa,Berman:2014hna} and which will be presented in this thesis include the special form of the equations of motion which always appear under a projector and the singularity structure of the wave and monopole solutions in extended spaces. Some unpublished ideas and material on the effects of localizing these solutions in winding space and corresponding worldsheet instanton corrections to the geometry will also be discussed.

Solutions in exceptional field theory play a crucial role in \cite{Berman:2015rcc} where the $\Slt$ EFT with $D=2$ is constructed and it is shown that it in fact provides an action for F-theory. The relevant EFT is a twelve-dimensional theory with manifest $\Slt$ symmetry. To identify it with F-theory \cite{Vafa:1996xn}, amongst other steps the relations between brane solutions in ten, eleven and twelve dimensions are analyzed. In some sense the $\Slt$ EFT is a minimal exceptional field theory since it introduces only a single extra coordinate and the corresponding exceptionally extended geometry is too small to include any effects of geometrizing the gauge fields.  

Nevertheless it provides the means to place F-theory together with M-theory in a unified geometric framework. In future it will be interesting to see how this minimal EFT can be included in the viewpoint of the higher rank groups (with larger $D$). We will comment on a potential hierarchy of EFTs and the inclusion of solution generating transformations in higher dimensional symmetry groups towards the end of this thesis.

\section{Overview of Bibliography}
So far many important references to the key results in the development of double field theory and the exceptional field theories have been given. Here we will give a more systematic overview of standard references which will be referred to throughout this thesis. 

Following the initial endeavours by Duff \cite{Duff:1989tf} on string dualities, Tseytlin formulated a version of the string in a doubled space \cite{Tseytlin:1990nb,Tseytlin:1990va}. The new geometry to describe a duality covariant version of supergravity was introduced by Siegel \cite{Siegel:1993xq,Siegel:1993th}. With the insights of generalized geometry \cite{Hitchin:2004ut,Gualtieri:2003dx}, Hull then constructed the double sigma model in \cite{Hull:2004in} and developed more of the ideas which eventually led to the conception of double field theory which was established with the seminal paper \cite{Hull:2009mi} by Hull and Zwiebach. Since DFT allows for dynamics in all the doubled dimensions, it goes beyond the duality covariant formulation of supergravity. There were two key groups that have developed DFT further, one of Hohm, Hull and Zwiebach \cite{Hull:2009zb,Hohm:2010jy,Hohm:2010pp,Hohm:2010xe,Hohm:2011nu} and the other of Jeon, Lee and Park \cite{Jeon:2010rw,Jeon:2011cn,Jeon:2011vx,Jeon:2011sq,Jeon:2012hp}. Extensions and further developments of DFT can be found in \cite{Blair:2013noa,Berman:2013uda,Blair:2014kla,Wu:2013sha,Wu:2013ixa,Ma:2014ala} whereas a relaxation of the section conditon is discussed in \cite{Aldazabal:2011nj,Geissbuhler:2011mx,Dibitetto:2012rk,Grana:2012rr,Geissbuhler:2013uka,Berman:2013cli} and \cite{Lee:2015qza}. Three excellent reviews of DFT and related developments are \cite{Aldazabal:2013sca, Berman:2013eva, Hohm:2013bwa}.

In the duality manifest formulation of M-theory there was again initial work by Duff \cite{Duff:1990hn} and then Hull \cite{Hull:2007zu}, Hillmann \cite{Hillmann:2009ci} and Waldram \etal \cite{Pacheco:2008ps, Coimbra:2011ky, Coimbra:2012af}, followed  by Berman, Perry, \etal \cite{Berman:2010is, Berman:2011pe,Berman:2011kg,Berman:2011cg, Berman:2011jh,Berman:2012vc}. Further contributions to the duality invariant theories are \cite{Park:2013gaj, Cederwall:2013oaa,Cederwall:2013naa,Strickland-Constable:2013xta,Aldazabal:2013via,Park:2014una}. The full non-truncated form of exceptional field theory was provided by Hohm and Samtleben \cite{Hohm:2013pua} and spelled out for the various exceptional groups in \cite{Hohm:2013vpa,Hohm:2013uia,Hohm:2014fxa,Abzalov:2015ega,Musaev:2015ces,Hohm:2015xna} and \cite{Berman:2015rcc}. Extensions to EFT such as supersymmetry and further developments can be found in \cite{Godazgar:2014nqa,Musaev:2014lna,Wang:2015hca}.

All of this is related to the long standing $E_{11}$ program of West and collaborators which anticipated some of these developments, see for example \cite{West:2001as, Englert:2003zs, West:2003fc, Kleinschmidt:2003jf, West:2004kb,West:2012qm,Tumanov:2015iea}. For quantum aspects of the duality manifest string see \cite{Berman:2007vi,Berman:2007xn,Berman:2007yf,Hohm:2013jaa,Betz:2014aia}. In addition, there have been a whole host of fascinating recent results, for a representative but by no means complete sample one may start with \cite{Berman:2013uda, Blair:2013noa, Blair:2013gqa, Lee:2014mla, Lee:2013hma, Hull:2014mxa, Coimbra:2014qaa, Rosabal:2014rga, Hohm:2014xsa, Hohm:2014sxa, Cederwall:2014kxa, Cederwall:2014opa, Blair:2014zba}.

More specifically, large (finite) gauge transformations are studied in \cite{Hohm:2012gk,Park:2013mpa,Berman:2014jba,Rey:2015mba,Chaemjumrus:2015vap} (with caveats concerning global issues discussed in \cite{Papadopoulos:2014mxa,Papadopoulos:2014ifa}), non-geometric fluxes are treated by \cite{Chatzistavrakidis:2012qj,Chatzistavrakidis:2013qca} and non-commutativity and non-associativity aspects of the geometry are discussed in \cite{Mathai:2004qq,Bouwknegt:2004ap,Mathai:2005fd,Blumenhagen:2010hj,Lust:2010iy,Blumenhagen:2011ph,Condeescu:2012sp,Andriot:2012vb}. 

The study of solutions of DFT and EFT which is the main topic of this thesis began with \cite{Berkeley:2014nza,Berman:2014jsa,Berman:2014hna} by the author, his supervisor and collaborators. This is related to work on ``exotic branes'' \cite{deBoer:2010ud,deBoer:2012ma,Kimura:2013fda,Kimura:2013zva,Kimura:2013khz,Kimura:2014upa,Kimura:2014bxa,Kimura:2014aja}. When studying classic supergravity solutions such as the wave, the string, the membrane, the fivebrane and the monopole, as well as reviewing concepts like T-duality, Kaluza-Klein reductions and smearing, the book by Ortin \cite{Ortin:2004ms} was an invaluable reference.

\section{Outline of Thesis}
This thesis is structured in the following way. Chapters \ref{ch:DFT} and \ref{ch:DFTsol} are devoted to double field theory. The first one gives a technical introduction to the most important features of DFT while the second one presents the wave and monopole solutions of the theory whose reduction to supergravity shed some light on how T-duality is viewed in the doubled formalism. Chapters \ref{ch:EFT} and \ref{ch:EFTsol} deal with exceptional field theories in their truncated and full form. Again we first give an overview of the formalism (which has a lot of parallels with DFT) to then construct various solutions to the theory which are related to known supergravity objects.

In Chapter \ref{ch:aspects} we look at some interesting aspects and extensions of the DFT and EFT solutions presented, including their singularity structure and localizing them in winding space. Chapter \ref{ch:SL2} outlines the construction of another EFT, the one with duality group $\Slt$. Here the emphasis is on the relation to F-theory for which it provides an action. We will finish with some concluding remarks in Chapter \ref{ch:outro}. The appendices \ref{ch:appDFT}, \ref{ch:appE7}, \ref{ch:appEFT}, \ref{ch:apptorus} and \ref{ch:appSL2} contain various detailed calculations and extra material while Appendix \ref{ch:glossary} provides a concise glossary of all the standard supergravity solutions referred to in the main text.

\section{Notation}
Each piece of notation is introduced as required throughout the chapters of this thesis. An attempt was made to keep the notation as consistent as possible, but some sections use their own. Here we want to summarize some of the notation which is standard throughout, especially those objects which are common to double field theory and the exceptional field theories. 

The \textbf{dimension} of the underlying space which gets extended is denoted by $D$. In double field theory the ordinary spacetime is thus $D$-dimensional and the extended space has $2D$ dimensions. In the exceptional field theories $D$ is the dimension of the ``internal'' space which gets extended to $\dim R_1$. Here $\dim R_1$ is the dimension of the coordinate representation $R_1$ in Table \ref{tab:GH}. The ``external'' space has dimension $11-D$.

The \textbf{generalized coordinate vectors} of the doubled and exceptional extended spaces are usually denoted by $X^M$ where  $M=1,\dots,\dim R_1$. The generalized coordinate is composed of a spacetime coordinate $x^\mu$ where $\mu=1,\dots,D$ and one or more winding/wrapping coordinates. These are denoted differently for each theory, \eg in DFT we use $\tx_\mu$ or $\tx^\bmu$ for the winding coordinates whereas for the $\Slf$ invariant theory we use $y_{\mu\nu}$ (the index pair is antisymmetric) corresponding to membrane wrappings. Note that for the full non-truncated EFTs the generalized coordinate is $Y^M$ which is composed of $y^m$ or $y^\alpha,y^s$ whereas the external coordinate is $x^\mu$.

The \textbf{generalized metric} is written as $\HH_{MN}$ and $\MM_{MN}$ for DFT and EFT respectively. It is parametrized by the metric of the underlying spacetime $g_{\mu\nu}$ or $g_{mn}$ and the $p$-form fields $B_2$, $C_3$ and $C_6$, depending on the framework. The generalized metric is constrained to live in the coset G/H. In the exceptional theories, a frequently occurring object is $g_{\mu\nu,\rho\sigma} = g_{\mu[\rho}g_{\sigma]\nu} = \frac{1}{2}(g_{\mu\rho}g_{\sigma\nu}-g_{\mu\sigma}g_{\rho\nu})$ and similar for the inverse $g^{\mu\nu,\rho\sigma}$. They are used to lower and raise antisymmetrized index pairs such as $y_{\mu\nu}$.

\chapter{Doubled Field Theory}
\label{ch:DFT}

In the introduction a first glimpse at double field theory was provided. In this chapter we will give a more technical introduction to the basics of DFT. We begin with an overview of the most important features and elements of the doubled geometry with its $\Odd$ structure. Then we take a closer look at the global and local symmetries, the corresponding algebra and the crucial constraints that need to be imposed. This is followed by the presentation of the action of DFT together with its equations of motion. 

This chapter aims to cover just the basic ingredients of DFT which are needed in the presentation of the solutions in the next chapter. References to the original results in the literature are given throughout and the interested reader should consult them for a more detailed exposition. Recommended reviews of the wider field include \cite{Aldazabal:2013sca,Hohm:2013bwa,Berman:2013eva}.

\section{The Doubled Geometry with $\Odd$ Structure}
\label{sec:DFTintro}

Two of the key achievements of double field theory are a manifestly T-duality invariant formulation of supergravity and the geometric unification of the fields and local symmetries of the bosonic sector, that is the metric and the B-field (which usually mix under T-duality). Both of these feats are accomplished by constructing an extended tangent space which is the combination of the usual tangent space (for momentum modes) plus the cotangent space (for winding modes), and then supplementing the spacetime coordinates (conjugate to momentum) by novel coordinates conjugate to the winding charges. Thereby the target space is also doubled in dimension and one arrives at a truly doubled geometry.

Note that a setup of an extended tangent bundle but without introducing extra coordinates is called \emph{generalized geometry} and was developed by Hitchin \cite{Hitchin:2004ut} and Gualtieri \cite{Gualtieri:2003dx}. Generalized geometry provides some powerful tools to study the geometric unification of symmetries by combining diffeomorphisms and gauge transformation into generalized diffeomorphism of the extended tangent space. Only doubling the underlying space though enables the formulation of a duality covariant action and therefore this extra step is necessary to arrive at double field theory \cite{Hohm:2010pp}.

Throughout the DFT chapters of this thesis we will denote by $D$ the number of dimensions of ordinary supergravity and hence the doubled space of DFT has $2D$ dimensions. Generally one can think of $D=10$, the critical dimension of superstring theory, and thus the doubled space being $20$-dimensional. But everything holds for general $D$ and indeed for certain toy models it helps to look at the simplest case of $D=1$ with the structure group being $O(1,1)$. 

The standard spacetime coordinates are denoted by $x^\mu$ with $\mu=1,\dots,D$, they are conjugate to momentum and are associated with momentum modes. We now introduce a dual set of coordinates\footnote{Occasionally it will be useful to utilize the alternative notation $\tx^\bmu$ for the winding coordinates in order to differentiate between winding and inverse components of tensors.}, $\tx_\mu$ which are conjugate to winding and are associated with winding modes. Hence they are called \emph{winding coordinates}. Vectors are sections of the tangent space $TM$ of the spacetime manifold $M$ while one-forms naturally live in the cotangent space $T^*M$. The idea is now to have the space and the dual space present at the same time, \ie an extended tangent bundle is constructed as
\begin{equation}
E = TM \oplus T^*M 
\end{equation}
whose sections are the formal sum of a vector and a one-form. 

The next step is to introduce coordinates for the doubled space which are denoted by $X^M$ where $M=1,\dots 2D$. The doubled coordinates consist of the spacetime coordinates and the winding coordinates
\begin{equation}
X^M = \begin{pmatrix} x^\mu \\ \tx_\mu \end{pmatrix} \, .
\end{equation}
This doubled space is equipped with an invariant $\Odd$ structure $\eta$. In the above coordinates it can be written as
\begin{equation}
	\eta_{MN} = 
	\begin{pmatrix}
		0 & 1 \\ 1 & 0
	\end{pmatrix}
\label{eq:eta}
\end{equation}
where the off-diagonal blocks are $D\times D$ identity matrices reflecting the standard inner product between $TM$ and $T^*M$. This structure can be seen as an invariant metric to form an inner product for doubled vectors and it provides a pairing between spacetime and winding coordinates. Its inverse $\eta^{MN}$ is given by the same matrix. 

After setting the scene by introducing the coordinates of the doubled space, we now turn to the fields in the theory. In its simplest incarnation DFT is formulated for the bosonic NSNS sector of supergravity, \ie it contains the spacetime metric $g_{\mu\nu}$, the Kalb-Ramond B-field $B_{\mu\nu}$ and the dilaton $\phi$. To construct a theory which is manifestly T-duality invariant, these fields need to be repackaged into tensor representations of $\Odd$, the T-duality group. This leads to the metric and B-field to be treated on an equal footing by combining them into a \emph{generalized metric} $\HH_{MN}$ on the doubled space
\begin{equation}
\HH_{MN} = 
\begin{pmatrix}
g_{\mu\nu} - B_{\mu\rho}g^{\rho\sigma}B_{\sigma\nu} & B_{\mu\rho}g^{\rho\nu} \\
-g^{\mu\sigma}B_{\sigma\nu} & g^{\mu\nu} 
\end{pmatrix}
\label{eq:DFTmetric}
\end{equation}
where $g^{\mu\nu}$ is the inverse of $g_{\mu\nu}$. This generalized metric parametrizes the symmetric coset space $G/H=\Odd/O(D)\times O(D)$ and satisfies
\begin{equation}
	\HH_{MK}\eta^{KL}\HH_{LN} = \eta_{MN} \, .
\end{equation}
Therefore one can construct a generalized vielbein ${\EE^A}_M$ such that
\begin{equation}
\HH_{MN} = {\EE^A}_M{\EE^B}_N \HH_{AB} \, .
\label{eq:DFTvielbein}
\end{equation}
Here $M,N$ are \emph{curved} doubled spacetime indices and $A,B$ are \emph{flat} doubled tangent space indices. The tangent space metric $\HH_{AB}$ can conveniently be taken as
\begin{equation}
\HH_{AB} = 
	\begin{pmatrix}
		1 & 0 \\ 0 & 1 
	\end{pmatrix}
\end{equation}
where the diagonal blocks are $D\times D$ identity matrices. Now if the vielbein of the spacetime metric $g_{\mu\nu}$ is taken to be ${e^a}_\mu$ then the generalized vielbein can be parametrized as
\begin{equation}
{\EE^A}_M = 
\begin{pmatrix}
{e^a}_\mu & 0 \\
-{e_a}^\nu B_{\nu\mu} & {e_a}^\mu
\end{pmatrix} \, .
\end{equation}
In choosing the lower-triangular form, the local $H$-transformations have partially been fixed. 

Let us briefly comment on the signatures of the metrics involved here. In the above exposition the Euclidean theory was presented, \ie the maximal compact subgroup $H$ is $O(D)\times O(D)$. The Lorentzian version of the theory takes the same form with $H=O(1,D-1)\times O(1,D-1)$ and the $D$-dimensional identity matrices in the tangent space metric $\HH_{AB}$ replaced by the $D$-dimensional Minkowski metric.

The remaining field of string theory to be rewritten as an $\Odd$ object is the dilaton. It gets shifted or rescaled by the determinant $g$ of the spacetime metric to form an $\Odd$ scalar which we will call the DFT dilaton
\begin{equation}
	d= \phi - \frac{1}{4}\ln g \qorq e^{-2d} = \sqrt{g}e^{-2\phi} \, .
\label{eq:DFTdilaton}
\end{equation}
The form of the generalized metric and rescaled dilaton here essentially encodes the Buscher rules for T-duality \cite{Buscher:1987sk,Buscher:1987qj} which describe how the fields $g_{\mu\nu}$, $B_{\mu\nu}$ and $\phi$ change under T-duality. In the language here, the objects $\HH_{MN}$ and $d$ simply undergo an $\Odd$ transformation. This will be explained in more detail below.

In general the DFT fields $\HH$ and $d$ are functions of the doubled space, \ie $\HH(X)$ and $d(X)$. We will see later how this coordinate dependence needs to be restricted, but first let us turn to the symmetries of the fields and the corresponding algebra.

\section{Generalized Lie Derivative and Diffeomorphisms}
\label{sec:DFTgenlie}

This section follows the clear and concise exposition of \cite{Berman:2013uda}. In general relativity  we have the global symmetry group $G=GL(D)$ with maximal compact subgroup $H=SO(D)$ (in the Euclidean case) for the local symmetry transformations. The transformation of a vector $V^\mu$ under a infinitesimal diffeomorphism generated by a vector field $U^\mu$ is described by the Lie derivative which is given by the commutator -- the Lie bracket -- of the two vectors $V^\mu$ and $U^\mu$
\begin{equation}
L_U V^\mu = [U,V]^\mu = U^\nu\partial_\nu V^\mu - V^\nu\partial_\nu U^\mu \, .
\end{equation}
The first term here is the transportation of the vector field itself while the second term is a $GL(D)$ transformation of the vector $V^\mu$. This defines the algebra of diffeomorphisms
\begin{equation}
[L_U,L_V] = L_{[U,V]} \, ,
\end{equation}
that is the commutator of two Lie derivatives is the Lie derivative with respect to the Lie bracket of the two vector fields. This definition of the Lie derivative can be extended to arbitrary tensors by requiring it satisfies the Leibniz rule. For a scalar field $S$ one simply has the transport term
\begin{equation}
L_U S = U^\mu\partial_\mu S \, .
\end{equation}

A similar story can be told for generalized geometry where we now have $G=\Odd$ and the local symmetries are in $H=O(D)\times O(D)$. The \emph{generalized Lie derivative} of a generalized vector $V^M$ with weight $\omega(V)$ is defined as \cite{Hull:2009mi,Hull:2009zb,Berman:2012vc}
\begin{equation}
\LL_U V^M = U^N\partial_N V^M - \left(V^N\partial_N U^M - \eta^{MK}\eta_{NL}V^N\partial_K U^L\right) 
			+ \omega(V) \partial_N U^N V^M \,.
\label{eq:DFTgenlie}
\end{equation}
As for the ordinary Lie derivative there is the transport term but now the second term (the term in parenthesis) is an $\Odd$ transformation. For completeness the weight term has been included in the definition here as it will be needed later. This definition can be extended to arbitrary tensors by requiring it satisfies the Leibniz rule. For a scalar $S$ with weight $\omega(S)$ the generalized Lie derivative is 
\begin{equation}
\LL_U S = U^N\partial_N S + \omega(S) \partial_N U^N V^M
\end{equation}
An important property of the generalized Lie derivative is that it preserves the $\Odd$ structure $\eta$
\begin{equation}
\LL_U \eta_{MN} = 0 \, .
\end{equation}

In generalized geometry, any object whose transformation under infinitesimal generalized diffeomorphisms is given by the generalized Lie derivative 
\begin{equation}
\delta_U T  = \LL_U T 
\label{eq:gaugetransformation}
\end{equation}
is a proper generalized tensor (where $T$ is an arbitrary tensor). Unlike in general relativity ($\delta_\xi\partial_\mu S=L_\xi\partial_\mu S$), the partial derivative of a scalar, $\partial_MS$ is not a tensorial quantity since we have
\begin{equation}
\delta_U(\partial_MS) = \LL_U\partial_MS + \eta_{MN}\eta^{KL}\partial_KU^N \partial_LS \,.
\end{equation}

The generalized Lie derivative defines the algebra of gauge symmetries. The commutator of two generalized Lie derivatives acting on a generalized vector $W^M$ is
\begin{equation}
[\LL_U,\LL_V] W^M = \LL_{[U,V]_\sfC}W^M + T^M(U,V,W)
\label{eq:genliecom}
\end{equation}
where we introduce the $\sfC$-bracket \cite{Hull:2009zb} of two generalized vectors $U^M$ and $V^M$ as
\begin{equation}
[U,V]_\sfC^M = U^N\partial_NV^M - V^N\partial_NU^M 
				- \frac{1}{2}\eta^{MN}\eta_{KL}\left(U^K\partial_NV^L - V^K\partial_NU^L \right) \, .
\label{eq:Cbracket}
\end{equation}
Due to the presence of the extra term $T^M(U,V,W)$ which is given by
\begin{equation}
\begin{aligned}
T^M(U,V,W) &= \eta^{PQ}\eta_{KL}\left[\vphantom{\frac{1}{2}}
				\left(\partial_PU^K\partial_QV^M - \partial_PU^M\partial_QV^K\right)W^L \right.\\
			&\hspace{2.2cm} \left.
			 	+\frac{1}{2}\left(U^L\partial_QV^K - V^L\partial_QU^K\right)\partial_PW^M\right]\, ,
\end{aligned}
\end{equation}
the generalized Lie derivative does not produce a closed algebra for arbitrary vectors $U^M$ and $V^M$. Furthermore, the $\sfC$-bracket does not satisfy the Jacobi identity. 

In order to have a consistent theory with a closed algebra, a constraint needs to be introduced. This takes the form of the so called \emph{physical section condition} which will be discussed in more detail in the next section. It requires that
\begin{equation}
\eta^{MN}\partial_M \otimes \partial_N = 0
\label{eq:DFTsection}
\end{equation}
where the differential operators act on any field or products of fields. The extra term $T^M$ in \eqref{eq:genliecom} vanishes under the section condition since each term contains two derivatives $\partial_P,\partial_Q$ contracted with $\eta^{PQ}$. So if the constraint is imposed, the algebra of generalized diffeomorphism closes.

The generalized Lie derivative has some interesting properties under the section condition. For example, there are ``trivial'' gauge parameters of the form
\begin{equation}
	U^M =\eta^{MN}\partial_N S
\label{eq:trivial}
\end{equation}
where $S$ is an arbitrary scalar. Such a parameter does not generate a gauge transformation via \eqref{eq:gaugetransformation} once the section condition is imposed since the result is proportional to $\eta^{MN}\partial_M\partial_N S$. The Jacobiator\footnote{The Jacobiator for any bracket is defined as $J(U,V,W) = [[U,V],W] + [[V,W],U + [[W,U],V]$. If the Jacobiator vanishes, the Jacobi identity is satisfied.} of the $\sfC$-bracket precisely takes such a form of a trivial parameter. To see this, consider the $\sfD$-bracket \cite{Hohm:2010pp}. In the same way the ordinary Lie derivative can be defined in terms of the Lie bracket, the generalized Lie derivative is given in terms of the $\sfD$-bracket
\begin{equation}
[U,V]_\sfD^M = \LL_U V^M \, .
\end{equation}
This bracket is not antisymmetric, but using the definition \eqref{eq:Cbracket} one can show that it differs from the $\sfC$-bracket by a term of the form \eqref{eq:trivial} which is symmetric in the two arguments
\begin{equation}
	[U,V]_\sfD^M = [U,V]_\sfC^M + \frac{1}{2}\eta^{MN}\partial_N(\eta^{KL}U_KV_L) \, .
\end{equation}
Therefore under the section condition, $[U,V]_\sfC^M$ and $[U,V]_\sfD^M$ generate the same generalized Lie derivative. Using this and the algebra \eqref{eq:genliecom} it is straightforward to show that the $\sfD$-bracket satisfies the Jacobi identity. So while the $\sfC$-bracket is antisymmetric but does not satisfy the Jacobi identity, the $\sfD$-bracket does so but is not antisymmetric.

Combining all this, it can be shown that the Jacobiator of the $\sfC$-bracket is given by\footnote{The object inside the parenthesis is called the Nijenhuis operator.}
\begin{equation}
	J(U,V,W)^M = \eta^{MN}\partial_N\left(
		\eta_{KL}[U,V]^K W^L + \eta_{KL}[V,W]^K U^L + \eta_{KL}[W,U]^K V^L\right)
\end{equation}
which is a trivial parameter of the form \eqref{eq:trivial}. Therefore the algebra of diffeomorphisms given by the $\sfC$-bracket closes and the Jacobi identity is satisfied when the section condition is imposed. The $\sfC$-bracket and $\sfD$-bracket reduce to the Courant bracket \cite{Courant:1990} and Dorfmann bracket \cite{Dorfman:1987} respectively under the section condition.

To conclude, the section condition \eqref{eq:DFTsection} is a crucial ingredient to the doubled geometry of DFT to ensure a consistent theory with a properly defined gauge algebra. We will now have a closer look at the physical implications of the section condition.

\section{The Physical Section Condition}
\label{sec:DFTsection}

In the introduction we have seen that the section condition has its string theory origin in the level-matching condition of the closed string \eqref{eq:levelmatching}. If the oscillator modes $N$ and $\tilde{N}$ are equal or zero the product of momentum and winding has to vanish for each direction. Expanding the section condition \eqref{eq:DFTsection} in terms of ordinary and winding coordinates by using $\eta^{MN}$ as given in \eqref{eq:eta} leads to
\begin{equation}
	\partial_\mu \otimes \tpartial^\mu = 0 
\end{equation}
where $\partial_\mu = \frac{\partial}{\partial x^\mu}$ and $\tpartial^\mu = \frac{\partial}{\partial \tx_\mu}$. Therefore this is equivalent to the expression $p_\mu w^\mu = 0$ in the conjugate variables. The $\Odd$ invariant structure $\eta$ thus provides a pairing between a normal and a winding coordinate for each value of the index $\mu$. 

In order to satisfy the section condition, the fields and gauge parameters of the theory can only depend on either $x$ or $\tx$, but not both. Therefore the simplest way of solving the section condition is to only allow for a coordinate dependence on the spacetime coordinates $x^\mu$ with everything independent of the winding coordinates $\tx_\mu$. The theory then reduces to a formulation of supergravity along the lines of generalized geometry where the tangent space is doubled but the underlying space itself is not. The opposite choice where everything depends on the $\tx_\mu$ but not the $x^\mu$ gives a description of the T-dual picture. One can also pick a mixture of normal and winding coordinates, but for each pair matched by $\eta_{MN}$ only one or the other is allowed. 

Therefore, imposing the section condition picks out a maximally isotropic subspace of half-maximal dimension. One can think of this as a $D$-dimensional slice or section through the $2D$-dimensional doubled space which forms the physical spacetime of supergravity. The orientation of the subspace, \ie which of the $2D$ coordinates are part of the section, can be rotated by an $\Odd$ transformation with different orientations (or polarizations) corresponding to different T-duality frames. This perspective of T-duality and the identification of spacetime as a null subspace determined by the $\Odd$ structure $\eta$ was described first in \cite{Hull:2004in}. 

At this point it is useful to clarify a slight ambiguity in the statement of the section condition in \eqref{eq:DFTsection} where $\otimes$ means that the differential operators act on any field or product of fields. It is better to state these two conditions separately as
\begin{equation}
	\eta^{MN}\partial_M\partial_N \Phi = 0 \qandq \eta^{MN}\partial_M \Phi \partial_N \Psi = 0 
\label{eq:weakstrong}
\end{equation}
where $\Phi$ and $\Psi$ are any fields or gauge parameters of DFT. The first form is called the \emph{weak constraint} while the second form is the \emph{strong constraint}. The former is the one with its origin in the level-matching condition as explained above while the latter one is equivalent to the section condition, \ie demanding that fields only depend on half the number of coordinates. Together these two constraints are sufficient for the closure of the gauge algebra and to render to whole formulation of DFT gauge invariant (see below). However, they need not to be necessary.

Since imposing both the weak and the strong constraint gives an elegant formulation which makes T-duality a manifest symmetry but has no extra physical content beyond supergravity, it would be interesting to consider relaxations of the strong constraint and study theories for backgrounds which depend on both the normal and dual coordinates.

Important developments in this direction \cite{Aldazabal:2011nj,Geissbuhler:2011mx,Dibitetto:2012rk,Grana:2012rr,Geissbuhler:2013uka,Berman:2013cli} consider compactifications of DFT with a Scherk-Schwarz reduction. In such a scenario the reduction of DFT leads to gauged supergravities in lower dimensions. This approach also provides higher dimensional origins for gauged supergravities with non-geometric fluxes which become purely geometric in the doubled space. These methods and results are reviewed in great detail in \cite{Aldazabal:2013sca}. A entirely different approach towards a weakly constrained formulation of double field theory can be found in \cite{Lee:2015qza}. 

In this thesis we will always impose the strong form of the constraint and in general use the terms (physical) ``section condition'' and ``strong constraint'' synonymously. Only in Section \ref{sec:Tduality} we will make a subtle distinction between these two expressions when discussing how T-duality emerges in double field theory.

\section{The Action and Projected Equations of Motion}
\label{sec:DFTaction}

Having studied the generalized geometry of the doubled space and its generalized gauge structure we are finally in a position to present one of the key results of double field theory: a manifestly $\Odd$ invariant action for the bosonic NSNS-sector fields of supergravity. The action integral \cite{Hohm:2010pp} may be written in an Einstein-Hilbert-like form as
\begin{equation}
S = \int \dd^{2D}X e^{-2d} \RR
\label{eq:DFTaction}
\end{equation}
with the scalar $\RR$ given by
\begin{equation}
\begin{aligned}
\RR 	&= \frac{1}{8}\HH^{MN}\partial_M\HH^{KL}\partial_N\HH_{KL} 
		- \frac{1}{2}\HH^{MN}\partial_M\HH^{KL}\partial_K\HH_{NL} \\
	&\quad+ 4\HH^{MN}\partial_M\partial_N d - \partial_M\partial_N\HH^{MN}
		-4\HH^{MN}\partial_M d \partial_N d + 4\partial_M\HH^{MN}\partial_N d \\
	&\quad+ \frac{1}{2}\eta^{MN}\eta^{KL}\partial_M{\EE^A}_K\partial_N{\EE^B}_L\HH_{AB} \, . 
\end{aligned}
\label{eq:DFTscalarR}
\end{equation}
Recall that $M,N$ are curved doubled spacetime indices and $A,B$ are flat doubled tangent space indices with the generalized vielbein defined in \eqref{eq:DFTvielbein}. The object $\RR$ is expressed in a fully covariant manner solely in terms of the $\Odd$ tensor $\HH_{MN}$ and scalar $d$. It can be shown \cite{Hohm:2010pp} that $\RR$ is itself an $\Odd$ scalar\footnote{Note that due to the last line in \eqref{eq:DFTscalarR} $\RR$ is only a weakly constrained $\Odd$ scalar. This last line is not present in the original formulation of DFT \cite{Hull:2009mi} but is required when considering Scherk-Schwarz reduction to reproduce the correct gauged supergravity actions \cite{Grana:2012rr}.}. Furthermore it is also a gauge scalar under generalized diffeomorphisms. 

When imposing the strong constraint by setting $\tpartial = 0$ the last line in \eqref{eq:DFTscalarR} drops out and this DFT action reduces (after integrating by parts) to the standard action for the common NSNS-sector of supergravity
\begin{equation}
S = \int \dd^{D}x \sqrt{-g} e^{-2\phi} \left[ R + 4(\partial\phi)^2 - \frac{1}{12}H^2 \right] \, . 
\end{equation}
Here $R$ is the usual Ricci scalar corresponding to the spacetime metric $g_{\mu\nu}$ with determinant $g$, $\phi$ is the dilaton and $H=\dd B$ is the three-form field strength of the two-form B-field.

The DFT action \eqref{eq:DFTaction} is a great success since it is manifestly $\Odd$ invariant and reduces to the correct low-energy action of supergravity. However, it is not constructed in a geometric manner \`a la general relativity, \ie it is not expressed in terms of a curvature or a  generalization thereof. There has been a number of different approaches \cite{Jeon:2010rw,Jeon:2011cn,Hohm:2010pp,Hohm:2011si,Berman:2013uda} to construct an appropriate connection and curvature tensor for the doubled geometry of DFT. These constructions are all legitimate and each has its advantages, but all come with some drawbacks (only being partially covariant, connection not completely determined, no curvature but torsion). As of yet, there is no complete and consistent picture of how to describe the doubled geometry and its curvature. 

We will now turn to the equations of motion of the DFT fields $\HH$ and $d$. The equation of motion for the dilaton is easily obtained by varying the action
\begin{equation}
\delta S = \int \dd^{2D}X e^{-2d} (-2\RR)\delta d
\end{equation}
which has to vanish for any variation $\delta d$ and thus gives
\begin{equation}
\RR=0
\label{eq:DFTeomDilaton}.
\end{equation}
(Note that $\delta \RR/\delta d=0$ up to total derivatives.) To find the equation of motion for the generalized metric we have to be a bit more careful. Varying the action with the generalized metric gives
\begin{equation}
\delta S = \int \dd^{2D}X e^{-2d} K_{MN}\delta \HH^{MN}
\label{eq:DFTvaraction}
\end{equation}
where $K_{MN}$ is  given by
\begin{equation}
\begin{aligned}
K_{MN} &= \frac{1}{8}\partial_M\HH^{KL}\partial_N\HH_{KL} + 2\partial_M\partial_N d \\
	&\quad +(\partial_L-2\partial_L d) 		
		\left[\HH^{KL}\left(\partial_{(M}\HH_{N)K}
		- \frac{1}{4}\partial_K\HH_{MN}\right)\right] \\
	&\quad  + \frac{1}{4}\left(\HH^{KL}\HH^{PQ}-2\HH^{KQ}\HH^{LP}\right)
		\partial_K\HH_{MP}\partial_L\HH_{NQ} \\
	&\quad - \eta^{KL}\eta^{PQ}\left(\partial_K d\partial_L{\EE^A}_P 
		- \frac{1}{2}\partial_K\partial_L{\EE^A}_P\right)\HH_{(N|R}{\EE^R}_A\HH_{|M)Q} \, . 
\end{aligned}
\end{equation}
The last term uses the variation of the vielbein with respect to the metric
\begin{equation}
\delta{\EE^A}_M = \frac{1}{2}\HH^{AB}{\EE^N}_B\delta\HH_{MN} \, .
\end{equation} 

The expression in \eqref{eq:DFTvaraction} does not have to vanish for any $\delta\HH^{MN}$ since the generalized metric is constrained to parametrize the coset space $O(D,D)/O(D)\times O(D)$ (see \ref{sec:DFTintro} above). This means the generalized metric can be parametrized by $g_{\mu\nu}$ and $B_{\mu\nu}$ as written in \eqref{eq:DFTmetric}. Thus deriving the equations of motion is a little more complicated. This was first done in \cite{Hohm:2010pp}. We will re-derive the equations of motion here using a slightly different method because this method will be more readily applicable to the cases of extended geometry with the exceptional groups that we discuss later. The basic idea is that rather than varying with respect to the generalized metric one varies with respect to the spacetime metric and the B-field and then make the result $\Odd$ covariant.

By applying the chain rule, the action can be varied with respect to $\delta g_{\mu\nu}$ and $\delta B_{\mu\nu}$ separately. Making use of
\begin{align}
\frac{\delta g_{\mu\nu}}{\delta g_{\rho\sigma}} &= {\delta_\mu}^{(\rho}{\delta_\nu}^{\sigma)}, &
\frac{\delta g^{\mu\nu}}{\delta g_{\rho\sigma}} &= -g^{\mu(\rho}g^{\sigma)\nu}, &
\frac{\delta B_{\mu\nu}}{\delta B_{\rho\sigma}} &= {\delta_\mu}^{[\rho}{\delta_\nu}^{\sigma]}
\end{align}
leads to
\begin{align}
\delta S &= \int \dd^{2D}X e^{-2d} K_{MN}\left[
		\frac{\delta \HH^{MN}}{\delta g_{\rho\sigma}}\delta g_{\rho\sigma}
		+ \frac{\delta \HH^{MN}}{\delta B_{\rho\sigma}}\delta B_{\rho\sigma}\right] \\
		&= \int \dd^{2D}X e^{-2d} \left\{
			\left[- K_{\mu\nu}g^{\mu(\rho}g^{\sigma)\nu}
					+ 2{K_\mu}^\nu g^{\mu(\rho}g^{\sigma)\tau}B_{\tau\nu} 
					\vphantom{\left({\delta_\mu}^{(\rho}\right)}\right.\right. \notag\\
		& \left.\left. \hspace{4cm} 	+ K^{\mu\nu}\left({\delta_\mu}^{(\rho}{\delta_\nu}^{\sigma)} 
					+ B_{\mu\tau}g^{\tau(\rho}g^{\sigma)\lambda}B_{\lambda\nu}
					\right)\right]\delta g_{\rho\sigma} \right. \\
		& \left. \hspace{3.5cm} + 
			\left[- 2{K_\mu}^\nu g^{\mu\tau}{\delta_\tau}^{[\rho}{\delta_\nu}^{\sigma]}
					- 2K^{\mu\nu}B_{\mu\tau}g^{\tau\lambda}
						{\delta_\lambda}^{[\rho}{\delta_\nu}^{\sigma]} \right]\delta B_{\rho\sigma} 	
			\right\}\, .\notag
\end{align}
Now the $g$'s and $B$'s are re-expressed in terms of $\HH$, the symmetrizing brackets are dropped and the antisymmetrizing ones are expanded
\begin{align}
\delta S &= \int \dd^{2D}X e^{-2d} \left\{\vphantom{\frac{1}{2}}
	\left[- K_{\mu\nu}\HH^{\mu\rho}\HH^{\sigma\nu}
			+ 2{K_\mu}^\nu \HH^{\mu\rho}{\HH^{\sigma}}_\nu 
			+ K^{\mu\nu}\left({\delta_\mu}^{\rho}{\delta_\nu}^{\sigma} 
					- {\HH_\mu}^{\rho}{\HH^{\sigma}}_\nu\right)\right]\delta g_{\rho\sigma}\right. \notag\\
		& \left. \hspace{3.5cm} -2 
			\left[{K_\mu}^\nu \HH^{\mu\tau} + K^{\mu\nu}{\HH_\mu}^\tau\right]
			\frac{1}{2}\left({\delta_\tau}^{\rho}{\delta_\nu}^{\sigma} 
				- {\delta_\tau}^{\sigma}{\delta_\nu}^{\rho}\right) \delta B_{\rho\sigma} 	
			\right\}.
\end{align}
The crucial step is to then re-covariantize the indices by using $\eta_{MN}$ given in \eqref{eq:eta}
\begin{equation}
\begin{aligned}
\delta S &= \int \dd^{2D}X e^{-2d} \left\{
	K_{KL} \left(\eta^{K\rho}\eta^{\sigma L} - \HH^{K\rho}\HH^{\sigma L}\right) 
		\delta g_{\rho\sigma}\right. \\
	& \left. \hspace{3cm} - K_{KL}\left(\HH^{KP}\eta_{PM}\eta^{LN} - \HH^{KP}{\delta_P}^N{\delta_M}^L\right)
	\eta^{M\rho}{\delta^\sigma}_N\delta B_{\rho\sigma} \right\} 
\end{aligned}
\end{equation}
which reproduces the previous line once the doubled indices are expanded and summed over. In a final step the terms inside the brackets are brought into a form corresponding to a projected set of equations as follows
\begin{align}
\delta S &= \int \dd^{2D}X e^{-2d} \Big\{
	K_{KL} \left({\delta_M}^K{\delta_N}^L - \HH^{KP}\eta_{PM}\eta_{NQ}\HH^{QL}\right) 	
		\eta^{M\rho}\eta^{\sigma N}\delta g_{\rho\sigma} \notag\\
	&  \hspace{3cm} - K_{KL}\left(\HH^{KP}\eta_{PM}\eta^{LQ}\HH_{QR} - \HH^{KP}{\delta_P}^Q{\delta_M}^L\HH_{QR}\right) \notag\\
	&  \hspace{9cm} \HH^{RN}\eta^{M\rho}{\delta^\sigma}_N\delta B_{\rho\sigma} \Big\} \notag\\
	&= \int \dd^{2D}X e^{-2d} 2{P_{MN}}^{KL}K_{KL}
		\left(\eta^{M\rho}\eta^{\sigma N}\delta g_{\rho\sigma}
			+ \eta^{M\rho}\HH^{\sigma N}\delta B_{\rho\sigma}\right)
\end{align}
where we have introduced the projector
\begin{equation}
{P_{MN}}^{KL} = \frac{1}{2}\left[{\delta_M}^{(K}{\delta_N}^{L)} 
	- \HH_{MP}\eta^{P(K}\eta_{NQ}\HH^{L)Q}\right] \, .
\end{equation}
which is symmetric in both $MN$ and $KL$. 

The variation of the action has to vanish for \emph{any} $\delta g_{\mu\nu}$ and $\delta B_{\mu\nu}$ independently, therefore the equations of motion are given by  
\begin{equation}
{P_{MN}}^{KL}K_{KL} = 0
\label{eq:DFTeom}
\end{equation}
and not $K_{MN}=0$, the naive equations expected from setting \eqref{eq:DFTvaraction} to zero. 

This equation of motion was derived in a slightly different way in \cite{Hohm:2010pp} by using the constraint equation $\HH^t\eta\HH=\eta$ which ensures $\HH$ is an element of $\Odd$. The result is 
\begin{equation}
\frac{1}{2}(K_{MN} - \eta_{MK}\HH^{KP}K_{PQ}\HH^{QL}\eta_{LN}) = {P_{MN}}^{KL}K_{KL} = 0
\end{equation}
in agreement with ours. We wish to emphasize the point of re-deriving these equations is just so that we can use this method in the exceptional case later. 

Also note that the expression for $K_{MN}$ found in the literature, especially in \cite{Hohm:2010pp}, differs from the one given here. This difference arises as one can use either the invariant $\Odd$ metric $\eta$ or the generalized metric $\HH$ to raise and lower indices in the derivation of $K_{MN}$. Both methods are valid and the discrepancy disappears once the projector acts. In a way, the projector enforces the constraint that $\HH$ parametrizes a coset space. When using $\eta$, this constraint is taken into account automatically, but when using $\HH$ the constraint needs to be imposed by the projector. Since $K_{MN}$ appears in the equations of motion only with the projector acting on it, it does not matter which version is used.

The importance for the presence of the projector can be seen by counting degrees of freedom. The symmetric spacetime metric has $\frac{1}{2}D(D+1)$ degrees of freedom and the antisymmetric B-field contributes $\frac{1}{2}D(D-1)$ for a total of $D^2$ independent components. The dimension of the doubled space is $2D$, therefore $K_{MN}$ has $2D^2+D$ components. Of these, $D^2+D$ are in the kernel of the projector and are therefore eliminated, leaving $D^2$ degrees of freedom as desired. This can be shown by computing the characteristic polynomial and all the eigenvalues of the projector $P$.

\chapter{Solutions in DFT}
\label{ch:DFTsol}

Having set up the basics of double field theory and considered the equations of motion for its constituent fields, we will now turn our attention to constructing solutions to these equations and interpreting them in terms of known supergravity objects. This will lead to a better understanding of the novel coordinates -- the winding directions introduced -- and some interesting insights into how T-duality emerges in the doubled picture. 

In the introduction the parallels between double field theory and Kaluza-Klein theory were highlighted. In particular, a pure momentum mode such as the wave solution gives rise to a fundamental charge in the reduced picture of the KK-ansatz. A good example for this is the wave in M-theory which gives the D0-brane in string theory. In other words, the D0-brane is a momentum mode along the eleventh direction, the M-theory circle. Its mass and charge are given by the wave momentum in that direction and hence satisfy a BPS condition. It was this example which led us to consider momentum modes in the dual space of DFT. We will therefore begin by considering a null wave propagating in the dual space. 

Before we start, there is a caveat concerning the generalized metric. It is an open question if the generalized metric is an actual metric tensor on the doubled space or something different, in which case the term ``metric'' is a misnomer. For the purpose of this thesis it is sufficient that the generalized metric transforms under generalized diffeomorphisms (generated by the generalized Lie derivative) and for a given solution satisfies the DFT equations of motion. Nevertheless, for convenience we will encode the matrix $\HH_{MN}$ in terms of a ``line element'' $\dd s^2 = \HH_{MN}\dd X^M \dd X^N$ which provides a concise way of presenting the components of $\HH_{MN}$. It is not necessary to us that this line element defines an actual metric tensor in the doubled space.

\section{DFT Wave}
\label{sec:DFTwave}

We seek a solution for the generalized metric corresponding to a null wave whose momentum is pointing the $\tz$ direction. The ansatz will be that of a pp-wave in usual general relativity \cite{Aichelburg:1970dh}. This has no compunction to be a solution of DFT. As we have seen the equations of motion of the generalized metric in DFT are certainly not the same as the equations of motion of the metric in relativity. Let us immediately remove any source of confusion the reader may have, the pp-wave as a solution for $g_{\mu\nu}$ may of course, by construction, be embedded as a solution in DFT by simply inserting the pp-wave solution for $g_{\mu \nu}$ into $\HH_{MN}$. Here we will consider a pp-wave (that is the usual pp-wave ansatz \cite{Aichelburg:1970dh}) not for $g_{\mu\nu}$ but for the doubled metric $\HH_{MN}$  itself and then determine its interpretation in terms of the usual metric $g_{\mu\nu}$ and two-form $B_{\mu\nu}$.

The following is a solution to DFT in $2D$ dimensions given by a constant DFT dilaton \eqref{eq:DFTdilaton}, $d=const.$, and the generalized metric $\HH_{MN}$ with line element
\begin{equation}
\begin{aligned}
\dd s^2 &= \HH_{MN}\dd X^M \dd X^N \\
	&= (H-2)\left[\dd t^2 - \dd z^2\right] + \delta_{mn}\dd y^m\dd y^n \\
	&\quad + 2(H-1)\left[\dd t\dd\tz + \dd\ttt\dd z\right] \\
	&\quad - H\left[\dd\ttt^2 - \dd\tz^2\right] + \delta^{mn}\dd\ty_m\dd\ty_n
\label{eq:DFTwave}	
\end{aligned}
\end{equation}
where the generalized coordinates are split as
\begin{equation}
X^M = (x^\mu,\tx_\mu)=(t,z,y^m;\ttt,\tz,\ty_m)
\end{equation}
and a tilde denotes a dual coordinate as introduced above. This generalized metric and rescaled dilaton solve the equations of motion of the DFT derived in Section \ref{sec:DFTintro}. The appendix \ref{sec:wavecheck} provides the details demonstrating it is indeed a solution.

Since it is exactly of the same form as the usual pp-wave solution, the natural interpretation is of a pp-wave in the doubled geometry. One therefore imagines a massless mode which propagates and therefore carries momentum in the $\tz$ direction. In \cite{Blair:2015eba,Park:2015bza,Naseer:2015fba} the conserved charges of DFT were constructed (by defining an analogue to the ADM-mass and setting up generalized Komar integrals respectively) and it was indeed shown that the DFT wave is massless but has momentum in the dual $\tz$ direction.

In this solution, $H$ is taken to be a harmonic function of the usual transverse coordinates\footnote{The range of the transverse index is $m=1,\dots,D-2$.} $y^m$ (but not of their duals $\ty_m$) and as such is annihilated (up to delta function sources) by the Laplacian operator in these directions, \ie $\delta^{mn}\partial_m\partial_n H=0$. In DFT language, it is required (at least naively) that $H$ satisfies the strong constraint and therefore it is not a function of any of the dual coordinates. The fact that the harmonic function $H$ is taken to only depend on $y^m$ and not the dual transverse directions implies that the wave solution is \emph{smeared} in these $\ty_m$ directions. One can think of it as a plane wave front extending along the dual directions described by coordinates $\ty_m$ but with momentum in the $\tz$ direction. An explicit form of $H$ is
\begin{equation}
H = 1 + \frac{h}{r^{D-4}} \qquad \mathrm{with} \qquad r^2 = y^my^n\delta_{mn}
\end{equation}
where $h$ is a constant proportional to the momentum carried and $r$ is the radial coordinate of the transverse space.

We will now use the form of the doubled metric $\HH_{MN}$ in terms of $g_{\mu\nu}$ and $B_{\mu\nu}$ to rewrite this solution in terms of $D$-dimensional quantities, effectively reducing the dual dimensions. This is like in Kaluza-Klein theory, writing a solution of the full theory in terms of the reduced metric and vector potential using the ansatz 
\begin{align}
\dd s^2 &= (g_{\mu\nu} - B_{\mu\rho}g^{\rho\sigma}B_{\sigma\nu})\dd x^\mu \dd x^\nu
	+ 2B_{\mu\rho}g^{\rho\nu}\dd x^\mu \dd\tx_\nu + g^{\mu\nu}\dd\tx_\mu\dd\tx_\nu \, .
\label{eq:KKforDFT}  
\end{align}

By comparing \eqref{eq:KKforDFT} with \eqref{eq:DFTwave}, the fields of the reduced theory with coordinates $x^\mu=(t,z,y^m)$ can be computed. We find the metric and its inverse to be
\begin{equation}
g_{\mu\nu} = \diag [-H^{-1}, H^{-1}, \delta_{mn}] \qandq 
g^{\mu\nu} = \diag [-H, H, \delta^{mn}]
\end{equation}
whereas the only non-zero component of the B-field is given by
\begin{equation}
B_{tz} = -B_{zt} = -(H^{-1}-1) \, .
\end{equation}
From the definition $e^{-2d} = \sqrt{-g}e^{-2\phi}$ of the rescaled dilaton $d$ (which is a constant here) it follows that the dilaton $\phi$ is given by ($\phi_0$ is another constant)
\begin{equation}
e^{-2\phi} = H e^{-2\phi_0}
\end{equation}
since $g=-H^{-2}$. The corresponding line element is 
\begin{equation}
\dd s^2 = -H^{-1}(\dd t^2-\dd z^2)+\delta_{mn}\dd y^m\dd y^n 
%\label{eq:string}
\end{equation}
which together with the B-field and the dilaton $\phi$ gives the fundamental string solution extended along the $z$ direction \cite{Dabholkar:1990yf}. We have thus shown that the solution \eqref{eq:DFTwave} which carries momentum in the $\tz$ direction in the doubled space corresponds to the string along the $z$ direction from a reduced point of view. In other words, the fundamental string is a massless wave in the doubled space with momentum in a dual direction.

As mentioned above, this follows the logic of usual Kaluza-Klein theory. In the doubled formalism the solution is a massless wave with $P_MP_N\HH^{MN}=0$ (where the $P^M$ are some generalized momenta), but from a the reduced normal spacetime point of view the string has a mass or rather tension $T$ and charge $q$ which are obviously given by the momenta in the dual directions with a resulting BPS equation
\begin{equation}
T = |q| \, .
\end{equation}

Of course this is no surprise from the point of view of T-duality. Momentum and string winding exchange under T-duality. It is precisely as expected that momentum in the dual direction corresponds to a string. What is more surprising is when one views this from the true DFT perspective. There are null wave solutions that can point in any direction. When we analyze these null waves from the reduced theory we see them as fundamental strings or as usual pp-waves. It is a simple $\Odd$ rotation of the direction of propagation that takes one solution into the other. We will come back to this view on T-duality in DFT in the last section of this chapter.

\subsection{Goldstone Modes of the Wave Solution}
\label{sec:goldstones}
In the previous section we presented a solution to the equations of motion of DFT which reduces to the fundamental string. It will be  interesting to analyze the Goldstone modes of this solution in double field theory. Especially since the advent of M-theory, it was understood that branes are dynamical objects and that when one finds a solution of the low energy effective action one can learn about the theory by examining the dynamics of the Goldstone modes. For D-branes in string theory this was done in \cite{Adawi:1998ta} and for the membrane and fivebrane in M-theory, where such an analysis was really the only way of describing brane dynamics, this was done in  \cite{Kaplan:1995cp, Adawi:1998ta}. We will follow the excellent exposition and the method described in \cite{Adawi:1998ta} as closely as possible.

In DFT, the diffeomorphisms and gauge transformations are combined into generalized diffeomorphisms generated by a generalized Lie derivative. We will consider small variations in the generalized metric, $h_{MN}$ and the dilaton, $\lambda$ generated by such transformations as follows
\begin{align}
h_{MN} &= \delta_\xi \HH_{MN} = \LL_\xi \HH_{MN} \, ,  &
\lambda &= \delta_\xi d = \LL_\xi d  \, .
\end{align}
The generalized Lie derivative of DFT was introduced in \eqref{eq:DFTgenlie}. Acting on the generalized metric which is a symmetric tensor, it gives
\begin{equation}
\LL_\xi \HH_{MN} =	\xi^L\partial_L \HH_{MN} 
		+ 2\HH_{L(M}\partial_{N)}\xi^L 
		- 2\eta^{LP}\eta_{Q(M}\HH_{N)L}\partial_P\xi^Q \, .
\label{eq:genLieMetric}
\end{equation}
If the metric $\HH_{MN}$ and the transformation parameter $\xi^{M}=(\xi^\mu,\tilde{\xi}_\mu)$ both satisfy the section condition, then the vector part $\xi^\mu$ generates a coordinate transformation while the one-form part $\tilde{\xi}_\mu$ gives a gauge transformation of the B-field. 

The generalized Lie derivative of the dilaton contains just the transport term plus a term for $d$ being a tensor density
\begin{align}
\LL_\xi d &=	\xi^M\partial_M d - \frac{1}{2}\partial_{M}\xi^M \, .
\label{eq:genLieDilaton}
\end{align}

For the purpose of this analysis, the space spanned by the coordinates $\{t,z\}$ of the wave solution is treated like the worldvolume of an extended object. All remaining coordinates are treated as transverse in the extended space. The solution clearly breaks translation symmetry and so one naturally expects scalar zero-modes. One immediate puzzle would be to ask about the number of degrees of freedom of the Goldstone modes. Given that the space is now doubled one would naively image that any solution which may be interpreted as a string would have $2D-2$ degrees of freedom rather than the expected $D-2$. We will answer this question and show how the Goldstone modes have the correct number of degrees of freedom despite the solution living in a $2D$ dimensional space. The projected form of the equations of motion are crucial in making this work out.

To carry out the analysis it will be useful to split up the space into parts longitudinal and transverse to the wave. One collects the worldvolume coordinates $t$ and $z$ into $x^a$ and similarly for their duals\footnote{In what follows we will use the alternative notation $\tx^\bmu$ for the dual coordinates to avoid confusion between inverse and dual parts of the metric.} $\tx^\ba = (\ttt,\tz)$ such that the generalized coordinates are $X^M=(x^a,y^m,\tx^\ba,\ty^\bm)$. This allows the non-zero components of the metric and its inverse to be written as
\begin{equation}
\begin{aligned}
\HH_{ab} &= (2-H)\II_{ab} 					&	\HH^{ab} &= H\II^{ab} \\
\HH_{\ba\bb} &= H\II_{\ba\bb} 				&	\HH^{\ba\bb} &= (2-H)\II^{\ba\bb}\\
\HH_{a\bb} &= \HH_{\bb a} = (H-1)\JJ_{a\bb}	&	\HH^{a\bb} &= \HH^{\bb a} = (H-1)\JJ^{a\bb} \\
\HH_{mn} &= \delta_{mn}, \quad \HH_{\bm\bn} = \delta_{\bm\bn}		
					&	\HH^{mn} &= \delta^{mn}, \quad\HH^{\bm\bn} = \delta^{\bm\bn}
\end{aligned}
\end{equation}
where the constant symmetric $2\times 2$ matrices $\II$ and $\JJ$ are defined as
\begin{equation}
\II = \begin{pmatrix} -1 & 0 \\ 0 & 1 \end{pmatrix}
\qandq
\JJ = \begin{pmatrix} 0 & 1 \\ 1 & 0 \end{pmatrix} \, .
\label{eq:IJmatrix}
\end{equation}
For later use also define their (antisymmetric) product
\begin{equation}
\KKK = \II \cdot \JJ = - \JJ\cdot\II = \begin{pmatrix} 0 & -1 \\ 1 & 0 \end{pmatrix} \, .
\label{eq:Kmatrix}
\end{equation}

Following \cite{Adawi:1998ta}, we now pick a transformation parameter $\xi^M$ with non-zero components only in the transverse directions, but with no transformation along the worldvolume directions (and the directions dual to the worldvolume). This transformation may then be described by the DFT vector field
\begin{equation}
\xi^M = (0,H^\alpha\hphi^m, 0, H^\beta\htphi^\bm)
\end{equation}
where $\hphi^m$ and $\htphi^\bm$ are the constant vectors that later will become the Goldstone modes once we allow them to have dependence on the worldvolume coordinates, $H$ is the harmonic function given above and $\alpha,\beta$ are constants that are to be determined by demanding that the Goldstone modes become normalizable. Using \eqref{eq:genLieMetric}, we can compute the components of $h_{MN}$ in terms of $\hphi^m,\htphi^\bm$. Recall that both the metric and the transformation parameter only depend on $y$ through the harmonic function $H$. Therefore $\partial_m$ is the only derivative that gives a non-zero contribution. We find
\begin{equation}
\begin{aligned}
h_{ab} &= -\hphi^m (H^\alpha\partial_m H)  \II_{ab} 
	& h_{mn} &= 2\hphi^q\delta_{q(m}{\delta_{n)}}^p\partial_pH^\alpha \\
h_{\ba\bb} &= \hphi^m (H^\alpha\partial_m H)  \II_{\ba\bb} 
	& h_{\bm\bn} &= -2\hphi^q\delta_{q(\bm}{\delta_{\bn)}}^p\partial_p H^\alpha \\
h_{a\bb} &= h_{\bb a} = \hphi^m (H^\alpha\partial_m H)  \JJ_{a\bb}
	& h_{m\bn} &= h_{\bn m} = -2\htphi^\bq\delta_{\bq[m}{\delta_{\bn]}}^p\partial_p H^\beta
\end{aligned}
\label{eq:h}
\end{equation}
and all terms with indices mixing $a,\ba$ with $m,\bm$ vanish. For the dilaton there is no contribution from the transport term as $d$ is a constant for our solution. This leaves the density term which gives
\begin{equation}
\lambda =  - \frac{1}{2}\hphi^m\partial_{m} H^\alpha \, .
\label{eq:lambda}
\end{equation}

Once we have these equations, the next step is to allow the moduli to have dependence on the worldvolume coordinates, 
\begin{equation}
\hphi^m\rightarrow\phi^m(x) \, , \qquad \htphi^\bm\rightarrow \tphi^\bm(x)  \label{eq:zeromodes}
\end{equation}
 and the hats are removed. These are the zero-modes.

We now determine their equations of motion by inserting \eqref{eq:zeromodes} into \eqref{eq:h} and \eqref{eq:lambda} and then subsequently into the equations of motion for DFT, \eqref{eq:DFTeomDilaton} and \eqref{eq:DFTeom}. As usual we keep only terms with two derivatives and first order in $h_{MN}$ and $\lambda$ themselves. (It would certainly be interesting to move beyond this expansion and compare with a Nambu-Goto type action but we will not do so here.) This gives
\begin{align}
K_{MN} &= \HH^{LK}\partial_L\partial_{(M} h_{N)K} 	
	- \frac{1}{4}\HH^{LK}\partial_L\partial_K h_{MN}
	+ 2\partial_M\partial_N \lambda \\
R &= 4 \HH^{MN}\partial_M\partial_N \lambda
		- \partial_M\partial_N h^{MN}.
\end{align}
For convenience we will define $\Box = H\II^{ab}\partial_a\partial_b$ and $\Delta = \delta^{kl}\partial_k\partial_l$. Inserting $h_{MN}$ from \eqref{eq:h}, we find
\begin{align}
K_{ab} &= -(1+\alpha H^{-1})\partial_a\partial_b\phi^m (H^\alpha\partial_m H) 
	+ \frac{1}{4}\II_{ab}\Box\phi^m(H^\alpha\partial_m H) \notag \\
K_{\ba\bb} &= -\frac{1}{4}\II_{\ba\bb}\Box\phi^m(H^\alpha\partial_m H) \notag \\
K_{a\bb} &= K_{\bb a} = \frac{1}{2}{\KKK^c}_\bb\partial_c\partial_a\phi^m 	
	(H^\alpha\partial_m H) - \frac{1}{4}\JJ_{a\bb}\Box\phi^m(H^\alpha\partial_m H) \notag \\	
K_{mn} &= - \frac{\alpha}{2}\Box\phi^p\delta_{p(m}{\delta_{n)}}^q(H^\alpha\partial_qH) \notag \\
K_{\bm\bn} &= \frac{\alpha}{2}\Box\phi^p\delta_{p(\bm}{\delta_{\bn)}}^q 
	(H^\alpha\partial_qH) \notag \\
K_{m\bn} &= \delta K_{\bn m} = \frac{\beta}{2}\Box\tphi^\bp
	\delta_{\bp[m}{\delta_{\bn]}}^q(H^\beta\partial_q H) \notag \\
K_{am} &= K_{ma} = \frac{1}{2}\partial_a\phi^n\left[
	\delta_{mn}\Delta H^\alpha - \partial_m\partial_n H^\alpha 
	- \partial_m(H^\alpha\partial_n H)\right] \notag \\
K_{\ba m} &= K_{m\ba} = \frac{1}{2}{\KKK^b}_\ba\partial_b\phi^n
	\partial_m(H^\alpha\partial_n H) \notag \\
K_{a\bm} &= K_{\bm a} = \frac{1}{2}\partial_a\tphi^\bn{\delta_\bn}^k{\delta_\bm}^l
	\left[\delta_{kl}\Delta H^\beta - \partial_k\partial_l H^\beta \right] \notag \\
K_{\ba\bm} &= K_{\bm\ba} = 0
\end{align}
where $\KKK$ was defined in \eqref{eq:Kmatrix}. Further, inserting $\lambda$ from \eqref{eq:lambda} gives the dilaton equation
\begin{equation}
R = -H^{-1}(2\alpha + 1) \Box\phi^m (H^\alpha\partial_m H) = 0.
\end{equation}
It is straight forward to see that the dilaton equation is solved by $\Box\phi = 0$. For the other equations we have to work a bit harder. The full equations of motion for the generalized metric are the projected equations \eqref{eq:DFTeom} which contain $D^2$ linearly independent equations
\begin{equation}
\begin{aligned}
K_{mn} &= {\delta_m}^\bk {\delta_n}^\bl K_{\bk\bl}  \\
K_{m\bn} &= {\delta_m}^\bk {\delta_\bn}^l K_{\bk l}
\end{aligned}
\label{eq:DFTblock1}
\end{equation}
\begin{equation}
\begin{aligned}
K_{mt} &= (H-1){\delta_m}^\bn K_{\bn z} - (2-H){\delta_m}^\bn K_{\bn\bt} \\
K_{mz} &= (H-1){\delta_m}^\bn K_{\bn t} + (2-H){\delta_m}^\bn K_{\bn\bz} \\
K_{m\bt} &= (H-1){\delta_m}^\bn K_{\bn\bz} - H{\delta_m}^\bn K_{\bn t} \\
K_{m\bz} &= (H-1){\delta_m}^\bn K_{\bn\bt} + H{\delta_m}^\bn K_{\bn z}
\end{aligned}
\label{eq:DFTblock2}
\end{equation}
\begin{equation}
\begin{aligned}
0 &= (H-1)(K_{\bt\bt} - K_{\bz\bz}) + H(K_{t\bz} + K_{z\bt}) \\
0 &= (H-1)(K_{tt} - K_{zz}) + (2-H)(K_{t\bz} + K_{z\bt}) \\
0 &= (H-1)(K_{t\bz} - K_{z\bt}) - HK_{zz} + (2-H)K_{\bz\bz} \\
0 &= (H-1)(K_{t\bt} - K_{z\bz}) + HK_{tz} + (2-H)K_{\bt\bz}.
\end{aligned}
\label{eq:DFTblock3}
\end{equation}
Inserting for $K_{MN}$ from above yields the equations of motion for the zero modes. The first two read
\begin{equation}
\begin{aligned}
-\alpha\Box\phi^p\delta_{p(m}{\delta_{n)}}^q(H^\alpha\partial_qH) &= 0 \\
\beta\Box\tphi^\bq\delta_{\bq[m}{\delta_{\bn]}}^p(H^\beta\partial_q H) &= 0 
\end{aligned}
\end{equation}
and can be solved by $\Box\phi=0$ and $\Box\tphi=0$ respectively. The next block of equations \eqref{eq:DFTblock2} can be re-covariantized by using
\begin{equation}
-\II_{ac}\epsilon^{cb}=
-\begin{pmatrix} -1 & 0 \\ 0 & 1 \end{pmatrix}\begin{pmatrix} 0 & 1 \\ -1 & 0 \end{pmatrix}
=\begin{pmatrix}0 & 1 \\ 1 & 0 \end{pmatrix}
\end{equation}
which leads to
\begin{equation}
\begin{aligned}
\partial_a\phi^n\left[\delta_{mn}\Delta H^\alpha - \partial_m\partial_n H^\alpha 
	\right. & \left.- \partial_m(H^\alpha\partial_n H)\right] \\
	&= -\II_{ac}\epsilon^{cb}\partial_b\tphi^\bn{\delta_\bn}^p
	(H-1)\left[\delta_{pm}\Delta H^\beta - \partial_p\partial_m H^\beta \right] \\
\partial_a\phi^n\partial_m(H^\alpha\partial_nH) 
	&= \II_{ac}\epsilon^{cb}\partial_b\tphi^\bn{\delta_\bn}^p
	H\left[\delta_{pm}\Delta H^\beta - \partial_p\partial_m H^\beta \right].
\end{aligned}
\end{equation}
Adding these two equations gives
\begin{equation}
\partial_a\phi^n W_{mn}^{(\alpha)} =  \II_{ac}\epsilon^{cb}\partial_b\tphi^\bn{\delta_\bn}^n W_{mn}^{(\beta)}
\end{equation}
where for $\gamma=\alpha,\beta$ we have $W_{mn}^{(\gamma)}=\delta_{mn}\Delta H^\gamma - \partial_m\partial_n H^\gamma$. If $\alpha=\beta$ we have the same object $W_{mn}$ on both sides which can be shown to be invertible. The equation can thus be reduced to a duality relation between $\phi$ and $\tphi$
\begin{equation}
\partial_a\phi^m = \II_{ab}\epsilon^{bc}\partial_c\tphi^\bn\delta_\bn^m 
\qquad \mathrm{or} \qquad
\dd\phi^m = \star\dd\tphi^\bn\delta_\bn^m.  
\label{eq:dualityzeromodes}
\end{equation}
This equation implies both $\Box\phi=0$ and $\Box\tphi=0$ as can be seen by acting with a contracted derivative on the equation. If $\phi^m$ and $\tphi^\bm$ are placed in a generalized vector $\Phi^M=(0,\phi^m,0,\tphi^\bm)$ this can be written as a self-duality relation
\begin{equation}
\HH_{MN}\dd\Phi^M = \eta_{MN}\star\dd\Phi^N 
\end{equation}
and precisely matches the result in \cite{Duff:1989tf} for the duality symmetric string.

The final block of equations of motion \eqref{eq:DFTblock3} are either trivial or are also of the form $\Box\phi^m(H^\alpha\partial_mH)=0$ provided $\alpha=-1$. If one was not concerned by normalization of the modes then this also provides a way of constraining the value of $\alpha$. The consistent choice of $\alpha=-1$ is fortunately the choice that also leads to normalizable modes. This may be seen by examining the case $\alpha=-1$ and integrating over the transverse space. This exactly mirrors the situation described in \cite{Adawi:1998ta}. The Goldstone modes are really the normalizable modes corresponding to broken gauge transformations. Where for gravity the gauge transformations are ordinary diffeomorphisms, in the case of DFT it is generated by the generalized Lie derivative. 

One can now turn equation \eqref{eq:dualityzeromodes} into a (anti-)chiral equation for a linear combination of $\phi$ and  $\tphi$ as follows. Introducing $\psi_\pm$ to be given by
\begin{equation}
\psi_\pm = \phi \pm \tphi
\end{equation}
and inserting them into \eqref{eq:dualityzeromodes} and its Hodge dual gives the familiar (anti-)self-dual left- and right-movers
\begin{equation}
\dd\psi_\pm = \pm\star\dd\psi_\pm 
\end{equation}
of the Tseytlin-string \cite{Tseytlin:1990nb,Tseytlin:1990va}. Thus the dynamics of the Goldstone modes of the wave solution reproduce the duality symmetric string in doubled space. The number of physical degrees of freedom are not doubled but just become rearranged in terms of chiral and anti-chiral modes on the world-sheet.

\section{DFT Monopole}
\label{sec:DFTmonopole}

In the previous section a null wave in the doubled space was shown to reduce to a fundamental string or a pp-wave when viewed from the ordinary supergravity point of view. The interpretation of the solution in terms of the normal supergravity theory associated to the reduction of DFT was determined by the direction the null wave was traveling in. If the DFT solution carries momentum in a spacetime direction $z$ it reduces to a wave. But if it carries momentum in a dual winding direction $\tz$ it gives the string whose mass and charge are determined by the momentum in that dual direction. 

Instead of the wave we will now consider the Kaluza-Klein monopole solution also known as the Sorkin-Gross-Perry monopole \cite{Sorkin:1983ns,Gross:1983hb}
\begin{equation}
\begin{aligned}
\dd s^2 &= H^{-1}\left[\dd z + A_i\dd y^i\right]^2 + H\delta_{ij}\dd y^i\dd y^j \\
H &= 1 + \frac{h}{|\vec{y}_{(3)}|} \, ,  \qquad
\partial_{[i}A_{j]} = \frac{1}{2}{\epsilon_{ij}}^k\partial_k H 
\end{aligned}
\end{equation} 
where $H$ is a harmonic function and $A_i$ a vector potential with $i=1,2,3$. This solution has the following non-trivial topology. The $z$-direction is taken to be compact, \ie a circle $S^1$. The three $y$-directions span $\mathbb{R}^3$ which becomes $\mathbb{R}^+\times S^2$ when expressed in polar coordinates. Locally, we can thus identify $S^2\times S^1$ as the Hopf fibration of $S^3$ where the first Chern class is proportional to the magnetic charge.

If this solution is supplemented by some trivial world volume directions, it can be turned into something known as a KK-brane. The low energy limit of string theory is ten-dimensional supergravity. Thus, embedding the monopole solution (which is four-dimensional) requires adding six trivial dimensions (one of which is timelike) which would then produce a KK6-brane solution as follows
\begin{equation}
\dd s^2 = -\dd t^2 + \dd\vec{x}_{(5)}^{\, 2} 
				+ H^{-1}\left[\dd z + A_i\dd y^i\right]^2 + H\dd\vec{y}_{(3)}^{\, 2}
\end{equation}
where $H$ and $A_i$ are the same as above.\footnote{By adding seven trivial dimensions, the monopole solution can be embedded in M-theory (eleven-dimensional supergravity) where it takes the form of a KK7-brane. Then from the point of view of Type IIA supergravity -- which is the theory that emerges upon Kaluza-Klein reduction of M-theory -- the KK7-brane becomes either the Type IIA D6- or KK6-brane, depending on which direction is reduced.  All of this is part of the usual supergravity story relating solutions of eleven-dimensional supergravity to those of the Type IIA theory \cite{Townsend:1995kk}.}

Now let us consider a monopole-type solution in double field theory which we call the DFT monopole. Appendix \ref{sec:monopolecheck} shows that the following is a solution and satisfies the DFT equations of motion. The monopole solution is described by the generalized metric $\HH_{MN}$ 
\begin{equation}
\begin{aligned}
\dd s^2 &= \HH_{MN}\dd X^M \dd X^N \\
	&= H(1+H^{-2}A^2) \dd z^2 + H^{-1} \dd \tz^2 
			+ 2H^{-1} A_i[\dd y^i \dd \tz - \delta^{ij} \dd \ty_j \dd z ]\\
	&\quad + H(\delta_{ij}+H^{-2}A_iA_j) \dd y^i \dd y^j 
			+ H^{-1} \delta^{ij} \dd \ty_i \dd \ty_j \\
	&\quad +\eta_{ab}\dd x^a \dd x^b + \eta^{ab}\dd \tx_a \dd \tx_b 
\end{aligned}
\label{eq:DFTmonopole}
\end{equation}
and the rescaled dilaton of DFT (defined as $e^{-2d}=\sqrt{-g}e^{-2\phi}$) 
\begin{equation}
e^{-2d} = He^{-2\phi_0} 
\end{equation}
where $\phi_0$ is a constant. The generalized coordinates are
\begin{equation}
X^M = (z,\tz,y^i,\ty_i,x^a,\tx_a)
\label{eq:DFTmonopolecoords}
\end{equation}
where $i=1,2,3$ and $a=1,\dots,6$. The last line in \eqref{eq:DFTmonopole} uses the Minkowski metric $\eta_{ab}$, \ie $x^1=t$ and $\tx_1=\ttt$ are timelike, the signature being mostly plus. 

Here $H$ is a harmonic function of the $y^i$ only; it is annihilated (up to delta function sources) by the Laplacian in the $y$-directions and is given by
\begin{equation}
H(r) = 1 + \frac{h}{r}\, , \qquad r^2 = \delta_{ij}y^iy^j
\end{equation} 
with $h$ an arbitrary constant that is related to the first Chern class and hence the magnetic charge. The vector $A_i$ also obeys the Laplace equation, is divergence-free and its curl is given by the gradient of $H$
\begin{equation}
\vec{\nabla}\times\vec{A} = \vec{\nabla} H 
\qquad\mathrm{or}\qquad
\partial_{[i}A_{j]} = \frac{1}{2}{\epsilon_{ij}}^k\partial_k H \, .
\label{eq:DFTAH}
\end{equation}

This doubled solution is to be interpreted as a KK-brane of DFT. It can be rewritten to extract the spacetime metric $g_{\mu\nu}$ and the Kalb-Ramond two-form $B_{\mu\nu}$ in ordinary spacetime with coordinates $x^\mu=(z,y^i,x^a)$. We will show explicitly that the ``reduced'' solution is in fact an infinite periodic array of NS5-branes smeared along the $z$ direction.

One can also show that if $\tz$ is treated as a normal coordinate and $z$ as a dual coordinate the reduced solution is the string theory monopole introduced above. This means the (smeared) NS5-brane is the same as a KK-monopole with the KK-circle in a dual (winding) direction.

One might be concerned about the presence of $A_i$ in the generalized metric since for the monopole picture to make sense, $A_i$ must transform as a one-form gauge field. (Below we show how this one-form is a component of the two-form $B_{\mu \nu}$). Crucially, the generalized metric transforms under the generalized Lie derivative. When the generating double vector field of the generalized Lie derivative points in the dual space directions it generates the gauge transformations of the B-field. When we have an additional isometry, the $z$ direction of this solution, then this generalized Lie derivative generates the correct gauge transformations of a one-form field $A_i$. (This requires the gauge parameters to also be independent of $z$).

We will now use the same KK-ansatz \eqref{eq:KKforDFT} as for the DFT wave to rewrite the DFT monopole solution \eqref{eq:DFTmonopole} in terms of ten-dimensional non-doubled quantities. By Comparing \eqref{eq:KKforDFT} with \eqref{eq:DFTmonopole} the reduced fields can be computed. The spacetime metric $g_{\mu\nu}$ and the non-vanishing components of the B-field $B_{\mu\nu}$ are given by
\begin{equation}
\begin{aligned}
\dd s^2 &= -\dd t^2 + \dd\vec{x}_{(5)}^{\; 2} + H[\dd z^2 + \dd\vec{y}_{(3)}^{\; 2}] \\
B_{iz} &= A_i \, .
\label{eq:NSfivebrane}
\end{aligned}
\end{equation} 
The determinant of this metric is $-H^4$ and therefore the string theory dilaton becomes
\begin{equation}
e^{-2\phi} = (-g)^{-1/2} e^{-2d} = H^{-2}He^{-2\phi_0} = H^{-1}e^{-2\phi_0} \, .
\end{equation}
This solution is the NS5-brane solution of string theory \cite{Ortin:2004ms}, more precisely it is the NS5-brane smeared along the $z$ direction. Usually the harmonic function of the NS5-brane depends on all four transverse directions, that is $y^i$ and $z$. By smearing it over the $z$ direction the brane is no longer localized in $z$ and so the $z$-dependence is removed from the harmonic function.  

Smearing the solution along $z$ has also consequences for the field strength $H_{\mu\nu\rho}$. The NS5-brane comes with an H-flux whose only non-zero components are in the transverse directions $y^i$ and $z=y^4$. The field strength is written as
\begin{equation}
H_{mnp} = 3\partial_{[m}B_{np]} 
	= {\epsilon_{mnp}}^q\partial_q \ln H(r,z) 
\label{eq:DFTfieldstrength}
\end{equation}
where  $m=(i,z)=1,\dots,4$. We then note that the non-trivial part of the metric is $g_{mn}=H\delta_{mn}$ so that $g=\det g_{mn}=H^4$. This then allows us to write the field strength as
\begin{equation}
\begin{aligned}
H_{mnp} &= \sqrt{g}\tilde{\epsilon}_{mnpq}g^{qs}\partial_s \ln H \\
		&= H^2\tilde{\epsilon}_{mnpq}H^{-1}\delta^{qs}H^{-1}\partial_s H
		= \tilde{\epsilon}_{mnp}{}^q\partial_q H
\end{aligned}
\end{equation}
where the epsilon tensor has been converted to the permutation symbol (a tensor density) in order to make contact with the epsilon in a lower dimension. If the solution then is smeared along $z$, $H$ no longer depends on this coordinate. Therefore $H_{ijk} = 0$ and
\begin{equation}
\begin{aligned}
H_{ijz} &= 2\partial_{[i}B_{j]z} = \tilde{\epsilon}_{ijzk}\delta^{kl} \partial_l H \\
	&= \tilde{\epsilon}_{ijk}\delta^{kl} \partial_l H 
	= {\epsilon_{ij}}^k \partial_k H = 2\partial_{[i}A_{j]} \, .
\end{aligned}
\end{equation}
Thus the only non-zero component of the B-field (up to a gauge choice) of the smeared NS5-brane is $B_{iz}=A_i$. This then shows how the flux of the smeared NS5-brane is just the same as the usual magnetic two-form flux from a magnetic monopole for the electromagnetic field.

In conclusion, the smeared NS5-brane solution \eqref{eq:NSfivebrane} can be extracted from the DFT monopole \eqref{eq:DFTmonopole} using \eqref{eq:KKforDFT}. If $z$ and $\tz$ are exchanged, the same procedure recovers the KK-monopole of string theory. Since the monopole and the NS5-brane are T-dual to each other in string theory and DFT makes T-duality manifest, this should not come as a surprise. 

In order to identify the NS5-brane with the KK-monopole, it needed to be smeared along the $z$ direction. Any monopole type solution is expected to need more than a single patch to describe it (and in fact the topological charge may be viewed as the obstruction to a global description). In \cite{Papadopoulos:2014mxa} the problems of constructing a full global solution containing NSNS magnetic flux, with patching between different local descriptions in DFT, are discussed in detail. So have we resolved those issues here?

Not really, in the case described above, because of the additional isometry in the transverse directions, the three-form flux is completely encoded in a two-form flux. (This is non-trivial and can be constructed in the usual way, \`a la Dirac). In other words because of the additional isometry $H_3=F_2\wedge dz$, so that although the $H_3$ flux is an element of the third cohomology it is really completely given by the second cohomology of which $F_2$ is a non-trivial representative.

One can now ask the question if it is possible to localize the monopole and remove this additional smearing. We will look at this in Section \ref{sec:localization} where we consider solutions localized in winding space.

\section{T-duality in DFT}
\label{sec:Tduality}

In this chapter we have constructed and discussed two solutions to the DFT equations of motion, the DFT wave and the DFT monopole. Both are solutions in the doubled space which -- depending on their orientation -- reduce to a pair of T-duality related objects in the supergravity picture, \ie after removing the dual directions. The DFT wave gives either the F1-string or a pp-wave, the DFT monopole leads to the smeared NS5-brane or a KK-monopole. By considering these solutions and their relations, we can gain an interesting insight into how T-duality is viewed in DFT. For this we will make a distinction between the \emph{strong constraint} and the \emph{section condition}. 

Naively speaking, the strong constraint \eqref{eq:weakstrong} always needs to be imposed as explained in Section \ref{sec:DFTsection} to have a consistent theory which can be reduced to supergravity. Essentially it restricts the dependence of fields on half of the coordinates. For each pair $(x^\mu,\tx_\mu)$ of a normal and a winding coordinate one has to pick which coordinate the fields depend on and impose either $\partial_\mu = 0$ or $\tpartial^\mu = 0$ for the other coordinate.

The usual spacetime manifold of supergravity is defined by picking out a maximally isotropic subspace of the doubled space of DFT. It is the section condition (via the $\Odd$ structure $\eta$ which provides the pairing of $x^\mu$ and $\tx_\mu$) which selects the set of coordinates that form the physical spacetime slice or section within the doubled space. So normally by imposing the strong constraint the dependence on half of the coordinates is removed and the remaining coordinates are identified with the coordinates of spacetime via the section condition. In the absence of isometries the strong constraint and the section condition are therefore the same and so far we have used the expressions synonymously.

But what if there is an isometry? When there is an isometry then there is indeed an ambiguity in how one identifies the physical spacetime within the doubled space. In other words, due to the isometry fields can depend on fewer coordinates than required by the strong constraint. Then there is a choice of how to solve the section condition, \ie how to pick the coordinates for spacetime. 

This can be illustrated with the two DFT solutions we have encountered so far. The harmonic function and thus all fields of the DFT wave as presented in \eqref{eq:DFTwave} only depends on the $D-2$ transverse coordinates $y^m$. Therefore the strong constraint is satisfied since the solution depends on only half (actually less than half) the coordinates of the doubled space. But the section condition requires us to pick $D$ out of the $2D$ coordinates to form the physical spacetime. Obviously one has to pick the $D-2$ $y^m$ and it is also natural to include $t$, the time coordinate. But for the last coordinate there is a choice between $z$ and $\tz$. Since the solution depends on neither, the strong constraint remains satisfied by either choice. 

As we have seen, picking $z$ leads to the fundamental string solution while the opposite choice gives the pp-wave. Therefore the T-duality between these two supergravity solutions arises from the ambiguity in picking the physical spacetime. It is this choice of selecting either $z$ or $\tz$ which translates to the choice of (T-dual) frames in the supergravity picture. Thus from the DFT perspective, traditional T-duality comes from an isometry in the doubled space and hence an ambiguity in how one defines the half-dimensional subspace corresponding to a supergravity solution. The same holds for the DFT monopole \eqref{eq:DFTmonopole} where again there is a choice between $z$ and $\tz$ in addition to the $y^m$ and $x^a$ coordinates.

We will pick up this discussion again in Section \ref{sec:localization} where solutions localized in winding space which therefore have fewer isometries are considered. For now we will ask a different question. We have seen how T-duality related solutions descend from the same object in DFT. What about an electric/magentic duality or S-duality relation between the fundamental object, the F1-string, and the solitonic object, the NS5-brane?

Double field theory only makes T-duality of string theory a manifest symmetry. In order to incorporate S-duality, we have to move over to exceptional field theory which makes the U-duality of M-theory manifest. Since S-duality (together with T-duality) is part of U-duality, we will see how the same results for the wave and monopole in an extended space hold, but in addition they can also be directly related to each other. Therefore we will study exceptional field theory and its solutions next.

\chapter{Exceptional Field Theory}
\label{ch:EFT}

The primary idea behind exceptional field theory \cite{Hillmann:2009ci, Berman:2010is, Berman:2011cg, Berman:2011jh, Hohm:2013pua} is to make the exceptional symmetries of eleven-dimensional supergravity manifest. The appearance of the exceptional groups in dimensionally reduced supergravity theories was first discussed in \cite{Cremmer:1997ct, Cremmer:1998px}. In EFT one first performs a decomposition of eleven-dimensional supergravity but with no reduction or truncation into an $(11-D) + D$ split. That is one takes the eleven dimensions of supergravity to be
\begin{equation}
M^{11} = M^{11-D} \times M^D \, .
\end{equation}
Then one supplements the $D$ so called ``internal'' directions of $M^D$ with additional coordinates to linearly realize the exceptional symmetries and form an extended space $M^{\dim R_1}$ whose dimension is the dimension of the relevant\footnote{For $D=6,7,8$ this is the fundamental representation. For other groups a different representation might be needed for the construction, \eg for $D=4$ where $E_4=SL(5)$ it is the {\bf{10}} (\cf Table \ref{tab:GH} in the Introduction and Table \ref{tab:Ytensor} below).} representation $R_1$ of the exceptional group $E_D$. The introduction of novel extra dimensions therefore leads to an extension of the eleven-dimensional space to
\begin{equation}
M^{11} \longrightarrow M^{11-D} \times M^{\dim R_1}
\end{equation}
where $M^{\dim R_1}$ is a coset manifold that comes equipped with the coset metric of ${E_D}/{H}$ (where $H$ is the maximally compact subgroup of $E_D$ given in Table \ref{tab:GH}). This ``exceptional extended geometry'' has been constructed for several U-duality groups but was originally restricted to truncations of the eleven-dimensional theory \cite{Hull:2007zu, Pacheco:2008ps, Hillmann:2009ci, Berman:2010is, Coimbra:2011ky, Coimbra:2012af} where the ``external'' metric was taken to be flat and off-diagonal terms (the ``gravi-photon'') were set to zero. Furthermore, coordinate dependence was restricted to the internal extended coordinates.

The full, non-truncated exceptional field theory allows for a dependence on all coordinates, external, internal and extended. This allows for eleven-dimensional supergravity to be embedded into a theory that is fully covariant under the exceptional groups $E_D$. So far EFTs have been developed for $2\leq D\leq8$: $SL(2)\times\mathbb{R}^+$ \cite{Berman:2015rcc}, $SL(3) \times SL(2)$ \cite{Hohm:2015xna}, $SL(5)$ \cite{Musaev:2015ces}, $SO(5,5)$ \cite{Abzalov:2015ega}, $E_6$ \cite{Hohm:2013vpa}, $E_7$ \cite{Hohm:2013uia} and $E_8$ \cite{Hohm:2014fxa}, including the supersymmetrisation of the $E_6$ and $E_7$ cases \cite{Musaev:2014lna,Godazgar:2014nqa}. In addition, fairly complete constructions have been developed for the general field content of these theories \cite{Cederwall:2013naa,Wang:2015hca}. 

It is worthwhile at this stage to describe how the U-duality groups become related to the embedding of the eleven dimensions in the extended space. The combination of p-form gauge transformations and diffeomorphism give rise to a continuous local $E_D$ symmetry. This however is not U-duality which is a global discrete symmetry that only occurs in the presence of isometries. (See \cite{Berman:2014jba} for the equivalent discussion for DFT). Crucially however there is also a {\it{physical section condition}} that provides a constraint in EFT that restricts the coordinate dependence of the fields to a subset of the dimensions and thus there naturally appears a physical submanifold which we identify as usual spacetime. When there are no isometries present this section condition constraint produces a canonical choice of how spacetime is embedded in the extended space.  However, in the presence of isometries there is an ambiguity in how one identifies the submanifold in the extended space. This ambiguity is essentially the origin of U-duality with different choices of spacetime associated to U-duality related descriptions. (This is discussed in detail for the case of DFT in Section \ref{sec:Tduality} above and \cite{Berman:2014jsa}).

Another remark concerns the nature of the dimensionality $D$. Originally in double field theory only $D$ of the dimensions were taken to be compact which gives rise to the group $\Odd$. By doubling these $D$ coordinates, the T-duality in these directions could be made manifest. The remaining ``external'' dimensions were not doubled. Then $D$ was extended to include all dimensions of string theory, \ie everything was doubled. In EFT, the $E_D$ duality groups only arise if the eleven dimensions are split and $D$ dimensions are singled out. These $D$ directions are extended to form the exceptionally extended space while the remaining external dimension are not extended. The $E_{11}$ programme \cite{West:2001as, Englert:2003zs, West:2003fc, Kleinschmidt:2003jf, West:2004kb,West:2012qm,Tumanov:2015iea} states that all the EFTs with a given $D$ are included in a theory based on $E_{11}$, \ie by setting $D=11$ a split of the dimensions is avoided and all directions become extended. The problem here is that $E_{11}$ is of infinite dimension and thus the exceptionally extended space would have to include infinitely many novel coordinates to realize the symmetry.

In this thesis we will focus on $D=2,4$ and $7$. We will first give an overview of the exceptionally extended geometry, its structure and algebra, with an emphasis on $\Slf$ and $\Es$. Then as a warm-up the truncated theories for these two groups are presented. Only then do we move on to look at the full, non-truncated theory for $\Es$. The $\Slt$ EFT will be considered in Chapter \ref{ch:SL2} with a focus on its relation to F-theory.

\section{The Exceptional Extended Geometry of EFTs}
\label{sec:EEG}

All the defining features of double field theory such as an extended tangent space, a generalized Lie derivative with corresponding gauge algebra and bracket, and the section condition are also found in the exceptional extended geometries of exceptional field theories. In what follows, we will define these objects in a general manner, independent of $D$ and thus the relevant exceptional group $E_D$. To this end we will introduce the Y-tensor ${Y^{MN}}_{KL}$ \cite{Berman:2012vc}. It can be constructed for each symmetry group G and captures the deviation of the extended geometry from the usual Riemannian geometry of general relativity.

To illustrate this, we can cast some of the definitions for the doubled geometry from Chapter \ref{ch:DFT} into this general form by using the Y-tensor for DFT which is made from the $\Odd$ invariant structure $\eta$
\begin{equation}
	{Y^{MN}}_{KL} = \eta^{MN}\eta_{KL} \, .
\end{equation}
Then the generalized Lie derivative of a doubled vector $V^M$ can be written as (\cf \eqref{eq:DFTgenlie})
\begin{equation}
\begin{aligned}
	\LL_U V^M &= L_U V^M + {Y^{MN}}_{KL}\partial_NU^KV^L \\
			&= U^N\partial_N V^M - V^N\partial_N U^M + \eta^{MN}\eta_{KL}\partial_N U^K V^L
				+ \omega(V) \partial_N U^N V^M\, ,
\end{aligned}
\end{equation}
\ie it is the ordinary Lie derivative $L_U V^M$ plus a correction term for the doubled geometry. Similarly the $\sfC$-bracket \eqref{eq:Cbracket} can be expressed in terms of the Y-tensor and the extra term in the commutator of two generalized Lie derivatives is proportional to ${Y^{MN}}_{KL}$. Hence when imposing the section condition \eqref{eq:DFTsection}
\begin{equation}
	{Y^{MN}}_{KL}\partial_M \otimes \partial_N = \eta^{MN}\eta_{KL}\partial_M \otimes \partial_N= 0
\end{equation}
the extra term vanishes and the algebra closes. 

\begin{table}[h!]
\centering
\begin{tabular}{|c|c|rl|c|}
\hline
$D$ & $G$ 						& \multicolumn{2}{c|}{Y-tensor} & dim $R_1$ \\ \hline
$2$ & $SL(2)\times\mathbb{R}+$ 	& ${Y^{\alpha s}}_{\beta s}$ =  & $\delta^\alpha_\beta$ & $3$ \\
$3$ & $SL(3)\times SL(2)$		& ${Y^{i\alpha,j\beta}}_{k\gamma,l\delta} = $ & $4\delta^{ij}_{kl}\delta^{\alpha\beta}_{\gamma\delta}$ & $6$ \\
$4$ & $SL(5)$ 					& ${Y^{MN}}_{KL} =$ & $\epsilon^{iMN}\epsilon_{iKL}$ & $10$ \\
$5$ & $SO(5,5)$ 				& ${Y^{MN}}_{KL} =$ & $\frac{1}{2}(\Gamma^i)^{MN}(\Gamma_i)_{KL}$ & $16$ \\
$6$ & $E_6$ 					& ${Y^{MN}}_{KL} =$ & $10d^{MNP}d_{KLP}$ & $27$ \\
$7$ & $E_7$ 					& ${Y^{MN}}_{KL} =$ & $12{c^{MN}}_{KL} + \delta^{(M}_K\delta^{N)}_L + \frac{1}{2}\Omega^{MN}\Omega_{KL}$ & $56$ \\ 
$D$ & $O(D,D)$ 					& ${Y^{MN}}_{KL} =$ & $\eta^{MN}\eta_{KL}$ & $2D$ \\ \hline
\end{tabular}
\caption[The Y-tensor]{The Y-tensor for exceptional extended geometries with $2\leq D \leq 7$. The last column gives the dimensions of the extended space\protect\footnotemark\ (and thus the range of the index $M$ in each case). The indices and elements of the $Y$-tensor are the following: in the first two cases $s$ is a singlet index, $\alpha,\beta$ are $SL(2)$ indices and $i,j$ are $SL(3)$ indices; next $\epsilon_{iMN}=\epsilon_{iabcd}$ is the $SL(5)$ alternating tensor ($i=1,\dots,5$, $a=1,\dots,4$); then ${(\Gamma^i)^M}_N$ are the $16\times 16$ Majorana-Weyl representation of the $SO(5,5)$ Clifford algebra ($i=1,\dots,10$); $d^{MNK}$ is the symmetric invariant tensor of $E_6$; $c^{MNKL}$ is the symmetric tensor of $E_7$ and $\Omega^{MN}$ is the symplectic invariant tensor of its $\mathbf{56}$. For comparison the DFT Y-tensor made from the $\Odd$ invariant structure $\eta^{MN}$ has been included in the last line. In all cases except $E_7$, the Y-tensor is symmetric in both the upper and lower indices.}
\label{tab:Ytensor}
\end{table}

Therefore, by considering the Y-tensor for the various symmetry groups G, the corresponding extended geometry can be constructed. The Y-tensor for $2\leq D \leq 7$ is given in Table \ref{tab:Ytensor} \cite{Berman:2012vc}. While the extended coordinates $X^M$ are in the representation $R_1$ (usually the fundamental representation of G) of Table \ref{tab:GH} given in the introduction, the Y-tensor can be thought of a projector onto $R_2$ (usually the adjoint).

The generalized diffeomorphism which is the relevant gauge transformation with gauge parameter $U^M$ of a generalized vector $V^M$ can now be defined for any of the exceptional groups as
\begin{equation}
	\delta_UV^M = \LL_U V^M = L_U V^M + {Y^{MN}}_{KL}\partial_NU^KV^L \, .
	\label{eq:EFTgenlie}
\end{equation}
The gauge algebra closes and the Jacobi identity is satisfied once the section condition
\begin{equation}
	{Y^{MN}}_{KL}\partial_M \otimes \partial_N = 0 
	\label{eq:EFTsection}
\end{equation}
is imposed. The algebra can be written in terms of the $\sfE$-bracket \cite{Hohm:2013pua} which is the analogue of the the $\sfC$-bracket of DFT
\begin{equation}
	[\LL_U,\LL_V]W^M = \LL_{[U,V]_\sfE}W^M
\end{equation}
where the $\sfE$-bracket is given by
\begin{equation}
	[U,V]_\sfE^M = [U,V]^M	+ \frac{1}{2}{Y^{MN}}_{KL}\left(\partial_NU^KV^L - \partial_NV^KU^L \right) \, .
	\label{eq:Ebracket}
\end{equation}
Here $[U,V]^M$ is the ordinary Lie bracket and the extra term is proportional to the Y-tensor.
With this setup various further identities can be shown to hold, \eg that the Y-tensor is an invariant tensor
\begin{equation}
	\LL_U {Y^{MN}}_{KL} = 0 \, .
\end{equation}

\footnotetext{This is the dimension of the representation $R_1$ of the extended coordinates given in Table \ref{tab:GH} in the Introduction.}

With these general definitions for the exceptional extended geometries in mind, we can proceed to construct the theories for specific groups. First we will consider the truncated theory where the external sector is trivial and there are no cross-terms between external and internal sector. This setting allows us to gain some understanding of the extended geometries upon which the full theory can then be built.

\section{The truncated Theory}
\label{sec:truncatedEFT}

The $\Slf$ invariant theory \cite{Berman:2010is} was the first theory constructed which applied the ideas of generalized geometry and double field theory to the exceptional groups that arise in M-theory truncations. We will therefore start by giving an outline of this theory together with its action and equations of motion. After a brief intermezzo on the general form of the equations of motion of the truncated theories, we will move on to present the truncated $\Es$ theory. This should be a good preparation for the next section.

Note that the work on the truncated theories \cite{Hull:2007zu, Pacheco:2008ps, Hillmann:2009ci, Berman:2010is, Coimbra:2011ky, Coimbra:2012af} predates the inception of the full EFT formalism \cite{Hohm:2013pua,Hohm:2013vpa,Hohm:2013uia,Hohm:2014fxa}. Therefore certain formulations and calculations might seem superfluous in hindsight.

\subsection{The $SL(5)$ Duality Invariant Theory}
\label{sec:truncatedSL5}
Let us start by examining the extended geometry of the $SL(5)$ duality invariant theory. This arises from the full eleven-dimensional theory by splitting the dimensions into $4+7$. The U-duality group acts on the four dimensions and can be made manifest by including the six dual dimensions corresponding to membrane wrappings. There is then a (4+6)-dimensional extended space with manifest $SL(5)$ invariance and no dependence on the remaining seven dimensions. 

The extended tangent bundle of the generalized geometry for $\Slf$ combines the tangent space with the space of two-forms
\begin{equation}
	E = TM \oplus \Lambda^2T^*M
\end{equation}
where $M$ is the four-manifold in the $4+7$ split. This combines the diffeomorphisms on the four-manifold with the two-form gauge transformations of the three-form C-field. Therefore the two terms in the extended tangent bundle correspond to momentum modes and membrane wrapping modes respectively. 

The next step is to introduce generalized coordinate vectors (sections of $E$) for the extended tangent bundle. The four spacetime coordinates $x^\mu$ are combined with the six wrapping coordinates $y_{\mu\nu}$ to form the extended coordinate $X^M$ in the $\mathbf{10}$ of $\Slf$
\begin{equation}
	X^M = \begin{pmatrix} x^\mu \\ y_{\mu\nu} \end{pmatrix} \, .
\end{equation}
In the spirit of DFT and going beyond generalized geometry, not only the tangent bundle is extended but also the underlying space. The $\Slf$ theory is formulated on this (4+6)-dimensional extended space.

Referring to the $E_{11}$ decomposition into $SL(5)\times GL(7)$, schematically a generalized metric of such an (10+7)-dimensional space can be written as (see \cite{Malek:2012pw})
\begin{equation}
\HH = {\det g_{11}}^{-1/2}
\begin{pmatrix}
\tMM & 0 \\ 0 & g_7
\end{pmatrix}
\end{equation} 
where $\tMM$ is the generalized metric on the extended space and $g_7$ is the metric on the remaining seven dimensions. The conformal factor up front is important as it relates these two otherwise independent sectors, it is given in terms of the determinant of $g_{11}$, the metric of the full eleven-dimensional space.

This $\tMM_{MN}$ is the generalized metric as first given in \cite{Berman:2010is}. It parametrizes the coset $SL(5)/SO(5)$ in terms of the spacetime metric $g_{\mu\nu}$ and the form field $C_{\mu\nu\rho}$
\begin{equation}
\tMM_{MN} = 
\begin{pmatrix}
g_{\mu\nu} + \frac{1}{2}C_{\mu\rho\sigma}g^{\rho\sigma,\lambda\tau}C_{\lambda\tau\nu} 
		& \frac{1}{\sqrt{2}}C_{\mu\rho\sigma}g^{\rho\sigma,\lambda\tau} \\
\frac{1}{\sqrt{2}}g^{\rho\sigma,\lambda\tau}C_{\lambda\tau\nu} & g^{\rho\sigma,\lambda\tau}
\end{pmatrix}
\label{eq:SL5metric}
\end{equation}
for coordinates $X^M = (x^\mu,y_{\mu\nu})$ in the $\mathbf{10}$ of $SL(5)$ and with $g^{\mu\nu,\rho\sigma}=\frac{1}{2}(g^{\mu\rho}g^{\nu\sigma}-g^{\mu\sigma}g^{\nu\rho})$ which is used to raise an antisymmetric pair of indices. Note that there is no overall factor in front, this metric has a determinant of $g^{-2}$ where $g$ is the determinant of the four-metric $g_{\mu\nu}$. Therefore in this form it is actually an element of $GL(5)$, not $SL(5)$. This can be remedied by considering the following.

The theory contains a scaling symmetry for the $GL(5)$ which can be used to rescale $\tMM_{MN}$ by $g$, \eg $\MM_{MN} = g^{1/5}\tMM_{MN}$ (this particular rescaling leads to a generalized metric with unit determinant, \ie $\det \MM_{MN}=1$). Noting that $\det g_{11} = g\det g_7$ and assuming a simple form\footnote{For example when considering the compactification of the seven dimensions on a seven-torus with equal radius $R$ this is just $g_7=R\delta_7$ and thus $V=R^7$.} for the seven-metric such that $\det g_7 = V$ we have
\begin{equation}
\HH = 
\begin{pmatrix}
V^{-1/2}g^{-1/2}g^{-1/5}\MM & 0 \\ 0 & V^{-5/14}g^{-1/2}\delta_7
\end{pmatrix} =
\begin{pmatrix}
e^{-\Delta}\MM & 0 \\ 0 & e^{-5\Delta/7}\delta_7
\end{pmatrix} \, .
\end{equation} 
Under an $SL(5)$ transformation the seven-sector should remain unchanged, therefore we have the following $SL(5)$ scalar density
\begin{equation}
e^\Delta = V^{1/2}g^{7/10}
\end{equation}
which we will us to write down the correctly weighted action for the extended theory. In terms of the generalized metric $\MM_{MN}$ with unit determinant and the volume factor $\Delta$ the action reads
\begin{equation}
S = \int \dd^{10} X e^\Delta R
\label{eq:SL5action}
\end{equation}
where the scalar $R$ is given by
\begin{equation}
\begin{aligned}
R &= \frac{1}{12}\MM^{MN}\partial_M\MM^{KL}\partial_N\MM_{KL}
		-\frac{1}{2}\MM^{MN}\partial_M\MM^{KL}\partial_L\MM_{KN} \\
	&\qquad  + \partial_M\MM^{MN}\partial_N\Delta 
		+ \frac{1}{7}\MM^{MN}\partial_M\Delta\partial_N\Delta \, .
\end{aligned}
\label{eq:SL5R}
\end{equation}
The first two terms reproduce the Einstein-Hilbert and Maxwell term upon imposing section condition. The last two terms are kinetic terms for $\Delta$. The equations of motion for $\Delta$ can be found by varying the action and are given up to total derivatives by $R=0$.

On the other hand, varying the action with respect to the generalized metric and integrating by parts gives
\begin{align}
\delta S = \int \dd^{10} X e^{\Delta} &\left[\frac{1}{12}\left(\partial_M \MM^{KL}\partial_N \MM_{KL}
		- 2 \partial_K \MM^{KL}\partial_L \MM_{MN}  
		- 2 \MM^{KL} \partial_K \partial_L \MM_{MN} \right.\right. \notag \\
	&\quad\left.\left.
		+ 2 \MM^{KL}\MM^{PQ}\partial_K \MM_{MP} \partial_L \MM_{NQ} 
		- 2 \MM^{KL}\partial_K\Delta \partial_L \MM_{MN}\right) \right. \notag \\
	&\quad\left. -\frac{1}{2}\left(\partial_M \MM^{KL}\partial_L \MM_{KN}
		- 2 \partial_L \MM^{KL}\partial_M \MM_{KN} 
		- 2 \MM^{KL} \partial_L\partial_M \MM_{KN} \right.\right. \notag \\
	&\quad\left.\left.
		+ 2 \MM^{KP}\MM^{LQ}\partial_{(K} \MM_{M)Q} \partial_L \MM_{NP} 
		- 2 \MM^{KL} \partial_K \Delta \partial_M \MM_{LN}\right) \right. \notag \\
	&\quad\left. -\partial_M\partial_N\Delta - \frac{6}{7} \partial_M\Delta\partial_N\Delta	
		\right] \delta \MM^{MN} \, . 
\end{align}
Note that there is no term for varying $e^{\Delta}$. This factor contains information about the determinant of $\MM_{MN}$ but does not change if the metric is varied as it is fixed to have unit determinant. We will denote everything inside the brackets by $K_{MN}$
\begin{equation}
\delta S = \int \dd^{10} X e^\Delta K_{MN} \delta \MM^{MN} \, .
\label{eq:SL5varaction}
\end{equation}

As in the case of DFT, \eqref{eq:SL5varaction} does not have to vanish for any variation $\delta \MM^{MN}$ since the generalized metric is constrained to parametrize a coset space. This gives rise to a projector to eliminate the additional degrees of freedom. To impose this constraint and find this projector, one has to use the chain rule. In order to vary the generalized metric with respect to the spacetime metric and the C-field, it will be useful to use indices $a = \{\mu,5\}$ in the $\mathbf{5}$ of $SL(5)$. The coordinates are then
\begin{equation}
X^M = X^{ab} = 
\begin{cases}
X^{\mu 5} &= x^\mu \\
X^{\mu\nu} &= \frac{1}{2}\epsilon^{\mu\nu\rho\sigma}y_{\rho\sigma}
\end{cases}
\label{eq:SL5coords}
\end{equation}
where $\epsilon^{\mu\nu\rho\sigma}$ is the permutation symbol in four dimensions, a tensor density. The generalized metric and its inverse take the form
\begin{equation*}
\MM_{ab,cd} =
\begin{pmatrix}
\MM_{\mu 5,\nu 5} & \MM_{\mu 5,\lambda\tau} \\ \MM_{\rho\sigma,\nu 5} & \MM_{\rho\sigma,\lambda\tau}
\end{pmatrix}
= g^{1/5}
\begin{pmatrix}
g_{\mu\nu} + \frac{1}{2}C_{\mu\rho\sigma}g^{\rho\sigma,\lambda\tau}C_{\lambda\tau\nu}
	 & \frac{-1}{2\sqrt{2}} C_{\mu\rho\sigma} g^{\rho\sigma,\alpha\beta} 
	 		\epsilon_{\alpha\beta\lambda\tau} \\
	\frac{-1}{2\sqrt{2}} \epsilon_{\rho\sigma\alpha\beta}
			g^{\alpha\beta,\lambda\tau}C_{\lambda\tau\nu}
	 & g^{-1}g_{\rho\sigma,\lambda\tau}
\end{pmatrix}
\end{equation*}
\begin{equation}
\MM^{ab,cd} = g^{-1/5}
\begin{pmatrix}
g^{\mu\nu}  & \frac{1}{2\sqrt{2}}g^{\mu\nu}C_{\nu\alpha\beta}\epsilon^{\alpha\beta\lambda\tau} \\
		\frac{1}{2\sqrt{2}}\epsilon^{\rho\sigma\alpha\beta}C_{\alpha\beta\mu}g^{\mu\nu} 
	& gg^{\rho\sigma,\lambda\tau} + \frac{1}{8}\epsilon^{\rho\sigma\alpha\beta}C_{\alpha\beta\mu} 
		g^{\mu\nu}C_{\nu\gamma\delta}\epsilon^{\gamma\delta\lambda\tau}
\end{pmatrix}
\end{equation}
with $g^{\mu\nu,\alpha\beta}g_{\alpha\beta,\rho\sigma} = \frac{1}{2}(\delta^\mu_\rho\delta^\nu_\sigma - \delta^\mu_\sigma\delta^\nu_\rho)$. Note the factor of $g^{1/5}$ up front since this is the rescaled metric with unit determinant. Using the chain rule and varying the metric in \eqref{eq:SL5varaction} with respect to $\delta g_{\mu\nu}$ and $\delta C_{\mu\nu\rho}$ gives
\begin{align}
\delta S &= \int \dd^{10} X e^\Delta K_{MN} 
	\left[\frac{\delta \MM^{MN}}{\delta g_{\mu\nu}}\delta g_{\mu\nu} 
		+ \frac{\delta \MM^{MN}}{\delta C_{\mu\nu\rho}}\delta C_{\mu\nu\rho}\right] \\
	&= \int \dd^{10}X e^\Delta g^{-1/5} \notag \\
	&\hspace{1cm}\bigg\{\left[-K_{\alpha 5,\beta 5} g^{\alpha(\mu}g^{\nu)\beta} 
		-2 K_{\alpha 5,\beta\beta'}\frac{1}{2\sqrt{2}}g^{\alpha(\mu}g^{\nu)\alpha'}
			C_{\alpha'\gamma\gamma'}\epsilon^{\gamma\gamma'\beta\beta'}\right. \notag\\
	&\left.\hspace{1.8cm} 
		+ K_{\alpha\alpha',\beta\beta'}\left(\vphantom{\frac{1}{8}}
		gg^{\mu\nu}g^{\alpha\alpha',\beta\beta'} 
		- gg^{\alpha(\mu}g^{\nu)[\beta}g^{\beta']\alpha'}
		- gg^{\alpha[\beta}g^{\beta'](\mu}g^{\nu)\alpha'} \right.\right. \notag\\
	&\left.\left.\hspace{4cm}- \frac{1}{8}\epsilon^{\alpha\alpha'\gamma\gamma'}
			C_{\gamma\gamma'\sigma}g^{\sigma(\mu}g^{\nu)\sigma'}
			C_{\sigma'\lambda\lambda'}\epsilon^{\lambda\lambda'\beta\beta'}\right) \right. \notag\\
	&\left.\hspace{1.8cm}- \frac{1}{5}g^{1/5}K_{MN}\MM^{MN}g^{\mu\nu}
		\right]\delta g_{\mu\nu} \notag\\
	&\hspace{1cm} + \left[
		2K_{\alpha 5,\beta\beta'}\frac{1}{2\sqrt{2}}g^{\alpha\alpha'}
			\delta_\gamma^{[\mu}\delta_{\gamma'}^\nu\delta_\sigma^{\rho]}
			\epsilon^{\gamma\gamma'\beta\beta'} \right. \notag\\
	&\left.\hspace{1.8cm} + 2K_{\alpha\alpha',\beta\beta'}
			\frac{1}{8}\epsilon^{\alpha\alpha'\gamma\gamma'}
			\delta_\gamma^{[\mu}\delta_{\gamma'}^\nu\delta_\sigma^{\rho]}
			g^{\sigma\sigma'}C_{\sigma'\lambda\lambda'}\epsilon^{\lambda\lambda'\beta\beta'}
			\right]\delta C_{\mu\nu\rho}	\bigg\}
\end{align}
where the term $\frac{1}{5}K_{MN}\MM^{MN}g^{\mu\nu}\delta g_{\mu\nu}$ arises from varying the determinant factor. After cleaning up and dropping the symmetrizing and antisymmetrizing brackets, the $g$'s and $C$'s are re-expressed in terms of $\MM$ (factors of $g^{1/5}$ have to be accounted for carefully)
\begin{equation}
\begin{aligned}
\delta S &= \int \dd^{10}X e^\Delta  \\
	&\hspace{1cm}\bigg\{ g^{1/5} \left[\vphantom{\frac{1}{\sqrt{2}}} 
		-K_{\alpha 5,\beta 5} \MM^{\alpha 5,\mu 5}\MM^{\nu 5,\beta 5}
		-2 K_{\alpha 5,\beta\beta'}\MM^{\alpha 5,\mu 5}\MM^{\nu 5,\beta\beta'} 
			\right. \\
	&\left. \hspace{2.3cm}
		+ K_{\alpha\alpha',\beta\beta'}\left(g^{-1/5}\MM^{\mu 5,\nu 5}gg^{\alpha\alpha',\beta\beta'}
			- \MM^{\alpha\alpha',\mu 5}\MM^{\nu 5,\beta\beta'}\right)  \right. \\
	&\left.\hspace{2.3cm} - \frac{1}{5} K_{MN}\MM^{MN}\MM^{\mu 5,\nu 5}\right]\delta g_{\mu\nu} \\
	&\hspace{1cm} +\frac{1}{\sqrt{2}}\left[
		  K_{\alpha 5,\beta\beta'} \MM^{\alpha 5,\mu 5}\epsilon^{\nu\rho\beta\beta'}
		+ K_{\alpha\alpha',\beta\beta'} \MM^{\alpha\alpha',\mu 5}
			\epsilon^{\nu\rho\beta\beta'}\right]\delta C_{\mu\nu\rho}	\bigg\}
\end{aligned}
\end{equation}
Now the indices can be re-covariantized to be expressed as
\begin{equation}
\begin{aligned}
\delta S &= \int \dd^{10}X e^\Delta \left\{\vphantom{\frac{1}{\sqrt{2}}}
		g^{1/5}K_{KL}\left(\MM^{M, \mu 5}\MM^{\nu 5, N}\MM_{MP}	
			\frac{1}{4}\epsilon^{aPK}\epsilon_{aNQ}\MM^{QL} 
		- \MM^{K, \mu 5}\MM^{\nu 5, L}  \right.\right. \\
	&\left.\left.\hspace{4.5cm}	- \frac{1}{5}\MM^{KL}\MM^{\mu 5,\nu 5}\right)\delta g_{\mu\nu} 
	 	+ \frac{1}{\sqrt{2}}K_{KL}\MM^{K, \mu 5}\epsilon^{\nu\rho L5}\delta C_{\mu\nu\rho}
	 	\right\}
\end{aligned}
\label{eq:projderivation}
\end{equation}
which reproduces the previous line if the extended indices are expanded and summed over. In a final step these expressions can be written in terms of a projected set of equations
\begin{equation}
\delta S =\int \dd^{10}X e^\Delta (-3) \PP{M}{N}{K}{L}K_{KL}
	\left(g^{1/5}\MM^{M, \mu 5}\MM^{\nu 5, N}\delta g_{\mu\nu}
	 - \frac{1}{2\sqrt{2}}\MM^{M, \mu 5}\epsilon^{\nu\rho N5}\delta C_{\mu\nu\rho}\right)
\end{equation}
where the projector is given by
\begin{equation}
\PP{M}{N}{K}{L} = \frac{1}{3}\left({\delta_M}^{(K}{\delta_N}^{L)}
		+ \frac{1}{5}\MM_{MN}\MM^{KL}
		- \frac{1}{4}\MM_{MP}\epsilon^{aP(K}\epsilon_{aNQ}\MM^{L)Q} \right)
\end{equation}
which is symmetric in both $MN$ and $KL$ as can be seen from the contraction with the symmetric $\delta g_{\mu\nu}$ and $K_{KL}$ respectively. Note that the term containing $\delta C_{\mu\nu\rho}$ does not impose any symmetry property on the projector.

The variation of the action has to vanish for \emph{any} $\delta g_{\mu\nu}$ and $\delta C_{\mu\nu\rho}$ independently, therefore the equations of motion are given by  
\begin{equation}
\PP{M}{N}{K}{L}K_{KL} = 0
\label{eq:SL5eom}
\end{equation}
with $K_{MN}$ defined in \eqref{eq:SL5varaction}.

\subsection{Divertimento: Equations of Motion with a Projector}
\label{sec:divertimento}

In general, the dynamics of the extended geometry can be described using a projected equation of motion. In the truncated theories, \ie with trivial external sector and vanishing cross-terms, the action of the internal sector is given by
\begin{equation}
S = \int \dd^{N}X \LL
\end{equation}
where the Lagrangian $\LL$ includes the integration measure for the extended space and $N$ is the dimension of the extended space (that is $2D$ for $\Odd$ and $10$ for $\Slf$ for example -- see Table \ref{tab:Ytensor} where it is called dim $R_1$). Setting the variation of the action to zero gives
\begin{equation}
\delta S = \int \dd^{N}X K_{MN}\delta \MM^{MN} = 0 
\end{equation}
where $K_{MN}=\delta\LL/\delta\MM^{MN}$ is the variation of the Lagrangian with respect to the generalized metric. The integrand does not have to vanish for any $\delta\MM^{MN}$ since the generalized metric is constraint to parametrize the coset space $G/H$. This constraint gives rise to a projector in the equations of motion
\begin{equation}
{P_{MN}}^{KL}K_{KL} = 0.
\label{eq:genEoM}
\end{equation}
Following the method for $\Odd$ in Section \ref{sec:DFTaction} and $\Slf$ above where we use a chain rule type argument, we see that the projector may be written in a standard form using only the generalized metric and the $Y$-tensor
\begin{equation}
{P_{MN}}^{KL} = \frac{1}{a}\left( {\delta_M}^{(K}{\delta_N}^{L)}
		+ b \MM_{MN}\MM^{KL}	- \MM_{MP}{Y^{P(K}}_{NQ}\MM^{L)Q} \right) \, ,
\label{eq:genProj}
\end{equation}
together with the constants $a$ and $b$ which depend on the dimension of the extended space $N$ and thus the U-duality group. In \cite{Berkeley:2014nza} the projectors have been constructed for $\Odd$, $\Slf$ and $SO(5,5)$ and the constants $(a,b)$ where found to be $(2,0)$, $(3,\frac{1}{5})$ and $(4,\frac{1}{4})$ respectively.

Our ${P_{MN}}^{KL}$ is a genuine projector in the sense that $P^2=P$ and its eigenvalues are either $0$ or $1$. The eigenvectors with eigenvalue $0$ span the kernel of the projector. Those parts of $K_{MN}$ proportional to these eigenvectors are projected out and eliminated from the equations of motion. 

The multiplicity of the eigenvalues $0$ and $1$ are called nullity (dimension of the kernel) and rank of the projector respectively. They add up to $\frac{1}{2}N(N+1)$, the dimension of the vector space of eigenvectors. We have not shown that this is true beyond the groups mentioned above since the calculations have been done just by brute force. However,  given the structure of the exceptional geometric theories, in that the theories up to $E_7$ are completely determined by the generalized metric and the $Y$-tensor (along with a few dimensionally dependent constants), then we expect this projector to be true at least up to $E_7$ with only the constants $a$ and $b$ to be determined.

Note, the object $K_{MN}$ is symmetric and thus has $\frac{1}{2}N(N+1)$ independent components in a generalized space with $N$ dimensions. The bosonic degrees of freedom of the theories under consideration are given by the metric tensor $g_{\mu\nu}$ and the form fields $B_{\mu\nu}$ or $C_{\mu\nu\rho}$ (plus one for the dilaton $\phi$ in DFT and the volume factor $\Delta$ in the $SL(5)$ theory). One equation of motion is needed for each of those degrees of freedom. The projector reduces the components of the equation $K_{MN}=0$ such that the right number of independent equations remain.

\subsection{The $\Es$ Duality Invariant Theory}
\label{sec:truncatedE7}

Let's now turn to the extended geometry of th $\Es$ duality invariant theory. We consider the case where the eleven-dimensional theory is a direct product of $M^4 \times M^7$, the U-duality group acting on the seven-dimensional space $M^7$ is $E_7$. We will truncate the theory to ignore all dependence on the $M^4$ directions and will not allow any excitations of fields with mixed indices such as the gravi-photon. 

The exceptional extended geometry is constructed by extending the tangent bundle from seven to 56 dimensions
\begin{equation}
E = TM \oplus \Lambda^2 T^*M \oplus \Lambda^5 T^*M \oplus (T^*M \otimes \Lambda^7 T^*M) \, .
\label{eq:E7tangentbundle}
\end{equation}
where $M$ is the seven-space. The terms in the sum correspond to brane charges: momentum, membrane, fivebrane and KK-monopole charge. Details of this construction and the resulting theory are described in \cite{Hillmann:2009ci,Hull:2007zu,Pacheco:2008ps,Coimbra:2011ky,Coimbra:2012af,Berman:2010is,Berman:2011jh,Berman:2012vc}. The algebra is $E_7\otimes GL(4)$ with the $E_7$ acting along the seven spacetime dimensions of the extended space. The generators of the associated motion group are
\begin{equation}
P_\mu, Q^{\mu\nu},Q^{\mu_1\dots\mu_5},Q^{\mu_1\dots\mu_7,\nu}\qandq P^\alpha
\end{equation}
where $\mu=1,\dots,7$ and $\alpha=1,\dots,4$. The first four generate the $\mathbf{56}$ representation of $E_7$ and the last one generates translations in the remaining four directions, the $GL(4)$. For convenience, the following dualization of generators is used
\begin{equation}
\tQ_{\mu\nu}=\frac{1}{5!}\epsilon_{\mu\nu\rho_1\dots\rho_5}Q^{\rho_1\dots\rho_5} 
\qandq
\tQ^{\mu}=\frac{1}{7!}\epsilon_{\nu_1\dots\nu_7}Q^{\nu_1\dots\nu_7,\mu} \, .
\end{equation}
For the $E_7$ generators we can now introduce generalized coordinates\footnote{For this subsection and Section \ref{sec:E7monopole} which are both concerned with the truncated $\Es$ EFT we will use this alternative notation for the extended coordinates.} 
\begin{equation}
\XX^M = (X^\mu,Y_{\mu\nu},Z^{\mu\nu},W_\mu)
\end{equation}
to form the extended 56-dimensional space. Note that an index pair $\mu\nu$ is antisymmetric and we thus have indeed $7+21+21+7=56$ coordinates. These are the coordinates conjugate to the brane charges mentioned above which can be seen as brane \emph{wrapping} coordinates in analogy to the string \emph{winding} coordinates of DFT.

The generalized metric $\MM_{MN}$ of this extended space can be constructed from the vielbein given in \cite{Hillmann:2009ci,Hull:2007zu,Pacheco:2008ps,Coimbra:2011ky,Coimbra:2012af,Berman:2011jh}. The full expression is quite an unwieldy structure, so we will introduce it in several steps.

The underlying structure of $\MM_{MN}$ can be seen clearly if the M-theory potentials $C_3$ and $C_6$ are turned off. Then the only field present is the spacetime metric $g_{\mu\nu}$ and the line element of the extended space\footnote{As for the generalized metric in DFT, we utilize a line element to present the components of the matrix $\MM_{MN}$ and the coordinates of the extended space in a concise form. We do not wish to imply that the generalized metric is an actual metric tensor on the extended space.} reads
\begin{equation}
\begin{aligned}
\dd s^2 &= \MM_{MN}\dd\XX^M\dd\XX^N \\
		&= g^{-1/2}\Big\{
			g_{\mu\nu}\dd X^\mu X^\nu
			+ g^{\rho\sigma,\lambda\tau}\dd Y_{\rho\sigma}\dd Y_{\lambda\tau} \\
		&\qquad\qquad 
			+ g^{-1} g_{\rho\sigma,\lambda\tau}\dd Z^{\rho\sigma}\dd Z^{\lambda\tau}  
			+ g^{-1}g^{\mu\nu}\dd W_\mu \dd W_\nu \Big\} \, .
\end{aligned}
\label{eq:genmetric}
\end{equation}
Here the determinant of the spacetime metric is denoted by $g=\det g_{\mu\nu}$ and the four-index objects are defined by $g_{\mu\nu,\rho\sigma}=\frac{1}{2} \left(g_{\mu\rho}g_{\nu\sigma} - g_{\mu\sigma}g_{\nu\rho}\right)$ and similarly for the inverse. 

The generalized metric has a scaling symmetry and can be rescaled by a power of its determinant which in turn is just a power of $g$. The bare metric, \ie without the factor of $g^{-1/2}$ upfront, has $\det \MM_{MN} = g^{-28}$. One could choose to rescale by including a factor of $g^{1/2}$ which would then lead to $\det\MM_{MN}=1$, an often useful and desirable property. Here the factor $g^{-1/2}$ is included. It arises completely naturally from the $E_{11}$ programme, see \cite{Berman:2011jh}, and interestingly gives solutions in the Einstein frame when rewritten by a KK-ansatz (\ie no further rescaling is necessary).

If the gauge potentials are non-zero, there are additional terms for the ``diagonal'' entries of \eqref{eq:genmetric} and also ``cross-terms'' mixing the different types of coordinates. For what follows we will not need to use the full generalized metric with both potentials present at the same time. We will just need to consider the two special cases where either the $C_3$ potential or the $C_6$ potential vanishes. 

In the first case with no three-form, the six-form is dualized and encoded as
\begin{equation}
U^\mu = \frac{1}{6!}\epsilon^{\mu\nu_1\dots\nu_6}C_{\nu_1\dots\nu_6}
\label{eq:defU}
\end{equation}
which allows the line element to be written as
\begin{equation}
\begin{aligned}
\dd s^2 &= g^{-1/2}\left\{ \left[g_{\mu\nu} 
	+ \frac{1}{2}(g_{\mu\nu}U^\rho U_\rho - U_\mu U_\nu)\right]\dd X^\mu X^\nu \right.
	+ \frac{2}{\sqrt{2}}g^{-1/2}g_{\mu[\lambda}U_{\tau]}\dd X^\mu \dd Z^{\lambda\tau} \\
	&\qquad 
	+ \left[g^{\rho\sigma,\lambda\tau} 
		- \frac{1}{2}U^{[\rho}g^{\sigma][\lambda}U^{\tau]}\right] 
			\dd Y_{\rho\sigma}\dd Y_{\lambda\tau} 
	+ \frac{2}{\sqrt{2}}g^{-1/2}U^{[\rho}g^{\sigma]\nu} \dd Y_{\rho\sigma}\dd W_\nu \\
	&\qquad
	+ g^{-1} g_{\rho\sigma,\lambda\tau}\dd Z^{\rho\sigma}\dd Z^{\lambda\tau}  
		+ g^{-1}g^{\mu\nu}\dd W_\mu \dd W_\nu \vphantom{\frac{1}{2}} 
		\left. \vphantom{\frac{1}{2}} \right\} \, .
\end{aligned}
\label{eq:genmetric1}
\end{equation}

In the second case with no six-form, the three-form components are encoded in $C, V$ and $X$ (see \cite{Berman:2011jh}). We will concentrate on the special case where 
\begin{equation}
V^{\mu_1\dots\mu_4} = 
	\frac{1}{3!}\epsilon^{\mu_1\dots\mu_4\nu_1\dots\nu_3}C_{\nu_1\dots\nu_3}\neq 0
\quad\mathrm{but}\quad 
{X_\mu}^{\rho\sigma} = C_{\mu\lambda\tau}V^{\lambda\tau\rho\sigma} = 0 \, .
\label{eq:defV}
\end{equation}
Then the line element for the generalized metric is then given by
\begin{equation}
\begin{aligned}
\dd s^2 &= g^{-1/2}\left\{\left[g_{\mu\nu} 
		+ \frac{1}{2}C_{\mu\nu\rho}g^{\rho\sigma,\lambda\tau}
			C_{\lambda\tau\nu} \right]\dd X^\mu \dd X^\nu \right. \\ 
	&\qquad 	
		+ \left[g^{\mu_1\mu_2,\nu_1\nu_2} + \frac{1}{2}V^{\mu_1\mu_2\rho\sigma}
			g_{\rho\sigma,\lambda\tau}V^{\lambda\tau\nu_1\nu_2}\right] 
				\dd Y_{\mu_1\mu_2}\dd Y_{\nu_1\nu_2} \\
	&\qquad  + g^{-1} \left[g_{\mu_1\mu_2,\nu_1\nu_2} 
		+ \frac{1}{2}C_{\mu_1\mu_2\rho}g^{\rho\sigma}C_{\sigma\nu_1\nu_2}\right]
				\dd Z^{\mu_1\mu_2}\dd Z^{\nu_1\nu_2} \\ 
	&\qquad	
		+ g^{-1}g^{\mu\nu}\dd W_\mu \dd W_\nu 
		+ \frac{2}{\sqrt{2}}g^{-1/2}C_{\mu\rho\sigma}g^{\rho\sigma,\lambda\tau}
			\dd X^\mu \dd Y_{\lambda\tau} \\ 
	&\qquad	
		+ \frac{2}{\sqrt{2}}g^{-1/2}V^{\mu_1\mu_2\rho\sigma}g_{\rho\sigma,\nu_1\nu_2} 
			\dd Y_{\mu_1\mu_2}\dd Z^{\nu_1\nu_2}  \\ 
	&\qquad	
		+ \left.\frac{2}{\sqrt{2}}g^{-1/2}C_{\mu_1\mu_2\rho}g^{\rho\nu}
			\dd Z^{\mu_1\mu_2}\dd W_\nu  \right\} \, .
\end{aligned}
\label{eq:genmetric2}
\end{equation}

The action for the $E_7$ theory can be constructed as in \cite{Hillmann:2009ci,Hull:2007zu,Pacheco:2008ps,Coimbra:2011ky,Coimbra:2012af,Berman:2011jh}. The corresponding equations of motion can be derived as explained in the previous subsection and will appear under a projector to take into account that the generalized metric is constrained to be in the $E_7/SU(8)$ coset.

\section{The full non-truncated Theory}
\label{sec:fullEFT}

By constructing the generalized metrics on the exceptional extended spaces of $\Slf$ and $\Es$ we have encountered some of the essential parts of exceptional field theory. But so far the formalism was restricted to the ``extended internal'' sector, the external space (with $11-D$ dimensions) was taken to be trivial and cast aside together with the cross-terms which would have mixed -- external and internal -- indices. Also the coordinate dependence of the fields was restricted to the extended internal sector. 

The next step is therefore to find a formalism which includes all sectors, is fully covariant and makes the $E_D$ duality group a manifest symmetry. This allows for the embedding of full eleven-dimensional supergravity without a truncation\footnote{Note that when we speak of ``non-truncated'' theories, we simply mean that the external sector and the cross-terms between the external and internal extended sectors are included. This does not mean that there are no other truncations of any kind to the theories.}. Furthermore, there is no restriction on the coordinate dependence of fields. We will demonstrate this by studying the full EFT for the duality group $\Es$ \cite{Hohm:2013uia}.

The $E_7$ exceptional field theory lives in a $(4+56)$-dimensional spacetime. The four dimensional external space has coordinates $x^\mu$ and metric $g_{\mu\nu}={e_\mu}^\ua{e_\nu}^\ub\eta_{\ua\ub}$ which may be expressed in terms of a vierbein. The 56-dimensional extended internal space has coordinates $Y^M$ which are in the fundamental representation of $E_7$. This exceptional extended space is equipped with a generalized metric $\MM_{MN}$ which parametrizes the coset $E_7/SU(8)$ and takes the same form as the one constructed for the truncated version of the theory. The 56-dimensional exceptional extended geometry with the extended tangent bundle \eqref{eq:E7tangentbundle} was already given in the previous section. Just note the different notation $Y^M$ as opposed to $\XX^M$ for the generalized coordinates. 

In addition to the external metric $g_{\mu\nu}$ and the generalized metric $\MM_{MN}$, EFT also requires a generalized gauge connection ${\AAA_\mu}^M$ and a pair of two-forms $B_{\mu\nu\ \alpha}$ and $B_{\mu\nu\ M}$ to describe all degrees of freedom of eleven-dimensional supergravity (their corresponding field strengths ${\FF_{\mu\nu}}^M$, $\HH_{\mu\nu\rho\ \alpha}$ and $\HH_{\mu\nu\rho\ M}$ are described below). Here $\alpha=1,\dots,133$ labels the adjoint and $M=1,\dots,56$ the fundamental representation of $E_7$. For more on the nature of these two-forms see \cite{Hohm:2013uia}. For the main part of this chapter and the next they will both be zero and not play a role in what follows though they are of course crucial for the consistency of the theory.

Thus, the field content of the $E_7$ exceptional field theory is
\begin{equation}
\left\{ g_{\mu\nu}, \MM_{MN}, {\AAA_\mu}^M, B_{\mu\nu\ \alpha}, B_{\mu\nu\ M}\right\} \, .
\label{eq:E7fieldcontent}
\end{equation}
All these fields are then subjected to the physical section condition which picks a subspace of the exceptional extended space. This section condition can be formulated in terms of the $E_7$ generators $(t_\alpha)^{MN}$ and the invariant symplectic form $\Omega_{MN}$ of $Sp(56)\supset E_7$ as
\begin{align}
(t_\alpha)^{MN}\partial_M\partial_N \Phi &= 0 \, , &
(t_\alpha)^{MN}\partial_M\Phi\partial_N \Psi &= 0 \, , &
\Omega^{MN}\partial_M\Phi\partial_N \Psi &= 0
\label{eq:E7section}
\end{align}
where $\Phi,\Psi$ stand for any field and gauge parameter. This form of the section condition is equivalent to the form given in \eqref{eq:EFTsection} in terms of the Y-tensor. 

The equations of motion describing the dynamics of the fields can be derived from the following action which now is an integral over the full $(4+56)$-dimensional spacetime
\begin{equation}
\begin{aligned}
S = \int \dd^4x \dd^{56}Y e&\left[\hR 
		+ \frac{1}{48}g^{\mu\nu}\DD_\mu\MM^{MN}\DD_\nu\MM_{MN} \right. \\
		&\quad \left.	- \frac{1}{8}\MM_{MN}\FF^{\mu\nu\ M}{\FF_{\mu\nu}}^N
		- V(\MM_{MN},g_{\mu\nu}) + e^{-1}\LL_{\mathrm{top}}\right] \, .
\end{aligned}
\label{eq:action}
\end{equation}
The first term is a covariantized Einstein-Hilbert term given in terms of the spin connection $\omega$ of the vierbein ${e_\mu}^\ua$ (with determinant $e$)
\begin{equation}
\LL_{\mathrm{EH}}=e\hR =e {e_\ua}^\mu{e_\ub}^\nu\hR_{\mu\nu}{}^{\ua\ub} \qquad \mathrm{where} \qquad
\hR_{\mu\nu}{}^{\ua\ub} \equiv R_{\mu\nu}{}^{\ua\ub}[\omega] + {\FF_{\mu\nu}}^M e^{\ua\rho}\partial_M{e_\rho}^\ub \, .
\end{equation}
The second term is a kinetic term for the generalized metric $\MM_{MN}$ which takes the form of a non-linear gauged sigma model with target space $E_7/SU(8)$. The third term is a Yang-Mills-type kinetic term for the gauge vectors ${\AAA_\mu}^M$ which are used to define the covariant derivatives $\DD_\mu$. The fourth term is the ``potential'' $V$ built from internal extended derivatives $\partial_M$
\begin{equation}
\begin{aligned}
V = &-\frac{1}{48}\MM^{MN}\partial_M\MM^{KL}\partial_N\MM_{KL}
		+ \frac{1}{2}\MM^{MN}\partial_M\MM^{KL}\partial_L\MM_{NK} \\
	& - \frac{1}{2}\g^{-1}\partial_M\g\partial_N\MM^{MN} 
		- \frac{1}{4}\MM^{MN}\g^{-1}\partial_M\g\g^{-1}\partial_N\g
		- \frac{1}{4}\MM^{MN}\partial_M g^{\mu\nu}\partial_N g_{\mu\nu}
\end{aligned}
\end{equation}
where $\g = e^2 = \det g_{\mu\nu}$. The last term is a topological Chern-Simons-like term which is required for consistency. 

All fields in the action depend on all the external and extended internal coordinates. The derivatives $\partial_M$ appear in the non-abelian gauge structure of the covariant derivative and together with the two-forms $B_{\mu\nu}$ in the field strengths ${\FF_{\mu\nu}}^M$. Note that this action can be reduced to the action of the truncated theory presented in the previous section by setting ${\AAA_\mu}^M$ and the $B_{\mu\nu}$ to zero and restricting the coordinate dependence to $Y^M$, \ie $\partial_\mu=0$. Then only the potential term containing the generalized metric remains. 

The gauge connection ${\AAA_\mu}^M$ allows for the theory to be formulated in a manifestly invariant way under generalized Lie derivatives. The \emph{covariant derivative} for a vector of weight $\lambda$ is given by
\begin{equation}
\begin{aligned}
\DD_\mu V^M &= \partial_\mu V^M - {\AAA_\mu}^K\partial_K V^M
	+ V^K\partial_K {\AAA_\mu}^M 
	+ \frac{1-2\lambda}{2} \partial_K{\AAA_\mu}^K V^M \\
	&\quad + \frac{1}{2}\left[24(t^\alpha)^{MN}(t^\alpha)_{KL}+\Omega^{MN}\Omega_{KL}\right]
		\partial_N{\AAA_\mu}^K V^L \, .
\end{aligned}
\end{equation}
The associated non-abelian field strength of the gauge connection, defined as
\begin{equation}
\begin{aligned}
{F_{\mu\nu}}^M &\equiv 2\partial_{[\mu}{\AAA_{\nu]}}^M 
	- 2{\AAA_{[\mu}}^N\partial_N{\AAA_{\nu]}}^M \\
	&\quad - \frac{1}{2}\left[24(t^\alpha)^{MN}(t^\alpha)_{KL}-\Omega^{MN}\Omega_{KL}\right]
		{\AAA_{[\mu}}^K\partial_N{\AAA_{\nu]}}^L \, ,
\end{aligned}
\end{equation}
is not covariant with respect to vector gauge transformations. In order to form a properly covariant object we extend the field strength with St\"uckelberg-type couplings to the compensating two-forms $B_{\mu\nu\ \alpha}$ and $B_{\mu\nu\ M}$ as follows
\begin{equation}
{\FF_{\mu\nu}}^M = {F_{\mu\nu}}^M - 12(t^\alpha)^{MN}\partial_N B_{\mu\nu\ \alpha} 
		- \frac{1}{2}\Omega^{MN}B_{\mu\nu\ N}  \, .
\label{eq:genfieldstrength}
\end{equation}
For a detailed derivation and explanation of this we refer to \cite{Hohm:2013uia}. The Bianchi identity for this generalized field strength is
\begin{equation}
3\DD_{[\mu}{\FF_{\nu\rho]}}^M = -12(t^\alpha)^{MN}\partial_N\HH_{\mu\nu\rho\ \alpha}
	- \frac{1}{2}\Omega^{MN}\HH_{\mu\nu\rho\ N}
\end{equation}
which also defines the three-form field strengths $\HH_{\mu\nu\rho\ \alpha}$ and $\HH_{\mu\nu\rho\ M}$. The final ingredient of the theory are the \emph{twisted self-duality} equations for the 56 EFT gauge vectors ${\AAA_\mu}^M$
\begin{equation}
{\FF_{\mu\nu}}^M = \frac{1}{2}e \epsilon_{\mu\nu\rho\sigma} \Omega^{MN} \MM_{NK} \FF^{\rho\sigma\ K} 
\label{eq:selfduality}
\end{equation}
which relate the 28 ``electric'' vectors to the 28 ``magnetic'' ones. This self-duality relation is a crucial property of the $E_7$ EFT and is essential for the results presented here. In fact this sort of twisted self-duality equation has been described many years ago in the seminal work of \cite{Cremmer:1998px}. 

To conclude this brief overview of the full exceptional field theory, we note that the bosonic gauge symmetries uniquely determine the theory. They are given by the generalized diffeomorphisms of the external and extended internal coordinates. For more on the novel features of the generalized diffeomorphisms in exceptional field theory see \cite{Hohm:2013uia}.

An immediate simplification to the above equations presents itself when the coordinate dependence of fields and gauge parameters is restricted. In Chapter \ref{ch:EFTsol} we will consider a solution of EFT which only depends on external coordinates. Thus any derivative of the internal extended coordinates, $\partial_M$, vanishes trivially. Furthermore, our solution comes with zero two-form fields $B_{\mu\nu\ \alpha}$ and $B_{\mu\nu\ M}$, thus simplifying the gauge structure further. The upshot of this is a drastic simplification of the theory: covariant derivatives $\DD_\mu$ reduce to ordinary partials $\partial_\mu$, the generalized field strength ${\FF_{\mu\nu}}^M$ is simply given by $2\partial_{[\mu}{\AAA_{\nu]}}^M$, the covariantized Einstein-Hilbert term reduces to the ordinary one and the potential $V$ of the generalized metric vanishes. Finally the Bianchi identity reduces to the usual $\dd \FF^M =0$.

\subsection{Embedding Supergravity into EFT}
\label{sec:embedding}

Having outlined the main features of the $E_7$ EFT, we proceed by showing how eleven-dimensional supergravity can be embedded in it (again following \cite{Hohm:2013uia} closely). Applying a specific solution of the section condition \eqref{eq:E7section} to the EFT produces the dynamics of supergravity with its fields rearranged according to a $4+7$ Kaluza-Klein coordinate split.

The appropriate solution to the section condition is related to a decomposition of the fundamental representation of $E_7$ under its maximal subgroup $GL(7)$
\begin{equation}
\mathbf{56} \rightarrow 7 + 21 + 7 + 21
\end{equation}  
which translates to the following splitting of the extended internal coordinates
\begin{equation}
Y^M = (y^m, y_{mn}, y_m, y^{mn})
\label{eq:gencoordM}
\end{equation}
where $m=1,\dots,7$ and the pair $mn$ is antisymmetric. We thus have indeed $7+21+7+21=56$ coordinates. The section condition is solved by restricting the coordinate dependence of fields and gauge parameters to the $y^m$ coordinates. We thus have
\begin{equation}
\begin{aligned}
\partial^{mn} &\rightarrow 0 \, , & 
\partial^m &\rightarrow 0 \, , & 
\partial_{mn} &\rightarrow 0 \\
{B_{\mu\nu}}^{mn} &\rightarrow 0 \, , & 
{B_{\mu\nu}}^m &\rightarrow 0 \, , & 
{B_{\mu\nu}}_{mn} &\rightarrow 0 
\end{aligned}
\end{equation}
where the second line is the necessary consequence for the compensating two-form ${B_{\mu\nu}}^M$.

The complete procedure to embed supergravity into EFT can be found in \cite{Hohm:2013vpa,Hohm:2013uia}, here we will focus on those aspects relevant to our results. The Kaluza-Klein decomposition of the eleven-dimensional spacetime metric takes the following form
\begin{equation}
\hg_{\hmu\hnu} = 
			\begin{pmatrix}
				\hg_{\mu\nu} & \hg_{\mu n} \\
				\hg_{m\nu} & \hg_{mn}
			\end{pmatrix} = 
			\begin{pmatrix}
				g_{\mu\nu} + {A_\mu}^m{A_\nu}^ng_{mn} & {A_\mu}^mg_{mn} \\
				g_{mn}{A_\nu}^n & g_{mn}
			\end{pmatrix}
\end{equation}
where hatted quantities and indices are eleven-dimensional. The four-dimensional external sector with its metric $g_{\mu\nu}$ is carried over to the EFT. The seven-dimensional internal sector is extended to the 56-dimensional exceptional space and the internal metric $g_{mn}$ becomes a building block of the generalized metric $\MM_{MN}$. The KK-vector ${A_\mu}^m$ becomes the $y^m$-component of the EFT vector ${\AAA_\mu}^M$. 

The gauge potentials $C_3$ and $C_6$ of supergravity are also decomposed under the $4+7$ coordinate split. Starting with the three-form, there is the purely external three-form part $C_{\mu\nu\rho}$ which lives in the external sector. The purely internal scalar part $C_{mnp}$ is included in $\MM_{MN}$. The one-form part $C_{\mu\ mn}$ is the $y_{mn}$-component of ${\AAA_\mu}^M$. The remaining two-form part $C_{\mu\nu\ m}$ gets encoded in the compensating two-form $B_{\mu\nu\ \alpha}$. Similarly for the six-form, the purely internal scalar $C_{m_1\dots m_6}$ is part of $\MM_{MN}$. The one-form $C_{\mu\ m_1\dots m_5}$ is dualized on the internal space and forms the $y^{mn}$-component of ${\AAA_\mu}^M$. The remaining components of $C_6$ with a mixed index structure (some of which need to be dualized properly) are again encoded in the two-form $B_{\mu\nu\ \alpha}$.

In the next chapter we will work with supergravity solutions where the gauge potentials only have a single non-zero component which will be of the one-form type under the above coordinate split. There will not be any internal scalar parts or other mixed index components. The above embedding of supergravity fields into EFT can therefore be simply summarized as follows. The spacetime metric $g_{\mu\nu}$ of the external sector is carried over; the generalized metric $\MM_{MN}$ of the extended internal sector is given in terms of the internal metric $g_{mn}$ by (\cf \eqref{eq:genmetric})
\begin{equation}
\MM_{MN}(g_{mn}) = g^{1/2}\diag [ g_{mn}, g^{mn,kl}, g^{-1}g^{mn}, g^{-1}g_{mn,kl}]
\label{eq:genmetricSUGRA}
\end{equation}
where the determinant of the internal metric is denoted by $g=\det g_{mn}$, the four-index objects are defined by $g_{mn,kl}=g_{m[k}g_{l]n}$ and similarly for the inverse; and the components of the EFT vector potential ${\AAA_\mu}^M$ are
\begin{align}
{\AAA_\mu}^m 	&= {A_\mu}^m \, , &
\AAA_{\mu\ mn}	&= C_{\mu\ mn}  \, , &
{\AAA_\mu}^{mn}	&= \frac{1}{5!}\epsilon^{mn\ m_1\dots m_5} C_{\mu\ m_1\dots m_5} \, .
\label{eq:vecpotSUGRA}
\end{align}
The final component, $\AAA_{\mu\ m}$, is related to the dual graviton and has no appearance in the supergravity picture, see \cite{Hohm:2013uia}.

It is also possible to embed the Type II theories in ten dimensions into EFT. The Type IIA embedding follows from the above solution to the section condition by a simple reduction on a circle. In contrast, the Type IIB embedding requires a different, inequivalent solution to the section condition \cite{Hohm:2013uia}. Both Type II embeddings are presented in Appendix \ref{ch:appEFT}

We are now equipped with the tools to relate exceptional field theory to eleven-dimensional supergravity and the Type IIA and Type IIB theory in ten dimensions. This will be useful when analyzing the EFT solution we are presenting next.

\chapter{Solutions in EFT}
\label{ch:EFTsol}

In this chapter we look at solutions such as the wave and the monopole for the exceptional field theories introduced in the previous chapter. First the wave solution in the truncated $\Slf$ theory is considered which gives the membrane. By working in the larger extended space of the truncated $\Es$ theory, one can construct a monopole solution which corresponds to the fivebrane in the reduced picture. These two results mirror the DFT wave and monopole discussions. Then we move on the the full, non-truncated $\Es$ theory to present a single, self-dual solution which gives all 1/2 BPS branes in ten- and eleven-dimensional supergravity.

\section{The Wave in the truncated $SL(5)$ EFT}
\label{sec:SL5wave}
In double field theory we have seen how a null wave, \ie a momentum mode, in a dual direction of the doubled space gives a fundamental string. Now we want to demonstrate the same holds for the membrane in M-theory. It will be given by a null wave along one of the ``wrapping'' directions of an extended geometry. A simple scenario to start with is the truncated $SL(5)$ exceptional field theory where the extended space consists of four ordinary and six dual directions (here $\mu=1,\dots,4$)
\begin{equation}
	X^M = (x^\mu, y_{\mu\nu}) \, .
\end{equation}

The wave solution for this case is given by a generalized metric $\MM_{MN}$ with line element
\begin{equation}
\begin{aligned}
\dd s^2 &= \MM_{MN}\dd X^M \dd X^N \\
		&= (H-2)\left[(\dd x^1)^2 - (\dd x^2)^2 - (\dd x^3)^2 \right] 
			+ (\dd x^4)^2 \\
		&\quad + 2(H-1)\left[\dd x^1\dd y_{23} 
			+ \dd x^2\dd y_{13} - \dd x^3\dd y_{12} \right] \\
		&\quad	- H\left[(\dd y_{13})^2 + (\dd y_{12})^2 - (\dd y_{23})^2 \right]
		 	+ (\dd y_{34})^2  + (\dd y_{24})^2 - (\dd y_{14})^2.
\end{aligned}
\label{eq:SL5ppwave}
\end{equation}
This generalized metric solves the equations of motion of the $SL(5)$ theory derived in Section \ref{sec:truncatedSL5} (see Appendix \ref{sec:SL5check}). It can be interpreted as a pp-wave in the extended geometry which carries momentum in the directions dual to $x^2$ and $x^3$ \ie combinations of $y_{12}, y_{13}$ and $y_{23}$. Since it is a pp-wave it has no mass or charge and the solution is pure metric, there is no form field it couples to. As before, $H$ is a harmonic function of the transverse coordinate $x^4$: $H=1+h\ln x^4$. It is smeared in the remaining dual directions. 

A Kaluza-Klein ansatz suitable for the extended geometry here that allows us to rewrite the solution in terms of four-dimensional quantities and reducing the dual directions is
\begin{equation}
\begin{aligned}
\dd s^2 	&= \left(g_{\mu\nu} + e^{2\phi}C_{\mu\lambda\tau}g^{\lambda\tau,\rho\sigma}
			C_{\rho\sigma\nu}\right)\dd x^\mu \dd x^\nu \\
	&\quad + 2e^{2\phi}C_{\mu\lambda\tau}g^{\lambda\tau,\rho\sigma}
		\dd x^\mu \dd y_{\rho\sigma} 
	+ e^{2\phi}g^{\lambda\tau,\rho\sigma}\dd y_{\lambda\tau}\dd y_{\rho\sigma}.
\end{aligned}
\label{eq:KKforSL5}
\end{equation}
The factor $e^{2\phi}$ is a scale factor and needs to be included for consistency. This decomposition of the generalized metric into the usual metric and C-field  resembles the form of the generalized metric \eqref{eq:SL5metric} as in the DFT case.

By comparing \eqref{eq:KKforSL5} with \eqref{eq:SL5ppwave}, the fields of the reduced system with coordinates $x^\mu$ can be computed. From the diagonal terms we find
\begin{equation}
g_{\mu\nu} = \diag (-H^{-1}, H^{-1}, H^{-1}, 1) \qandq 
g^{\mu\nu,\rho\sigma} = e^{-2\phi}\diag (-H, -H, -1, H, 1, 1)
\end{equation}
and since $g^{\mu\nu,\rho\sigma}$ is given by $g^{\mu\nu}$, the inverse of $g_{\mu\nu}$, we need $e^{2\phi}=H^{-1}$ for consistency. The corresponding line element is
\begin{equation}
\dd s^2 = -H^{-1}\left[(\dd x^1)^2 - (\dd x^2)^2 - (\dd x^3)^2 \right] + (\dd x^4)^2.
\label{eq:membraneKK}
\end{equation}
The off-diagonal terms give the antisymmetric C-field whose only non-zero component is 
\begin{equation}
C_{123} = -(H^{-1}-1).
\end{equation}
This metric and C-field look like the membrane in M-theory. To complete this identification, \eqref{eq:membraneKK} has to be rescaled to be expressed in the Einstein frame.

The standard rescaling procedure (in four dimensions) gives
\begin{equation}
g_{\mu\nu} = \Omega^{-2}\tg_{\mu\nu} = H^{-3/2}\tg_{\mu\nu}
\end{equation}
where 
\begin{equation}
\Omega^2 = \sqrt{|\det e^{2\phi} g^{\mu\nu,\rho\sigma}|} = H^{3/2}.
\end{equation}
Therefore the rescaled metric reads $\tg_{\mu\nu}=H^{3/2}g_{\mu\nu}$ and the full solution in the Einstein frame is\footnote{The C-field is unaffected by the rescaling, only its field strength obtains a different factor in the action.}
\begin{equation}
\dd s^2 = -H^{-1/2}\left[(\dd x^1)^2-(\dd x^2)^2-(\dd x^3)^2 \right]+H^{3/2}(\dd x^4)^2 
%\label{eq:membrane}
\end{equation}
which is indeed the M2-brane in four dimensions in the Einstein frame. The membrane is extended in the $x^2-x^3$ plane. We have thus shown that the solution \eqref{eq:SL5ppwave} which carries momentum in the directions dual to $x^2$ and $x^3$ in the extended geometry corresponds to a membrane stretched along these directions from a reduced point of view. By similar arguments as in the string case, the mass and charge of the M2-brane are given by the momenta in the dual directions.

\section{The Monopole in the truncated $\Es$ EFT}
\label{sec:E7monopole}

So far it was not only shown how the wave in DFT gives rise to the fundamental string but also that a null wave in the truncated $\Slf$ EFT reduces to the membrane in ordinary spacetime. The same is true for the truncated $\Es$ EFT. A null wave propagating along a membrane wrapping direction gives rise to the M2-brane. 

Furthermore, due to the larger extended space, it is now also possible to consider a wave traveling in a fivebrane wrapping direction. Unsurprisingly, this reduces to the M5-brane in ordinary spacetime. We will demonstrate this explicitly and for completeness reproduce the membrane result. 

In DFT, the section condition is easily solved by reducing the coordinate dependence to half the doubled space. Thus each pair of solutions related by an $\Odd$ transformation, such as the wave and string or the monopole and fivebrane, can be presented in a straightforward fashion. In contrast in the exceptional extended geometry, the solutions to the section condition are more complex since a much larger extended space has to be dealt with. In the case of $\Es$, the section condition takes one from 56 to seven dimensions. We thus present the solutions step by step and relate them ``by hand'' rather than constructing the different solutions to the section condition explicitly.

\subsection{The M2- and M5-brane as a Wave}
\label{sec:M2M5wave}
Consider the following solution for an extended $\Es$ theory built from a seven-dimensio-nal spacetime with coordinates $X^\mu=(t,x^m,z)\rightarrow\XX^M$ with $m=1,\dots,5$ (in the coordinate system of the classic supergravity solutions given in Appendix \ref{sec:classicsol}, reduce on $x^3,x^4,x^5$ and $x^6$ and collect the remaining transverse directions $x^1,x^2$ and $y^i$ into $x^m$). The generalized metric is given by\footnote{The delta with four indices is defined as $\delta_{mn,kl}=\frac{1}{2}\left(\delta_{mk}\delta_{nl}-\delta_{ml}\delta_{nk}\right)$ and similarly for the inverse.}
\begin{equation}
\begin{aligned}
\dd s^2 &= (2-H)\left[-(\dd X^t)^2 + \delta^{mn}\dd Y_{mz} \dd Y_{nz} 
						+ \delta_{mn}\dd Z^{tm} \dd Z^{tn} - (\dd W_z)^2 \right] 
						- (\dd Y_{tz})^2\\
		&\qquad + H\left[(\dd X^z)^2 - \delta^{mn}\dd Y_{tm} \dd Y_{tn} 
						- \delta_{mn}\dd Z^{mz} \dd Z^{nz} + (\dd W_t)^2 \right]	
						+ (\dd Z^{tz})^2\\	
		&\qquad + 2(H-1)\left[\dd X^t \dd X^z - \delta^{mn}\dd Y_{tm} \dd Y_{nz} 
						+ \delta_{mn}\dd Z^{tm} \dd Z^{nz} - \dd W_t\dd W_z \right] \\	
		&\qquad + \delta_{mn} \dd X^m \dd X^n + \delta^{mn,kl}\dd Y_{mn} \dd Y_{kl} 
				- \delta_{mn,kl}\dd Z^{mn} \dd Z^{kl} - \delta^{mn} \dd W_m \dd W_n \, .
\end{aligned}
\label{eq:E7wave}
\end{equation}
This is a massless, uncharged null wave carrying momentum in the $X^z=z$ direction and $H=1+\frac{h}{|\vec{x}_{(5)}|^3}$ is a harmonic function of the transverse coordinates $x^m$. The solution is smeared over all other directions and thus there is no coordinate dependence on them. If the extra wrapping dimensions are reduced by using a Kaluza-Klein ansatz based on \eqref{eq:genmetric}, one recovers the pp-wave in M-theory in seven dimensions.

If the wave is rotated to travel in a different direction, the momentum it carries becomes the mass and charge of an extended object in the reduced picture. The different M-theory solutions obtained upon a KK-reduction of the extended wave solution pointing in various directions are summarized in Table \ref{tab:E7wave}.

\begin{table}[ht]
\begin{center}
\begin{tabular}{|c|c|}
\hline
\begin{tabular}[c]{@{}c@{}}direction of \\ propagation\end{tabular} & \begin{tabular}[c]{@{}c@{}}supergravity \\ solution\end{tabular} \\ \hline
$X\in TM$							& pp-wave 	\\
$Y\in \Lambda^2T^*M$				& M2-brane	\\
$Z\in\Lambda^5 T^*M$				& M5-brane	\\
$W\in(T^*M\otimes\Lambda^7 T^*M)$	& KK-monopole \\ 
\hline                                                    
\end{tabular}
\end{center}
\caption{The wave in exceptional extended geometry can propagate along any of the  extended directions giving the various classic solutions when seen from a supergravity perspective.}
\label{tab:E7wave}
\end{table}

The rotation that points the wave in the $Z^{tz}$ direction is achieved by the following swap of coordinate pairs in the above solution
\begin{equation}
\begin{aligned}
X^z &\longleftrightarrow Z^{tz} 	&
W_z &\longleftrightarrow Y_{tz} \\
X^{m} &\longleftrightarrow Z^{tm} &
W_{m} &\longleftrightarrow Y_{tm} \, .
\end{aligned}
\label{eq:wave2fivebrane}
\end{equation}
The rotated wave solution can now be rewritten by using a KK-ansatz based on the line element given in \eqref{eq:genmetric1} to remove the extra dimensions. This gives the M5-brane solution \eqref{eq:classicfivebrane} reduced to seven dimensions (and smeared over the reduced directions)
\begin{equation}
\begin{aligned}
\dd s^2 &= H^{1/5}\left[-\dd t + \dd\vec{x}_{(5)}^{\, 2} + H \dd z^2\right] \\
\tC_{tx^1x^2x^3x^4x^5} &= -(H^{-1}-1)  \\
H &= 1 + \frac{h}{z}  \, .
\end{aligned}
\end{equation}
The details of this calculation can be found in Appendix \ref{sec:appWave}.

It can also be shown that the wave in the $E_7$ extended theory pointing along one of the $Y$-directions gives the membrane from a reduced point of view. The key steps of this calculation are given here.

Start by splitting the transverse coordinates $x^m$ into $x^a$ and $y^i$ with $a=1,2$ and $i=1,2,3$ as before so that the extended space is given by $X^\mu=(t,x^a,y^i,z)\rightarrow\XX^M$. Then the wave can be rotated to point in the $Y_{x^1x^2}$ direction. This is achieved by the mapping
\begin{equation}
\begin{aligned}
X^z &\longleftrightarrow Y_{x^1x^2}  	&  
W_z &\longleftrightarrow Z^{x^1x^2}  \\
X^a &\longleftrightarrow  \epsilon^{ab}Y_{bz} & 
W_a &\longleftrightarrow  \epsilon_{ab}Z^{bz} \\
Y_{ij} &\longleftrightarrow \epsilon_{ijk}Z^{tk} & 
Z^{ij} &\longleftrightarrow \epsilon^{ijk}Y_{tk} 
\end{aligned}
\label{eq:wave2membrane}
\end{equation}
while leaving the remaining coordinates unaltered. The extended solution \eqref{eq:E7wave} then reads (recall that $x^1=u$ and $x^2=v$)
\begin{align}
\dd s^2 &= (2-H)\left[-(\dd X^t)^2 + \delta_{ab} \dd X^a \dd X^b 
				+ \delta^{ij}\dd Y_{iz} \dd Y_{jz} \right. \notag \\
		&\hspace{3cm} \left. 
				+ \delta_{ab}\dd Z^{ta} \dd Z^{tb} 
				+ \delta^{ij,kl}\dd Y_{ij} \dd Y_{kl} 
				- (\dd Z^{uv})^2 \right] - (\dd Y_{tz})^2 \notag \\
		&\quad + H\left[(\dd Y_{uv})^2 
				- \delta^{ab}\dd Y_{ta} \dd Y_{tb} 
				- \delta_{ij,kl}\dd Z^{ij} \dd Z^{kl} \right. \notag \\
		&\hspace{3cm} \left. 	
				- \delta^{ab} \dd W_a \dd W_b 
				- \delta_{ij}\dd Z^{iz} \dd Z^{jz}	
				+ (\dd W_t)^2 \right] + (\dd Z^{tz})^2   \label{eq:E7membrane}\\ 	
		&\quad + 2(H-1)\left[\dd X^t \dd Y_{uv} 
				- \dd X^u\dd Y_{tv} + \dd X^v\dd Y_{tu}
				- {\epsilon_{ij}}^k\dd Z^{ij} \dd Y_{kz} \right. \notag \\
		&\hspace{3cm} \left. 
				+ {\epsilon^{ij}}_k\dd Y_{ij} \dd Z^{kz} 
				+ \dd W_u \dd Z^{tv} - \dd W_v \dd Z^{tu}
				- \dd W_t\dd Z^{uv} \right]  \notag \\	
		&\quad + \delta^{ab}\dd Y_{az} \dd Y_{bz} + \delta_{ij} \dd X^i \dd X^j  
				+ (\dd X^z)^2 + \delta_{ij}\dd Z^{ti} \dd Z^{tj} 
				+ \delta^{ab}\delta^{ij} \dd Y_{ai} \dd Y_{bj}  \notag \\
		&\quad  - (\dd W_z)^2 
				- \delta^{ij}\dd Y_{ti} \dd Y_{tj}
				- \delta_{ab}\delta_{ij} \dd Z^{ai} \dd Z^{bj}  
				- \delta_{ab}\dd Z^{az} \dd Z^{bz} - \delta^{ij} \dd W_i \dd W_j \, .	 \notag		
\end{align}
The KK-ansatz to reduce this metric is based on the line element given in \eqref{eq:genmetric2} and is spelled out explicitly in the appendix in \eqref{eq:KKansatzC}. The procedure is the same as in the reduction calculation that yielded the fivebrane and gives 
\begin{equation}
\begin{aligned}
\dd s^2 &= H^{-2/5}\left[-\dd t + \dd\vec{x}_{(2)}^{\, 2} 
			+ H(\dd\vec{y}_{(3)}^{\, 2} + \dd z^2)\right]   \\
C_{tx^1x^2} &= -(H^{-1}-1)\\
H &= 1 + \frac{h}{\vec{y}_{(3)}^{\, 2} + z^2}
\end{aligned}
\end{equation}
which is the M2-brane solution reduced to seven dimensions (with the harmonic function smeared accordingly).

Hence, both the M2 and the M5 can be obtained from the same wave solution in the exceptional extended geometry and all branes in M-theory are just momentum modes of a null wave in the extended theory. The direction of the wave determines the type of brane (from the reduced perspective) or indeed gives a normal spacetime wave solution. From this point of view the duality transformations between the various solutions are just rotations in the extended space.

\subsection{The M5-brane as a Monopole}
\label{sec:M5monopole}
In Section \ref{sec:DFTmonopole} we showed that the NS5-brane of string theory was the monopole solution of DFT. In this section we want to show something similar for the M5-brane in exceptional extended geometry. 

If the KK-circle of the monopole in the $E_7$ extended theory is not along a usual spacetime direction but instead along one of the novel $Y$-directions, then this produces a smeared fivebrane solution.

First, a slightly different extended space has to be constructed. Starting from eleven dimensions and reducing on $x^3,x^4,x^5$ and $t$ allows for a construction of the monopole solution in the extended space with coordinates $X^\mu=(x^a,w,y^i,z)\rightarrow\XX^M$ (where $w=x^6$) and potential $A_i$. The generalized metric is given by
\begin{align}
\dd s^2 &= (1+H^{-2}A^2)\left[\delta^{ab}\dd Y_{az}\dd Y_{bz} 
		+ (\dd Y_{wz})^2 + H^{-2}(\dd W_z)^2\right] \notag \\
	&\quad + (1+H^{-2}A_1^2)\left[(\dd X^1)^2 + H^{-2}\delta_{ab}\dd Z^{a1}\dd Z^{b1} 
		+ H^{-2}(\dd Z^{w1})^2\right] \notag\\
	&\qquad + (1+H^{-2}A_2^2)\left[\dots\right]+ (1+H^{-2}A_3^2)\left[\dots\right] \notag\\
	&\quad + (1+H^{-2}A_1^2+H^{-2}A_2^2)
		\left[H^{-1}(\dd Y_{3z})^2 + H^{-1}(\dd Z^{12})^2\right] \notag\\
	&\qquad + (1+H^{-2}A_1^2+H^{-2}A_3^2)	\left[\dots\right]
		+ (1+H^{-2}A_2^2+H^{-2}A_3^2)\left[\dots\right] \notag\\
	&\quad + 2H^{-2}A_1A_2\left[\dd X^1 \dd X^2 - H^{-1}\dd Y_{1z}\dd Y_{2z}
		+ H^{-1}\dd Z^{13}\dd Z^{23} \right. \notag\\
	&\hspace{3cm} \left. + H^{-2}\delta_{ab}\dd Z^{a1}\dd  Z^{b2} 
		+ H^{-2}\dd Z^{w1}\dd Z^{w2}\right] \notag\\
	&\qquad + 2H^{-2}A_1A_3\left[\dots\right] + 2H^{-2}A_2A_3\left[\dots\right] \notag\\
	&\quad + 2H^{-1}A_1\left[H^{-1}(\dd X^1\dd X^z - \delta^{ab}\dd Y_{az}\dd Y_{b1}
		 - \dd Y_{wz}\dd Y_{w1})\right. \notag\\
	&\hspace{2.6cm}\left. + H^{-2}(\dd Y_{12}\dd Y_{2z} + \dd Y_{13}\dd Y_{3z}
		- \dd Z^{12}\dd Z^{2z} - \dd Z^{13}\dd Z^{3z}) \right. \notag\\
	&\hspace{2.6cm}\left. + H^{-3}(\delta_{ab}\dd Z^{a1}\dd Z^{bz}
		+ \dd Z^{w1}\dd Z^{wz} - \dd W_1\dd W_z)\right] \notag\\
	&\qquad + 2H^{-1}A_2\left[\dots\right] + 2H^{-1}A_3\left[\dots\right] \notag\\
	&\quad  + H^{-1}\left[\delta_{ab}\dd X^a\dd X^b + (\dd X^w)^2
		+ \delta^{ab} \dd Y_{aw}\dd Y_{bw} + \delta^{ab,cd}\dd Y_{ab}\dd Y_{cd}\right] \notag\\
	&\quad + H^{-2}\left[(\dd X^z)^2  + \delta^{ab}\delta^{ij}\dd Y_{ai}\dd Y_{bj}
		+ \delta^{ij} \dd Y_{wi}\dd Y_{wj}\right] \notag\\
	&\quad + H^{-3}\left[\delta^{ab}\dd W_a\dd W_b + (\dd W_w)^2  
		+ \delta_{ab} \dd Z^{aw}\dd Z^{bw} + \delta_{ab,cd}\dd Z^{ab}\dd Z^{cd} \right. \notag\\
	&\hspace{2cm} \left. + \delta^{ij,kl}\dd Y_{ij}\dd Y_{kl}
		+ \delta_{ij} \dd Z^{iz}\dd Z^{jz} \right] \notag\\
	&\quad + H^{-4}\left[\delta_{ab}\dd Z^{az}\dd Z^{bz} + (\dd Z^{wz})^2
		+ \delta^{ij} \dd W_i\dd W_{j}\right] \label{eq:E7monopole}
\end{align}
where $A^2=A_iA^i=A_1^2+A_2^2+A_3^2$. The ellipsis denotes the same terms as in the line above, with the obvious cycling through the $i$ index. The harmonic function $H$ is a function of the three $y$'s and is given by $H=1+\frac{h}{|\vec{y}_{(3)}|}$. The relation between the harmonic function and the vector potential are the same as before given in \eqref{eq:DFTAH}.

This is a monopole with the KK-circle in the $X^z=z$ direction. The solution as before may be rotated such that this ``special'' direction is of a different kind. If the KK-circle is along $Y_{wz}$, a membrane wrapping direction, the solution reduces to a M5-brane smeared along $z$. This rotation is achieved by the following map (recall that $x^1=u$ and $x^2=v$)
\begin{equation}
\begin{aligned}
X^z &\longleftrightarrow -Y_{wz} 	&
W_z &\longleftrightarrow Z^{wz} \\
X^w &\longleftrightarrow Y_{uz} &
W_w &\longleftrightarrow Z^{uz} \\
Y_{uv} &\longleftrightarrow -Y_{vz} &
Z^{uv} &\longleftrightarrow Z^{vz} \\
Y_{ui} &\longleftrightarrow -Y_{iz} &
Z^{ui} &\longleftrightarrow Z^{iz} \\
Y_{vi} &\longleftrightarrow \frac{1}{2}\epsilon_{ijk}Z^{jk} &
Z^{vi} &\longleftrightarrow  \frac{1}{2}\epsilon^{ijk}Y_{jk} \, .
\end{aligned}
\label{eq:monopole2fivebrane}
\end{equation}
Using \eqref{eq:genmetric1} to read off the fields, the exceptional extended geometry monopole reduces to the M5-brane solution 
\begin{equation}
\begin{aligned}
\dd s^2 &= H^{-3/5}[\dd\vec{x}_{(2)}^{\, 2} 
	+ H(\dd w^2 + \dd\vec{y}_{(3)}^{\, 2} + \dd z^2)]  \\
C_{izw} &= A_i  \\
H &= 1 + \frac{h}{|w^2 + \vec{y}_{(3)}^{\, 2} + z^2|^{3/2}} \, .
\end{aligned}
\end{equation}
The fivebrane is given in terms of its magnetic potential, \ie to the dual gauge potential $C_3$ given in \eqref{eq:classicfivebrane}. The full calculation is shown explicitly in Appendix \ref{sec:appMonopole}.  

We have thus demonstrated how a monopole with its KK-circle along a membrane wrapping direction is identified with a (smeared) fivebrane. This is the analogous result to the KK-monopole/NS5-brane identification in DFT shown in Section \ref{sec:DFTmonopole}.

\subsection{The Situation for the M2-brane}
In theory the same story should be true for the membrane. In the previous sections the wave was shown not only to give the membrane but also the fivebrane. From the same reasoning the monopole should not only give the fivebrane, but also the membrane. 

The problem is that this cannot be shown as simply as for the fivebrane in the $E_7$ truncated theory. To obtain the membrane from the monopole one has to consider its magnetic potential $C_6$ given in \eqref{eq:classicmembrane}. But this six-form has non-zero components with indices $C_{izx^3x^4x^5x^6}$, \ie in directions which are truncated in order to construct the exceptional extended geometry. 

More technically, if the electric $C_3$ of the membrane is dualized in seven dimensions, this gives a two-form. This means that only some part of the above six-form lives in the seven-space that gets extended, the remainder lives in the other four directions. Thus it is not possible to describe the membrane this way and stay in the truncated space. This is simply a problem with the tools at our disposal, \ie the truncated version of the $E_7$ exceptional field theory. By looking at all the relations we have built between the solutions in the extended space, it seems natural that a monopole with its KK-circle in a fivebrane wrapping direction gives a membrane. This problem then is demanding the full non-truncated EFT \cite{Hohm:2013uia} and we will turn to this next.

% \newpage
\section{A Self-dual Solution in the full $E_7$ EFT}
\label{sec:selfdual}

In order to construct a solution which has both wave and monopole aspects and thus can be simultaneously related to the membrane and the fivebrane requires the full non-truncated EFT outlined in \ref{sec:fullEFT}. The coordinates of the $4+56$-dimensional theory are $(x^\mu,Y^M)$, the field content was given in \eqref{eq:E7fieldcontent} and the crucial twisted self-duality relation is equation \eqref{eq:selfduality}.

Now consider the following set of fields. We take the external sector with coordinates $x^\mu$ to be four-dimensional spacetime with one timelike direction $t$ and three spacelike directions $w^i$ with $i=1,2,3$. The external metric is that of a point-like object, given in terms of a harmonic function of the transverse coordinates by
\begin{equation}
g_{\mu\nu} = \diag [-H^{-1/2},H^{1/2}\delta_{ij}]\, , \qquad H(r) = 1 + \frac{h}{r}
\label{eq:exmetric}
\end{equation}
where $r^2=\delta_{ij}w^iw^j$ and $h$ is some constant (which will be interpreted later). 

The 56-dimensional extended internal sector uses the coordinates $Y^M$ given in \eqref{eq:gencoordM}. The EFT vector potential ${\AAA_\mu}^M$ of our solution has ``electric'' and ``magnetic'' components (from the four-dimensional spacetime perspective) that are given respectively by
\begin{equation}
{\AAA_t}^M 	= \frac{H-1}{H} a^M \qquad\mathrm{and}\qquad
{\AAA_i}^M	= A_i \ta^M \, ,
\label{eq:vecpot}
\end{equation}
where $A_i$ is a potential of the magnetic field. The magnetic potential obeys a BPS-like condition where its curl is given by the gradient of the harmonic function that appears in the metric
\begin{equation}
\vec{\nabla}\times\vec{A} = \vec{\nabla} H 
\qquad\mathrm{or}\qquad
\partial_{[i}A_{j]} = \frac{1}{2}{\epsilon_{ij}}^k\partial_k H \, .
\label{eq:AH}
\end{equation}
The index $M$ in ${\AAA_\mu}^M$ labels the 56 vectors, only two of which are non-zero for our solution. The vector $a^M$ in the extended space (a scalar form a spacetime point of view) points in one of the 56 extended directions. Later we will interpret this direction as the direction of propagation of a wave or momentum mode. The dual vector $\ta^M$ denotes the direction dual to $a^M$ given approximately by $a^M \sim \Omega^{MN}\MM_{NK}\ta^K$. This sense of duality between directions of the extended space will be formalized in Section \ref{sec:selfduality}.

Using the relation between $H$ and $A_i$, one can immediately check that ${\AAA_\mu}^M$ satisfies the twisted self-duality equation \eqref{eq:selfduality}. Loosely speaking, the duality on the external spacetime via $\epsilon_{\mu\nu\rho\sigma}$ exchanges electric ${\AAA_t}^M$ and magnetic ${\AAA_i}^M$ components of the potential. The symplectic form $\Omega_{MN}$ acts on the extended internal space and swaps $a^M$ with its dual $\ta^M$. If one goes through the calculation carefully, one sees that minus signs and factors of powers of $H$ only work out if both actions on the external and extended internal sector are carried out simultaneously. We will show this explicitly in Section \ref{sec:selfduality}.

The generalized metric of the extended internal sector, $\MM_{MN}$, is a diagonal matrix with just four different entries, $\{ H^{3/2},H^{1/2},H^{-1/2},H^{-3/2}\}$. The first and last one appear once each, the other two appear 27 times each. The precise order of the 56 entries of course depends on a coordinate choice, but once this is fixed it characterizes the solution together with the choice of direction for $a^M$.

For definiteness, let's fix the coordinate system and pick a direction for $a^M$ which we call $z$, i.e. $a^M=\delta^{Mz}$. The dual direction is denoted by $\tz$ and we have $\ta^M=\delta^{M\tz}$. Then $\MM_{zz}=H^{3/2}$ and $\MM_{\tz\tz}=H^{-3/2}$. For completeness, the full expression for the generalized metric for the coordinates in \eqref{eq:gencoordM} is\footnote{Here $\delta_n$ denotes an $n$-dimensional Kronecker delta.}
\begin{equation}
\MM_{MN} = \diag[ H^{3/2}, H^{1/2}\delta_6, H^{-1/2}\delta_6, H^{1/2}\delta_{15},
				H^{-3/2}, H^{-1/2}\delta_6, H^{1/2}\delta_6, H^{-1/2}\delta_{15} ] \, .
\label{eq:SDSgenmetric}
\end{equation}
The second 28 components are the inverse of the first 28 components, reflecting the split of the EFT vector ${\AAA_\mu}^M$ into 28 ``electric'' and 28 ``magnetic'' components.

To get the fields for any other direction, one simply has to perform a rotation in the extended space which is a duality transformation. The rotation matrix $R \in E_7$ rotates $a^M$ in the desired direction ${a'}^M$ and at the same time transforms $\MM_{MN}$ according to
\begin{align}
{a'}^M &= {R^M}_N a^N \, , & 
\MM'_{MN} &= {R_M}^K\MM_{KL}{R^L}_N \, .
\label{eq:rot}
\end{align}
Since the action and the self-duality equation is invariant under such a transformation, the fields can freely be rotated in the extended space.

The remaining fields of the theory, namely the two-form gauge fields $B_{\mu\nu\ \alpha}$ and $B_{\mu\nu\ M}$, are trivial. Also the external part of the three-form potential, $C_{\mu\nu\rho}$, vanishes for our solution. This will eventually restrict somewhat the possible supergravity solutions obtained from this EFT solution. Dropping these restrictions would be interesting and it would provide a technical challenge to repeat this section and include other fluxes such as those on the external space.

To recap, the fields $g_{\mu\nu}$, ${\AAA_\mu}^M$ and $\MM_{MN}$ as given in equations \eqref{eq:exmetric}, \eqref{eq:vecpot} and \eqref{eq:SDSgenmetric} (together with \eqref{eq:AH}) form our solution to EFT. They satisfy the self-duality equation and their respective equations of motion.

Note that all fields directly or indirectly depend on the harmonic function $H$ which in turn only depends on the external transverse coordinates $w^i$. There is thus no coordinate dependence on any of the internal or extended coordinates.  This solution therefore is de-localized and smeared over all the internal extended directions. It is an interesting open question to look at solutions localized in the extended space. In theory, EFT allows for coordinate dependencies on \emph{all} coordinates, even the extended ones. We leave this for future work.

\subsection{Interpreting the Solution}

How do we interpret this solution in exceptional field theory? Before we do this let us return to how solutions in the truncated theory may be interpreted. A wave whose momentum is in a {\it{winding}} direction describes a brane associated with that winding direction, \eg a wave with momentum along $y_{12}$ describes a membrane extended over the $y^1,y^2$ directions. A monopole-like solution --- by which we mean a Hopf fibration --- where the $S^1$ fibre is a winding direction describes the S-dual brane to that winding direction, \eg if the fibre of the monopole is $y_{12}$ then the solution describes a fivebrane. Thus in the extended (but truncated) theory branes can have either a description as monopole or as a wave. These statements were the conclusions of \cite{Berkeley:2014nza,Berman:2014jsa} and presented in the chapters so far. 

Now because of the truncation it was not possible to describe a given solution in both ways within the same description of spacetime. The key point of EFT is that there is no truncation and so such things are possible. The self-duality relation is simply the Kaluza-Klein description of a solution that has both momentum and non-trivial Hopf fibration, \ie it is simultaneously electric and magnetic from the point of view of the KK-gravi-photon. These are not just solutions to some linear abelian theory but full solutions to the gravitational theory (or in fact EFT). As such they are exact self-dual solutions to the non-linear theory though are charged with respect to some $U(1)$ symmetry that is given by the existence of the $S^1$ in extended space. Our intuition should be shaped by this experience with Kaluza-Klein theory and the solution thought of as simultaneously a wave and a monopole whose charge is equal to the wave's momentum.

Let us look at the moduli of the solution. The solution is specified by two pieces of data, the vector $a^M$ and the constant $h$ that appears in the harmonic function. The vector specifies the direction the wave is propagating in. That is, it gives the direction along which there is momentum. The constant $h$ in the harmonic function of the solution is then proportional to the amount of momentum carried. 

In addition, the solution comes with a monopole-like structure, whose fibre is in the direction dual to the direction of propagation of the wave and whose base is in the external spacetime. In the case of the smeared solution studied in this chapter this fibration may be classified by its first Chern class which is $h$. (See Section \ref{sec:localization}  for a discussion of the localized non-smeared solution.)

To give a non-trivial first Chern class the fibre must be an $S^1$ and then the {\it{magnetic charge}} $h$ is integral. This is essentially Dirac quantization but now our theory also requires self-duality which in turn implies that the momentum in the dual direction to the fibre is quantized. The presence of quantized momentum in this direction then implies that this direction itself must also be an $S^1$. Let us examine this quantitatively.

The electric charge of the solution is related to the radius of the circle by
\begin{equation}
q_e = \frac{n}{R_e}  \qquad  {\rm{with}} \, \, n \in \mathbb{Z}
\end{equation}
and the magnetic charge is related to the radius of the fibre by
\begin{equation}
q_m = m  R_m  \qquad  {\rm{with}} \, \, m \in \mathbb{Z}   \, .
\end{equation}
Now the twisted self-duality relation implies 
\begin{equation}
q_e=q_m \qquad \implies \qquad n/m =  R_e R_m \, .
\end{equation}
From examining the norms of the $E_7$ vectors that specify the solution we can determine $R_e$ and $R_m$ as
\begin{equation}
R_e=|a^M| \qandq R_m=|\ta^M| \, .
\end{equation}
Calculating these norms using the metric of the solution (see the next section) then reveals 
\begin{equation}
R_e=H^{3/4}   \, , \quad R_m= H^{-3/4}  \qquad {\rm{thus}} \qquad 
R_e R_m =1   \quad {\rm{and}} \quad  n=m.
\end{equation}
So the $E_7$ related radii are duals and the electric and magnetic quantum numbers are equal. Note that the harmonic function $H$ (and thus the radii) is a function of $r$, the radial coordinate of the external spacetime. This will lead to interesting insights when we analyze the solution close to its core or far away from it in Section \ref{sec:singularities}.

The actual direction $a^M$ that one chooses determines how one interprets the solution in terms of the various usual supergravity descriptions. That is we can interpret this single solution in terms of the brane solutions in eleven dimensions or the Type IIA and Type IIB brane solutions in ten dimensions. We will show this in detail in Section \ref{sec:reduction}.

Finally let us a add a comment about the topological nature of these solutions. The more mathematically minded reader will note that brane solutions like the NS5-brane are not classified by the first Chern class which in cohomology terms is given by $H^2(M;\mathbb{Z})$ but instead by the Dixmier-Douady class, \ie $H^3(M;\mathbb{Z})$. For the smeared solution these two are related since $H^3(S^2\times S^1;\mathbb{Z}) = H^2(S^2;\mathbb{Z})\times H^1(S^1;\mathbb{Z})$. Thus for the smeared branes there is no issue. The question of the global structure of the localized solutions where one has a genuine $H^3$ is however an important open question that has recently received some attention \cite{Papadopoulos:2014mxa, Papadopoulos:2014ifa, Hull:2014mxa}.

\subsection{Twisted Self-duality }
\label{sec:selfduality}
The EFT gauge potential ${\AAA_\mu}^M$ presented above satisfies the twisted self-duality equation \eqref{eq:selfduality}. This can be checked explicitly by looking at the components of the equation and making use of the relation between the harmonic function $H$ and the spacetime vector potential $A_i$ given in \eqref{eq:AH}.

First though, we will look at the relation between the two vectors $a^M$ and $\ta^M$ that define the directions of ${\AAA_t}^M$ and ${\AAA_i}^M$. The duality relation between them can be made precise by normalizing the vectors using the generalized metric $\MM_{MN}$. The unit vectors
\begin{equation}
\ha^M = \frac{a^M}{|a|} = \frac{a^M}{\sqrt{a^Ka^L\MM_{KL}}} \qandq
\hta^M = \frac{\ta^M}{|\ta|} = \frac{\ta^M}{\sqrt{\ta^K\ta^L\MM_{KL}}}
\end{equation}
are related via the symplectic form $\Omega$ by
\begin{equation}
\ha^M = \Omega^{MN}\MM_{NK}\hta^K\, , \qquad 
\Omega_{MN}=\begin{pmatrix} 0 & 1 \\ -1 & 0 \end{pmatrix} \, .
\label{eq:vectorduality}
\end{equation}
If the vectors are not normalized the metric in the duality relation introduces extra factors. For the specific directions given above we have $\ha^M=H^{-3/4}\delta^{Mz}$ and $\hta^M=H^{3/4}\delta^{M\tz}$ which indeed satisfy \eqref{eq:vectorduality} for the $\MM_{MN}$ given in \eqref{eq:SDSgenmetric}.

Let's now turn to the self-duality of the field strength. We begin by computing the field strength ${\FF_{\mu\nu}}^M$ of ${\AAA_\mu}^M$ as given in \eqref{eq:genfieldstrength}, recalling the simplifications our solution provides. There are two components which read
\begin{equation}
\begin{aligned}
{\FF_{it}}^M &= 2\partial_{[i}{\AAA_{t]}}^M 
	= -\partial_i(H^{-1}-1)a^M = H^{-2}\partial_i H \delta^{Mz} \\
{\FF_{ij}}^M &= 2\partial_{[i}{\AAA_{j]}}^M 
	= 2\partial_{[i}A_{j]}\ta^M ={\epsilon_{ij}}^k\partial_k H \delta^{M\tz} \, .
\end{aligned}
\label{eq:F}
\end{equation}
The spacetime metric $g_{\mu\nu}$ is given in \eqref{eq:exmetric} and has determinant $e^2=|\det g_{\mu\nu}|=H$. This can be used to rewrite the self-duality equation \eqref{eq:selfduality} as
\begin{align}
{\FF_{\mu\nu}}^M &= \frac{1}{2}H^{1/2} \epsilon_{\mu\nu\rho\sigma}  g^{\rho\lambda}g^{\sigma\tau}\Omega^{MN} \MM_{NK}{\FF_{\lambda\tau}}^K
\end{align}
where the spacetime metric is used to lower the indices on $\FF^M$. Now we can look at the components of the equation. Starting with
\begin{equation}
{\FF_{ij}}^M = H^{1/2} \epsilon_{ijkt}  g^{kl}g^{tt}\Omega^{MN} \MM_{NK}{\FF_{lt}}^K
\end{equation}
and inserting for the spacetime metric and the field strength gives
\begin{align}
{\FF_{ij}}^M &= -H^{1/2} \epsilon_{tijk}H^{-1/2}\delta^{kl}(-H^{1/2})
		\Omega^{MN} \MM_{NK}H^{-2}\partial_l H a^K \notag\\
	&= H^{-3/2}({\epsilon_{ij}}^k\partial_k H) \Omega^{MN} \MM_{NK} \delta^{Kz}
\end{align}
where the extra minus sign in the first line comes from permuting the indices on the four-dimensional epsilon which is then turned into a three-dimensional one. In the next step we make use of \eqref{eq:AH} and the components of $\Omega$ and $\MM$ that are picked out by the summation over indices are substituted
\begin{align}
{\FF_{ij}}^M &= H^{-3/2}2\partial_{[i}A_{j]}\Omega^{Mz} \MM_{zz} \notag \\
	&= H^{-3/2}2\partial_{[i}A_{j]}\delta^{M\tz}H^{3/2} 
	= 2\partial_{[i}A_{j]}\ta^M 
\end{align}
and we obtained the expected result. Similarly, the other component of the self-duality equation reads
\begin{equation}
{\FF_{it}}^M = \frac{1}{2}H^{1/2} \epsilon_{itjk} g^{jp}g^{lq} 
		\Omega^{MN} \MM_{NK}{\FF_{pq}}^K \, .
\end{equation}
Going through the same steps as before leads to 
\begin{align}
{\FF_{it}}^M &= -\frac{1}{2}H^{1/2} \epsilon_{tijk}H^{-1/2}\delta^{kp}H^{-1/2}\delta^{lq}
		\Omega^{MN} \MM_{NK}2\partial_{[p}A_{q]}\ta^K \notag\\
	&= -H^{-1/2}({\epsilon_i}^{jk}\partial_j A_k) \Omega^{MN} \MM_{NK} \delta^{K\tz} \, .
\end{align}
Again substituting for $\Omega$ and $\MM$ gives the expected result	
\begin{align}
{\FF_{it}}^M 	&= -H^{-1/2}\partial_i H\Omega^{M\tz} \MM_{\tz\tz} \notag \\
	&= -H^{-1/2}\partial_i H(-\delta^{Mz})H^{-3/2} = H^{-2}\partial_i Ha^M 
\end{align}
to match with \eqref{eq:F}.

Thus the components of the field strength of the EFT vector ${\AAA_\mu}^M$ given in \eqref{eq:F} satisfy the self-duality condition. It is also possible to satisfy an anti-self-duality equation.  If the magnetic charge of our solution is taken to be minus the electric charge, this has the effect of modifying the magnetic component of the EFT vector by an extra minus sign, ${\AAA_i}^M=-A_i\ta^M$. The above calculation then works exactly the same but the extra minus sign ensures that the field strength is anti-self-dual. This choice would then be consistent with the original EFT paper \cite{Hohm:2013uia} (of course the choice of self-dual or anti-self-dual is ultimately related to how supersymmetry is represented).

\subsection{Reductions to Supergravity Solutions}
\label{sec:reduction}

The self-dual EFT solution presented in the previous section gives rise to the full spectrum of 1/2 BPS branes in eleven-dimensional supergravity and the Type IIA and Type IIB theories in ten dimensions. We will now show how applying the appropriate solution to the section condition and rotating our solution in a specific direction of the exceptional extended space leads to the wave solution, the fundamental, solitonic and Dirichlet p-branes, the KK-branes which are extended monopoles, and an example of an intersecting brane solution. All these extracted solutions together with their Kaluza-Klein decomposition can be found in Appendix \ref{ch:glossary} for easy referral.

\subsubsection{Supergravity Solutions in Eleven Dimensions}
We start by looking at the EFT solution from an eleven-dimensional supergravity point of view. Using the results of Section \ref{sec:embedding} in reverse, the supergravity fields can be extracted from the EFT solution. Recall that the resulting supergravity fields will be rearranged according to a $4+7$ Kaluza-Klein coordinate split. 

First, the extended coordinates $Y^M$ are decomposed into $y^m$, $y_{mn}$ and so on as given in \eqref{eq:gencoordM}. Then by comparing the expression for the generalized metric of the internal extended space, $\MM_{MN}$, of our solution in \eqref{eq:SDSgenmetric} to \eqref{eq:genmetricSUGRA}, one can work out the seven-dimensional internal metric $g_{mn}$. The components of the EFT vector potential ${\AAA_\mu}^M$ given in \eqref{eq:vecpot} can be related to the KK-vector of the decomposition and the $C_3$ and $C_6$ form fields respectively according to \eqref{eq:vecpotSUGRA}. Finally, the external spacetime metric $g_{\mu\nu}$ in \eqref{eq:exmetric} is simply carried over to the 4-sector of the KK-decomposition. 

As mentioned before, the EFT solution is characterized by the direction of the vector $a^M$ and a corresponding ordering in the diagonal entries of $\MM_{MN}$. If the procedure of extracting a supergravity solution just described is applied to the EFT solution as presented in Section \ref{sec:selfdual}, \ie with the direction of the $y^m$-type, $a^M=\delta^{Mz}$ where we now identify $z$ with $y^1$, the first of the ordinary $y^m$ directions, the \emph{pp-wave} solution of supergravity can be extracted. From $\MM_{MN}$, the internal metric is given by
\begin{equation}
g_{mn} = \diag [H, \delta_6] 
\label{eq:intmetricWM}
\end{equation}
where $\delta_6$ is a Kronecker delta of dimension six. These are the remaining six directions of $y^m$. The ``electric'' part of the EFT vector, ${\AAA_t}^z = -(H^{-1}-1)$, becomes the cross-term in the supergravity metric. The ``magnetic'' part ${\AAA_i}^{\tz}=A_i$ is like a dual graviton and does not appear in the supergravity picture. Note that the dual direction to $z$ is $y_1=\tz$. See Appendix \ref{sec:SUGRAsolutions} for the supergravity wave decomposed under a $4+7$ split. Since our self-dual EFT solution is interpreted as a wave now propagating in the ordinary direction $y^1=z$, it is not too surprising to recover the supergravity wave once the extra exceptional aspects are removed.

As shown in previous chapters and \cite{Berkeley:2014nza,Berman:2014jsa}, we know that a wave in an exceptional extended geometry can also propagate along the novel dimensions such as $y_{mn}$ or $y^{mn}$. If our solution is rotated to propagate in those directions, \eg $a^M={\delta^M}_{12}$ or $a^M=\delta^{M\ 67}$, the \emph{membrane} and \emph{fivebrane} solutions of supergravity are recovered. For the former, the membrane is stretched along $y^1$ and $y^2$, for the latter, the fivebrane is stretched along the complimentary directions to $y^6$ and $y^7$, \ie $y^1,y^2,y^3,y^4$ and $y^5$. This result is obtained by an accompanying rotation of the generalized metric according to \eqref{eq:rot} and extracting the internal metrics for the M2 and the M5 (\cf Appendix \ref{sec:SUGRAsolutions})
\begin{align}
g_{mn} &= H^{1/3}\diag [H^{-1}\delta_2, \delta_5] \, , &
g_{mn} &= H^{2/3}\diag [H^{-1}\delta_5, \delta_2] \, .
\label{eq:intmetricM2M5}
\end{align}
The masses and charges of the branes are provided by the momentum in the extended directions. The electric potential is given by ${\AAA_t}^M$ which encodes the $C_3$ for the M2 and the $C_6$ for the M5. The magnetic potential is given by ${\AAA_i}^M$ which gives their duals, \ie the $C_6$ for the M2 and the $C_3$ for the M5. We will explain this procedure of obtaining the membrane and fivebrane from the EFT solution in more detail below.

In Section \ref{sec:E7monopole} we have previously hinted at the idea that the wave in EFT along $y_m$, the fourth possible direction, should correspond to a monopole-like solution in supergravity. Since we are now working with a self-dual solution, we can show that this is indeed the case. If the direction of $a^M$ is of the $y_m$-type, \eg $a^M=\delta^{M\tz}$, and thus $\ta^M$ along $y^m$ (essentially swapping $a^M$ and $\ta^M$ of the pp-wave), the \emph{KK-monopole} is obtained. Again performing the corresponding rotation of the generalized metric, the internal metric can be extracted
\begin{equation}
g_{mn} = \diag [H^{-1}, \delta_6] \, .
\label{eq:intmetricKKM}
\end{equation}
The ``magnetic'' part of the EFT vector, ${\AAA_i}^z = A_i$, becomes part of the KK-monopole metric in supergravity. The ``electric'' part ${\AAA_t}^{\tz} = -(H^{-1}-1)$ now has the nature of a dual graviton and does not contribute in the supergravity picture. This is the opposite scenario to the pp-wave described above, underlining the electric-magnetic duality of these two solutions.

The four supergravity solutions we have extracted from our EFT solution all have the same external spacetime metric $g_{\mu\nu}$ under the KK-decomposition,
\begin{equation}
g_{\mu\nu} = \diag [-H^{-1/2},H^{1/2}\delta_{ij}]
\end{equation}
which has the character of a point-like object (in four dimensions). The four solutions only differ in the internal metric $g_{mn}$, the KK-vector of the decomposition and of course the C-fields. But these elements are just rearranged in ${\AAA_\mu}^M$ and $\MM_{MN}(g_{mn})$ and are all the same in EFT, up to an $\Es$ rotation of the direction $a^M$ of the solution.

\subsubsection{From Wave to Membrane}
Let's pause here briefly and take a closer look at a specific example of such a rotation. The EFT solution presented in Section \ref{sec:selfdual} with the choice for the vector $a^M$ given there and the generalized metric $\MM_{MN}$ in \eqref{eq:SDSgenmetric} for a fixed coordinate system directly reduces to the pp-wave in eleven dimensions. 

We now want to demonstrate how this can also give the M2-brane at the same time by simply picking a different duality frame, that is choosing a different section of the extended space to give the physical spacetime. This new duality frame is obtained by rotating the fields of the solution according to \eqref{eq:rot}. 

As explained above, if the EFT solution is propagating along a $y_{mn}$ direction, say $y_{12}$, it gives the membrane. Thus the vector $a^M=\delta^{M1}$ has to be rotated into ${a'}^M={\delta^M}_{12}$. This has the effect of exchanging $y^1$ with $y_{12}$ and their corresponding components in the metric, \ie $\MM_{1\,1} \leftrightarrow \MM^{12\, 12}$. This should not come as a surprise since here momentum and winding directions are exchanged which is exactly what is expected in relating the wave and the membrane via duality.

Besides $y^1 \leftrightarrow y_{12}$, the frame change also swaps the following pairs of coordinates and the corresponding components of the metric (here the index $a$ takes the values 3 to 7)
\begin{equation}
\begin{aligned}
y^2 &\leftrightarrow y_2 \, , &
y_{ab} &\leftrightarrow y^{ab} \, , &
y_{1a} &\leftrightarrow y^{1a} \, .
\end{aligned}
\label{eq:pp-M2}
\end{equation}
These are simple exchanges between dual pairs of coordinates that reflect the new duality frame.

After the rotation, the generalized metric reads (still in the coordinate system given by \eqref{eq:gencoordM})
\begin{equation}
\begin{aligned}
\MM_{MN} = \diag[&H^{-1/2}\delta_2, H^{1/2}\delta_5, H^{3/2}, H^{1/2}\delta_{10}, H^{-1/2}\delta_{10}, \\
&H^{1/2}\delta_2, H^{-1/2}\delta_5, H^{-3/2}, H^{-1/2}\delta_{10}, H^{1/2}\delta_{10}] \, .
\end{aligned}
\label{eq:genmetricM2}
\end{equation}
This can now be compared to \eqref{eq:genmetricSUGRA} to read off the internal metric in the reduced, eleven-dimensional picture as described above, and gives \eqref{eq:intmetricM2M5}, the M2-brane stretched along $y^1$ and $y^2$. Similar rotation procedures can be applied to relate the pp-wave or the membrane to the fivebrane and the monopole.

Our self-dual EFT wave solution with attached monopole-structure thus unifies the four classic eleven-dimensional supergravity solutions and provides the so-far missing link in the duality web of exceptionally extended solutions.

\subsubsection{The Membrane / Fivebrane Bound State}
The self-dual EFT solution does not only give the standard 1/2 BPS branes of supergravity but also bound states. Such solutions were first mentioned in \cite{Gueven:1992hh} and then interpreted by Papadopoulos and Townsend in \cite{Papadopoulos:1996uq}. As an illustrative example we will show how the dyonic M2/M5-brane solution of \cite{Izquierdo:1995ms} can be obtained from our EFT solution. 

Before we find this bound state of a membrane and a fivebrane, it is useful to see how to pick a duality frame such that the EFT solution reduces to the (pure) fivebrane. Above we have just seen how to rotate the frame to get the (pure) membrane instead of the wave. If we rotate further to have the solution propagate in the $y^{67}$ direction, \ie ${a''}^M=\delta^{M\, 67}$, then we get the fivebrane. 

Starting from the membrane frame of the previous subsection with ${a'}^M={\delta^M}_{12}$ and the generalized metric in \eqref{eq:genmetricM2}, the new frame rotation exchanges the membrane direction $y_{12}$ with the fivebrane direction $y^{67}$ and their corresponding components in the metric, $\MM^{12\, 12} \leftrightarrow \MM_{67\, 67}$. Again it is very natural to exchange a membrane coordinate $y_{mn}$ with a fivebrane coordinate $y^{mn}$ in this kind of duality transformation.

\begin{table}[h]
\centering
\begin{tabular}{|l|cccc|ccccccc|} \hline
sector		& \multicolumn{4}{c|}{external} & \multicolumn{7}{c|}{internal} \\ \hline
\multirow{2}{*}{coordinate}	 &      & \multicolumn{3}{|c|}{$w^i$} & \multicolumn{2}{c}{$y^a$} & \multicolumn{3}{|c|}{$y^A$} & \multicolumn{2}{c|}{$y^\alpha$} \\
    & \multicolumn{1}{l|}{$t$} & $w^1$ & $w^2$ & $w^3$ & $y^1$ & $y^2$ & \multicolumn{1}{|l}{$y^3$} &  $y^4$  & \multicolumn{1}{l|}{$y^5$}   & $y^6$   & $y^7$     \\ \hline
membrane    & $\circ$   &  -    &  -    &  -    & $\circ$     & $\circ$     & -      & -   &   - & -   & - \\
fivebrane    & $\circ$   & -     &  -    &  -    & $\circ$     & $\circ$  & $\circ$   &  $\circ$  & $\circ$  & -   & -    \\ \hline
\end{tabular}
\caption{The coordinates of the membrane and fivebrane are either in the external or internal sector of the KK-decomposition. A circle $\circ$ denotes a worldvolume direction of the brane while a dash $-$ indicates a transverse direction.}
\label{tab:coords}
\end{table}
In what follows, it will be useful to split the index of the coordinate $y^m$ into $m=(a,A,\alpha)$. Here $y^a$ with $a=1,2$ are the two worldvolume directions of the membrane or two of the worldvolume directions of the fivebrane. The $y^A$ with $A=3,4,5$ are the remaining three worldvolume directions of the fivebrane, they are transverse directions for the membrane. And finally the $y^\alpha$ with $\alpha=6,7$ are transverse directions to both the membrane and fivebrane. Table \ref{tab:coords} shows the worldvolume (a circle $\circ$) and transverse (a dash $-$) directions of the membrane and fivebrane together with the coordinate labels and sector under the KK-decomposition.

Besides $y_{ab}=y_{12} \leftrightarrow y^{\alpha\beta}=y^{67}$, the frame change also swaps some other pairs of coordinates. This can now be neatly written as 
\begin{equation}
\begin{aligned}
y^{ab} &\leftrightarrow y_{\alpha\beta} \, , &
y_{a\alpha} &\leftrightarrow y^{a\alpha} \, , \\
y^A &\leftrightarrow y_{BC} \, , &
y_A &\leftrightarrow y^{BC} \, .
\end{aligned}
\label{eq:M2-M5}
\end{equation}
The first line contains further exchanges between membrane directions and fivebrane directions. The second line is a result of going from the membrane frame to the fivebrane frame. Once the corresponding components of the metric have been exchanged as well, it reads
\begin{equation}
\begin{aligned}
\MM_{MN} = \diag[&H^{-1/2}\delta_5, H^{1/2}\delta_2, H^{1/2}\delta_{10}, H^{-1/2}\delta_{10}, H^{-3/2}, \\
&H^{1/2}\delta_5, H^{-1/2}\delta_2, H^{-1/2}\delta_{10}, H^{1/2}\delta_{10},  H^{3/2}]\, .
\end{aligned}
\label{eq:genmetricM5}
\end{equation}
which can be reduced to give the internal fivebrane metric \eqref{eq:intmetricM2M5}. Inserting the rotated vector ${a''}^M$ into \eqref{eq:vecpot} then gives the corresponding C-form field as explained above.

Now that it is clear how to obtain both the M2-brane and the M5-brane from our self-dual EFT solution, we can attempt to obtain the dyonic M2/M5 bound state of \cite{Izquierdo:1995ms}. To achieve this, we will again start from the membrane duality frame. This time though, we do not rotate the frame all the way into the fivebrane frame but introduce a parameter $\xi$ which interpolates between a purely electric M2-brane and a purely magnetic M5-brane. For $\xi=0$ the transformation gives the fivebrane whereas for $\xi=\pi/2$ the membrane is recovered\footnote{This choice of $\xi$ -- and not one shifted by $\pi/2$ -- might be counter-intuitive but has been made to match the $\xi$ in \cite{Izquierdo:1995ms}.}. Therefore a vector of the form
\begin{equation}
a^M_{\mathrm{(M2/M5)}} = \sin\xi \, a^M_{\mathrm{(M2)}} + \cos\xi \, a^M_{\mathrm{(M5)}} \\
\label{eq:aM2M5}
\end{equation}
points the EFT solution in the direction which gives the M2/M5-brane (here $a^M_{\mathrm{(M2)}}={\delta^M}_{12}$ and $a^M_{\mathrm{(M5)}}=\delta^{M\, 67}$ from above). If this vector is inserted into the EFT vector potential, one obtains both the $C_3$ and the $C_6$ (together with their duals) of the membrane and fivebrane, each modulated by $\sin\xi$ or $\cos\xi$. Since we are dealing with a dyonic solution, both an electric and a magnetic potential are expected. 

Having found the new EFT vector, the above rotation now needs to be applied to the generalized metric. Comparing the metric for the M2 in \eqref{eq:genmetricM2} and the M5 in \eqref{eq:genmetricM5}, one finds that the components which get exchanged in \eqref{eq:M2-M5} differ by a factor of $H$. In most cases $H^{1/2}$ becomes $H^{-1/2}$ or vice versa, \eg $\MM_{A\, B} = H^{1/2}\delta_{AB}$ and $\MM^{AB\, CD} = H^{-1/2}\delta^{AB,CD}$ are exchanged. The only exceptions are for the $ab=12$ and $\alpha\beta=67$ components where $H^{\pm 3/2}$ becomes $H^{\pm 1/2}$. The partial, $\xi$-dependent rotation now introduces factors of $\sin\xi$ and $\cos\xi$ into the metric components and generates off-diagonal entries. To see how they arise, one has to consider the effect of the rotation on the coordinates.

The coordinate pairs which get rotated into each other are the same as in \eqref{eq:M2-M5}, but now superpositions are formed instead of exchanging them completely. One can think of each pair as a 2-vector acted on by 
\begin{equation}
R_2=\begin{pmatrix}\sin\xi & \cos\xi \\ -\cos\xi & \sin\xi \end{pmatrix}
\end{equation}
which is the $2\times 2$ submatrix of the full rotation matrix $R$ in \eqref{eq:rot}. Then the new coordinate pair is schematically\footnote{More formally, one can introduce epsilon symbols so that the index structure works out, \eg ${y'}^A = \sin\xi\, y^A + \cos\xi\ \epsilon^{ABC}y_{BC}$.} given by, for example 
\begin{equation}
\begin{pmatrix}
{y'}^A \\ y'_{BC}
\end{pmatrix} =
R_2 \begin{pmatrix}
y^A \\ y_{BC}
\end{pmatrix} =  
\begin{pmatrix}
\sin\xi\ y^A + \cos\xi\ y_{BC} \\ 
\sin\xi\ y_{BC} - \cos\xi\ y^A
\end{pmatrix} \, .
\label{eq:rotationbyxi}
\end{equation}
The other coordinate pairs which are acted on by copies of $R_2$ are
\begin{equation}
\begin{pmatrix}
y_{ab} \\ y^{\alpha\beta}
\end{pmatrix} , \ 
\begin{pmatrix}
y^{ab} \\ y_{\alpha\beta}
\end{pmatrix} , \
\begin{pmatrix}
y_{a\alpha} \\ y^{b\beta}
\end{pmatrix} \quad\mathrm{and}\quad
\begin{pmatrix}
y_A \\ y^{BC}
\end{pmatrix} .
\label{eq:coordsM2M5}
\end{equation}
These rotations have quite non-trivial consequences for the corresponding components of the generalized metric. Conjugating the $2\times 2$ blocks of the metric with $R_2$ gives for our example
\begin{equation}
\begin{pmatrix}
\MM'_{A\, B} & {\MM'_A}^{EF} \\ {{\MM'}^{CD}}_{B} & {\MM'}^{CD\, EF}
\end{pmatrix} = 
R_2
\begin{pmatrix}
\MM_{A\, B} & 0 \\ 0 & \MM^{CD\, EF}
\end{pmatrix}
R_2^{\, -1}
\end{equation}
and similarly for all the other metric components which are rotated into each other. The essential action of this rotation becomes clearest when the indices are suppressed. The result, which is the same for all the blocks, is
\begin{equation}
R_2
\begin{pmatrix}
H^{1/2} & 0 \\ 0 & H^{-1/2}
\end{pmatrix}
R_2^{\, -1} =
\begin{pmatrix}
H^{1/2}\sin^2\xi + H^{-1/2}\cos^2\xi & -H^{-1/2}(H-1)\sin\xi\cos\xi \\ 
-H^{-1/2}(H-1)\sin\xi\cos\xi & H^{-1/2}\sin^2\xi + H^{1/2}\cos^2\xi
\end{pmatrix} \, .
\end{equation}
This transformation produces additional off-diagonal terms in the generalized metric. In the ordinary supergravity picture these extra terms reduce to components of the C-field in the internal sector of the KK-decomposition which are of the form $C_{mnk}$. These terms are not present for the pure membrane and fivebrane, they only occur in the bound state solution. The $E_7$ generalized metric with cross-terms due to non-vanishing internal C-field was constructed in general form in \cite{Berman:2011jh} and the appropriate reduction ansatz (which has the same form as the metric) for our concrete scenario was spelled out in \cite{Berman:2014jsa} and can be found in Section \ref{sec:truncatedE7}. 

The next step is thus to bring the above matrix into the standard coset form of a generalized metric or a KK-reduction ansatz which can be done by using some trigonometric identities and introducing the shorthand
\begin{equation}
\Xi = \sin^2\xi + H \cos^2\xi \, .
\label{eq:Xi}
\end{equation}
Then the new metric components read
\begin{equation}
\begin{pmatrix}
H^{1/2}\Xi^{-1}\left[1+\frac{(H-1)^2}{H}\sin^2\xi\cos^2\xi\right] & 
-H^{-1/2}\Xi \frac{H-1}{\Xi}\sin\xi\cos\xi \\ 
-H^{-1/2}\Xi \frac{H-1}{\Xi}\sin\xi\cos\xi &
H^{-1/2}\Xi 
\end{pmatrix} 
\end{equation}
which is of the desired form. It is interesting to see that it is actually possible to rewrite the transformed metric in a coset form. The underlying reason for this is that the original matrix was already in coset form, just without any off-diagonal terms, \ie without a C-field. 

Now this matrix can be compared to a suitable reduction ansatz to extract the (components of) the metric and the C-field in supergravity. Such an ansatz -- adapted to our coordinates here -- takes the form
\begin{equation}
g^{1/2}
\begin{pmatrix}
g_{AB} + C_{ACD}g^{CD,EF}C_{EFB} & C_{ACD}g^{CD,EF} \\
g^{CD,EF}C_{EFB} & g^{CD,EF}
\end{pmatrix}
\end{equation}
where $g = \det g_{mn}$ is the determinant of the internal metric. Comparing these two matrices leads to the following components (note that of course to find the determinant all blocks have to be taken into account, not just those corresponding to $y^A$ and $y_{BC}$)
\begin{equation}
\begin{aligned}
g_{AB} &= H^{1/3}\Xi^{-2/3} \delta_{AB} & 
g &= H^{1/3}\Xi^{-2/3}  \\
g^{CD,EF} &= H^{-2/3}\Xi^{4/3} \delta^{CD,EF} &
C_{ABC} &= -\frac{H-1}{\Xi}\sin\xi\cos\xi\ \epsilon_{ABC}\, .
\end{aligned}
\end{equation}

The same procedure also works for the other pairs of indices that need to be transformed and their corresponding metric components. The only difference is for $(y_{12}, y^{67})$ where the metric has an extra factor of $H$, \ie $\MM^{12,12} = H^{3/2}$ and $\MM_{67,67} = H^{1/2}$. But this factor is just carried through and does not affect the calculation presented above. Similarly, for $(y^{12}, y_{67})$ there is an extra factor of $H^{-1}$ in the metric.

Also note that for some cross-terms in the reduction ansatz one needs to define $V^{m_1\dots m_4} = \frac{1}{3!}\epsilon^{m_1\dots m_4n_1\dots n_3}C_{n_1\dots n_3}$, see Section \ref{sec:truncatedE7} for more details. Since the only non-zero component (in the internal sector) of $C_3$ is $C_{ABC}$, the only component of this $V$ that does not vanish is $V^{ab\alpha\beta}=-\frac{H-1}{\Xi}\sin\xi\cos\xi\epsilon^{ab\alpha\beta}$.

Once the transformation of each index pair and the corresponding metric component together with the reduction to supergravity is performed, the dyonic M2/M5-brane solution 
is obtained in the usual 4+7 Kaluza-Klein split. Its internal metric $g_{mn}$ (and its determinant $g$) recovered form the generalized metric together with the external metric $g_{\mu\nu}$ which is just carried over from the external sector of the EFT solution are given by
\begin{equation}
\begin{aligned}
g_{mn} &= H^{1/3}\Xi^{1/3}\diag[H^{-1}\delta_{ab},\Xi^{-1}\delta_{AB},\delta_{\alpha\beta}] \, , & 
g &=H^{1/3}\Xi^{-2/3}  \\
g_{\mu\nu} &= \diag[-H^{-1/2},H^{1/2}\delta_{ij}]  \, .
\end{aligned}
\end{equation}
with $\Xi$ as defined in \eqref{eq:Xi}. Reversing the KK-decomposition, finally gives the eleven-dimensional spacetime metric of the solution as in \cite{Izquierdo:1995ms}
\begin{equation}
\begin{aligned}
\dd s^2 &= H^{-2/3}\Xi^{1/3}[-\dd t^2 + \delta_{ab}\dd y^a \dd y^b]
			+ H^{1/3}\Xi^{-2/3}[\delta_{AB}\dd y^A \dd y^B] \\
		&\qquad	+ H^{1/3}\Xi^{1/3}[\delta_{ij}\dd w^i \dd w^j 
			+ \delta_{\alpha\beta}\dd y^\alpha \dd y^\beta] \, .
\end{aligned}
\end{equation}
The harmonic function $H=H(r)$ here only depends on the three $w^i$ where $r^2=\delta_{ij}w^iw^j$. It is smeared over the remaining transverse coordinates $y^A$ and $y^\alpha$. In \cite{Izquierdo:1995ms} the solution is only delocalized in the three $y^A$ since it is constructed in eight dimensions and then lifted to eleven dimensions by including the $y^A$. Simply delocalizing it in $y^\alpha$ allows for a complete identification with the solution here. Furthermore, in the reference a multi-brane solution is constructed whereas here only a single source is considered. The result can of course be extended to take several identical brane sources into account.

It can be checked that setting $\xi$ to $0$ or $\pi/2$ and thus either $\Xi=H$ or $\Xi=1$ reproduces the pure M5-brane \eqref{eq:M5} and the pure M2-brane \eqref{eq:M2} respectively.

The components of the three-form gauge potential which have one external and two internal indices, \ie $C_{\mu\ mn}$ were obtained from the EFT vector potential ${\AAA_\mu}^M$. The component $C_{mnk}$ which is entirely in the internal sector was extracted from the generalized metric $\MM_{MN}$. Together they read
\begin{equation}
\begin{aligned}
C_{tab} &= \frac{H-1}{H}\sin\xi\ \epsilon_{ab} \\
C_{i\alpha\beta} &= A_i\cos\xi\ \epsilon_{\alpha\beta} \\
C_{ABC} &= -\frac{H-1}{\Xi}\sin\xi\cos\xi\ \epsilon_{ABC} 
\end{aligned}
\label{eq:M2M5C3}
\end{equation}
where $A_i$ is defined as before. These are exactly the C-field components of the bound state solution (in \cite{Izquierdo:1995ms} they are given in terms of their field strengths). Again one can check that in the pure cases where either $\cos\xi=0$ or $\sin\xi=0$, the three-form potential only has a single component as given in \eqref{eq:M2} and \eqref{eq:M5} respectively. The third component above vanishes in the two pure cases. 

An interesting observation is that $\Xi=\sin^2\xi + H \cos^2\xi$ goes to $1$ far away from the brane solution since $H\rightarrow 1$ for $r\rightarrow\infty$ while near the core where $\Xi\sim H$ the fivebrane geometry prevails.

In summary, it has been shown that the self-dual EFT solution contains the dyonic M2/M5-brane solution. Therefore, the EFT solution does not only give the standard supergravity branes but in fact also the brane bound states. The standard ones are the objects obtained by pointing the vector along one of the axes of our 56-dimensional exceptional extended coordinate space. But any combination of directions is possible, thus giving rise to dyonic bound states of branes. Furthermore, the solutions of the Type II theories in ten dimensions are also included. We will look at this aspect next.

\subsubsection{Type IIA Solutions}
In Appendix \ref{ch:appEFT} it is shown how the ten-dimensional Type IIA theory can directly be embedded into EFT without an intermediate step to the eleven-dimensional theory. Applying this procedure in reverse, the EFT solution can be viewed from a Type IIA point of view.

In the case of extracting the eleven-dimensional solutions, the internal extended coordinate $Y^M$ was decomposed into four distinct subsets \eqref{eq:gencoordM} such as $y^m$ or $y_{mn}$. Having the EFT wave propagating along those four kinds of directions gave rise to the four different solutions in supergravity with the four components of the EFT vector potential \eqref{eq:vecpotSUGRA} providing the KK-vector and C-fields. Now in the ten-dimensional Type IIA case, the generalized coordinate splits into eight separate sets of directions \eqref{eq:gencoordIIA} and we can thus expect to get eight different solutions, one for each possible orientation of the EFT solution (together with the eight types of components of the EFT vector \eqref{eq:vecpotIIA}). 

Let us first obtain the WA-solution, the pp-wave spacetime in Type IIA. The generalized metric has to be slightly reshuffled to accommodate our new choice of coordinates, its precise form can be found in the appendix. To obtain the wave, the EFT solution is made to propagate along one of the ordinary directions $y^\bm$, say $y^1=z$. Using the ansatz \eqref{eq:genmetricIIA} for $\MM_{MN}$ in Type IIA and comparing it to the (rotated) generalized metric of the EFT solution gives the dilaton $e^{2\phi}$ and the internal 6-metric $\bg_{\bm\bn}$ of the WA-solution under the $4+6$ KK-decomposition. The corresponding EFT vector component is ${\AAA_t}^z=-(H^{-1}-1)$ which provides the KK-vector of the decomposition and combines with the internal and external metrics to form the ten-dimensional metric of the wave. The other component of the vector potential, ${\AAA_i}^\tz = A_i$ is related to the dual graviton which does not appear in the ten-dimensional picture. The KK-decomposition of the wave and other Type II solutions can be found in Appendix \ref{sec:stringsolutions}.

If instead the EFT solution is chosen to propagate along the compact circle $y^\theta$, the same procedure as above leads to the D0-brane. The RR-one-form $C_1$ it couples to can be extracted from the EFT vector ${\AAA_t}^\theta$. The dual seven-form $C_7$ is derived from the other component, ${\AAA_i}^\ttheta$.

The picture should be clear by now. The EFT solution, that is the generalized metric $\MM_{MN}$ and the vector potential ${\AAA_\mu}^M$, are rotated in a specific direction. Depending on the nature of that direction, different solution in the ten-dimensional theory arise. The F1-string and NS5-brane solution can be extracted if the EFT solution propagates along one of the $y_{\bm\theta}$ and $y^{\bm\theta}$ directions respectively. The corresponding EFT vector provides the NSNS-two-form $B_2$ and dual NSNS-six-form $B_6$ for the string and vice versa for the fivebrane. Similarly, if the directions are $y_{\bm\bn}$ and $y^{\bm\bn}$, the D2- and D4-branes with the corresponding set of dual RR-three-form $C_3$ and  RR-five-form $C_5$ are obtained. 

The last two directions the EFT solution can be along are $y_\bm$ and $y_\theta$. These are the dual directions to $y^\bm$ and $y^\theta$ and hence provide the solutions dual to WA and D0, that is the KK6A-brane and the D6-brane. For the KK6A-brane, essentially the KK-monopole of the Type IIA theory, if we choose $y_1=\tz$ as the direction, the EFT vector ${\AAA_i}^\tz=A_i$ gives the KK-vector for the ten-dimensional metric and the dual ${\AAA_t}^z=-(H^{-1}-1)$ is the dual graviton for that solution. For the D6-brane, the EFT vector provides the RR-seven-form $C_7$ it couples to together with the dual one-form $C_1$.

We have thus outlined how eight different Type IIA solutions can all be extracted from a single self-dual solution in EFT. The fundamental wave and string, the solitonic monopole and fivebrane, and the four $p$-even D-branes all arise naturally by applying the Type IIA solution to the section condition to the EFT wave rotated in the appropriate direction. A summery of all the possible orientations and corresponding solutions can be found in Table \ref{tab:solutions} at the end of this section.

\subsubsection{Type IIB Solutions}
Along the same lines as above, using the ansatz for embedding the Type IIB theory into EFT allows for further solutions to be extracted from the EFT wave. The generalized coordinate $Y^M$ is now split into five distinct sets according to \eqref{eq:gencoordIIB} which gives five possible directions to align the EFT solution (together with five types of components in the EFT vector \eqref{eq:vecpotIIB}). 

As before, the entries of the generalized metric $\MM_{MN}$ have to be rearranged to accommodate the choice of coordinates (see Appendix \ref{ch:appEFT}). Comparing the Type IIB ansatz for $\MM_{MN}$ in \eqref{eq:genmetricIIBapp} to the (rotated) generalized metric leads to the six-dimensional internal metric $\bg_{\bm\bn}$ together with the $SL(2)$ matrix $\gamma_{ab}$. If the direction of choice is of the $y^\bm$ type, the WB-solution can be extracted. This is the pp-spacetime of the Type IIB theory which is identical to the WA-solution. The procedure is exactly the same as before with the ${\AAA_\mu}^M$ providing the KK-vector for the ten-dimensional metric (and the dual graviton which plays no role).

The dual choice of direction, \ie $y_\bm$, gives the dual solution, that is the KK6B-brane, the KK-monopole of the Type IIB theory. Again the EFT vector contributes the KK-vector and dual graviton. The KK6B-brane is identical to the KK6A-brane.

\begin{table}[h]
\centering
\begin{tabular}{|c|c|clcl|cc|}
\hline
theory & solution & \multicolumn{2}{c}{orientation}
 & \multicolumn{2}{c|}{\begin{tabular}[c]{@{}c@{}}EFT\\vector\end{tabular}} 
 & ${\AAA_t}^M$ & ${\AAA_i}^M$ \\ \hline
\multirow{5}{*}{$D=11$} 
 & WM 	&\ \ \ & $y^m$ 		&& ${\AAA_\mu}^m$ 	& KK-vector 		& dual graviton 	\\
 & M2 	&& $y_{mn}$ 		&& $\AAA_{\mu\ mn}$ 		& $C_3$ 			& $C_6$ 			\\
 & M2/M5 && 	\ *	&&		\ *	&  $C_3\oplus C_6$ 	&		$C_6\oplus C_3$		\\
 & M5 	&& $y^{mn}$ 		&& ${\AAA_\mu}^{mn}$ 	& $C_6$ 			& $C_3$ 			\\
 & KK7 	&& $y_m$ 		&& $\AAA_{\mu\ m}$ 		& dual graviton 	& KK-vector 		\\ \hline
\multirow{8}{*}{\begin{tabular}[c]{@{}c@{}}$D=10$\\Type IIA\end{tabular}} 
 & WA 	&& $y^\bm$ 			&& ${\AAA_\mu}^\bm$			& KK-vector & dual graviton 	\\
 & D0 	&& $y^\theta$ 		&& ${\AAA_\mu}^\theta$		& $C_1$ 			& $C_7$ 		\\
 & D2 	&& $y_{\bm\bn}$ 		&& $\AAA_{\mu\ \bm\bn}$		& $C_3$ 			& $C_5$ 		\\
 & F1 	&& $y_{\bm\theta}$ 	&& $\AAA_{\mu\ \bm\theta}$	& $B_2$ 			& $B_6$ 		\\
 & KK6A && $y_\bm$ 			&& $\AAA_{\mu\ \bm}$			& dual graviton 	& KK-vector 	\\ 
 & D6 	&& $y_\theta$ 		&& $\AAA_{\mu\ \theta}$		& $C_7$ 			& $C_1$ 		\\
 & D4 	&& $y^{\bm\bn}$		&& ${\AAA_\mu}^{\bm\bn}$		& $C_5$ 			& $C_3$ 		\\
 & NS5 	&& $y^{\bm\btheta}$	&& ${\AAA_\mu}^{\bm\theta}$	& $B_6$ 			& $B_2$ 		\\\hline
\multirow{5}{*}{\begin{tabular}[c]{@{}c@{}}$D=10$\\ Type IIB\end{tabular}} 
 & WB 		&& $y^\bm$ 			&& ${\AAA_\mu}^\bm$ 			& KK-vector 		& dual graviton \\
 & F1 / D1 	&& $y_{\bm\ a}$ 		&& $\AAA_{\mu\ \bm \ a}$ 		& $B_2$ / $C_2$ 	& $B_6$ / $C_6$ 	\\
 & D3 		&& $y_{\bm\bn\bk}$ 	&& $\AAA_{\mu\ \bm\bn\bk}$ 	& $C_4$ 			& $C_4$ 			\\
 & NS5 / D5 	&& $y^{\bm\ a}$ 		&& ${\AAA_\mu}^{\bm\ a}$		& $B_6$ / $C_6$ 	& $B_2$ / $C_2$ 	\\
 & KK6B 		&& $y_\bm$			&& $\AAA_{\mu\ \bm}$			& dual graviton 	& KK-vector 		\\ 
\hline
\end{tabular}
\caption{This table shows all supergravity solutions in ten and eleven dimensions discussed in this section. The orientation indicates the type of direction along which the EFT solution propagates to give rise to each of the supergravity solutions. It also determines the nature of the components of the EFT vector in the supergravity picture. * The orientation of the M2/M5 bound state requires a superposition of a membrane direction $y_{mn}$ and a fivebrane direction $y^{mn}$. Therefore the EFT vector in that hybrid direction gives both the $C_3$ and the $C_6$ since it is a dyonic solution.}
\label{tab:solutions}
\end{table}

A more interesting choice of direction is to rotate the EFT solution along one of the $y_{\bm\ a}$. This produces the Type IIB S-duality doublet of the F1-string and D1-brane. They couple to a two-form which carries an additional $SL(2)$ index $a$ to distinguish between the NSNS-field $B_2$ and the RR-field $C_2$. From the generalized metric $\MM_{MN}$ the internal metric $\bg_{\bm\bn}$ and the $SL(2)$ matrix $\gamma_{ab}$ containing the dilaton $e^{2\phi}$ ($C_0$ vanishes for this solution) can be extracted. The EFT vector $\AAA_{\mu\ \bm\ a}$ provides the two-form (and also the dual six-form).

Similarly, the EFT solution along one of the $y^{\bm\ a}$ gives rise to the other S-duality doublet of the Type IIB theory, the NS5-brane and the D5-brane. They couple to a six-form  which also carries an $SL(2)$ index to distinguish the NSNS- and RR-part, $B_6$ and $C_6$ respectively. The six-form is encoded in the electric part of the EFT vector ${\AAA_t}^{\bm\ a}=-(H^{-1}-1)$ (and the dual two-form is encoded in the magnetic part $\AAA_{i\ \bm\ a}=A_i$) upon dualization on the internal coordinate. 

Finally, having the EFT solution along the fifth direction from a Type IIB point of view, $y_{\bm\bn\bk}$, leads to the self-dual D3-brane together with its self-dual four-form $C_4$ encoded in $\AAA_{\mu\ \bm\bn\bk}$. 

As in the Type IIA theory, the fundamental wave and string, the solitonic monopole and fivebrane, and three $p$-odd D-branes, can all be extracted from the EFT solution by applying the Type IIB solution to the section condition and rotating the fields appropriately. All the obtained solutions are summarized in Table \ref{tab:solutions}, together with the orientation the EFT solution, \ie its direction of propagation.

In theory it should also be possible to obtain the D-instanton (the D(-1)-brane) and its dual, the D7-brane, from the EFT solution. The reason why this is not as straightforward as for all the other D-branes is that the instanton, as the name implies, does not have a time direction, it is a ten-dimensional Euclidean solution. Therefore the EFT solution has to be set up in such a way that the time coordinate is not in the external sector but in the internal sector of the KK-decomposition. Then being part of the exceptional extended space it can be rotated and ``removed'' when taking the section back to the physical space, leaving a solution without a time direction.

The issue for the D7-brane is that it only has two transverse directions, so it cannot fully be accommodated by our KK-decomposition which places time plus \emph{three} transverse direction in the external sector and the the world volume (with the remaining transverse bits, if there are any) in the internal sector. This clearly does not work for the D7-brane.

Both of these reasons are not fundamental shortcomings of the EFT solution, they are just technical issues arising from the way we have set everything up in this chapter. In Chapter \ref{ch:SL2} where the EFT for the duality group $\Slt$ is constructed we will revisit some of the above solutions and also look at D7-branes. Before doing this we will discuss some more properties and aspects of the DFT solutions of Chapter \ref{ch:DFTsol} and the EFT solutions of this chapter next.

\chapter{Aspects and Discussion of Solutions}
\label{ch:aspects}

In this chapter we will look at some aspects of the DFT and EFT solutions presented so far in more detail and analyze them further. Specifically we want to study the presence or absence of singularities at the core of the solutions and see if there is a naturally preferred frame choice to avoid the encounter with a singularity. Furthermore we want to extend some of the solutions we have found by localizing them in the dual space, \ie give them a dependence on a winding coordinate. This leads to some interesting insights into worldsheet instanton corrections. To some extend this has been done before \cite{Gregory:1997te,Harvey:2005ab}, but it is much more natural to formulate such a setup in the language of double field theory.

\section{Singularities and Preferred Frame Choices}
\label{sec:singularities}

In this section we wish to enter into some speculation that has motivated some of the work at the heart of this thesis. In particular we want to comment on the issue of the singularity structure of supergravity solutions\footnote{We are grateful to Michael Duff for discussions on this issue.}. 

Having constructed the self-dual wave solution with monopole-structure in EFT (Section \ref{sec:selfdual}) and shown how it relates to the known solutions of supergravity, we can analyze it further. The fields of the solution, that is the metrics $g_{\mu\nu}$ and $\MM_{MN}$ and the vector potential ${\AAA_\mu}^M$, are all expressed in terms of the harmonic function $H(r)$ with $r^2=\delta_{ij}w^iw^j$ where $r$ is the radial coordinate of the transverse directions in the external sector. This leads to the immediate question of what happens to the solution when $r$ goes to zero or infinity, \ie what happens close to the core of the solution or far away from it? 

As is well known from the works of Duff and others \cite{Duff:1991pe,Duff:1994an}, the nature of singularities at the core of brane solutions depends on the duality frame that one uses. Given that EFT provides a formalism unifying different duality frames then one would imagine that solutions in EFT maybe be singularity free at the core. Since the solution lives in an extended space, the additional dimensions help to smooth the singularity at the core. This is not the case for the fundamental string, for example.

A good way to think about DFT or EFT is as a Kaluza-Klein-type theory. The space is extended and the reduction of the theory through use of the section condition gives supergravity. Let's recall some basic properties of ordinary Kaluza-Klein theory that will be useful for our intuition. The reduced theory is gravity plus electromagnetism (and a scalar field which will not be relevant here). One typically allows various singular solutions such as electric sources which have delta function-type singularities and magnetic sources which also are singular. Then the Kaluza-Klein lift of these solutions smooths out the singularities. The electric charges are just waves propagating around the KK-circle and the magnetic charges come from fibering the KK-circle to produce a total space describing an $S^3$. Thus the singularities inherent from the abelian charges become removed when one considers the full theory and the $U(1)$ is just a subgroup of some bigger non-abelian group, in this case five-dimensional diffeomorphisms. 

A similar process happens when one considers the 't\,Hooft-Polyakov monopole where in the low energy effective field theory the gauge group is broken to $U(1)$ and the monopole is a normal Dirac monopole (with a singularity at the origin). Near the core of the monopole however, the low energy effective description breaks down, and the full non-abelian theory becomes relevant. The non-abelian interactions smooth out the core of the monopole and the singularity is removed. This intuition is exactly what we wish to evoke when thinking about DFT and EFT. Solutions become smoothed out by the embedding in a bigger theory, $U(1)$ charges in particular are simply the result of some reduction and the singularities are non-existent in the full theory \cite{Harvey:1996ur}.

So can one show that the EFT solutions described here are free of singularities at their core? To see how this works, we will first return to a simpler example, the DFT wave of Section \ref{sec:DFTwave} which gives the fundamental string when the direction of propagation is a winding direction. After this short digression we will return to the EFT solution and find a similar result.

\subsection{The Core of the DFT Wave}
It is well known that the fundamental string of string theory has a singularity at its core (essentially there are delta function sources required by the solution). The existence of this singularity can easily be inferred from looking at the Ricci scalar or sending a probe towards the core of the string and looking at the proper time it takes the probe to do so \cite{Duff:1991pe,Duff:1994an}.

The other fundamental object in string theory, the T-dual of the string, is the wave. The wave is clearly non-singular as a straightforward calculation of the Riemann curvature shows. If one takes a closer look at the string and the wave, it is the cross-terms in the wave metric that ensure that the curvature remains finite and does not develop a singularity. The T-duality (essentially the Buscher rules) that turns the wave into the string moves these cross-terms into the B-field of the string. The curvature of the string then becomes singular as $r$ goes to zero. 

In double field theory T-duality simply is a rotation in the doubled space. In DFT there is a single fundamental solution, the DFT wave. Depending on its orientation in the doubled space, either the string or the wave solution of string theory can be extracted when seen from an un-doubled point of view.

The DFT solution is non-singular everywhere. The notion of curvature in DFT is a slightly ambiguous concept. By non-singular we mean that the generalized Ricci tensor defined by varying the DFT action with respect to the generalized metric vanishes everywhere for the DFT wave. Of course the equations of motion for DFT dictate that this must vanish in the absence of any RR or fermionic sources since the NSNS sector is contained in the generalized geometry. What is significant is that one might have allowed delta function sources as one does for the Schwarzschild solution in ordinary relativity. None are required by the wave solution in DFT. The lack of a singularity at the core may also be argued from analogy with ordinary relativity. The solution in question is the wave solution for DFT and the wave solution in ordinary GR is free from singularities therefore we expect the DFT wave to also be singularity free. Thus looking at the solution from the perspective of the doubled space eliminates the singularity at the core of the fundamental string. 

One can ask further how the solution behaves closes to the core or far away, \ie what happens when $r$ is small or large? Is there is a natural choice of picking the coordinates that form the physical spacetime? To answer this question we have to take a closer look at the DFT wave. Recall the generalized metric of the DFT wave solution \eqref{eq:DFTwave} 
\begin{equation}
\begin{aligned}
\dd s^2 &= \HH_{MN}\dd X^M \dd X^N \\
	&= (H-2)\left[\dd t^2 - \dd z^2\right] + \delta_{mn}\dd y^m\dd y^n \\
	&\quad + 2(H-1)\left[\dd t\dd\tz + \dd\ttt\dd z\right] \\
	&\quad - H\left[\dd\ttt^2 - \dd\tz^2\right] + \delta^{mn}\dd\ty_m\dd\ty_n
\end{aligned}
\end{equation}
where the doubled coordinates are $X^M=(t,z,\ttt,\tz,y^m,\ty_\bm)$. The doubled space has dimension $2D$ and the transverse coordinates are labeled by $m,\bm=1,\dots,D-2$. The harmonic function here (for $D>4$) is given by $H = 1 + \frac{h}{r^{D-4}}$ with $r^2=\delta_{mn}y^my^n$. In order to study the behaviour of this solution we will take all the relevant doubled coordinates to be compact such that $(t,z,\ttt,\tz)$ are periodic coordinates.

In order to use our Kaluza-Klein inspired intuition we will first examine the behaviour of DFT on a simple $2D$-dimensional torus. The $2D$ DFT torus is decomposed into a $D$-torus with volume $R^D$ and a dual torus with volume $(\alpha'/R)^D$. This decomposition is always possible due to the presence of the invariant $\Odd$ structure $\eta$ that provides the doubled space with a polarization. Double field theory comes equipped with a coupling 
\begin{equation}
G_{DFT}= e^{2d} \,  \alpha'^{D-1} 	
\end{equation}
where $e^{2d}$ is the DFT dilaton and acts as a dimensionless coupling for DFT in the same way as the usual dilaton does for supergravity. The DFT coupling $G_{DFT}$ is to be compared with the usual Newton's constant in $D$-dimensional supergravity given by
\begin{equation}
G_N=e^{2\phi} \alpha'^{\frac{D-2}{2}} \, .
\end{equation}
It is known (almost by construction) that reducing DFT on the \emph{dual} $D$-torus \`a la Kaluza-Klein gives supergravity in $D$ dimensions and thus we can relate the coupling for the reduced theory to the DFT coupling, $G_{DFT} = \frac{\alpha'^D}{R^D}G_N $, to get
\begin{equation}
e^{2\phi}= \frac{R^D}{\alpha'^{D/2}} e^{2d} \, .
\end{equation}
Equivalently, we may instead reduce on the $D$-torus, to give the T-dual supergravity picture via $G_{DFT} = R^D G_N$ which gives the relation
\begin{equation}
e^{2\phi_{{\rm dual}}}=\frac{\alpha'^{D/2}}{R^D} e^{2d} \, .
\end{equation}
These three couplings then potentially provide a hierarchy that is governed by $R^D$, the volume of the torus. This analysis gives the completely intuitive result that for $R$ much larger than the string scale, the appropriate description is DFT reduced on the dual torus since its coupling, $e^{2\phi}$, is greatest. For small $R$ the appropriate description is for the theory reduced on the torus itself, as its coupling, $e^{2\phi_{{\rm dual}}}$, is greatest. For a circle whose radius is of order $\alpha'$ there is no preferred reduction and the hierarchy breaks down. Thus the total doubled space should be taken into account without reduction. This is as it should be, for tori near the string scale we need to include both ordinary modes and the winding modes simultaneously. This somewhat pedestrian analysis is just slightly formalizing the notion that on a compact space there is a natural T-duality frame that is picked out by the one where the volume is largest.{\footnote{The reader may prefer that this argument should be expressed in terms of energy scales in which case simply convert the hierarchy of gravitational couplings to effective Planck masses and examine the theory that dominates the low energy effective action (\cf the discussion in the Introduction).}} 

 We now wish to apply this lesson to our DFT wave solution and determine which dimensions become large and thus pick the T-duality frame. These large dimensions will be those that then become identified with the spacetime of supergravity. This is non-trivial in the sense that the volume of the compact space will be a function of $r$, the distance from the core of the solution. One should think of the solution as a toroidal fibration with a one-dimensional base space with coordinate $r$ and the toroidal fibre being given by the generalized metric. For the solution at hand this is space described by the $4\times 4$ generalized metric for the coordinates $X^A=(t,z,\ttt,\tz)$
\begin{equation}
\HH_{AB} = \begin{pmatrix}
H-2		& 0		& 0   		& H-1 	\\
0		& 2-H	& H-1   	& 0 		\\
0		& H-1	& -H	  	& 0		\\
H-1		& 0		& 0 		& H
\end{pmatrix} \, .
\end{equation}
The function $H = 1 + \frac{h}{r^{D-4}}$ completely determines the geometry and is solely a function of $r$, the radial distance to the solution's core. In these coordinates there is no notion of one dimension being larger than another since the metric has off-diagonal components. In order to see which dimensions become large it is necessary to go to a choice of coordinates where the generalized metric is diagonal. To simplify the notation we set $\rho=r^{D-4}$. Then the four eigenvalues of $\HH_{AB}$ are $\lambda_A$. They are given together with their limits by
\begin{equation}
\begin{array}{lccc}
& & r \rightarrow 0 & r \rightarrow \infty  \\ 
\lambda_1 = \phantom{-}\displaystyle\frac{h-f}{\rho}	& \quad\longrightarrow\quad	& 0	& -1 \\
\lambda_2 = -\displaystyle\frac{h-f}{\rho}	& \longrightarrow	& 0	& +1  \\
\lambda_3 = -\displaystyle\frac{h+f}{\rho}	& \longrightarrow	& -\infty			& -1 \\
\lambda_4 = \phantom{-}\displaystyle\frac{h+f}{\rho}	& \longrightarrow	& +\infty & +1
\end{array}
\end{equation}
where $f=\sqrt{h^2+\rho^2}$. The corresponding (normalized) eigenvectors can be used to construct the diagonalizing matrix which in turn is utilized to find the new basis where $\HH_{AB}$ is diagonal with entries $\lambda_A$. This new basis is given by 
\begin{equation}
{X'}^A =  \sqrt{\frac{h}{2f}}
\begin{pmatrix}
\sqrt{\frac{f+\rho}{h}} t - \sqrt{\frac{h}{f+\rho}} \tz \\
\sqrt{\frac{h}{f+\rho}} \ttt  + \sqrt{\frac{f+\rho}{h}} z \\
\sqrt{\frac{h}{f-\rho}} \ttt - \sqrt{\frac{f-\rho}{h}} z  \\
\sqrt{\frac{f-\rho}{h}} t + \sqrt{\frac{h}{f-\rho}} \tz
\end{pmatrix} \, .
\end{equation}
If we now take the limit where $r$ either goes to zero or infinity (and thus $f$ goes to $h$ or infinity), the diagonalized coordinate basis looks like (recall that the original basis was $X^A=(t,z,\ttt,\tz)$)
\begin{align}
r \rightarrow 0:\quad	{X'}^A &= \frac{1}{\sqrt{2}}
\begin{pmatrix}
t - \tz \\ \ttt + z \\  \ttt - z \\    t + \tz
\end{pmatrix}\, , &
r \rightarrow \infty: \quad {X'}^A &= 
\begin{pmatrix}
t \\  z  \\ \ttt \\ \tz
\end{pmatrix} \, .
\end{align}

This can be interpreted as follows. The DFT wave is asymptotically flat, thus for large $r$ there is no preferred set of coordinates, \ie it does not matter which pair, $(t,z)$ or $(\ttt,\tz)$, we call ``dual" and which we take as being our usual spacetime. Different choices will just give T-duality related solutions. However, as one approaches the core of the solution and $r$ gets smaller, the space diagonalizes and takes the form of a sort of twisted light-cone. It is twisted in the sense that the light cone mixes the coordinates that at asymptotic infinity describe both the space and its dual. That is we have 
\begin{align}
u &= \frac{1}{\sqrt{2}}(t-\tz)\, , &
v &= \frac{1}{\sqrt{2}}(\ttt+z)\, , &
\tv &= \frac{1}{\sqrt{2}}(\ttt-z)\, , &
\tu &= \frac{1}{\sqrt{2}}(t + \tz) \, .
\end{align}

There is a clear hierarchy in the volumes between the two sets. One set, $\tv$ and $\tu$, comes with a large volume associated to those dimensions as can be seen from the eigenvalues $\lambda_3$ and $\lambda_4$ which diverge. On the other hand, the other set of coordinates, $u$ and $v$, is associated with a small volume as the eigenvalues $\lambda_1$ and $\lambda_2$ tend to zero. Thus one set is picked out and we must think of the core of the solution as being given by the space described by $\tv$ and $\tu$. Following these coordinates out of the core to asymptotic infinity these coordinates become just $\ttt$ and $\tz$. This fits in with the intuition that the dual description is suitable for describing the spacetime near the singular core. What may not have been apparent before DFT is that actually as one approaches the core these dual coordinates twist with the normal spacetime coordinates to give a twisted light-cone at the core of the fundamental string.

\subsection{Wave vs Monopole}
Now we return to the self-dual solution of EFT. To carry out the analysis, the solution is not treated as living in 4+56 dimensions but as a truly 60-dimensional solution. Thus the three constituents of the solution, the external metric $g_{\mu\nu}$, the extended internal metric $\MM_{MN}$ and the EFT vector potential ${\AAA_\mu}^M$ are combined in the usual Kaluza-Klein fashion to form a 60-dimensional metric
\begin{equation}
\HH_{\hat{M}\hat{N}} = 
\begin{pmatrix}
g_{\mu\nu} + {\AAA_\mu}^M{\AAA_\nu}^N\MM_{MN} & {\AAA_\mu}^M\MM_{MN} \\
\MM_{MN}{\AAA_\nu}^N & \MM_{MN}
\end{pmatrix}
\end{equation}
where $g_{\mu\nu}$ is given in \eqref{eq:exmetric}, ${\AAA_\mu}^M$ in \eqref{eq:vecpot} and $\MM_{MN}$ is taken from \eqref{eq:genmetric} to be
\begin{equation}
\MM_{MN} = \diag[H^{3/2},H^{1/2}\delta_{27},H^{-1/2}\delta_{27},H^{-3/2}] \, .
\end{equation}
We then insert the EFT monopole/wave solution to find\footnote{The resulting object has dimension 60 whereas the two blocks have dimension 29 and 31 respectively as indicated by the indices $\hat{M}$, $A$ and $\bar{A}$.}
\begin{equation}
\HH_{\hat{M}\hat{N}} = 
\begin{pmatrix}
H^{1/2} \HH_{AB}^\mathrm{wave} & 0 \\
0 & H^{-1/2} \HH_{\bar{A}\bar{B}}^\mathrm{mono}
\end{pmatrix}
\label{eq:60metric}
\end{equation}
where to top left block is simply the metric of a wave with 27 transverse dimensions and the bottom right block is the metric of a monopole, also with 27 transverse dimensions
\begin{align}
\HH_{AB}^\mathrm{wave} &= 
\begin{pmatrix}
H-2 & H-1 & 0 \\
H-1 & H & 0 \\
0 & 0 & \delta_{27}
\end{pmatrix} , \\
\HH_{\bar{A}\bar{B}}^\mathrm{mono} &= 
\begin{pmatrix}
H(\delta_{ij} + H^{-2}A_iA_j) & H^{-1}A_i & 0 \\
H^{-1}A_j & H^{-1} & 0 \\
0 & 0 & \delta_{27}
\end{pmatrix} .
\end{align}
In \eqref{eq:60metric} it is interesting to see that there is a natural split into a block diagonal form simply by composing the fields \`a la Kaluza-Klein. These two blocks come with prefactors of $H^{\pm1/2}$ with opposite power, so the geometry will change distinctly between large and small $r$. 

Close to the core of the solution, where $r$ is small, $H$ becomes large and the wave geometry dominates. Far away for large $r$, $H$ will be close to one (and thus $A_i$ vanishes) and neither the monopole nor wave dominates. Thus one would imagine asymptotically either description is valid and the different choices related through a duality transformation. However what is curious is the dominance of the wave solution in the small $r$ region. It appears according to this analysis that all branes in string and M-theory when thought of as solutions in EFT are wave solutions in the core. It is hoped to explore these questions in a more rigorous fashion in future work.

\section{Localization in Winding Space}
\label{sec:localization}

The supergravity solutions we have considered so far to construct DFT and EFT solutions were \emph{de-localized} and heavily smeared over many dimensions, especially the dual ones. This was not only done to keep things simple but also to relate the various solutions to each other via dualities which, at least in a first step, is easier done when there are isometries.

In this section we want to study solutions which have a dependence on a winding coordinate, \ie they are \emph{localized} in winding space. To put this scenario into context, it will be useful to recall the discussion on T-duality in Section \ref{sec:Tduality} which we will pick up here. There it was argued that the presence of an isometry in the doubled space leads to an ambiguity of how to pick the section which forms the physical spacetime. The different choices give rise to the different T-duality frames from the supergravity point of view. The crucial point was that either choice satisfies the strong constraint. 

Now we want to ask the question what happens when the solution depends on one of those coordinates, say $z$. Note that the solution is still independent of the dual coordinate $\tz$. In such a scenario the strong constraint is still satisfied since overall the solution only depends on half (or less than half) the coordinates. But now there is a canonical choice to take the $z$ coordinate as part of the physical spacetime. The reduction still gives a good supergravity solution, this time it is not smeared but localized in $z$. 

However, what happens if a non-canonical choice is made for the spacetime, \ie what if $\tz$ is chosen in order to end up with a solution in the T-dual frame? One obtains a supergravity solution with a dependence on the ``parameter'' $z$ which is not a spacetime coordinate anymore. To see this in more detail, it helps to look at a concrete example.

\subsection{The Monopole Solution Localized in Winding Space}
In order to appreciate the existence of a monopole solution localized in winding space, some background material is required which we will briefly review here. Superstring theory compactified on a circle contains Kaluza-Klein monopole solutions \cite{Sorkin:1983ns,Gross:1983hb} which have an isometry on the circle. T-duality along the circle transforms them into H-monopoles \cite{Banks:1988rj,Ooguri:1995wj} which can be seen as the NS5-branes \cite{Gauntlett:1992nn,Khuri:1992ww} of string theory. The KK-monopole is charged under the $U(1)$ arising from the KK-reduction ansatz for the spacetime metric while the H-monopole is charged under the $U(1)$ originating from the KK-reduction of the two-form B-field (whose field strength is the three-form H-flux, hence the name).

The NS5-brane naturally corresponds to a localized H-monopole geometry while the solution obtained by T-duality from the KK-monopole is smeared over the circle. This breaking of the isometry around the circle leads to a throat behaviour when probed by scattering strings at short distances. It was shown by Tong \cite{Tong:2002rq} that the smeared and localized versions of the NS5-brane can be reconciled by considering string worldsheet instantons. In the smeared background there are instanton corrections to the geometry to reproduce the localized solution. 

The presence of a duality means that the physics of the dual solutions is the same. The problem now is that the throat behaviour of the NS5-brane is qualitatively different to the T-dual KK-monopole solution. It was therefore argued by Gregory, Harvey and Moore \cite{Gregory:1997te} that a proper KK-monopole solution in string theory should be modified to have a throat behaviour near its core when being probed by scattering string winding states (as opposed to string momentum states as we are now in the T-dual picture). 

This was confirmed by Harvey and Jensen \cite{Harvey:2005ab} who used the same gauged linear sigma model techniques as Tong \cite{Tong:2002rq} to demonstrate that instanton effects give corrections to the low energy effective action which can be interpreted as changes to the effective geometry of the solution. In the same way the smeared H-monopole was turned into the localized NS5-brane, we should now think of the KK-monopole as becoming localized in winding space. The problem with such a result is to accommodate both the normal and winding coordinates such a solution requires in a single framework. This is where the doubled geometry of DFT comes into play.

Double field theory is a natural formalism to consider these background geometries since by doubling the target space and thus making T-duality manifest, solutions which depend on a dual coordinate can be analyzed \cite{Dabholkar:2005ve}. In \cite{Jensen:2011jna} Jensen used the gauged linear sigma model for a doubled target space background to study the KK-monopole in winding space and clarify some aspects of the worldsheet instanton calculation. Here we now want to complement this picture by giving the DFT solution which corresponds to the localized KK-monopole (from a target space as opposed to sigma model point of view).

To construct a monopole solution which is not smeared but localized in the $z$ direction, the harmonic function $H$ needs an explicit dependence on $z$. For now we will just denote this by $H=H(r,z)$ using the coordinates of the DFT monopole solution given in \eqref{eq:DFTmonopolecoords}. We will discuss the exact form of the localized harmonic function below. A first immediate effect of this extra coordinate dependence is that the field strength $H_{\mu\nu\rho}$ in \eqref{eq:DFTfieldstrength} of the NS5-brane has now two non-zero components
\begin{equation}
\begin{aligned}
H_{ijz} &= 2\partial_{[i}B_{j]z} = {\epsilon_{ij}}^k\partial_k H(r,z) = 2\partial_{[i}A_{j]} \\
H_{ijk} &= 3\partial_{[i}B_{jk]} = {\epsilon_{ijk}}\partial_z H(r,z) \, .
\end{aligned}
\end{equation}
The first one can be expressed in terms of the magnetic potential $A_i$ as before in the smeared case. The second one is new, as the $\partial_z$ derivative now does not vanish. The localized monopole solution of DFT then reads
\begin{equation}
\begin{aligned}
\dd s^2 	&= H(1+H^{-2}A^2) \dd z^2 + H^{-1} \dd \tz^2 \\
	&\quad + 2H^{-1} A_i\dd y^i \dd \tz - 2H^{-1} A^i \dd \ty_i \dd z
			+ 2H^{-1}{B_i}^j\dd y^i\dd \ty_j \\
	&\quad + H(\delta_{ij}+H^{-2}A_iA_j+H^{-2}{B_i}^kB_{kj}) \dd y^i \dd y^j 
			+ H^{-1} \delta^{ij} \dd \ty_i \dd \ty_j \\
	&\quad +\eta_{ab}\dd x^a \dd x^b + \eta^{ab}\dd \tx_a \dd \tx_b
\end{aligned}
\label{eq:DFTlocalized}
\end{equation}
where extra terms for $\dd y^2$ and $\dd y^i\dd \ty_j$ involving $B_{ij}$ arise as compared to \eqref{eq:DFTmonopole}. 

Upon rewriting this solution by using the ansatz \eqref{eq:KKforDFT}, one obtains the localized NS5-brane with its full field strength. If we carry out the simple operation of swapping the roles of $z$ and  $\tz$ in the reduction, then this gives the following result
\begin{equation}
\begin{aligned}
\dd s^2 &= -\dd t^2 + \dd\vec{x}_{(5)}^{\, 2} 
				+ H^{-1}\left[\dd \tz + A_i\dd y^i\right]^2 + H\dd\vec{y}_{(3)}^{\, 2} \\
H_{ijk} &= 3\partial_{[i}B_{jk]} = \epsilon_{ijk}\partial_z H(r,z) \, .
\end{aligned}
\end{equation}
This solution is the localized KK-monopole. The spacetime coordinates in this duality frame now include $\tz$, crucially though the harmonic function $H$ still depends on $z$, which is a dual coordinate in this frame. Since the monopole can be thought of as a purely gravitational solution, the field strength $H_{ijk}$ can be re-interpreted as torsion of the geometry. One thus concludes that this is the monopole solution localized in the dual winding space. This property is discussed in detail in \cite{Jensen:2011jna}. This is exactly the same result as blindly applying the Buscher rules \cite{Buscher:1987sk,Buscher:1987qj} (which would require an isometry) to the localized NS5-brane along the $z$ direction. It produces the monopole (which is indeed the T-dual of the fivebrane) but the solution is localized in the dual winding direction.

The alert reader will be aware that obviously one should not be allowed to use the Buscher rules to carry out a T-duality in the $z$ direction in the case where the NS5-brane is localized. The $z$ direction is not an isometry of the localized solution. This shows how the $\Odd$ symmetry in DFT (which is a local continuous symmetry) goes beyond the usual notion of T-duality (\`a la Buscher) since it is applicable to any background without any assumptions about the existence of isometries. This perspective was discussed for example in \cite{Berman:2014jba,Cederwall:2014opa} amongst other places.

The usual spacetime manifold is defined by picking out a maximally isotropic subspace of the doubled space. Normally this is done by solving the strong constraint which removes the dependence of fields on half of the coordinates. We then identify the remaining coordinates with the coordinates of spacetime. 

The DFT monopole is a single DFT solution which obeys the strong constraint; how we identify spacetime is essentially a choice of the duality frame. When the half-dimensional subspace which we call spacetime matches that of the reduction through the section condition, then we have a normal supergravity solution which, in the case described above, is the NS5-brane. Alternatively, one can pick the identification of spacetime not to be determined by the section condition, this then gives an alternative duality frame. Generically this will not have a supergravity description even though it is part of a good DFT solution. This is precisely the case described in this section. There is a localization in winding space and so this solution cannot be described through supergravity alone -- even though it maybe a good string background. In DFT it is just described by picking a spacetime submanifold that is not determined by the solution of the section condition.

\begin{table}[h]
\begin{center}
\begin{tabular}{|c|c|c|}
\hline
duality frame & \begin{tabular}[c]{@{}c@{}}DFT solution \\ with $H=H(r,z)$\end{tabular}          & \begin{tabular}[c]{@{}c@{}}DFT solution \\ with $H=H(r,\tz)$\end{tabular}       \\ \hline
A             & \begin{tabular}[c]{@{}c@{}}NS5-brane\\ localized in spacetime\end{tabular}       & \begin{tabular}[c]{@{}c@{}}NS5-brane \\ localized in winding space\end{tabular} \\ \hline
B             & \begin{tabular}[c]{@{}c@{}}KK-monopole\\ localized in winding space\end{tabular} & \begin{tabular}[c]{@{}c@{}}KK-monopole\\ localized in spacetime\end{tabular}    \\ \hline
\end{tabular}
\end{center}
\caption{In this table both DFT solutions are of the form \eqref{eq:DFTlocalized} but with different coordinate dependencies in the harmonic function. Each solution can be viewed in two different duality frames. In frame A the $z$ coordinate is a spacetime coordinate while $\tz$ is a dual winding coordinate. In frame B it is the other way round, $z$ is a dual winding coordinate while $\tz$ is a spacetime coordinate. The solutions extracted from the DFT solutions that are localized in spacetime have good supergravity descriptions while those that are localized in winding space have not.}
\label{tab:localized}
\end{table}

With this in mind, we come to the following conclusion. There are two different DFT solutions of the form \eqref{eq:DFTlocalized}, one with $H(r,z)$ and the other with $H(r,\tz)$ as harmonic function. Here by $z$ and $\tz$ we do not mean spacetime and winding coordinates \emph{a priori}, but just the coordinates as expressed in \eqref{eq:DFTlocalized}. For each of these two DFT solutions there is a choice of duality frames which are of course related by $\Odd$ rotations. In one frame, for clarity call it frame A, $z$ is a spacetime coordinate and $\tz$ is a dual winding coordinate. In another frame, say frame B, the role of $z$ and $\tz$ is exchanged, \ie $\tz$ is a spacetime coordinate and $z$ is dual. See Table \ref{tab:localized} for an overview.

In the case where $H$ is a function of $z$, the DFT solution rewritten in the duality frame A is the NS5-brane localized in spacetime. Its T-dual, found by going to frame B, is the KK-monopole localized in winding space which has no supergravity description as explained above. In the other case where $H$ is a function of $\tz$, the DFT solution rewritten in frame B gives the KK-monopole localized in spacetime while frame A gives the NS5-brane localized in winding space. Again this is a solution with no supergravity description but valid from a string theory point of view. 

The DFT solution listed in the first column of Table \ref{tab:localized} containing the winding localized monopole and spacetime localized NS5-brane was first given in the work by Jensen \cite{Jensen:2011jna}. The DFT solution described in the second column extend Jensen's ideas but are of course a natural consequence of the structure of DFT. We would also like to emphasize that one may interpret Jensen's solution as a DFT monopole as described here (this interpretation has not been made before). Thus DFT allows us to pick subspaces that do not match the section condition. This choice does not allow a spacetime interpretation but does have an interpretation from a string theory point of view.

As explained at the beginning of this section, in Tong's paper \cite{Tong:2002rq} and more recently in related works by Harvey and Jensen \cite{Harvey:2005ab, Jensen:2011jna} and Kimura \cite{Kimura:2013fda,Kimura:2013zva,Kimura:2013khz,Kimura:2014upa,Kimura:2014bxa,Kimura:2014aja} a gauged linear sigma model was used to describe the NS5-brane and related solutions. By ``related solutions'' we mean the KK-monopole and in fact also the exotic $5_2^2$ brane \cite{deBoer:2010ud,deBoer:2012ma}. These are all solutions in the same $\Odd$ duality orbit. The advantage of the gauged linear sigma model description is that one may examine the inclusion of world sheet instanton effects. As first shown in \cite{Tong:2002rq}, the inclusion of such world sheet instantons gives rise exactly to the localization in dual winding space we are describing above. Thus in some sense DFT knows about world sheet instantons. We will discuss this in the next section.

In terms of the topological questions raised by \cite{Papadopoulos:2014mxa}, the localized solution (which does not have the additional isometry) requires an appropriate patching to form a globally defined solution. Thus for this thesis we restrict ourselves to giving only descriptions in a local patch. What is hopeful is that the solution described here has very specific topology of the dual space since it is itself a monopole. It is hoped to carry out a detailed analysis of the global properties in the future.

\subsection{Worldsheet Instanton Corrections}

Let us now return to the harmonic function $H(r,z)$ of the localized solution above. First, consider the harmonic function of the NS5-brane in string theory. Since the fivebrane in ten dimensions has four transverse directions, it is expressed in terms of a harmonic function $H(y_{(4)})$. By \emph{smearing} the solution over one of these directions, it becomes de-localized (called the H-monopole above) and can be related to the KK-monopole via T-duality which requires an isometry. Schematically we have 
\begin{equation}
H_{\mathrm{NS5}} = 1 + \frac{h}{|\vec{y}_{(4)}|^2} \qquad\longrightarrow\qquad 
H_{\mathrm{smeared}} = 1 + \frac{h'}{|\vec{y}_{(3)}|} = H_{\mathrm{KK}} 
\end{equation}
where $h,h'$ are constants. Below we will set $r^2 = y^iy^j\delta_{ij}$ with $i=1,2,3$ and $y^4 = z$.

When working with localized solutions, we want to retain the coordinate dependence on all four transverse directions. But the monopole solution requires a compact direction for consistency (recall how it is given in terms of the Hopf fibration of $S^3$ which in a local patch is $S^2\times S^1$ where the $S^1$ -- the ``monopole circle'' -- is the compact direction), therefore the localization is performed on a circle of radius $R$ and the harmonic function on $\mathbb{R}^3 \times S^1$ (as opposed to $\mathbb{R}^4$ above) is
\begin{equation}
H_{\mathrm{localized}} = 1 + \frac{h}{2Rr}\frac{\sinh r/R}{\cosh r/R - \cos z/R} \, . 
\label{eq:Hlocalized}
\end{equation}
In the limit of $R\rightarrow\infty$, this reduces to the original function for the fivebrane on $\mathbb{R}^4$ where the circle is decompactified. On the other hand, for $r\gg R$ this gives the smeared solution in $\mathbb{R}^3$, \ie without a dependence on $z$. Note that $r$ is the radial coordinate of $\mathbb{R}^3=\mathbb{R}^+\times S^2$ expressed in polar coordinates.

The Hopf fibration of $S^3$ has vanishing first homotopy $\pi_1(S^3)=0$ and so there is no conserved string winding number. Therefore strings may dynamically unwind which was shown by Gregory, Harvey and Moore in \cite{Gregory:1997te}. This means that a string in the background of the localized monopole which carries winding charge can unwind itself. A zeromode can be identified which transfers the winding charge to the monopole background\footnote{Note that of course the winding charge can always be transfered by the string ``falling into'' the monopole. The observation here is that the string can unwind at an arbitrary distance away from the core.}. So in the same way a string momentum mode scattering off the NS5-brane sees a throat behaviour at short distances, a string winding mode scattering off the localized KK-monopole also experiences the throat behaviour and T-duality indeed relates dual solutions with the same physics.

This zeromode can be interpreted as an worldsheet instanton effect which correct the smeared geometry to give the localized solutions. This can be seen by expanding this localized harmonic function in Fourier modes 
\begin{equation}
H(r,z) = 1 + \frac{h}{2Rr}\sum_{k=-\infty}^{\infty} e^{-|k|\frac{r}{R}+ik\frac{z}{R}} \, ,
\label{eq:Hexpanded}
\end{equation}
where now $k$ can be related to the instanton number. This expression was used in the above cited works to find the corrective contribution to the low energy effective action of the instanton effect which can be labelled ``winding mode corrections''

\subsection{Localization on a Torus}

With the above discussion in mind, we now want to try something similar for M-theory backgrounds localized in the extended space. The equivalent solution in EFT to the DFT monopole is the self-dual solution presented in Section \ref{sec:selfdual} which contains both the M5-brane and the KK-monopole. These solutions have five transverse directions in eleven-dimensional supergravity. The harmonic function used for the EFT solution was smeared to only depend on three of them. Therefore we want to localize the fivebrane on two more directions, say $z_1$ and $z_2$. These need to be compact directions to relate to the monopole and thus have the topology of a torus $T^2=S^1_{R_1} \times S^1_{R_2}$. This will introduce the complex modulus $\tau$ which together with the volume $V$ encodes all the information about the torus. Both are given in terms of the two radii $R_1$ and $R_2$ and the angle between the circles. The hope is that once a harmonic function on $\mathbb{R}^3 \times T^2$ is found, it can be expanded (in terms of modular forms) and might provide some insights into membrane ``wrapping mode corrections''.

It is still an open question how this can be achieved due to the lack of a membrane sigma model. Here we would like to provide a stepping stone by presenting the harmonic function localized on $\mathbb{R}^3\times T^2$
\begin{equation}
\begin{aligned}
H(r,z,\bz;\tau,V) &= 1 + \frac{h}{r}\sum_{m,n}\sum_{\pm}\exp\left\{\frac{\pi}{\sqrt{\tau_2V}}
	\Big[z(m+\btau n) - \bz(m+\tau n) \right.\\ 
	& \left. \hspace{5cm}\pm i r \sqrt{2(m+n\tau)(m+\btau n)}\Big]\vphantom{\frac{\pi}{\sqrt{\tau_2V}}}\right\}
\end{aligned}
\end{equation}
where $h$ is a constant (see Appendix \ref{ch:apptorus} for a detailed derivation). The two coordinates $z^a$ have been turned into the complex coordinate $z=z^1+\tau z^2$ where $\tau=\tau_1+i \tau_2$ is the complex modulus of the torus. The torus can be seen as a doubly periodic lattice, each point is identified with $z \sim z + \sqrt{\frac{V}{\tau_2}}(m + \tau n)$ where $m,n$ are integers.

Although we currently lack a closed-form expression for this harmonic function, it can be used in the EFT solution of Section \ref{sec:selfdual} to provide a background which is localized on two of the 56 exceptionally extended internal directions. This geometry can now be probed by a membrane. In analogy to the expansion in \eqref{eq:Hexpanded}, it might be possible to identify extra modes and find additional terms to the low energy effective action which correct the geometry. This remains to be seen.

\chapter{$SL(2)\times\mathbb{R}^+$ EFT -- An Action for F-theory}
\label{ch:SL2}

Now we will slightly shift focus from the extended solutions and their properties back to exceptional field theory in its own right and what it can be used for. This chapter provides a shortened version of \cite{Berman:2015rcc} where the EFT for $D=2$ with duality group $\Slt$ was constructed and it was shown how this provides an action for F-theory. We begin by placing this achievement in a historical context.

It is 20 years since the study of non-perturbative string dynamics \cite{Witten:1995ex} and U-duality \cite{Hull:1994ys} led to the idea of M-theory. From its inception the low energy effective description of M-theory was known to be eleven-dimensional supergravity with the coupling of the Type IIA string promoted to the radius of the eleventh dimension. 

The natural extension of this idea to the Type IIB string gave rise to F-theory \cite{Vafa:1996xn} where the complex coupling in the Type IIB theory is taken to have its origin in the complex modulus of a torus fibred over the usual ten dimensions of the Type IIB string theory. Thus by definition, F-theory is the twelve-dimensional lift of the Type IIB string theory. The status of this twelve-dimensional theory has been somewhat different to that of its Type IIA spouse with no direct twelve-dimensional description in terms of an action and fields that reduce to the Type IIB theory. Indeed, there is no twelve-dimensional supergravity and thus no limit in which the twelve-dimensions can be taken to be ``large'', unlike in the M-theory case. The emphasis has thus largely been on using algebraic geometry to describe F-theory compactifications \cite{Morrison:1996na, Morrison:1996pp} such that now F-theory is synonymous with the study of elliptically fibred Calabi-Yau manifolds. 

The complex coupling of the Type IIB theory is naturally acted on by an $\mathrm{SL}(2)$ S-duality, which is a symmetry of the theory. F-theory provides a geometric interpretation of this duality. An idea which was raised swiftly after the introduction of F-theory was whether there exist similar geometrizations of the U-duality symmetries which one finds after descending in dimension \cite{Kumar:1996zx}. Here, one would hope to be able to associate the scalars of compactified supergravity with the moduli of some auxiliary geometric space. However, the scope for doing so turns out to be somewhat limited.

More recently, the idea of re-imagining duality in a geometric origin has resurfaced in the form of double field theory and exceptional field theory. Perhaps the key development that allows for the construction of these theories is that the new geometry is not a conventional one, but an ``extended geometry'' based on the idea of ``generalized geometry'' \cite{Hitchin:2004ut, Gualtieri:2003dx}. For instance, a key role is played by a ``generalized metric'', in place of the torus modulus, and one introduces an extended space with a novel ``generalized diffeomorphism'' symmetry. We aim to use these innovations to construct a twelve-dimensional theory which provides a local action for F-theory. 

Let us briefly recap the key steps in construction an exceptional field theory for general $D$ following the original line of thinking on the subject adapted to the notation of this chapter:
\begin{itemize}
\item Let us consider ourselves as starting with eleven-dimensional supergravity.
\item We decompose the eleven-dimensional diffeomorphism group $$\mathrm{GL}(11) \longrightarrow \mathrm{GL}(11-D) \times \mathrm{GL}(D).$$
\item We promote the $\mathrm{SL}(D) \subset \mathrm{GL}(D)$ symmetry to $\Edd$ by rewriting the bosonic degrees of freedom in terms of objects which fall naturally into representations of $\Edd$ (in order to do so, we may need to perform various dualizations).
\item We now take the step of introducing $N$ new coordinates such that the $D$ original coordinates and the $N$ extra (or ``dual'') coordinates fit into a certain representation of $\Edd$: $D + N = \dim R_1$.
\item Our fields - which are already in representations of $\Edd$ - are now taken to depend on all the coordinates. 
\item The local symmetries of the theory are likewise rewritten in such a manner that diffeomorphisms and gauge transformations unify into ``generalized diffeomorphisms''.
\item Invariance under these local symmetries may be used to fix the EFT action in the total $(11+N)$-dimensional space.
\item For consistency the theory must be supplemented with a constraint, quadratic in derivatives, known as the section condition. Inequivalent solutions (in the sense that they cannot be mapped into each other by an $\Edd$ transformation) of this constraint will impose a dimensional reduction of the $(11+N)$-dimensional theory down to eleven or ten dimensions. The reduction down to eleven we call the M-theory section and the reduction down to ten the IIB section for reasons that will become obvious. Note that although one may motivate and begin the construction by starting in either eleven dimensions, or ten, the full theory will automatically contain both M-theory and IIB subsectors \cite{Hohm:2013vpa, Blair:2013gqa}
\end{itemize}

Here we give the result of following this procedure to construct the EFT relevant to F-theory. The result will be a twelve-dimensional theory which may be reduced either to the maximal supergravity in eleven dimensions or Type IIB supergravity in ten. This theory is manifestly invariant under a $\G$ symmetry. The $SL(2)$ part of this is easily identified with the S-duality of Type IIB and after imposing the section condition one can see the emergence of a fibration structure as considered in F-theory. 

We note that the construction is a little different to the cases considered before in that the extended space that is introduced is too small to include any effects from geometrizing gauge field potentials such as the three-form of eleven-dimensional supergravity. In some sense, one may think of it as the minimal EFT. As such, it provides the easiest way of seeing exactly how F-theory fits in to the general EFT framework. By constructing the theory our goal is to bring into sharp focus the points of comparison between the EFT construction and what is normally thought of as F-theory. It will be of interest to embed this viewpoint into the higher rank groups already studied. 

In this chapter, after describing the action and symmetries of the $\G$ EFT in section \ref{sec:theory}, we begin the process of analyzing how it relates to F-theory. We can precisely show the identification of the fields of the $\G$ EFT with those of eleven-dimensional supergravity and ten-dimensional Type IIB supergravity, which we do in section \ref{sec:supergravity}. After this, in section \ref{sec:Frel} we discuss in general terms how one may view the IIB section - where there is no dependence on two coordinates - as an F-theory description. The dynamics of the extended space are encoded in a generalized metric whose degrees of freedom are precisely such as to allow us to recover the usual sort of torus fibration familiar in F-theory. 

We also discuss in detail how one sees M-theory/F-theory duality in the EFT framework. One can work out the precise maps between the M-theory and F-theory pictures by comparing the identifications of the EFT fields in the different sections. In this way one recovers the usual relationships relating eleven-dimensional supergravity on a torus to Type IIB string theory on a circle in nine dimensions \cite{Schwarz:1995jq}. We carry this out for a selection of familiar brane solutions in section \ref{sec:solutions}, paying attention to the different features that are seen in EFT depending on whether the brane wraps part of the extended space or not. For instance, 1/2 BPS brane solutions may be seen as simple wave/monopole type solutions in the EFT \cite{Berman:2014hna} as was shown for the $\Es$ case in Chapter \ref{ch:EFT}.

At the same time, we stress that the EFT construction is an extension of the usual duality relationships. From the point of view of the M-theory section, we introduce one additional dual coordinate $y^s$, which can be thought of as being conjugate to a membrane wrapping mode when we have a background of the form $M_9 \times T^2$. Alternatively, the IIB section sees two dual coordinates $y^\alpha$ related to winding modes of the F1 and D1 in the direction $y^s$. In EFT we treat all these modes on an equal footing - they are related by the symmetry $\G$.\footnote{Actually, in this case $y^s$ is a singlet of the $SL2(2)$ part of the group while $y^\alpha$ is a doublet.}

We hope that by presenting the twelve-dimensional $\G$ exceptional field theory we can begin to properly connect the EFT constructions to F-theory. %We conclude with a discussion of possible directions in which one can use the EFT formalism to address aspects of F-theory, and comment on the outlook to higher dimensional duality groups. 

\section{The Theory and its Action}
\label{sec:theory}

In this section, we describe the general features of the theory. First, we describe the setup in terms of an extended space and the field content. Then we shall give the form of the action. After this, we will briefly discuss the symmetries of the theory, which can be used to determine the action. Our goal is to provide an introduction to the main features. Thus, we refer to the original work for the technical details of the construction, some of which can be found in Appendix \ref{ch:appSL2}. Note that here we will restrict ourselves to only the bosonic sector of the theory. We expect that supersymmetrization will follow the form of other supersymmetrized EFTs \cite{Godazgar:2014nqa,Musaev:2014lna}. 

The theory we will describe may be thought of as a twelve-dimensional theory with a $9+3$ split of the coordinates, so that we have
\begin{itemize}
\item nine ``external'' coordinates, $x^\mu$,
\item three ``extended'' coordinates, $Y^M$ that live in the $\mathbf{2}_{1} \oplus \mathbf{1}_{-1}$ reducible representation of $\G$ (where the subscripts denote the weights under the $\mathbb{R}^+$ factor). To reflect the reducibility of the representation we further decompose the coordinates $Y^M = ( y^\alpha, y^s)$ where $\alpha=1,2$ transforms in the fundamental of $\mathrm{SL}(2)$, and $s$ stands, appropriately as we will see, for ``singlet'' or ``string''\footnote{The reducibility of the coordinate representation is not a feature of higher rank duality groups.}.
\end{itemize}
The fields and symmetry transformation parameters of the theory can in principle depend on all of these coordinates. However, as always happens in exceptional field theory and double field theory, there is a consistency condition which reduces the dependence on the extended coordinates. This condition is usually implemented as the section condition, which directly imposes that that the fields cannot depend on all extended coordinates. In our case, it takes the form
\begin{equation}
\partial_\alpha \otimes \partial_s = 0 \,,
\label{eq:constraint}
\end{equation}
with the derivatives to be thought of as acting on any field or pair of fields, so that we require both $\partial_\alpha \partial_s \Phi =0$ and $\partial_\alpha \Phi \partial_s \Psi + \partial_\alpha \Psi \partial_s \Phi = 0$. The origin of the section condition is the requirement that the algebra of symmetries closes, which we will review in the next subsection.

The field content of the theory is as follows. The metric-like degrees of freedom are
\begin{itemize}
\item an ``external'' metric, $g_{\mu \nu}$, 
\item a generalized metric, $\gM_{MN}$ which parametrizes the coset $(\G)/\mathrm{SO}(2)$. (From the perspective of the ``external'' nine dimensions, this metric will correspond to the scalar degrees of freedom.) The reducibility of the $Y^M$ coordinates implies that the generalized metric is reducible and thus may be decomposed as, \begin{equation} \gM_{MN}=\gM_{\alpha \beta} \oplus \gM_{ss} \, . \end{equation}
\end{itemize}
The coset $(\G)/\mathrm{SO}(2)$ implies we have just three degrees of freedom described by the generalized metric. This means that $\gM_{ss}$ must be related to $\det \gM_{\alpha \beta}$. One can thus define $\gM_{\alpha \beta}$ such that
\begin{equation}
\cH_{\alpha \beta} \equiv (\gM_{ss})^{3/4} \gM_{\alpha \beta} \,, \label{eq:unitdetgenmetric}
\end{equation}
has unit determinant. The rescaled metric $\cH_{\alpha \beta}$ and $\gM_{ss}$ can then be used as the independent degrees of freedom when constructing the theory. This unit determinant matrix $\cH_{\alpha \beta}$ will then appear naturally in the IIB/F-theory description. 

In addition, we have a hierarchy of gauge fields, similar to the tensor hierarchy of gauged supergravities \cite{deWit:2005hv,deWit:2008ta}. These are form fields with respect to the external directions, and also transform in different representations of the duality group:
\begin{equation}
\begin{array}{|c|ll|ll|} \hline
\mathrm{Representation} & & \mathrm{Gauge\ potential} & & \mathrm{Field}\,\, \mathrm{strength} \\ \hline
\mathbf{2}_1 \oplus \mathbf{1}_{-1} & & \Aa_\mu{}^M & & \Fa_{\mu \nu}{}^M \\
\mathbf{2}_0 & & \Ab_{\mu \nu}{}^{\alpha s} & & \Fb_{\mu \nu \rho}{}^{\alpha s} \\ 
\mathbf{1}_1 & & \Ac_{\mu \nu \rho}{}^{[\alpha \beta] s} & & \Fc_{\mu \nu \rho \sigma}{}^{[\alpha \beta] s} \\
\mathbf{1}_0 & & \Ad_{\mu \nu \rho \sigma}{}^{[\alpha \beta] ss } & & \Fd_{\mu \nu \rho \sigma \lambda}{}^{[\alpha \beta]ss} \\
\mathbf{2}_1 & & \Ae_{\mu \nu \rho \sigma \kappa}{}^{\gamma [\alpha \beta] ss } & & \Fe_{\mu \nu \rho \sigma \kappa\lambda}{}^{\gamma [\alpha \beta]ss} \\
\mathbf{2}_0 \oplus \mathbf{1}_2 & & \Af_{\mu \nu \rho \sigma \kappa \lambda}{}^{M} & & \mathrm{not}\,\,\mathrm{needed} \\  \hline
\end{array}
\label{eq:forms}
\end{equation}
These fields also transform under ``generalized diffeomorphisms'' and ``external diffeomorphisms'' which we describe in the next subsection, as well as various gauge symmetries of the tensor hierarchy which we describe in Appendix \ref{sec:tensorhier}. The field strengths are defined such that the fields transform covariantly under generalized diffeomorphisms, \ie according to their index structure and the rules given in the following subsection, and are gauge invariant under a hierarchy of interrelated gauge transformations given in detail in \ref{sec:tensorhier}. The expressions for the field strengths are schematically
\begin{equation}
\begin{aligned} 
\mathcal{F}_{\mu \nu}{} & = 2 \partial_{[\mu} A_{\nu]} +  \dots + \hd\Ab_{\1\2}\, \\
\Fb_{\1\2\3} &= 3\D_{[\1}\Ab_{\2\3]} +  \dots + \hd\Ac_{\1\2\3} \,, \\
\Fc_{\1\ldots\1\4} &= 4\D_{[\1}\Ac_{\2\3\4]}  + \dots + \hd\Ad_{\1\ldots\4} \,, \\
\Fd_{\1\ldots\5} &= 5\D_{[\1}\Ad_{\2\ldots\5]} + \dots + \hd\Ae_{\1\ldots\5} \,, \\
\Fe_{\1\ldots\6} &= 6\D_{[\1}\Ae_{\2\ldots\6]}  + \dots + \hd\Af_{\1\ldots\6} \,,
\end{aligned}
\end{equation}
where for the $p$-form field strengths the terms indicated by dots involve quadratic and higher order of field strengths.
We also see that there is always a linear term, shown, of the gauge field at next order, under a particular nilpotent derivative $\hd$ defined in \ref{sec:cartancalc}. The derivative, $\D_\mu$ that appears is a covariant derivative for the generalized diffeomorphisms, as described in the next section.
The detailed definitions of the field strengths are also in \ref{sec:tensorhier}. 

Crucially, the presence of the two kinds of diffeomorphism symmetries may be used to fix the action up to total derivatives. For the details of this calculation we refer to the original work \cite{Berman:2015rcc}. The resulting general form of the action, which is common to all exceptional field theories is given schematically as follows, 
\begin{equation}
S = \int \dd^{9} x \dd^3 Y \sqrt{g} \left( \hR + 
\LL_{skin} 
+ \LL_{gkin} 
+ \frac{1}{\sqrt{g}} \LL_{top} 
+ V 
\right) \,.
\label{eq:S} 
\end{equation}
The constituent parts are (omitting total derivatives\footnote{In the quantum theory the total derivatives are important and EFT will have a natural boundary term given by an $\Edd$ covariantization of the Gibbons-Hawking term \cite{Berman:2011kg}.}):
\begin{itemize}
\item the ``covariantized'' external Ricci scalar, $\hR$, which is 
\begin{equation}
\begin{aligned}
\hR &= 
\frac{1}{4} g^{\mu \nu} \D_\mu g_{\rho \sigma} \D_\nu g^{\rho \sigma} 
	- \frac{1}{2} g^{\mu \nu} \D_\mu g^{\rho \sigma} \D_\rho g_{\nu \sigma} \\
&\qquad + \frac{1}{4} g^{\mu \nu} \D_\mu \ln g \D_\nu \ln g 
	+ \frac{1}{2}  \D_\mu \ln g \D_\nu g^{\mu \nu} \, ,	
\end{aligned}
\end{equation} 
\item a kinetic term for the generalized metric (containing the scalar degrees of freedom)
\begin{equation}
\mathcal{L}_{skin}
= -\frac{7}{32} g^{\mu \nu} \D_\mu \ln \gM_{ss} \D_\nu \ln \gM_{ss} 
+ \frac{1}{4}  g^{\mu \nu} \D_\mu \cH_{\alpha \beta} \D_\nu \cH^{\alpha \beta} \, ,
\end{equation}

\item kinetic terms for the gauge fields
\begin{equation}
\begin{aligned}
\mathcal{L}_{gkin} & = 
- \frac{1}{2\cdot 2!} \gM_{MN} \mathcal{F}_{\mu \nu}{}^M \mathcal{F}^{\mu \nu N} 
 - \frac{1}{2\cdot 3!} \gM_{\alpha \beta} \gM_{ss} \mathcal{H}_{\mu \nu \rho}{}^{\alpha s} \mathcal{H}^{\mu \nu \rho \beta s} 
\\ & \qquad - \frac{1}{2\cdot 2! \cdot 4!} \gM_{ss} \gM_{\alpha \gamma} \gM_{\beta \delta} \mathcal{J}_{\mu \nu \rho \sigma}{}^{[\alpha \beta] s} \mathcal{J}^{\mu \nu \rho \sigma [\gamma \delta]s} 
\,.
\end{aligned}
\end{equation}
We do not include kinetic terms for all the form fields appearing in \eqref{eq:forms}. As a result, not all the forms are dynamical. We will discuss the consequences of this below. 
\item a topological or Chern-Simons like term which is not manifestly gauge invariant in 9+3 dimensions. In a standard manner however we may write this term in a manifestly gauge invariant manner in 10+3 dimensions as
\begin{equation}
 \begin{aligned}
  S_{top} &= \kappa \int d^{10}x\, d^3Y\, \varepsilon^{\1\ldots\mt}  \frac{1}{4}  \epsilon_{\alpha \beta} \epsilon_{\gamma \delta}  \left[ 
  \frac{1}{5}  \partial_s \Fd_{\mu_1 \dots \mu_5}{}^{\alpha \beta ss} \Fd_{\mu_6 \dots \mu_{10}}{}^{\gamma \delta ss} \right. \\ 
  & \qquad \left. - \frac{5}{2} \Fa_{\mu_1 \mu_2}{}^s \Fc_{\3\ldots\6}{}^{\alpha \beta s} \Fc_{\7\ldots\mt}{}^{\gamma \delta}  \right. \\
    & \qquad \left. + \frac{10}{3}2 \Fb_{\mu_1 \dots \mu_3}{}^{\alpha s}\Fb_{\4\ldots\6}{}^{\beta s} \Fc_{\7 \ldots \mt}{}^{\gamma \delta}  \right] \,.
 \end{aligned}\label{eq:ToptTerm}
\end{equation}
The index $\mu$ is treated to an abuse of notation where it is simultaneously ten- and nine-dimensional. (This extra dimension is purely a notational convenience and is unrelated to the extra coordinates present in $Y^M$.) The above term is such that its variation is a total derivative and so can be written again in the correct number of dimensions. For further discussion of this term, including an ``index-free'' description, see Appendix \ref{sec:topterm}. The overall coefficient $\kappa$ is found to be $\kappa = +\frac{1}{5! \cdot 48}$. % PLUS
\item a scalar potential
\begin{align}
V & = 
\frac{1}{4} \gM^{ss}\left(\partial_s \cH^{\alpha \beta} \partial_s \cH_{\alpha \beta} 
 	+ \partial_s g^{\mu \nu} \partial_s g_{\mu \nu} 
 	+ \partial_s \ln g \partial_s \ln g \right) \notag \\ 
&\quad  +  \frac{9}{32}  \gM^{ss} \partial_s \ln \gM_{ss} \partial_s \ln \gM_{ss}
	- \frac{1}{2}  \gM^{ss} \partial_s \ln \gM_{ss} \partial_s \ln g \notag \\ 
&\quad   + \gM_{ss}^{3/4} \Bigg[ 
	\frac{1}{4} \cH^{\alpha \beta} \partial_\alpha \cH^{\gamma \delta} \partial_\beta \cH_{\gamma \delta} 
	+ \frac{1}{2} \cH^{\alpha \beta} \partial_\alpha \cH^{\gamma \delta} \partial_\gamma \cH_{\delta \beta} \\
&\hspace{2cm} + \partial_\alpha \cH^{\alpha \beta} \partial_\beta \ln \left( g^{1/2} \gM_{ss}^{3/4} \right)  
	+ \frac{1}{4} \cH^{\alpha \beta} \left( \vphantom{\frac{1}{4}}\partial_\alpha g^{\mu \nu} 
		\partial_\beta g_{\mu \nu} \right. \notag \\
&\hspace{2cm} \left. + \partial_\alpha \ln g \partial_\beta \ln g 
	+ \frac{1}{4} \partial_\alpha \ln \gM_{ss} \partial_\beta \ln \gM_{ss} 
	+ \frac{1}{2} \partial_\alpha \ln g \partial_\beta \ln \gM_{ss} \right) 
\Bigg] \,. \notag
\end{align}
\end{itemize}

This theory expresses the dynamics of eleven-dimensional supergravity and ten-di-mensional Type IIB supergravity in a duality covariant way. In order to do so, we have actually increased the numbers of degrees of freedom by simultaneously treating fields and their electromagnetic duals on the same footing. This can be seen in the collection of form fields in \eqref{eq:forms}. For instance, although eleven-dimensional supergravity contains only a three-form, here we have additional higher rank forms which can be thought of as corresponding to the six-form field dual to the three-form. 

The action for the theory deals with this by not including kinetic terms for all the gauge fields. The field strength $\Fd_{\mu\nu\rho\sigma\kappa}$ of the gauge field $D_{\mu \nu \rho \sigma}$ only appears in the topological term \eqref{eq:ToptTerm}. The field $D_{\mu\nu\rho\sigma}$ in fact also appears in the definition of the gauge field $\Fc_{\mu \nu \rho}$, under a $\partial_M$ derivative. One can show that the equation of motion for this field is 
\begin{equation}
\partial_s \left( \frac{\kappa}{2} \epsilon^{\mu_1 \dots \mu_9} \epsilon_{\alpha \beta} 
\epsilon_{\gamma \delta} \Fd_{\mu_5 \dots \mu_9}{}^{\gamma \delta ss} 
-  e \frac{1}{48} \gM_{ss}  \gM_{\alpha \gamma} \gM_{\beta \delta} \Fc^{\mu_1 \dots \mu_4 \gamma \delta s} \right) = 0  \,.
\label{eq:Deom} 
\end{equation}
The expression in the brackets should be imposed as a duality relation relating the field strength $\Fd_{\mu \nu \rho \sigma \lambda}$ to $\Fc_{\mu \nu \rho \sigma}$, and hence removing seemingly extra degrees of freedom carried in the gauge fields which are actually just the dualizations of physical degrees of freedom. The above relation is quite important -- for instance the proof that the EFT action is invariant under diffeomorphisms is only obeyed if it is satisfied. 

As for the remaining two gauge fields, the equation of motion following from varying with respect to $E_{\mu \nu \rho \sigma \kappa}$ is trivially satisfied (it only appears in the field strength $\Fd_{\mu \nu \rho \sigma \kappa}$), while $F_{\mu \nu \rho \sigma \kappa \lambda}$ is entirely absent from the action.

\subsection{Local and Global Symmetries} 

It is clear that the above action has a manifest invariance under a global $\G$ symmetry, acting on the indices $\alpha,s$ in an obvious way. In addition, the exceptional field theory is invariant under a set of local symmetries.

Alongside the introduction of the extended coordinates $Y^M$ one constructs so called ``generalized diffeomorphisms". In the higher rank groups, these give a unified description of ordinary diffeomorphisms together with the $p$-form gauge transformations. Although the group $\G$ is too small for the $p$-form gauge transformations to play a role here, the generalized diffeomorphisms provide a combined description of part of the ordinary local symmetries of Type IIB and eleven-dimensional supergravity.

The generalized diffeomorphisms, generated by a generalized vector $\Lambda^M$, act as a local $\G$ action, called the generalized Lie derivative $\gL_\Lambda$. These act on a vector, $V^M$ of weight $\lambda_V$ in a form which looks like the usual Lie derivative plus a modification involving the Y-tensor (\cf Section \ref{sec:EEG})
\begin{equation}
\delta_\Lambda V^M \equiv
\mathcal{L}_\Lambda V^M 
= \Lambda^N \partial_N V^M 
- V^N \partial_N \Lambda^M 
+ Y^{MN}{}_{PQ} \partial_N \Lambda^P V^Q 
+ ( \lambda_V + \omega) \partial_N \Lambda^N V^M \,.
\label{eq:gld}
\end{equation}
This modification is universal for all exceptional field theory generalized Lie derivatives \cite{Berman:2012vc} and is built from the invariant tensors of the duality group. In the case of $\G$ \cite{Wang:2015hca}, it is symmetric on both upper and lower indices and has the only non-vanishing components
\begin{equation}
Y^{\alpha s}{}_{\beta s} = \delta^\alpha_\beta \, .
\label{eq:Y}
\end{equation}
There is also a universal weight term, $+ \omega \partial_N \Lambda^N V^M$. The constant $\omega$ depends on the number $n=11-D$ of external dimensions as $\omega = - \frac{1}{n-2}$ and for us $\omega = - 1/7$. The gauge parameters themselves are chosen to have specific weight $\lambda_\Lambda = 1/7$, which cancels that arising from the $\omega$ term. 

In conventional geometry, diffeomorphisms are generated by the Lie derivative and form a closed algebra under the Lie bracket. 
The algebra of generalized diffeomorphisms involves the $E$-bracket,
\begin{equation}
[ U,V]_E = \frac{1}{2} \left( \mathcal{L}_U V - \mathcal{L}_V U \right) \,.
\end{equation}
The condition for closure of the algebra is
\begin{equation}
\mathcal{L}_U \mathcal{L}_V - 
\mathcal{L}_V \mathcal{L}_U 
= \mathcal{L}_{[U,V]_E}  
\end{equation}
which does not happen automatically. A universal feature in all exceptional field theories is that we need to impose the section condition \cite{Berman:2012vc} so the algebra closes. This is the following constraint determined by the Y-tensor
\begin{equation}
Y^{MN}{}_{PQ} \partial_M \otimes \partial_N = 0 \,,
\end{equation}
which implies the form \eqref{eq:constraint} given before.

From the definition of the generalized Lie derivative \eqref{eq:gld} and the Y-tensor \eqref{eq:Y}, we can write down the transformation rules for the components $V^\alpha$ and $V^s$, which are 
\begin{equation}
 \begin{aligned}
  \gL_\Lambda V^\alpha &= \Lambda^M \partial_M V^\alpha - V^\beta \partial_\beta \Lambda^\alpha - \frac17 V^\alpha \partial_\beta \Lambda^\beta + \frac67 V^\alpha \partial_\beta \Lambda^\beta \,, \\
  \gL_\Lambda V^s &= \Lambda^M \partial_M V^s + \frac67 V^s \partial_\beta \Lambda^\beta - \frac87 V^s \partial_s \Lambda^s \,.
 \end{aligned}
\label{eq:gldcpts} 
\end{equation}
Then by requiring the Leibniz property for the generalized Lie derivative, we can derive the transformation rules for tensors in other representations of $\G$, such as the generalized metric $\gM_{MN}$. (The form fields must be treated separately, see Appendix \ref{sec:tensorhier}.)

In doing so, we also need to specify the weight $\lambda$ of each object. It is conventional to choose the generalized metric to have weight zero under generalized diffeomorphisms. Meanwhile, the sequence of form fields $\Aa, \Ab, \Ac,\dots$ are chosen to have weights $\lambda_\Aa = 1/7$, $\lambda_{\Ab} = 2/7$, $\lambda_{\Ac} = 3/7$ and so on. Finally, we take the external metric $g_{\mu \nu}$ to be a scalar of weight $2/7$. 

In the above we have only treated the infinitesimal, local $\G$ symmetry. This should be related to finite $\G$ transformations by exponentiation. The relation between the exponentiated generalized Lie derivative and the finite transformations are quite nontrivial due to the presence of the section condition. For double field theory there are now are series of works dealing with this issue \cite{Hohm:2012gk,Park:2013mpa, Berman:2014jba,Hull:2014mxa,Naseer:2015tia,Rey:2015mba} and recently the EFT case has been studied in \cite{Chaemjumrus:2015vap}.

The other diffeomorphism symmetry of the action consists of external diffeomorphisms, parametrized by vectors $\xi^\mu$. These are given by the usual Lie derivative
\begin{equation}
\delta_\xi V^\mu \equiv L_\xi V^\mu 
= \xi^\nu \D_\nu V^\mu - V^\nu \D_\nu \xi^\mu 
+ \hat\lambda_V \D_\nu \xi^\nu V^\mu \,,
\end{equation}
with partial derivatives replaced by the derivative $\D_\mu$ which is covariant under internal diffeomorphisms, and explicitly defined by
\begin{equation}
\D_\mu = \partial_\mu - \delta_{\Aa_\mu} \,.
\label{eq:Dmu} 
\end{equation}
The weight $\hat\lambda_V$ of a vector under external diffeomorphisms is independent of that under generalized diffeomorphisms.

For this to work, the gauge vector $\Aa_\mu$ must transform under generalized diffeomorphisms as
\begin{equation}
\delta_\Lambda \Aa_\mu{}^M= \D_\mu \Lambda^M\,.
\label{eq:deltaLambdaAa} 
\end{equation}
The external metrics and form fields then transform under external diffeomorphisms in the usual manner given by the Leibniz rule, while the generalized metric is taken to be a scalar, $\delta_\xi \gM_{MN} = \xi^\mu \D_\mu \gM_{MN}$. 

In addition, we need to consider the gauge transformations of the remaining form fields \eqref{eq:forms}. Deriving the correct gauge transformations and field strengths is a non-trivial exercise, and we defer the presentation to Appendix \ref{sec:tensorhier}. 

Each individual term in the general form of the action \eqref{eq:S} is separately invariant under generalized diffeomorphisms and gauge transformations. The external diffeomorphisms though mix the various terms and so by requiring invariance under these transformations one may then fix the coefficients of the action. 

An alternative derivation of the generalized Lie derivative is the following. 
We consider a general ansatz for the generalized Lie derivative acting on elements in the two-dimensional and singlet representation
\begin{equation}
 \begin{aligned}
  \gL_\Lambda V^\alpha &= \Lambda^M \partial_M V^\alpha - V^\beta \partial_\beta \Lambda^\alpha + a V^\alpha \partial_\beta \Lambda^\beta + b V^\alpha \partial_\beta \Lambda^\beta \,, \\
  \gL_\Lambda V^s &= \Lambda^M \partial_M V^s + c V^s \partial_\beta \Lambda^\beta + d V^s \partial_s \Lambda^s \,.
 \end{aligned}
\end{equation}
We can fix the coefficients $a$, $b$, $c$, $d$ as follows. We require a singlet $\Delta_s$ and $\epsilon_{\alpha\beta}$ to define an invariant, \ie
\begin{equation}
 \gL_\Lambda \left(\epsilon_{\alpha\beta} \Delta_s\right) = 0 \,.
\end{equation}
This property allows us to define the unit-determinant generalized metric \eqref{eq:unitdetgenmetric}. Furthermore, we require that the algebra of generalized Lie derivatives closes subject to a section condition. Requiring this to allow for two inequivalent solutions then fixes the coefficients $a$, $b$, $c$, $d$ above and -- up to a redefinition -- reproduces \eqref{eq:gldcpts}. This definition of the generalized Lie derivative fits in the usual pattern of generalized diffeomorphism in EFT described by the Y-tensor \cite{Berman:2012vc} deformation of the Lie derivative.

\section{Relationship to Supergravity}
\label{sec:supergravity} 

Our $\G$ exceptional field theory is equivalent to eleven-dimensional and ten-dimensional Type IIB supergravity, in a particular splitting inspired by Kaluza-Klein reductions. In this section, we present the details of this split and give the precise relationships between the fields of the exceptional field theory and those of supergravity.\footnote{In general, our actions will have the same relative normalizations as that in the book \cite{Ortin:2004ms} although we use the opposite signature. The procedure that we use is essentially the same as carried out in the exceptional field theory literature, see for instance \cite{Hohm:2013vpa, Baguet:2015xha} for detailed descriptions of the M-theory and Type IIB cases.}

\subsection{Metric Terms} 

Let us consider first the $(n+d)$-dimensional Einstein-Hilbert term
\be
S = \int \dd^{n+d}x \sqrt{G} R \,.
\ee
This discussion applies equally to eleven-dimensional supergravity and Type IIB supergravity. In both cases, $n$ is the number of ``external'' dimensions and $d$ the number of internal. So $n=9$ always but $d=2$ in the eleven-dimensional case and $d=1$ in the IIB case. 

The $(n+d)$-dimensional coordinates $x^{\hmu}$ are split into $x^\mu$, $\mu = 1,\dots,9$ and $y^m$, $m=1,\dots,d$. After splitting the $(n+d)$-dimensional flat coordinates $\ha$ into $n$-dimensional flat coordinates $a$ and $d$-dimensional flat coordinates $\bm$, we write the $(n+d)$-dimensional vielbein as
\be
E^{\ha}{}_{\hmu} = \begin{pmatrix}
 \phi^{\omega/2} e^a{}_\mu & 0 \\
 A_\mu{}^m \phi^{\bm}{}_m & 
 \phi^{\bm}{}_m 
\end{pmatrix} \,.
\ee
Fixing this form of the vielbein breaks the $\mathrm{SO}(1,n+d-1)$ Lorentz symmetry to $\mathrm{SO}(1,n-1) \times \mathrm{SO}(d)$. Note however that we continue to allow the fields of the theory to depend on all the coordinates, so at no point do we carry out a dimensional reduction.

Here we treat $e^a{}_\mu$ as the vielbein for the external metric $g_{\mu \nu}$ and can think of $\phi^{\bm}{}_m$ as the vielbein for the internal metric $\phi_{mn}$. We have also defined $\phi \equiv \det \phi_{mn}$.
The corresponding form of the metric is
\be
G_{\hmu \hnu} = 
\begin{pmatrix} \phi^{\omega} g_{\mu \nu} + A_\mu{}^p A_\nu{}^q \phi_{pq} & A_\mu{}^p \phi_{pn} \\
A_\nu{}^p \phi_{pm} & \phi_{mn} 
\end{pmatrix} \,.
\label{eq:GKK}
\ee
The constant $\omega$ is fixed in order to obtain the Einstein-Hilbert term for the metric $g_{\mu \nu}$ and is the same as in Section \ref{sec:theory}: $\omega=- \frac{1}{n-2}=-1/7$.

Diffeomorphisms $\xi^{\hmu}$ split into internal $\Lambda^m$ and external $\xi^\mu$ transformations. We define covariant derivatives $\D_\mu = \partial_\mu  - \delta_{A_\mu}$ which are covariant with respect to internal diffeomorphisms
\begin{equation}
\begin{aligned}
\D_\mu e^a{}_\nu &= \partial_\mu e^a{}_\nu - A_\mu{}^m \partial_m e^a{}_\nu + \omega \partial_n A_\mu{}^n e^a{}_\nu \,, \\
\D_\mu \phi^{\bm}{}_m &= \partial_\mu \phi^{\bm}{}_m - A_\mu{}^n \partial_n \phi^{\bm}{}_m - \partial_m A_\mu{}^n \phi^{\bm}{}_n \,.
\end{aligned}
\end{equation}
The use of the letter $\D_\mu$ here and also for the covariant derivatives \eqref{eq:Dmu} appearing in the EFT is no accident. Indeed, on solving the section condition, those in EFT reduce exactly to the expressions here.

One can think of $e^a{}_\mu$ as carrying density weight $-\omega$ under internal diffeomorphisms. The Kaluza-Klein vector appears as a connection for internal diffeomorphisms, the field strength is 
\be
F_{\mu \nu}{}^m = 2 \partial_{[\mu} A_{\nu]}{}^m - 2 A_{[\mu|}{}^n \partial_n A_{|\nu]}^m \,.
\ee 
To obtain external diffeomorphisms one must add a compensating Lorentz transformation of the vielbein. The resulting expressions are not covariant with respect to internal diffeomorphisms, but can be improved by adding a field dependent internal transformation (with parameter $-\xi^\nu A_\nu{}^m$) to each transformation rule. This leads to the definition of external diffeomorphisms in our split theory
\begin{equation}
\begin{aligned}
\delta_\xi e^a{}_\mu &= \xi^\nu \D_\nu e^a{}_\mu + \D_\mu \xi^\nu e^a{}_\nu \,, \\
\delta_\xi \phi^{\bm}{}_m &= \xi^\nu \D_\nu \phi^{\bm}{}_m \,, \\
\delta_\xi A_\mu{}^m &= \xi^\nu F_{\nu \mu}{}^m + \phi^{\omega} \phi^{mn} g_{\mu \nu} \partial_n \xi^\mu \,.
\end{aligned}
\end{equation}
It is convenient to also define a derivative 
\be
\D_m e^a{}_\mu = \partial_m e^a{}_\mu + \frac{\omega}{2} e^a{}_\mu  \partial_m \ln \phi \,.
\ee
One can then show that the Einstein-Hilbert term $S = \int \dd^{n+d}x\sqrt{G} R$ becomes
\begin{align}
\int \dd^n x \dd^d y \, \sqrt{|g|} 
\Bigg[ & \hat{R} - \frac{1}{4} \phi^{-\omega} F^{\mu \nu m} F_{\mu \nu m} \notag\\ 
	&	+ \frac{1}{4} g^{\mu \nu} \D_\mu \phi^{mn} \D_\nu \phi_{mn} 
		+ \frac{1}{4} \omega g^{\mu \nu} D_\mu \ln \phi D_\nu \ln \phi \notag \\
	&	-  \omega \left( \D_\mu \D^\mu \ln \phi + \frac{1}{2} \D_\mu \ln g  \D^\mu \ln \phi \right) \notag\\ 
	& 	+ \phi^{\omega} \left(R_{int} ( \phi) + \frac{1}{4} \phi^{mn} \D_m g^{\mu \nu} \D_n g_{\mu \nu} 
		+ \frac{1}{4} \phi^{mn} \D_m \ln g \D_n \ln g \right)\Bigg]  \notag\\
 	- \int \dd^n x \dd^d y &\partial_m  \left( \sqrt{|g|}  \phi^{\omega} \phi^{mn} \D_n \ln g \right)  \,. 
 	\label{eq:EHresult}
\end{align} 
Here $R_{int}(\phi)$ is the object given by the usual formula for the Ricci scalar applied to the internal metric $\phi_{mn}$, using only the $\partial_m$ derivatives (and not involving any determinant $e$ factors). Note that this vanishes when $d=1$ \ie for the IIB splitting. Meanwhile, $\hat{R}$ is the improved Ricci scalar in which all derivatives that appear are $\D_\mu$ rather than $\partial_\mu$.

\subsection{Splitting of eleven-dimensional Supergravity}

Now, we come to eleven-dimensional supergravity. We write the (bosonic) action as
\be
S_{11} = \int d^{11} x \sqrt{G}\left( 
R 
- \frac{1}{48} \hat F^{\hmu \hnu \hrho \hlambda} 
\hat F_{\hmu \hnu \hrho \hlambda} % was 1/12
+ \frac{1}{(144)^2} \frac{1}{\sqrt{G}} \varepsilon^{\hmu_1 \dots \hmu_{11}} % was 1/ 12*216 
\hat F_{\hmu_1 \dots \hmu_4} 
\hat F_{\hmu_5 \dots \hmu_8} 
\hat C_{\hmu_9 \dots \hmu_{11}} 
\right) \,.
\ee
We will slightly adapt our notation here. The index $\alpha = 1,2$ is used to denote internal indices, while we denote the internal components of the metric by $\gamma_{\alpha \beta}$, \ie with respect to the previous section $\phi_{mn} \rightarrow \gamma_{\alpha \beta}$. 

The four-form field strength is as usual
\be
\hat F_{\hmu \hnu \hrho \hlambda} = 4 \partial_{[\hmu} \hat C_{\hnu \hrho \hlambda]} \,.
\ee
The degrees of freedom arising from the metric are then the external metric $g_{\mu \nu}$, the Kaluza-Klein vector $A_\mu{}^\alpha$ with field strength
\be
F_{\mu \nu}{}^\alpha = 2 \partial_{[\mu} A_{\nu]}{}^\alpha - 2 A_{[\mu|}{}^\beta \partial_\beta A_{|\nu]}^\alpha \,,
\ee
and the internal metric $\gamma_{\alpha \beta}$. 

The three-form field gives $n$-dimensional forms $\hat C_{\mu \nu \rho}$, $\hat C_{\mu \nu \alpha}$ and $\hat C_{\mu \alpha \beta}$. In order to obtain fields which have better transformation properties under the symmetries in the split (both diffeomorphisms and gauge transformations), one redefines the form field components by flattening indices with $E_{\ha}{}^{\hmu}$ and then curving them with $E^{\ha}{}_\mu$, so that for instance $\bA_{\mu \alpha \beta} \equiv E^{\ha}{}_\mu E_{\ha}{}^{\hmu} \hat{C}_{\hmu \alpha \beta}$,
\be
\begin{split}
\bA_{\mu \alpha \beta } & = \hat C_{\mu \alpha \beta } \,,\\ 
\bB_{\mu \nu \alpha} & = \hat C_{\mu \nu \alpha} - 2 A_{[\mu}{}^\beta \hat C_{\nu] \alpha \beta } \,,\\ 
\bC_{\mu \nu \rho } & = \hat C_{\mu\nu \rho} 
 - 3 A_{[\mu}{}^\alpha \hat C_{\nu \rho] \alpha} 
+ 3 A_{[\mu}{}^\alpha A_{\nu}{}^\beta \hat C_{\rho] \alpha \beta  } \,.
\label{eq:11dABC}
\end{split}
\ee
The fields defined in this way are such that they transform according to their index structure under internal diffeomorphisms $\Lambda^m$ acting as the Lie derivative. 
The field strengths may be similarly redefined
\be
\begin{split}
\bF_{\mu \nu \alpha \beta } & = 2 \D_{[\mu } \bA_{\nu ] \alpha \beta} +2 \partial_{[\alpha} \bB_{|\mu \nu| \beta ]}\,,\\
\bF_{\mu \nu \rho \alpha} & = 3 \D_{[\mu} \bB_{\nu \rho] \alpha} + 3 F_{[\mu \nu}{}^\beta \bA_{\rho] \alpha \beta } 
- \partial_\alpha \bC_{\mu \nu \rho} \,,\\
\bF_{\mu \nu \rho \sigma} & = 4 \D_{[\mu } \bC_{\nu \rho \sigma]} 
+ 6 F_{[\mu \nu}{}^\alpha \bB_{\rho \sigma] \alpha } \,.
\end{split} 
\ee
Here we have the covariant derivative $\D_\mu = \partial_\mu - \delta_{A_\mu}$ introduced above. 

The kinetic terms for the gauge fields may be easily decomposed by going to flat indices and then using the above redefinitions. 
Including the Kaluza-Klein vector, one finds the total gauge kinetic terms
\be
\begin{split}
- \frac{1}{4} &\gamma^{1/7} \gamma_{\alpha \beta} F_{\mu \nu}{}^{ \alpha} F^{\mu \nu \beta}
	- \frac{1}{8} \gamma^{1/7} \gamma^{\alpha \beta} \gamma^{\gamma \delta} \bF_{\mu \nu \alpha \gamma} 
		\bF^{\mu \nu}{}_{\beta \delta} \\
&\qquad 
	- \frac{1}{12} \gamma^{2/7} \gamma^{\alpha \beta} \bF_{\mu \nu \rho}{}_{\alpha} \bF_{\mu \nu \rho \beta} 
	-\frac{1}{48} \gamma^{3/7} \bF^{\mu \nu \rho \sigma} \bF_{\mu \nu \rho \sigma} \, . 
\end{split}
\label{eq:11dgkin}
\ee
Finally, consider the Chern-Simons term. A very convenient way of treating this reduction is to use the trick of rewriting the Chern-Simons term as a manifestly gauge invariant term in one dimension higher (note this fictitious extra dimension has nothing to do with the extra coordinate introduced in EFT). Thus one writes
\be
S_{CS} = -  % \frac{1}{4 \cdot 12\cdot 216}
\frac{1}{4\cdot (144)^2} 
\int \dd^{12}x \varepsilon^{ \hmu_1 \dots \hmu_{12} } 
\hat F_{\hmu_1 \dots \hmu_4} 
\hat F_{\hmu_5 \dots \hmu_8} 
\hat F_{\hmu_9 \dots \hmu_{12}} \,.
\label{eq:MCSlift}  
\ee
The variation of this is
\be
%\frac{3}{12 \cdot 216} 
\frac{3}{(144)^2}
 \int \dd^{12}x
\partial_{\hmu_{12}} 
\left(
\varepsilon^{ \hmu_1 \dots \hmu_{12} } 
\hat F_{\hmu_1 \dots \hmu_4} 
\hat F_{\hmu_5 \dots \hmu_8} 
\delta \hat C_{\hmu_9 \dots \hmu_{11}}
\right) \,,
\ee
which is a total derivative as expected. This term can be decomposed according to the above splitting. One obtains
\be
\begin{split}
S_{CS} = 
- %\frac{1}{8} \frac{1}{216} 
\frac{1}{8 \cdot 12 \cdot 144} 
\int \dd^{10}x \dd^2 y \varepsilon^{\mu_1 \dots \mu_{10}} \varepsilon^{\alpha \beta} \Big(
&	3 \bF_{\alpha \beta \mu_1 \mu_2} 
		\bF_{\mu_3 \dots \mu_6} 
					\bF_{\mu_7 \dots \mu_{10}} \\
& - 8\bF_{\alpha \mu_1 \mu_2\mu_3} 
		\bF_{\beta \mu_4 \mu_5\mu_6} 
		\bF_{\mu_7 \dots \mu_{10}} \Big) \,.	
\end{split}
\label{eq:MCStermsplit} 
\ee

\subsection{The EFT/M-theory Dictionary}
\label{sec:reductionM}

Now, we take our $\G$ EFT described in section \ref{sec:theory} and impose the M-theory section condition, $\partial_s = 0$. Thus, the fields of our theory depend on the coordinates $x^\mu$ and $y^\alpha$, which are taken to be the coordinates of eleven-dimensional supergravity in the $9+2$ splitting described above. 

The metric-like degrees of freedom are easily identified. The external metric $g_{\mu \nu}$ used in the $\G$ EFT is simply that appearing in the splitting of eleven-dimensional supergravity. Meanwhile, the generalized Lie derivative tells us how the generalized metric should be parametrized in spacetime, by interpreting the transformation rules it gives in the M-theory section in terms of internal diffeomorphisms. For instance, one sees from \eqref{eq:gldcpts} that with $\partial_s = 0$
\be
\begin{split} 
\delta_\Lambda \gM_{\alpha \beta} & = \Lambda^\gamma \partial_\gamma \gM_{\alpha \beta} + \partial_\alpha \Lambda^\gamma \gM_{\gamma \beta} 
+ \partial_\beta \Lambda^\gamma \gM_{\alpha \gamma} 
+ \frac{2}{7} \partial_\gamma \Lambda^\gamma \gM_{\alpha \beta} \,,\\ 
\delta_\Lambda \gM_{ss} & = \Lambda^\gamma \partial_\gamma \gM_{ss} - \frac{12}{7} \partial_\gamma \Lambda^\gamma \gM_{ss} \,.
\end{split} 
\ee
which tells us that $\MM_{\alpha \beta}$ transforms as a rank two tensor of weight $2/7$ under internal diffeomorphisms while $\MM_{ss}$ is a scalar of weight $-12/7$, so that
\begin{align}
\MM_{\alpha\beta} &= \gamma^{1/7}\gamma_{\alpha\beta} \, , &
\MM_{ss} &= \gamma^{-6/7} \, .
\label{eq:genmetricM}
\end{align}
It is straightforward to check that after inserting this into the EFT action that one obtains the correct action resulting from the eleven-dimensional Einstein-Hilbert term, given by \eqref{eq:EHresult}. This verification involves the scalar potential, the scalar kinetic terms and the external Ricci scalar. The part of this calculation involving the external derivatives $\D_\mu$ works almost automatically after identifying the Kaluza-Klein vector $A_\mu{}^\alpha$ with the $\alpha$ component of the vector $A_\mu{}^M$, such that the derivatives $\D_\mu$ used in the EFT become those defined above in the splitting of supergravity. 

Now we come to the form fields of EFT. The ones that appear with kinetic terms in the action are $A_\mu{}^M$, $B_{\mu \nu}{}^{\alpha \beta s}$, $C_{\mu \nu \rho}{}^{\alpha \beta s s}$. These are related to the Kaluza-Klein vector $A_\mu{}^\alpha$ and the redefined fields \eqref{eq:11dABC} by the following 
\be
\begin{split} 
A_\mu{}^s & \equiv  \frac{1}{2} \epsilon^{\alpha \beta} \bA_{\mu \alpha \beta} 
\\ 
B_{\mu \nu}{}^{\alpha s} & \equiv  \epsilon^{\alpha \beta}  \bB_{\mu \nu \beta} -  \frac{1}{2} \epsilon^{\gamma \delta} A_{[\mu}{}^\alpha \bA_{\nu] \gamma \delta} 
\\ 
C_{\mu \nu \rho}{}^{\alpha \beta s} & \equiv  \epsilon^{\alpha \beta} \bC_{\mu \nu \rho} 
- 2 \frac{1}{2} \epsilon^{\gamma \delta} A_{[\mu}{}^\alpha A_\nu{}^\beta \bA_{\rho] \gamma \delta}
\end{split}
\label{eq:EFTABC}
\ee
These definitions are such that the field strength components are
\be
\begin{split} 
\Fa_{\mu \nu}{}^\alpha & = F_{\mu \nu}{}^\alpha \,,\\ 
\Fa_{\mu \nu}{}^s & =  \frac{1}{2}  \epsilon^{\alpha \beta}  \bF_{\mu \nu \alpha \beta} \,, \\
\Fb_{\mu \nu \rho}{}^{\alpha s} & =  \epsilon^{\alpha \beta} \bF_{\mu \nu \rho \beta} \,,\\ 
\Fc_{\mu \nu \rho \sigma}{}^{\alpha \beta s} & =  \epsilon^{\alpha \beta} \bF_{\mu \nu \rho \sigma} \,.
\end{split} 
\ee
We can then straightforwardly write down the gauge kinetic terms of the EFT action, which in this section and with the parametrization \eqref{eq:genmetricM} of the generalized metric are given by
\be
\begin{split}
 - \frac{1}{4} &\gamma^{1/7} \gamma_{\alpha \beta} \Fa_{\mu \nu}{}^\alpha \Fa^{\mu \nu \beta} 
- \frac{1}{4} \gamma^{1/7} \gamma^{-1} \Fa_{\mu \nu}{}^s \Fa^{\mu \nu s} 
- \frac{1}{12} \gamma^{2/7} \gamma^{-1} \gamma_{\alpha \beta} \Fb_{\mu \nu \rho}{}^{\alpha s} \Fb^{\mu \nu \rho \beta s} \\ &
- \frac{1}{4 \cdot 24} \gamma^{3/7} \gamma^{-1} \gamma_{\alpha \gamma} \gamma_{\beta \delta} \Fc_{\mu \nu \rho \sigma}{}^{[\alpha \beta ] s} 
\Fc^{\mu \nu \rho \sigma [ \gamma \delta ] s} \,,
\end{split} 
\ee
and show that this automatically reduces to those of eleven-dimensional supergravity, \eqref{eq:11dgkin}. 

Similarly, one can show that we obtain the correct Chern-Simons term \eqref{eq:MCStermsplit}. The remaining gauge fields that appear in EFT, which are a four-form, five-form and six-form are not dynamical. The action must always be complemented by a self-duality relation that relates $p$-form field strengths to their magnetic duals. This is a natural consequence of the formalism where we have included both electric and magnetic descriptions in the action simultaneously.

\subsection{Splitting of ten-dimensional Type IIB Supergravity} 

The (bosonic) (pseudo-)action of Type IIB can be written as
\be
\begin{split}
S & = \int \dd^{10} x \sqrt{G} \Big(R 
	+ \frac{1}{4} G^{\hmu \hnu} \partial_{\hmu} \cH_{\alpha \beta} \partial_{\hmu} \cH^{\alpha \beta} 
	- \frac{1}{12} \cH_{\alpha \beta} \hat F_{\hmu \hnu \rho}{}^{\alpha} \hat F^{\hmu \hnu \rho \beta} 
	- \frac{1}{480} \hat F_{\hmu_1 \dots \hmu_5}\hat F^{\hmu_1 \dots \hmu_5} \Big) \\ 
& \hspace{1cm}
	+ \frac{1}{24 \cdot 144} \int \dd^{10}x \epsilon_{\alpha \beta} \epsilon^{\hmu_1 \dots \hmu_{10}} 
		\hat C_{\hmu_1 \hmu_2 \hmu_3 \hmu_4} \hat F_{\hmu_5 \hmu_6\hmu_7}{}^\alpha \hat 
		F_{\hmu_8 \hmu_9 \hmu_{10}}{}^\beta \,.
\end{split} 
\ee
This action must be accompanied by the duality relation for the self-dual five-form
\be
\hat{F}_{\hmu_1 \dots \hmu_5} = \frac{1}{5!} \sqrt{G} \epsilon_{\hmu_1 \dots \hmu_{10}} \hat F^{\hmu_6 \dots \hmu_{10} }
\ee
The field strengths themselves are written as
\be
\hat F_{\hmu \hnu \rho}{}^\alpha = 3 \partial_{[\hmu} \hat C_{\hnu \rho]}{}^\alpha
\ee
and
\be
\hat F_{\hmu_1 \dots \hmu_5} = 5 \partial_{[\hmu_1} \hat C_{\hmu_2 \dots \hmu_5]} 
+ 5  \epsilon_{\alpha \beta} \hat C_{[\hmu_1 \hmu_2}{}^\alpha \hat F_{\hmu_3 \hmu_4 \hmu_5]}{}^{\beta} 
\ee
We carry out a $9+1$ split of the coordinates. In this section, we will denote the internal index by $s$ (for singlet), and the single internal metric component by $\phi_{ss} \equiv \phi$. The metric then gives the external metric, $g_{\mu\nu}$ and the Kaluza-Klein vector, $A_\mu{}^s$ with field strength
\be
F_{\mu \nu}{}^s = 2 \partial_{[\mu} A_{\nu]}{}^s - 2 A_{[\mu}{}^s \partial_s A_{\nu]}{}^s \,.
\ee
The scalars $\cH_{\alpha \beta}$ are trivially reduced using the decomposition of the metric to give
\be
+ \frac{1}{4} g^{\mu \nu} \D_\mu \cH_{\alpha \beta} \D_\nu \cH^{\alpha \beta} 
+ \frac{1}{4} \phi^{-8/7} \partial_s \cH_{\alpha \beta} \partial_s \cH^{\alpha \beta}\,.
\label{eq:Skinred}
\ee
From the two-form, using the standard trick involving contracting with $E^{\ha}{}_\mu E_{\ha}{}^{\hmu}$ to obtain appropriate decompositions of the forms, we find the components
\be
\begin{split}
\bC_{\mu s}{}^{\alpha} & \equiv \hat C_{\mu s}{}^\alpha \,,\\
\bC_{\mu \nu}{}^{\alpha}  & \equiv \hat C_{\mu \nu}{}^\alpha + 2 A_{[\mu}{}^s \hat C_{\nu] s}{}^\alpha \,,
\end{split} 
\ee
so that the field strengths are
\be
\begin{split}
\bar{F}_{\mu \nu s}{}^\alpha & \equiv \hat{F}_{\mu \nu s}{}^\alpha  \\
& = 2 \D_{[\mu} \bC_{\nu] s}{}^\alpha + \partial_s \bC_{\mu \nu}{}^\alpha \,,  \\
\bar{F}_{\mu \nu \rho}{}^\alpha & \equiv \hat F_{\mu \nu \rho}{}^\alpha - 3 A_{[\mu}{}^s \hat{F}_{\nu \rho]s} \\
 & = 3 \D_{[\mu} \bC_{\nu \rho]}{}^\alpha -  3 F_{[\mu \nu}{}^s \bC_{\rho]s}{}^\alpha \,.
\end{split} 
\ee
Note that the two-form doublet $\hat C_{\mu \nu}{}^\alpha$ consists of the NSNS-two-form $B_{\mu \nu}$ as its first component and the RR-two-form $C_{\mu \nu}$ as its second component. For the four-form one has similarly 
\be
\begin{split}
\bC_{\mu \nu \rho s} & \equiv \hat C_{\mu \nu \rho s}\,, \\
\bC_{\mu \nu \rho \sigma} & \equiv \hat C_{\mu \nu \rho \sigma}  + 4 A_{[\mu}{}^s \hat C_{\nu \rho \sigma] s} \,,
\end{split} 
\ee
with field strengths
\be
\begin{split}
\bF_{\mu \nu \rho \sigma s}  & \equiv \hat F_{\mu \nu \rho \sigma s}  \\
 & = 4 \D_{[\mu} \bC_{\nu \rho \sigma]s} + \partial_s \bC_{\mu \nu \rho \sigma} 
+ \epsilon_{\alpha \beta} \left( 2 \bC_{s[\mu}{}^\alpha \bF_{\nu \rho \sigma]}{}^\beta
+ 3 \bC_{[\mu \nu} \bF_{\rho \sigma]z}{}^\beta \right) \,, \\
\bF_{\mu \nu \rho \sigma \lambda}  & \equiv \hat F_{\mu \nu \rho \sigma \lambda }  
-5 A_{[\mu}{}^s \hat F_{\nu \rho \sigma \lambda]s} 
 \\
 & = 5 \D_{[\mu} \bC_{\nu \rho \sigma \lambda]} -20 F_{[\mu \nu}{}^s \bC_{\rho \sigma \lambda]s}
+ 5 \epsilon_{\alpha \beta} 
\bC_{[\mu \nu} \bF_{\rho \sigma \lambda]}{}^\beta \,.
\end{split} 
\ee 
In terms of these objects, the duality relation becomes 
\be
\bF_{\mu \nu \rho \sigma s} = \frac{1}{5!} \phi^{4/7} g^{1/2} \epsilon_{\mu \nu \rho \sigma }{}^{\nu_1 \dots \nu_5} 
\bF_{\nu_1 \dots \nu_5} \,.
\label{eq:IIBstarF}
\ee
These definitions lead to the following kinetic terms in the Lagrangian
\be
\begin{split}   
- \frac{1}{4} &\phi^{8/7} F_{\mu \nu}{}^s F^{\mu \nu s} 
- \frac{1}{4} \cH_{\alpha \beta} \phi^{-6/7} \bF_{\mu \nu s}{}^\alpha \bF^{\mu \nu}{}_s{}^\beta
- \frac{1}{12} \cH_{\alpha \beta} \phi^{2/7} \bF_{\mu \nu \rho}{}^{ \alpha} \bF^{\mu \nu \rho \beta} \\
& 
- \frac{1}{96} \phi^{4/7} \bF_{\mu \nu \rho \sigma s} \bF^{\mu \nu \rho \sigma}{}_s 
- \frac{1}{480} \phi^{4/7} \bF_{\mu_1 \dots \mu_5} \bF^{\mu_1 \dots \mu_5} \,.
\end{split} 
\label{eq:IIBFormAction}
\ee
Finally, consider the Chern-Simons term which can be written in one dimension higher as 
\be
\begin{split} 
- \frac{1}{5 \cdot 24 \cdot 144} 
& \int \dd^{11} x 
\epsilon_{\alpha \beta} \epsilon^{\hmu_1 \dots \hmu_{11}}
\hat F_{\hmu_1 \hmu_2\hmu_3}{}^\alpha 
\hat F_{\hmu_4 \hmu_5 \hmu_{6}}{}^\beta 5
\partial_{\hmu_{11}} \hat C_{\hmu_7 \hmu_8 \hmu_9 \hmu_{10} } 
=\\ & 
- \frac{1}{5\cdot 24 \cdot 144} 
\int \dd^{11} x 
\epsilon_{\alpha \beta} \epsilon^{\hmu_1 \dots \hmu_{11}}
\hat F_{\hmu_1 \hmu_2\hmu_3}{}^\alpha 
\hat F_{\hmu_4 \hmu_5 \hmu_{6}}{}^\beta 
\hat F_{\hmu_7 \hmu_8 \hmu_9 \hmu_{11} } \,.
\end{split}
\ee
Under the split, this becomes
\be
\begin{split}
- \frac{1}{5\cdot 24 \cdot 144} \int \dd^{10} x \dd y \epsilon_{\alpha \beta} \epsilon^{\mu_1 \dots \mu_{10}}
\Big( 
&	6 \bF_{\mu_1 \mu_2 s}{}^\alpha \bF_{\mu_3 \mu_4 \mu_{5}}{}^\beta \bF_{\mu_6 \dots  \mu_{10} } \\
&	+ 5 \bF_{\mu_1 \mu_2 \mu_3}{}^\alpha \bF_{\mu_4 \mu_5 \mu_{6}}{}^\beta \bF_{ \mu_7 \dots\mu_{10} s }
\Big)	 \,.
\end{split}
\label{eq:IIBCSsplit} 
\ee

\subsection{The EFT/Type IIB Dictionary} 
\label{sec:reductionF}

Now, we take our $\G$ EFT and impose the IIB section, $\partial_\alpha = 0$. The fields then depend on the coordinates $x^\mu$ and $y^s$, which become the coordinates of ten-dimensional Type IIB supergravity in the $9+1$ split we have described above. 

The external metric can be immediately identified. The components of the generalized metric are 
\begin{align}
\MM_{\alpha\beta} &= \phi^{-6/7} \cH_{\alpha\beta} \, ,  &
\MM_{ss} &= \phi^{8/7} \, .
\label{eq:genmetricIIB}
\end{align}
The Kaluza-Klein vector $A_\mu{}^s$ can be identified as the $s$ component of the gauge field $A_\mu{}^M$. One can then verify the reduction of the scalar potential, scalar kinetic terms and external Ricci scalar, and verify that they give the expected reduction of the Einstein-Hilbert term, \eqref{eq:EHresult} and scalar kinetic terms \eqref{eq:Skinred}. For completeness let us give here the parametrization of $\cH_{\alpha \beta}$ in terms of $\tau = C_0 + i e^{-\varphi}$: 
\begin{equation}
\cH_{\alpha\beta} = \frac{1}{\tau_2}
\begin{pmatrix}
1 & \tau_1 \\ \tau_1 & |\tau|^2
\end{pmatrix} = e^\varphi
\begin{pmatrix}
1 & C_0 \\ C_0 & C_0^2 + e^{-2\varphi}
\end{pmatrix} \, .
\end{equation}
For the remaining forms, we need the following definitions
\be
\begin{split}
A_\mu{}^\alpha & \equiv \bC_{\mu s}{}^\alpha \,,
\\ 
B_{\mu \nu}{}^{\alpha s} & \equiv  \bC_{\mu \nu}{}^{\alpha} - A_{[\mu}{}^s \bC_{\nu] s}{}^\alpha 
  \,, \\
C_{\mu \nu \rho}{}^{\alpha \beta s} 
& \equiv \epsilon^{\alpha \beta} \bC_{\mu \nu \rho}
+3 \bC_{[\mu|s|}{}^{[\alpha} \bC_{\nu \rho]}{}^{\beta]} 
+ 4 \bC_{[\mu|s|}{}^\alpha \bC_{\nu|s|}{}^\beta A_{\rho]}{}^s \\ 
D_{\mu \nu \rho \sigma}{}^{\alpha \beta ss} & \equiv 
\epsilon^{\alpha \beta} \bC_{\mu \nu \rho \sigma} 
+ 6 \bC_{[\mu \nu}{}^{[\alpha} \bC_{\rho |s|}{}^{\beta]} A_{\sigma]}{}^s \,,
\end{split} 
\ee
such that
\be
\begin{split}
\mathcal{F}_{\mu \nu}{}^{s} & = F_{\mu \nu}{}^s \,, \\ 
\mathcal{F}_{\mu \nu}{}^{\alpha} & = \bF_{\mu \nu s}{}^\alpha \,, \\ 
\Fb_{\mu \nu \rho}{}^{\alpha s} & = \bF_{\mu \nu \rho}{}^\alpha \,, \\ 
\Fc_{\mu \nu \rho \sigma}{}^{\alpha \beta s} & =  \epsilon^{\alpha \beta}  \bF_{\mu \nu \rho \sigma s} \,,\\ 
\Fd_{\mu \nu \rho \sigma \lambda}{}^{\alpha \beta ss} 
& =  \epsilon^{\alpha \beta} \bF_{\mu \nu \rho \sigma \lambda} \,.
\end{split}
\ee
The duality relation that one obtains from the EFT action by varying with respect to $\Delta D_{\mu \nu \rho \sigma}$ is (equation \eqref{eq:Deom}) 
\be
\partial_s \left( \frac{\kappa}{2} \epsilon^{\mu_1 \dots \mu_9} \epsilon_{\alpha \beta} 
\epsilon_{\gamma \delta} \Fd_{\mu_5 \dots \mu_9}{}^{\gamma \delta ss} 
- 2 e \frac{1}{96} \gM_{ss}  \gM_{\alpha \gamma} \gM_{\beta \delta} \Fc^{\mu_1 \dots \mu_4 \gamma \delta s} \right) = 0 
\ee
Here $\kappa$ is the overall coefficient of the topological term \eqref{eq:ToptTerm}. 
After some manipulation one sees that this is consistent with the duality relation imposed in supergravity, given in \eqref{eq:IIBstarF}, if
\be
\kappa = \frac{ 2 }{5!\cdot 96}\,, % PLUS 
\ee
which is satisfied by our coefficients. 

The kinetic terms of the EFT action are in this parametrization given by 
\be
\begin{split}   
- \frac{1}{4} &\phi^{8/7} \Fa_{\mu \nu}{}^s \Fa^{\mu \nu s} 
- \frac{1}{4} \cH_{\alpha \beta} \phi^{-6/7} \Fa_{\mu \nu}{}^\alpha \Fa^{\mu \nu\beta}
- \frac{1}{12} \cH_{\alpha \beta} \phi^{2/7} \Fb_{\mu \nu \rho}{}^{ \alpha s} \Fb^{\mu \nu \rho \beta s} \\
& 
- \frac{1}{96} \phi^{4/7} \cH_{\alpha \gamma} \cH_{\beta \delta} \Fc_{\mu \nu \rho \sigma}{}^{[\alpha \beta]s}  \Fc^{\mu \nu \rho \sigma [ \gamma \delta ] s} \,.
\end{split} 
\ee
Using the above definitions, we find that the first line of this (involving just the Kaluza-Klein vector and the components of the two-form) matches exactly the first line of \eqref{eq:IIBFormAction}. Before discussing the remaining term, we first consider the Chern-Simons term. It is convenient in the IIB section to rewrite this using the Bianchi identities \eqref{eq:Bianchi} given in the appendix. Then the topological term as given in the form \eqref{eq:ToptTermApp} can be written as
\be
\begin{split}
\kappa \int \dd^{10} x d y \epsilon^{\mu_1 \dots \mu_{10}} 
	\frac{1}{4} \epsilon_{\alpha \beta} \epsilon_{\gamma \delta} \Bigg[
&	\D_{\mu_1} \left( \Fc_{\mu_2 \dots \mu_5}{}^{\alpha \beta s} 
		\Fd_{\mu_6 \dots \mu_{10}}{}^{\gamma \delta ss} \right) \\
& 	+ 4  \Fa_{\mu_1 \mu_2}{}^\alpha  \Fb_{\mu_3 \mu_4 \mu_5}{}^{\beta s}
		\Fd_{\mu_6 \dots \mu_{10}}{}^{\gamma \delta ss} \\ 
&	+ \frac{10}{3} \Fb_{\mu_1 \mu_2 \mu_3}{}^{\alpha s} \Fb_{\mu_4 \mu_5 \mu_6}{}^{\beta s} 
		\Fc_{\mu_7 \dots \mu_{10}}{}^{\gamma \delta s} \Bigg] \,.
\end{split} 
\ee
The last two lines here give exactly the Chern-Simons term \eqref{eq:IIBCSsplit}. 

The remaining terms give kinetic terms for the field strength. The terms that one obtains differ from those that one obtains from a decomposition of the Type IIB pseudo-action by a multiplicative factor of $2$, that is the EFT gives 
\be
- \frac{1}{48} \phi^{4/7} \bF_{\mu \nu \rho \sigma s} \bF^{\mu \nu \rho \sigma}{}_s 
- \frac{1}{240} \phi^{4/7} \bF_{\mu_1 \dots \mu_5} \bF^{\mu_1 \dots \mu_5} \,,
\ee
to be compared with the coefficients of $1/96$ and $1/480$ in \eqref{eq:IIBFormAction}. This is as expected due to the use of the self-duality relation which is an equation of motion for the gauge field $D_{\mu \nu \rho \sigma}$. In ten dimensions one has the normalization $\frac{1}{4 \cdot 5!} (F_5)^2$ for the five-form field strength instead of the standard $\frac{1}{2 \cdot 5!}$ due to the fact that there are unphysical degrees of freedom which are eliminated by the self-duality relation after varying the Type IIB action. Here we see only the physical half after using the self-duality relation in the EFT action. The upshot is that strictly speaking the EFT is only equivalent to Type IIB at the level of the equations of motion, as would be expected given that the Type IIB action is only a pseudo-action.

\subsection{Summary} 

The above results display the mapping between the fields of the $\G$ EFT and those of supergravity in a certain Kaluza-Klein-esque split. It is straightforward to relate this back directly to the fields in eleven- and ten-dimensions themselves. 

For M-theory, one has for the degrees of freedom coming from the spacetime metric, with the Kaluza-Klein vector $A_\mu{}^\alpha = \gamma^{\alpha \beta} G_{\mu \beta}$,
\begin{equation}
\begin{split}
\cH_{\alpha\beta} &= \gamma^{-1/2} \gamma_{\alpha\beta} \,, \qquad
{\cal M}_{ss} = \gamma^{-6/7} \,, \\
g_{\mu\nu} & = \gamma^{1/7} \left( G_{\mu \nu} - \gamma_{\alpha \beta} A_\mu{}^\alpha A_\nu{}^\beta \right) \,,   \\
  A_\mu{}^s &= \frac12 \epsilon^{\alpha\beta} \hat{C}_{\mu\alpha\beta} \,, \\
  B_{\mu\nu}{}^{\alpha,s} &= \epsilon^{\alpha\beta} \hat{C}_{\mu\nu\beta} + \frac12 \epsilon^{\beta\gamma} A_{[\mu}{}^\alpha \hat{C}_{\nu]\beta\gamma} \,, \\
  C_{\mu\nu\rho}{}^{\alpha\beta,s} &= \epsilon^{\alpha\beta} \left( \hat{C}_{\mu\nu\rho} - 3 A_{[\mu}{}^{\gamma} \hat{C}_{\nu\rho]\gamma} + 2 A_{[\mu}{}^{\gamma} A_{\nu]}{}^{\delta} \hat{C}_{\rho\gamma\delta} \right) \,.
\end{split} \label{eq:MEFT}
\end{equation}
The inverse relationships, giving the eleven-dimensional fields in terms of those in our EFT are
\begin{equation}
\begin{split}
  \gamma_{\alpha\beta} &= \cH_{\alpha\beta} \left( {\cal M}_{ss} \right)^{-7/12} \,,\qquad
  \gamma = \left( {\cal M}_{ss}\right)^{-7/6} \,, \\
  % G_{\mu \alpha} & = \gamma_{\alpha \beta} A_\mu{}^\beta \,,\\
  G_{\mu\nu} &= g_{\mu\nu} \left( {\cal M}_{ss} \right)^{1/6} + \gamma_{\alpha \beta} A_\mu{}^\alpha A_\nu{}^\beta \,, \\
  \hat{C}_{\mu\alpha\beta} &= A_\mu{}^s \epsilon_{\alpha\beta} \,, \\
  \hat{C}_{\mu\nu\alpha} &= \epsilon_{\alpha\beta} \left( - B_{\mu\nu}{}^{\beta,s} + A_{[\mu}{}^{\beta} A_{\nu]}{}^s \right) \\
  \hat{C}_{\mu\nu\rho} &= \epsilon_{\alpha\beta} \left( \frac12 C_{\mu\nu\rho}{}^{\alpha\beta,s} + A_{[\mu}{}^{\alpha} A_{\nu}{}^{\beta} A_{\rho]}{}^s - 3 A_{[\mu}{}^{\alpha} B_{\nu\rho]}{}^{\beta,s} \right) \,.
\end{split} \label{eq:EFTM}
\end{equation}

Similarly, for IIB we have, with $\phi \equiv \phi_{ss}$ and the Kaluza-Klein vector $A_\mu{}^s = \phi^{-1} G_{\mu s}$
\begin{equation}
 \begin{split}
  g_{\mu\nu} &= \phi^{1/7} \left( G_{\mu\nu} - \phi A_\mu{}^s A_\nu{}^s \right)   \,, \qquad
  {\cal M}_{ss} = \phi^{8/7} \,, \\
  A_\mu{}^\alpha &= \hat{C}_{\mu s}{}^\alpha \,, \\
  B_{\mu\nu}{}^{\alpha, s} &= \hat{C}_{\mu\nu}{}^{\alpha} + A_{[\mu}{}^s \hat{C}_{\nu]s}{}^\alpha \,, \\
  C_{\mu\nu\rho}{}^{\alpha\beta, s} &= \epsilon^{\alpha\beta} \hat{C}_{\mu\nu\rho s} + 3 \hat{C}_{[\mu|s|}{}^{[\alpha} \hat{C}_{\nu\rho]}{}^{\beta]} - 2 \hat{C}_{[\mu|s}{}^{\alpha} \hat{C}_{\nu|s|}{}^{\beta} A_{\rho]}{}^s \,, \\
  D_{\mu\nu\rho\sigma}{}^{\alpha\beta, ss} &= \epsilon^{\alpha\beta} \left( \hat{C}_{\mu\nu\rho\sigma} + 4 A_{[\mu}{}^s \hat{C}_{\nu\rho\sigma]s} \right) + 6 \hat{C}_{[\mu\nu}{}^{[\alpha} \hat{C}_{\rho|s|}{}^{\beta]} A_{\sigma]}{}^s \,,
 \end{split} \label{eq:IIBEFT}
\end{equation}
and 
\begin{equation}
 \begin{split}
  G_{\mu\nu} &=  \left( {\cal M}_{ss} \right)^{-1/8} g_{\mu\nu} + \left(\gM_{ss} \right)^{7/8} A_\mu{}^s A_\nu{}^s \,, \qquad
  \phi = \left( {\cal M}_{ss}\right)^{7/8} \,, \\
  \hat{C}_{\mu s}{}^{\alpha} &= A_\mu{}^\alpha \,, \\
  \hat{C}_{\mu\nu}{}^{\alpha} &= B_{\mu\nu}{}^{\alpha,s} - A_{[\mu}{}^s A_{\nu]}{}^{\alpha} \,, \\
  \hat{C}_{\mu\nu\rho s} &= \frac12 \epsilon_{\alpha\beta} \left(C_{\mu\nu\rho}{}^{\alpha\beta,s} - 3 A_{[\mu}{}^{[\alpha} B_{\nu\rho]}{}^{\beta],s} - A_{[\mu}{}^{\alpha} A_{\nu}{}^{\beta} A_{\rho]}{}^s \right) \,, \\
  \hat{C}_{\mu\nu\rho\sigma} &= \frac12 \epsilon_{\alpha\beta} \left(D_{\mu\nu\rho\sigma}{}^{\alpha\beta,ss} + 6 B_{[\mu\nu}{}^{\alpha,s} A_{\rho}{}^{\beta} A_{\sigma]}{}^s - 4 A_{[\mu}{}^s C_{\nu\rho\sigma]}{}^{\alpha\beta,s} \right) \,.
 \end{split} \label{eq:EFTIIB}
\end{equation}

\section{Relationship to F-theory} 
\label{sec:Frel}

In the previous section, we have given the detailed rules for showing the equivalence of the $\G$ EFT to both eleven-dimensional supergravity and ten-dimensional Type IIB supergravity. Let us now elaborate on the connection to F-theory, rather than just Type IIB supergravity.

What is F-theory? Primarily we will take F-theory to be a twelve-dimensional lift of Type IIB supergravity that provides a geometric perspective on the $SL(2)$ duality symmetry \footnote{We thank Cumrun Vafa for discussions on how one should think of F-theory.}. It provides a framework for describing (non-perturbative) IIB vacua with varying $\tau$, in particular it is natural to think of sevenbrane backgrounds as monodromies of $\tau$ under the action of $SL(2)$. Equivalently, there is a process for deriving non-perturbative IIB vacua from M-theory compactifications to a dimension lower. Crucially, singularities of the twelve-dimensional space are related to D7-branes. We take this duality with M-theory to be the second key property of F-theory.

Let us now briefly comment on how these two properties are realized here before expanding in detail.
We usually view the twelve-dimensional space of F-theory as consisting of a torus fibration of ten-dimensional IIB. The group of large diffeomorphisms on the torus is then viewed as a geometric realization of the $SL(2)$ S-duality of IIB.

In the $\G$ EFT a similar picture arises. This is because we take the group of large generalized diffeomorphisms acting on the extended space to give the $\G$ duality group. See \cite{Hohm:2012gk,Park:2013mpa, Berman:2014jba,Hull:2014mxa,Naseer:2015tia,Rey:2015mba,Chaemjumrus:2015vap} for progress on understanding the geometry of these large generalized diffeomorphisms.

The EFT is subject to a single constraint equation, the section condition, with two inequivalent solutions. One solution of the constraint leads to M-theory or at least eleven-dimensional supergravity, and one leads to F-theory. Thus $\G$ EFT is a single twelve-dimensional theory containing both eleven-dimensional supergravity and F-theory, allowing us to naturally realize the M-theory / F-theory duality.

\begin{figure}[h!]
\begin{center}
\vspace{12pt}
\includegraphics[scale=0.5]{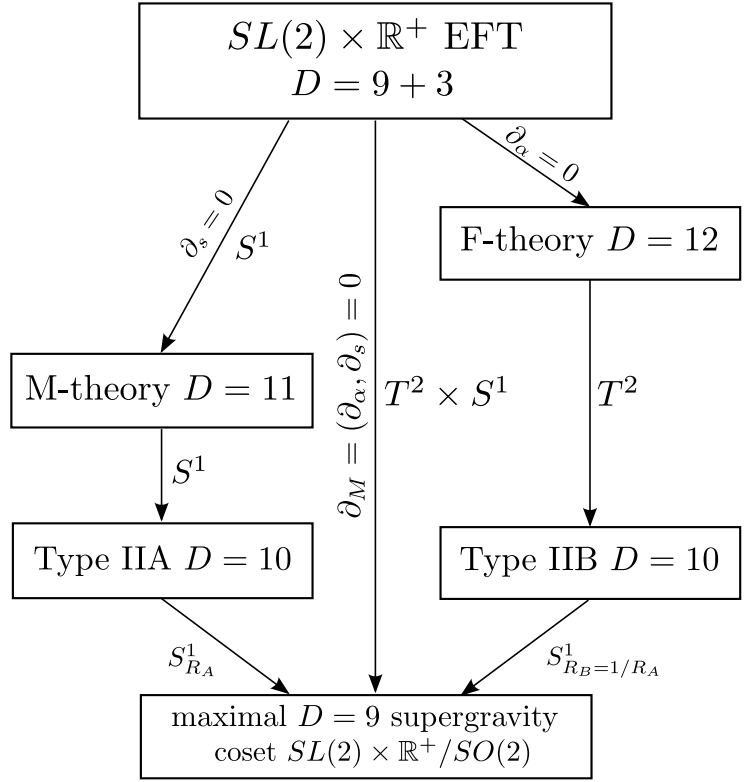}
\end{center}
\caption{The relation of supergravity theories in nine, ten, eleven and twelve dimensions. Note here $D$ denotes the overall dimensionality of the theory.}
\label{fig:SUGRAs}
\end{figure}

If we choose the IIB section, we can interpret any solution as being twelve-dimen-sional but with at least two isometries in the twelve-dimensional space. These two isometries lead to the two-dimensional fibration which in F-theory consists of a torus.

Finally, the fact that the generalized diffeomorphisms, not ordinary diffeomorphisms, play the key role here also allows one to use the section condition to ``dimensionally reduce'' the twelve-dimensional $\G$ to ten-dimensional IIB (as well as eleven-dimensional supergravity) as explained in Section \ref{sec:supergravity}. This explicitly shows how F-theory, interpreted as the $\G$ EFT, can be a twelve-dimensional theory, yet reduce to the correct eleven-dimensional and IIB supergravity fields. The relation between the twelve-dimensional theory and the supergravities in elven, ten and nine dimensions is depicted in Figure \ref{fig:SUGRAs}.

\subsection{M-theory/F-theory Duality} 
\label{sec:MF}

Before looking at the specifics of the $\G$ EFT, let us discuss in general how duality works in exceptional field theories or indeed double field theory. First consider the most familiar case of T-duality and how it comes about in DFT. The origin of T-duality is the ambiguity in identifying the $D$-dimensional spacetime embedded in the $2D$-dimensional doubled space. For the most generic DFT background that obeys the section condition constraint there is no T-duality. The section condition eliminates the dependence of the generalized metric on half the coordinates and so there is unique choice of how one identifies the $D$-dimensional spacetime embedding in the $2D$-doubled space.

However if there is an isometry then there is only a dependence on $D-1$ coordinates. Thus what we identify as the $D$-dimensional spacetime is ambiguous. This ambiguity is T-duality. This was discussed in detail in Section \ref{sec:Tduality}. (Note, even though DFT has a manifest $O(D,D;\mathbb{R})$ local symmetry this should not be confused with the global $O(n,n;\mathbb{Z})$ T-duality which only occurs when compactifying on a $T^n$.)

One has a similar situation in EFT but with some differences. In DFT a solution to the section condition will always provide a $D$-dimensional space. In EFT a solution to the $\Edd$ section condition will provide either a $D$-dimensional space or a $\left(D-1\right)$-dimensional space (where crucially the $D-1$ solution is not a subspace of the $D$-dimensional space). The two solutions are distinct (and not related by any element of $\Edd$). The $D$-dimensional solution is associated to the M-theory description and the $\left(D-1\right)$-dimensional solution is associated to the type IIB description. 

A completely generic solution that solves the section condition will be in one set or the other and one will be able to label it as an M or IIB solution. However, if there are \emph{two} isometries in the M-theory solution then again we have an ambiguity and one will be able interpret the solution in terms of the IIB section with \emph{one} isometry. This ambiguity gives the F-theory/M-theory duality. It is the origin of how M-theory on a torus is equivalent to Type IIB on a circle \cite{Schwarz:1995jq}. Thus in summary the F/M-duality is an ambiguity in the identification of spacetime that occurs when there are two isometries in an M-theory solution.

So the simplest case is where the eleven dimensional space is a $M_9 \times T^2$ and we take the two-torus to have complex structure $\tau$ and volume $V$. The reinterpretation in terms of a IIB section solution (which requires a single circle isometry) is given by
\be
R^{\mathrm{IIB}}_9 = V^{-3/4}
\eeq
where $R^{\mathrm{IIB}}_9$ is the radius of the IIB circle. This exactly recovers the well known M/IIB duality relations \cite{Schwarz:1995jq}.

Let us now further study the M-theory/F-theory duality with the following simple ansatz. Consider the nine-dimensional space to be of the form
\begin{equation}
 \dd s_{(9)}^2 = \dd s_{(1,2)}^2 + \dd s_{(6)}^2 \,,
\end{equation}
where $\dd s_{(6)}^2$ is the metric on some internal six-dimensional manifold $B_6$ and $\dd s_{(1,2)}^2$ is the metric of an effective three-dimensional theory.
We will consider the case where $y^\alpha$ parameterise a torus, as in F-theory. If for simplicity we ignore the one-form gauge potential $\Aa_\mu{}^M$, then using \eqref{eq:EFTM} we see that the M-theory section has the following metric
\begin{equation}
 \dd s_{(11)}^2 = \left({\cal M}_{ss}\right)^{1/6} \dd s_{(1,2)}^2 
 		+ \left[ \left( {\cal M}_{ss} \right)^{1/6} \dd s_{(6)}^2 
 		+ \left({\cal M}_{ss}\right)^{-7/12} {\cal H}_{\alpha\beta} \dd y^\alpha \dd y^\beta \right] \,.
\end{equation}
The internal manifold takes the form of a $T^2$-fibration over $B_6$. In the IIB section we instead have the Einstein-frame metric
\begin{equation}
 \dd s_{(10)}^2 = \left[ \left( {\cal M}_{ss} \right)^{-1/8} \dd s_{(1,2)}^2 
 	+ \left( {\cal M}_{ss} \right)^{7/8} (\dd y^s)^2 \right]
 	+ \left( {\cal M}_{ss} \right)^{-1/8} \dd s_{(6)}^2 \,.
\end{equation}
This is precisely the four-dimensional effective theory with six-dimensional internal space that we expect from F-theory, with the fourth direction $y^s$ becoming ``large'' in the small-volume limit, here given by ${\cal M}_{ss} \rightarrow \infty$. The dilaton and $C_0$ profile are given by ${\cal H}_{\alpha\beta}$ and are at this point arbitrary.

Let us mention another example of this M/IIB relationship which will be important to us later. Consider M-theory with a three-form $\hat{C}_{tx^1x^2}$. It is easy to show from the dictionary \eqref{eq:MEFT} -- \eqref{eq:EFTIIB} that in IIB this leads to a four-form tangential to the four-dimensional spacetime $\hat{C}_{tx_1x_2s}$.

\subsection{Sevenbranes}

In F-theory, a vital role is played by backgrounds containing sevenbranes. In this subsection we discuss some features of how one may view sevenbranes and their singularities in the context of the $\G$ EFT.

Sevenbrane solutions of type IIB supergravity have non-trivial metric and scalar fields $\tau$. From the point of view of EFT, all of these degrees of freedom are contained within the metric $g_{\mu \nu}$ and the generalized metric $\gM_{MN}$. Thus we may specify entirely a sevenbrane background by giving these objects. In the below, we will use the notation
\begin{align}
\dd s^2_{(9)} &= g_{\mu\nu}\dd x^\mu \dd x^\nu \, , &
\dd s^2_{(3)} &= \MM_{\alpha\beta}\dd y^\alpha \dd y^\beta + \MM_{ss} ( \dd y^s)^2\, ,
\end{align}
to specify the solutions. It is not obvious that one should view the generalised metric as providing a notion of line element on the extended space, so in a sense this is primarily a convenient shorthand for expression the solutions.

We consider a sevenbrane which is extended along six of the ``external directions'', denoted $\vec{x}_6$, and along $y^s$ which appears in the extended space. The remaining coordinates are time and the coordinates transverse to the brane which we take to be polar coordinates $(r,\theta)$. In this language, the harmonic function of the brane is $H \approx h \ln[r_0/r]$.\footnote{In the EFT point of view, one can consider the sevenbrane as having the form of a generalised monopole solution of the form given in section \ref{sec:monopole}, after smearing on one of the transverse directions. In this case $r_0$ is a cut-off that is introduced in this process and is related to the codimension-2 nature of the solution, \ie we expect it to be valid only up to some $r_0$ as the solution is not asymptotically flat.} The solution can be specified by
\begin{equation}
\begin{aligned}
\dd s^2_{(9)} &= -\dd t^2 + \dd \vec{x}_{(6)}^{\, 2} 
						+ H \left(\dd r^2 + r^2 \dd \theta^2 \right)  \\
\dd s^2_{(3)} &= H^{-1}\left[(\dd y^1)^2 + 2 h\theta \dd y^1\dd y^2 + K (\dd y^2)^2 \right]
						+ (\dd y^s)^2  \\
{\Aa_\mu}^M &= 0 \, , \qquad K = H^2 + h^2 \theta^2 \, .
\end{aligned}
\label{eq:sevenbrane}
\end{equation}
If one goes around this solution in the transverse space changing $\theta=0$ to $\theta=2\pi$ the $2\times 2$ block $\MM_{\alpha\beta}$ goes to
\begin{equation}
\MM \rightarrow \Omega^T\MM\Omega
\end{equation}
where the monodromy $\Omega$ is an element of $\mathrm{SL}(2)$ and is given by 
\begin{equation}
\Omega = 
\begin{pmatrix}
1 & 2\pi h \\ 0 & 1 
\end{pmatrix} \, .
\end{equation}

Reducing this solution to the IIB section gives the D7-brane. By using \eqref{eq:EFTIIB} one can extract the torus metric $\cH_{\alpha\beta}$ and the scalar $\phi$ of the $10=9+1$ split. From $\cH_{\alpha\beta}$ one then obtains the axio-dilaton, \ie $C_0$ and $e^\varphi$. The external metric is composed with $\phi$ to give the ten-dimensional solution
\begin{equation}
\begin{aligned}
\dd s^2_{(10)} &= -\dd t^2 + \dd \vec{x}_{(6)}^{\, 2} 
					+ H \left(\dd r^2 + r^2 \dd \theta^2 \right)	+ (\dd y^s)^2  \\
C_0 &= h\theta \, , \qquad e^{2\varphi} = H^{-2}			
\end{aligned}
\label{eq:D7}
\end{equation}
which is the D7-brane. Exchanging $y^1$ and $y^2$ and flipping the sign of the off-diagonal term (this is an $\mathrm{SL}(2)$ transformation of the $\MM_{\alpha\beta}$ block) leads to a solution which reduces to the S7-brane. On the M-theory section the solution \eqref{eq:sevenbrane} corresponds to a smeared KK-monopole which can be written as
\begin{equation}
\dd s^2_{(11)} = -\dd t^2 + \dd \vec{x}_{(6)}^{\, 2} 
					+ H \left[\dd r^2 + r^2 \dd \theta^2	+ (\dd y^1)^2\right]
					+ H^{-1}\left[\dd y^2 + h\theta\dd y^1\right]^2\, .
\label{eq:smearedKK}
\end{equation}
To see this more clearly, consider the usual KK-monopole in M-theory, which has three transverse and one isometric direction, the Hopf fibre. If this solution is smeared over one of the transverse directions to give another isometric direction one arrives at the above solution (where $(r,\theta)$ are transverse and $(y^1,y^2)$ are isometric). Therefore the M/IIB-duality between smeared monopole and sevenbrane relates the first Chern class of the Hopf fibration to the monodromy of the codimension-2 object. 

We have now seen how the $\mathrm{SL}(2)$ doublet of D7 and S7 is a smeared monopole with its two isometric direction along the $y^\alpha$ in the extended space. This can be generalized to give $pq$-sevenbranes in the IIB picture where the isometric directions of the smeared monopole correspond to the $p$- and $q$-cycles. The external metric is the same as above, the generalized metric now reads
\begin{equation}
\begin{aligned}
\dd s^2_{(3)} &= \frac{H^{-1}}{p^2+q^2}\bigg\{
					\left[p^2H^2 + (ph\theta - q)^2\right](\dd y^1)^2 
					+ \left[(p + qh\theta)^2 + q^2H^2\right](\dd y^2)^2  \\
	&\hspace{3cm} - 2\left[(p^2-q^2)h\theta + pq(K-1)\right] \dd y^1\dd y^2 \bigg\}
					+  (\dd y^s)^2  \, .
\end{aligned}
\label{eq:pqsevenbrane}
\end{equation}
The two extrema are $p=0$ which gives the D7 and $q=0$ which gives the S7. 

As for all codimension-2 objects, a single D7-brane should not be considered on its own. To get a finite energy density, a configuration of multiple sevenbranes needs to be considered. Introducing the complex coordinate $z=re^{i\theta}$ on the two-dimensional transverse space, such a multi-sevenbrane solution in EFT reads
\begin{equation}
\begin{aligned}
\dd s^2_{(9)} &= -\dd t^2 + \dd \vec{x}_{(6)}^{\, 2} + \tau_2|f|^2\dd z \dd \bz  \\
\dd s^2_{(3)} &= \frac{1}{\tau_2}\left[|\tau|^2 (\dd y^1)^2 
						+ 2\tau_1 \dd y^1\dd y^2 + (\dd y^2)^2 \right]	+(\dd y^s)^2 
\end{aligned}
\label{eq:multisevenbrane}
\end{equation}
where all the tensor fields still vanish. Instead of specifying a harmonic function on the transverse space, we now have the holomorphic functions $\tau(z)$ and $f(z)$. Their poles on the $z$-plane correspond to the location of the sevenbranes \cite{Greene:1989ya, Gibbons:1995vg}. One usually takes 
\be
\tau = j^{-1} \left( \frac{P(z)}{Q(z)} \right) \,,
\ee
where $P(z)$ and $Q(z)$ are polynomials in $z$ and $j(\tau)$ is the invariant $j$-function. The roots of $Q(z)$ will give singularities, which in the IIB section give the locations of the sevenbranes. The configuration in this case consists of the metric 
\be
\dd s^2_{(10)} = -\dd t^2 + \dd \vec{x}_{(6)}^{\, 2} + \dd y_s^2 + \tau_2|f|^2\dd z \dd \bz  
\ee
together with the scalar fields encoded by $\tau$. Meanwhile in the M-theory section, one finds a purely metric background,
\be
\dd s^2_{(11)}  = - \dd t^2 + \dd \vec{x}_{(6)}^{\, 2}
+ \tau_2|f|^2\dd z \dd \bz 
 + \tau_2 ( \dd y^1 )^2 
+ \frac{1}{\tau_2} \left( \dd y^2 + \tau_1 \dd y^1 \right)^2 \,.
\ee
This retains the singularities at the roots of $Q$, at which $\tau_2 \rightarrow i \infty$. 

A crucial point about singularities in F-theory is that they are seen as an origin for nonabelian gauge symmetries. These arise from branes wrapping vanishing cycles at the singularities. At this point we will not examine the details of this for our $\G$ EFT but instead we can point to interesting recent work in DFT. The authors of \cite{Aldazabal:2015yna} construct the full non-abelian gauge enhanced theory in DFT corresponding to the bosonic string at the self-dual point. One would hope that one could apply a simlar construction to produce a gauge enhanced EFT with non-abelian massless degrees of freedom coming from wrapped states on vanishing cycles.

\section{Solutions}
\label{sec:solutions} 

In this section, we discuss the embedding of supergravity solutions into the $\G$ EFT, in particular showing how one can use the EFT form of a configuration to relate eleven- and ten-dimensional solutions. 

\subsection{Membranes, Strings and Waves}
\label{sec:waves}

In higher rank EFT constructions, and double field theory, the fundamental string and the M2 solutions appear as generalized wave solutions in the extended space \cite{Berkeley:2014nza} as presented in Chapters \ref{ch:DFTsol} and \ref{ch:EFTsol}. This applies in the case when all worldvolume spatial directions of these branes lie in the extended space. In this subsection, we will describe first the form of such solutions in the $\G$ EFT, and then also look at the case where the extended space is transverse to an M2.

\subsubsection{Waves in extended space} 

The first family of solutions we are considering can be thought of as a null wave in EFT.\footnote{To properly think of a given solution as being a wave carrying momentum in a particular direction of the extended space, one should construct the conserved charges associated to generalized diffeomorphism invariance, as has been done in the DFT case \cite{Blair:2015eba , Park:2015bza , Naseer:2015fba}. } This can be reinterpreted on a solution of the section condition as branes whose worldvolume spatial directions are wholly contained in the extended space. We denote the spatial coordinates of the external space by $\vec{x}_8$, which become the transverse directions of the supergravity solutions. The harmonic function that appears is $H=1+\frac{h}{|\vec{x}_{(8)}|^6}$, $h$ a constant. As the fields only depend on $\vec{x}_{(8)}$ the section condition \eqref{eq:constraint} is automatically satisfied. The external metric and the one-form field for this set of solutions are
\begin{equation}
\begin{aligned}
\dd s^2_{(9)} &= H^{1/7}\left[-H^{-1}\dd t^2 + \dd \vec{x}_{(8)}^{\, 2}\right]  \\
{\Aa_t}^M &= -(H^{-1}-1)a^M \, .
\end{aligned}
\label{eq:waveexternal}
\end{equation}
Here $a^M$ is a unit vector in the extended space which -- depending on its orientation -- distinguishes between the different solutions in this family. For $a^M=\delta^M_s$, \ie a wave propagating along $y^s$, the generalized metric is
\begin{equation}
\dd s^2_{(3)} =   H^{-6/7}\left[\delta_{\alpha\beta}\dd y^\alpha \dd y^\beta 
						+ H^2(\dd y^s)^2 \right]  \, .
\label{eq:membrane}
\end{equation}
Upon a reduction to the M-theory section this corresponds to a membrane (M2) stretched along $y^1$ and $y^2$. On the IIB section one obtains a pp-wave propagating along $y^s$ . In other words, a membrane wrapping the two internal directions is a wave from a IIB point of view.

Explicitly, one sees that the internal two-metric $\gamma_{\alpha\beta}$ of the $11=9+2$ split of M-theory together with its determinant can be extracted from the above line element with the help of \eqref{eq:genmetricM} as
\begin{align}
\gamma_{\alpha\beta} &= H^{-2/3}\delta_{\alpha\beta} \, , & \gamma &= H^{-4/3} \, .
\end{align}
Using the dictionary \eqref{eq:EFTM} we obtain the membrane solution
\begin{equation}
\begin{aligned}
\dd s^2_{(11)} &= H^{-2/3}\left[-\dd t^2 + \delta_{\alpha\beta}\dd y^\alpha \dd y^\beta \right]
					+ H^{1/3} \dd \vec{x}_{(8)}^{\, 2} \\
C_{ty^1y^2} &= -(H^{-1}-1) \, .			
\end{aligned}
%\label{eq:M2}
\end{equation}
A similar procedure on the IIB section using \eqref{eq:genmetricIIB} and \eqref{eq:EFTIIB} leads to the pp-wave. This time the EFT vector ${\Aa_t}^s$ yields the KK-vector in the $10=9+1$ split, and one gets the pp-wave metric
\begin{equation}
\begin{aligned}
\dd s^2_{(10)} &= -H^{-1}\dd t^2 + H\left[\dd y^s -(H^{-1}-1) \dd t\right]^2
					+  \dd \vec{x}_{(8)}^{\, 2} \, .			
\end{aligned}
\label{eq:pp}
\end{equation}
If on the other hand the solution \eqref{eq:membrane} is oriented along one of the $y^\alpha$, say $y^1$ and thus $a^M=\delta^{M}_{1}$, the corresponding generalized metric is
\begin{equation}
\dd s^2_{(3)} = H^{9/14} \left[H^{1/2}(\dd y^1)^2 + H^{-1/2}(\dd y^2)^2 
					+ H^{-3/2} (\dd y^s)^2 \right] \, .
\label{eq:string}					
\end{equation}
This now corresponds to a fundamental string on the IIB section. The opposite orientation, \ie $a^M=\delta^{M}_{ 2}$ and swapping $y^1$ and $y^2$ in $\MM_{MN}$, gives the D1-brane. The reduction is performed as explained above, resulting in the 10-dimensional solutions
\begin{equation}
\begin{aligned}
\dd s^2_{(10)} &= H^{-3/4}\left[-\dd t^2 + (\dd y^s)^2 \right]
					+ H^{1/4} \dd \vec{x}_{(8)}^{\, 2} \\
\mathrm{F1:} \quad B_{ts} &= -(H^{-1}-1) \\
\mathrm{D1:} \quad C_{ts} &= -(H^{-1}-1) \\
e^{2\varphi} &= H^{\mp 1} \, .			
\end{aligned}
\label{eq:FD1}
\end{equation}
Here $e^{2\varphi}$ is the string theory dilaton. The fundamental string comes with the minus sign in the dilaton and couples to the NSNS-two-form $B_{ty^s}$. The D-string has the plus sign in the dilaton and couples to the RR-two-form $C_{ty^s}$.

On the M-theory section the solution \eqref{eq:string} corresponds to a pp-wave propagating along $y^1$ or $y^2$. Thus the $\mathrm{SL}(2)$ doublet of F1 and D1 is a wave along the $y^\alpha$ directions in the extended space. This can be generalized to give the $pq$-string in the IIB picture by orienting the wave in a superposition of the two $y^\alpha$ directions like $\frac{py^1+qy^2}{\sqrt{p^2+q^2}}$. Then the direction vector is given by
\begin{equation}
a^M = \frac{1}{\sqrt{p^2+q^2}}
\begin{pmatrix}
p \\ q \\ 0
\end{pmatrix}
\label{eq:pqvector}
\end{equation}
and the generalized metric for this configuration is
\begin{equation}
\begin{aligned}
\dd s^2_{(3)} &= H^{9/14}\left\{\frac{H^{-1/2}}{p^2+q^2}\left[(p^2H+q^2)(\dd y^1)^2 
					+ 2pq(H-1)\dd y^1 \dd y^2 \right.\right. \\
				&\hspace{4.5cm}\left.\left.
					+ (p^2+q^2H)(\dd y^2)^2 \right] + H^{-3/2} (\dd y^s)^2 
					\vphantom{\frac{H^{-1/2}}{p^2+q^2}}\right\}
\end{aligned}
\label{eq:pqstring}
\end{equation}
which reduces to the F1 for $q=0$ and the D1 for $p=0$. On the M-theory section this now corresponds to a wave propagating in a superposition of the two internal directions. 

The solutions \eqref{eq:membrane}, \eqref{eq:string} and \eqref{eq:pqstring} form a family which all look the same from the external point of view and only differ in the generalised metric (and the corresponding orientation in the extended space via $a^M$). The EFT wave solutions therefore nicely combines the membrane solution of M-theory and the F1, D1 and $pq$-string solutions of IIB into a single mode propagating in the extended space.

\subsubsection{M2 and D3} 

We have now seen an M2-brane wrapping the $y^\alpha$ space maps to a pp-wave in the IIB theory. In this case, as we mentioned, the worldvolume directions were aligned within the extended space and in particular this meant that we automatically had isometries in these directions. 

Now instead consider an M2-brane extending on two of the external directions plus time, so that its worldvolume extends along $t$ and $\vec{x}_{(2)}$. From the general considerations of the previous section, we expect this to get mapped to a D3-brane along the $t$, $\vec{x}_{(2)}$ and $y^s$ directions. Let's see this explicitly. In these coordinates, the supergravity solution of the M2-brane is given by
\begin{equation}
\begin{aligned}
\dd s^2_{(11)} &= H^{-2/3}\left[-\dd t^2 + \dd \vec{x}_{(2)}^{\, 2} \right]
					+ H^{1/3} \left[\dd \vec{w}_{(6)}^{\, 2} 
						+ \delta_{\alpha\beta}\dd y^\alpha \dd y^\beta \right] \\
\hat C_{tx^1x^2} &= -(H^{-1}-1) \, .			
\end{aligned}
\end{equation}
The harmonic function is a function of the transverse coordinates $H = H(\vec{w}_{(6)}, y^\alpha)$. We take the $y^\alpha$ to be compact so that the harmonic function must be periodic in $y^1$ and $y^2$. With respect to the nine-dimensional theory, the dependence on the $y^\alpha$ coordinates gives rise to massive fields with masses inversely proportional to the volume of the $y^\alpha$ space. In the usual ``F-theory limit'' where this volume is taken to zero, we can thus ignore the dependence on $y^1$ and $y^2$ and smear the solution over the $y^\alpha$ space. Using the dictionary \eqref{eq:MEFT}, we see that as an EFT solution this would correspond to
\begin{equation}
\begin{aligned}
\dd s^2_{(9)} &= H^{-4/7}\left[-\dd t^2 + \dd \vec{x}_{(2)}^{\, 2}\right] 
					+ H^{3/7} \dd \vec{w}_{(6)}^{\, 2}   \\
\dd s^2_{(3)} &= H^{3/7}\delta_{\alpha\beta}\dd y^\alpha \dd y^\beta	+ H^{-4/7} (\dd y^s)^2 \\
{\Ac_{tx^1x^2}}^{[\alpha\beta]s} &= \epsilon^{\alpha\beta}\hat C_{tx^1x^2} \, .
\end{aligned}
\end{equation}

We find that on the IIB section this corresponds to a D3-brane wrapped on the four-dimensional spacetime $\dd s_{(4)}^2 = -\dd t^2 + \dd \vec{x}_{(2)}^{\, 2} + (\dd y^s)^2$, as one would expect. In particular, we have
\begin{equation}
\begin{aligned}
\dd s^2_{(10)} &= H^{-1/2} \left[-\dd t^2 + \dd \vec{x}_{(2)}^{\, 2} + (\dd y^s)^2 \right] 
					+ H^{1/2} \dd \vec{w}_{(6)}^{\, 2}  \\
\hat{C}_{tx^1x^2s} &= -(H^{-1}-1) \,, \qquad e^{\varphi} = 1 \,.
\end{aligned}
\end{equation}

\subsection{Fivebranes and Monopoles}
\label{sec:monopole}

Similar to the relation between generalized null waves, strings and membranes above, the solitonic fivebrane and the M5 solutions can be seen to appear as generalized monopole solutions in the extended space (see \cite{Berman:2014jsa} and Chapters \ref{ch:DFTsol} and \ref{ch:EFTsol} above), in the case where the special isometry direction of the monopole lies in the extended space. We will first review this case, and then give the other solutions in which one obtains M5 branes (partially) wrapping the internal space.

\subsubsection{Monopole in extended space} 

The first set of solutions we consider takes the form of a monopole structure (Hopf fibration) in the extended space, so as stated above the special isometric direction of the monopole is one of the $Y^M$. The coordinates on the external space are denoted $x^\mu = ( t, \vec{x}_{(5)}, \vec{w}_{(3)})$. The harmonic function is $H=1+\frac{h}{|\vec{w}_{(3)}|}$ (satisfying the section condition) and the field configuration of the common sector is
\begin{equation}
\begin{aligned}
\dd s^2_{(9)} &= H^{-1/7}\left[-\dd t^2 + \dd \vec{x}_{(5)}^{\, 2} 
						+ H \dd \vec{w}_{(3)}^{\, 2}\right]  \\
{\Aa_i}^M &= A_ia^M \, , \qquad 2\partial_{[i}A_{j]} = {\epsilon_{ij}}^k\partial_kH \, .
\end{aligned}
\label{eq:monopoleexternal}
\end{equation}
Here $A_i$ is the magnetic potential obtained by dualizing the six-form ${\Af_{tx^1\dots x^5}}^M =$ $-(H^{-1}-1)a^M$ (and $w^i$ with $i=1,2,3$ are the transverse coordinates). As for the EFT wave, the unit vector $a^M$ distinguishes between the solutions in this family by specifying the isometric direction. If the monopole is oriented along $y^s$ and therefore $a^M=\delta^M_s$, the generalized metric is
\begin{equation}
\dd s^2_{(3)} = H^{6/7}\left[\delta_{\alpha\beta}\dd y^\alpha \dd y^\beta 
						+ H^{-2} (\dd y^s)^2 \right] \, .
\label{eq:fivebraneM}
\end{equation}
This corresponds to a fivebrane (M5) on the M-theory section. Note that the fivebrane is smeared over two of its transverse directions. Using the dictionary \eqref{eq:EFTM}, the internal two-metric extracted from this generalized metric combines with the external metric to give an eleven-dimensional solution with the $C_{iy^1y^2}$ component of the M-theory three-form provided by ${\Aa_i}^s$. This can be dualized to give the six-form the fivebrane couples to electrically. The full solution reads
\begin{equation}
\begin{aligned}
\dd s^2_{(11)} &= H^{-1/3}\left[-\dd t^2 + \dd \vec{x}_{(5)}^{\, 2} \right]
					+ H^{2/3} \dd \vec{w}_{(3)}^{\, 2} \\
C_{iy^1y^2} &= A_i \, , \qquad C_{tx^1\dots x^5} = -(H^{-1}-1) \, .			
\end{aligned}
%\label{eq:M5}
\end{equation}
If on the other hand \eqref{eq:fivebraneM} is considered on the IIB section, one obtains the KK-monopole in ten dimensions via \eqref{eq:EFTIIB} where the KK-vector for the $10=9+1$ split is encoded in  ${\Aa_i}^s$. The ten-dimensional solution is given by
\begin{equation}
\dd s^2_{(10)} = -\dd t^2 + \dd \vec{x}_{(5)}^{\, 2} + H^{-1}\left[\dd y^s + A_i\dd w^i\right]^2
					+ H \dd \vec{w}_{(3)}^{\, 2} \, .			
\label{eq:KK6B}
\end{equation}

The alternative choice for the isometric direction of the EFT monopole is one of the $y^\alpha$, say $y^1$ and thus $a^M=\delta^{M}_{ 1}$, the corresponding generalized metric is
\begin{equation}
\dd s^2_{(3)} = H^{-9/14}\left[H^{-1/2}(\dd y^1)^2 + H^{1/2}(\dd y^2)^2 
					+ H^{3/2} (\dd y^s)^2 \right] \, .
\label{eq:fivebraneIIB}
\end{equation}
This is the solitonic fivebrane (NS5) in the IIB picture. Note again that the resulting fivebrane is smeared over one of its transverse directions which corresponds to the isometric direction of the monopole. One can also pick the opposite choice for $a^M$, \ie $a^M=\delta^{M}_{ 2}$, which gives the D5-brane. The reduction procedure should be clear by now, the ten-dimensional solution is
\begin{equation}
\begin{aligned}
\dd s^2_{(10)} &= H^{-1/4}\left[-\dd t^2 + \dd \vec{x}_{(5)}^{\, 2} \right]
					+ H^{3/4} \left[\dd \vec{w}_{(3)}^{\, 2} + (\dd y^s)^2 \right] \\
\mathrm{NS5:} \quad B_{is} &= A_i   	\qquad 	B_{tx^1\dots x^5} = -(H^{-1}-1) \\
\mathrm{D5:} \quad C_{is} &= A_i 		\qquad  C_{tx^1\dots x^5} = -(H^{-1}-1) \\
e^{2\varphi} &= H^{\pm 1}			
\end{aligned}
\label{eq:NSD5}
\end{equation}
where the NS5 couples to the NSNS-B-fields and takes the positive sign in the dilaton while the D5 couples to the RR-C-fields and takes the negative sign in the dilaton. So the $\mathrm{SL}(2)$ doublet of NS5 and D5 is a monopole with its isometric direction along the $y^\alpha$ in the extended space. On the M-theory section the solution \eqref{eq:fivebraneIIB} is a KK-monopole with its isometric direction along one of the two internal dimensions.

As for the EFT wave, this can be generalized to give the $pq$-fivebrane of the IIB picture by having the isometric direction in a superposition of the two $y^\alpha$ directions like $\frac{py^1+qy^2}{\sqrt{p^2+q^2}}$. Then the $a^M$ is given again by \eqref{eq:pqvector} and the generalized metric for this configuration is
\begin{equation}
\begin{aligned}
\dd s^2_{(3)} &= H^{-9/14}\left\{\frac{H^{1/2}}{p^2+q^2}\bigg[(p^2H^{-1}+q^2)(\dd y^1)^2 
					+ 2pq(H^{-1}-1) \dd y^1\dd y^2 \right. \\
				&\hspace{4.7cm}\left.
					+ (p^2+q^2H^{-1})(\dd y^2)^2 \bigg] + H^{3/2} (\dd y^s)^2 
				\vphantom{\frac{H^{1/2}}{p^2+q^2}}\right\} \, .
\end{aligned}
\label{eq:pqfivebrane}
\end{equation}
For $q=0$ this is the NS5 and for $p=0$ this gives the D5 (both smeared). On the M-theory section this now corresponds to a KK-monopole with its isometric direction in a superposition of the two internal directions.

The EFT monopole combines the fivebrane solution of M-theory and the NS5, D5 and $pq$-fivebrane solutions of IIB into a single monopole structure with isometric direction in the extended space.

\subsubsection{M5 wrapped on internal space}

In the above we saw that M5-branes oriented completely along the external nine directions lead to KK-monopoles in the IIB picture. Let us now also study M5-branes (partially) wrapping the $y^\alpha$ space. We consider the setup in Section \ref{sec:MF} where our external spacetime consists of a three-dimensional part containing time and with spatial coordinates $\vec{x}_{(2)}$ and a six-dimensional part denoted $B_6$.

Let us begin with an M5-brane wrapping both $y^1$ and $y^2$, as well as wrapping a two-cycle $\Sigma_2$ in $B_6$ (along the two directions $\vec{z}_{(2)}$) and being extended along $t$ and $x^1$. This gives a D3-brane extended along $t$, $x^1$ and wrapping the two-cycle $\Sigma_2$, as expected. The twelve-dimensional EFT solution to which this corresponds is given by
\begin{equation}
 \begin{split}  
 \dd s_{(9)}^2 &= H^{-3/7} \left[-\dd t^2 + (\dd x^1)^2 + \dd \vec{z}_{(2)}^{\, 2} \right] 
  						+ H^{4/7} \left[(\dd x^2)^2 + \dd \vec{w}_{(4)}^{\, 2}\right] \,, \\
  \dd s_{(3)}^2 &= H^{-2/21} \delta_{\alpha\beta}\dd y^\alpha \dd y^\beta 
  						+ H^{4/7} (\dd y^s)^2 \,,
 \end{split}
\end{equation}
with non-trivial gauge field $\Ac_{\mu\nu\rho}{}^{[\alpha\beta]s}$. On the M-theory section this gives rise to the solution
\begin{equation}
\begin{aligned}
\dd s^2_{(11)} &= H^{-1/3}\left[-\dd t^2 + (\dd x^1)^2 + \delta_{\alpha\beta}\dd y^\alpha \dd y^\beta
			+	\dd \vec{z}_{(2)}^{\, 2}	  \right]
			+ H^{2/3} \left[(\dd x^2)^2 + \dd \vec{w}_{(4)}^{\, 2}\right] \, ,	
\end{aligned}
\end{equation}
with three-form $\hat{C}_3$ which couples magnetically to the M5. % and has legs in the four other directions of $B_6$.
In the IIB section we find the expected D4-brane
\begin{equation}
\dd s^2_{(10)} = H^{-1/2} \left[-\dd t^2 + (\dd x^1)^2 +	\dd \vec{z}_{(2)}^{\, 2}	 \right] 
			+ H^{1/2} \left[ (dy^s)^2 + (\dd x^2)^2 + \dd \vec{w}_{(4)}^{\, 2} \right] \,,
\end{equation}
with the magnetic four-form $\hat{C}_{\mu\nu\rho s}$ given by $\hat{C}_{\mu\nu\rho}$ and all other fields vanishing.

We can also consider an M5-brane which wraps a closed three-cycle $\Sigma_3$ of $B_6$ as well as one of the $y^{\alpha}$. In the IIB picture this should correspond to a NS5-brane or D5-brane, for $\alpha = 1$ or $2$. Let us denote the coordinates on $B_6$ as $\vec{z}_{(3)}$ for the $\Sigma_3$ and $\vec{w}_{(3)}$ for the other three coordinates. The explicit EFT solution is given by
\begin{equation}
 \begin{split}  
 ds_{(9)}^2 &= H^{-2/7} \left[-\dd t^2 + (\dd x^1)^2 + \dd \vec{z}_{(3)}^{\, 2} \right] 
 			+ H^{5/7} \left[ (\dd x^2)^2 + \dd \vec{w}_{(3)}^{\, 2} \right] \, , \\
  ds_{(3)}^2 &= H^{-2/7} (\dd y^1)^2 + H^{5/7} (\dd y^2)^2 + H^{-2/7} (dy^s)^2 \,, 
 \end{split}
\end{equation}
and gauge field $B_{\mu\nu}{}^{\alpha s} = \epsilon^{\alpha\beta} \hat{C}_{\mu\nu\beta}$ where $\hat{C}_{\mu\nu\alpha}$ will be specified shortly. As for the M2/D3 case, we can in the zero-volume limit take the harmonic functions to be smeared across the $y^1$ direction. On the M-theory section this gives the expected M5-brane
\begin{equation}
\dd s^2_{(11)} = H^{-1/3} \left[-\dd t^2 + (\dd x^1)^2 + (\dd y^1)^2 
					+ \dd \vec{z}_{(3)}^{\, 2} \right] 
					+ H^{2/3} \left[(\dd x^2)^2 + \dd \vec{w}_{(3)}^{\, 2} + (dy^s)^2 \right] \,.
\end{equation}
The gauge field has only non-zero components $\hat{C}_{z^iz^jy^2}$, where $i,j = 1, \ldots, 3$. On the other hand, on the IIB section we obtain a D5-brane along $t$, $x^1$ and wrapping $y^s$ as well as $\Sigma_3$
\begin{equation}
\dd s^2_{(10)} = H^{-1/4} \left[-\dd t^2 + (\dd x^1)^2  + (dy^s)^2 
					+ \dd \vec{z}_{(3)}^{\, 2}  \right]
					+ H^{3/4} \left[ (\dd x^2)^2 + \dd \vec{w}_{(3)}^{\, 2} \right] \,.	
\end{equation}
The six-form gauge field is then the dual to $\hat{C}_{z^{i}z^{j}} = \hat{C}_{z^{i} z^{j} y^2}$ and we find a D5-brane. By an $\mathrm{SL}(2)$ transformations we can also obtain NS5-branes this way.

\section{Comments} 

We saw above that the solutions described in Section \ref{sec:waves} in which one had waves in the extended space (corresponding to F1, D1 and M2 totally wrapping the internal space) essentially split into two categories, depending on whether the vector $a^M$ giving the wave's direction pointed in the $y^\alpha$ or $y^s$ directions. These different solutions were not related by duality.\footnote{However they can be related by a $\mathbb{Z}_2$ transformation on the generalized metric and the extended coordinates as in \cite{Malek:2015hma}. This transformation maps the common NS-NS sector of Type IIA / IIB into each other and would thus allow us to map the generalized metric containing the membrane into the F1-string in IIB.} However, as seen in previous chapters, in higher rank duality groups one finds that all branes whose worldvolumes are only spatially extended in the internal space are unified into a single solution of the EFT \cite{Berman:2014hna,Berman:2014jsa}. Similarly, as one increases the rank of the duality group considered, various of the M2 and M5 brane solutions will be unified as more of the worldvolume directions fall into the internal space. Thus, families of these solutions appear in a unified manner. With this lesson learned, one may hope that studying F-theory compactifications in bigger EFTs may allow one to more easily consider complicated set-ups.

\bigskip

We are aware that the approach of most practitioners in F-theory that has yielded so much success over a number of years has been through algebraic geometry. It is doubtful if the presence of this action can help in those areas where the algebraic geometry has been so powerful. We do hope though that it may provide some complementary techniques given that we now have a description in terms of twelve-dimensional degrees of freedom equipped with an action to determine their dynamics. 

One question people have tried to answer is to find the theory on a D3-brane when $\tau$ varies. This might be computable in this formalism using a Goldstone mode-type analysis similar to that in \cite{Berkeley:2014nza} where a Goldstone mode analysis was used to determine string and brane effective actions in DFT and EFT (\cf Section \ref{sec:goldstones}). A useful result from this formalism would be to show why elliptic Calabi-Yau are good solutions to the twelve-dimensional theory. This would likely involve the construction of the supersymmetric version of the $\G$ EFT in order to study the generalised Killing spinor equation. Another interesting area of investigation would be the Type II / heterotic duality, where we should then consider EFT on a K3 background and relate this to heterotic DFT \cite{Hohm:2011ex,Grana:2012rr}.

An interesting consequence of this work is that it shows how F-theory fits into a general picture of EFT with various $\Edd$ groups. One might then be inspired to consider far more general backgrounds with higher dimensional fibres and with monodromies in $\Edd$ and so one would not just have sevenbranes but more exotic objects (of the type described in \cite{deBoer:2012ma}). In fact this idea appeared early in the F-theory literature \cite{Kumar:1996zx}. More recent work in this direction has appeared where one takes the fibre to be $K3$ and then one has a U-duality group act on the $K3$ \cite{Braun:2013yla,Candelas:2014jma,Candelas:2014kma} in a theory sometimes called G-theory. Further, the dimensional reduction of EFTs has now been examined in some detail and in particular one can make use of Scherk-Schwarz-type reductions that yield gauged supergravities \cite{Aldazabal:2011nj,Geissbuhler:2011mx,Grana:2012rr,Berman:2012uy,Musaev:2013rq,Aldazabal:2013mya,Geissbuhler:2013uka,Berman:2013cli,Condeescu:2013yma,Aldazabal:2013via,Lee:2014mla,Baron:2014yua,Lee:2015xga}. This shows that one should perhaps consider a more general type of reduction than the simple fibrations described here. This means one could consider Scherk-Schwarz-type reductions on the F-theory torus. This makes no sense from the Type IIB perspective but it does from the point of view of the $\G$ EFT.

A further quite radical notion would be the EFT version of a T-fold where we only have a local choice of section so that the space is not globally described by Type IIB or M-theory. One could have a monodromy such that as one goes round a one-cycle in nine dimensions and then flips between the IIB section and M-theory section. This would exchange a wrapped membrane in the M-theory section with a momentum mode in the IIB section just as a T-fold swaps a wrapped string with a momentum mode. 
Note, this is not part of the $\G$ duality group and thus is not a U-fold. This is simply because with two isometries one has a $\mathbb{Z}_2$ choice of section that one can then twist.

\chapter{Conclusion}
\label{ch:outro}

Dualities lie at the heart of string theory. They relate seemingly different supergravity backgrounds and provide a unifying link to the various types of string theories. The presence of a duality or hidden symmetry often highlights a possible extension of the underlying theory. One is therefore led to find a formalism where the duality can be elevated to a manifest symmetry of the theory. Double field theory and exceptional field theory provide such a formalism for T-duality and U-duality respectively by recasting supergravity in a duality manifest way. 

In this thesis an outline of the basic features and underlying concepts of DFT and EFT was presented. The core idea is to extend the space by including dual coordinates which correspond to winding modes of the fundamental objects of string theory and M-theory. This leads to an doubled or exceptionally extended geometry which unifies the metric and the gauge field(s) in a geometric manner. Also their local symmetries are combined into general diffeomorphisms generated by a generalized Lie derivative. This provides an elegant reformulation of supergravity with the dualities turned into manifest symmetries of the action.

The main topic of this thesis was the study of solutions to the equations of motion of DFT and EFT. It was shown how supergravity objects related by a duality transformation have a common origin in a single solution in the extended theory. The simplest example is a null wave -- a pure momentum mode -- in the doubled space which gives the F1-string and pp-wave of string theory depending on its direction of propagation. Essentially the momentum carried in a dual direction gives a fundamental object with mass and charge in the reduced picture. Similarly we have seen how the DFT monopole leads to both the NS5-brane and the KK-monopole in string theory, \ie the monopole circle (the $S^1$ of the Hopf fibration) in a dual direction gives a solitonic object with magnetic charge. 

These two results can be combined in EFT where a single, self-dual solution was found which has both wave-like and monopole-like aspects. Depending on its orientation in the exceptionally extended space of the $\Es$ theory, it gives rise to the full spectrum of 1/2 BPS branes in both ten- and eleven-dimensional supergravity. It is the crucial feature of a twisted self-duality the $\Es$ EFT is equipped with which allows for this neat combination of electric and magnetic components into a single object\footnote{Note that strictly speaking we are not dealing with a \emph{dyonic} solution since the electric and magnetic charges arise from KK-reductions in different directions (although the directions are paired by the duality structure).}

Besides the EFTs for the symmetry groups $\Slf$ and $\Es$, we have also constructed the theory for $\Slt$ which subsequently was related to F-theory. It is a twelve-dimensional theory with manifest $\Slt$ symmetry. The two inequivalent ways of imposing the section condition reproduce M-theory and F-theory respectively. By a dimensional reduction to ten dimensions one arrives at Type IIA and Type IIB string theory. This EFT not only promotes the $SL(2)$ S-duality of the Type IIB sector to a symmetry, it also incorporates the M/F-theory duality via the two different solutions to the section condition. Furthermore, it provides an action for F-theory which can be reduced to the appropriate supergravity action in ten dimensions. 

By considering solutions in these extended field theories, it was established how the presence of isometries in the extended space gives rise to the dualities from a reduced point of view. Due to the isometries there is an ambiguity in how to pick the physical spacetime after imposing the section condition. It is this ambiguity which manifests itself as a duality between backgrounds arising from possible choices of section. For example, the isometry of the DFT wave solution leads to the T-duality relation between the F1-string and the pp-wave. The same can be said for other T-duality and S-duality relations of supergravity objects, in fact all the 1/2 BPS branes listed in Table \ref{tab:solutions} are connected by U-dualities which come from the isometries in the 56-dimensional exceptionally extended space of the $\Es$ EFT. This idea can also be extended to the duality between M-theory and F-theory where now two isometries are required.

%[recpap of goldstones, singularities, localization]

\bigskip

The fields of DFT and EFT are rapidly expanding and developing. Progress is made on a huge variety of fronts such as the study of large gauge transformations \cite{Hohm:2012gk,Park:2013mpa,Berman:2014jba,Rey:2015mba,Chaemjumrus:2015vap} and $\alpha'$-corrections \cite{Hohm:2013jaa,Coimbra:2014qaa,Bedoya:2014pma,Marques:2015vua,Hohm:2015mka,Hohm:2015doa,Hohm:2015ugy,Linch:2015lwa,Linch:2015fya,Linch:2015qva,Linch:2015fca}. On the solution side the extended theories provide a natural framework to deal with non-geometric backgrounds, exotic branes and their fluxes. The extended spaces virtually render everything geometric, albeit in a generalized sense. 

One idea the author would like to pursue in future concerns the relation between the various EFTs and to study the difference between symmetry transformations and solution generating transformations. The former is a subset of the latter, but this depends on the ``size'' of the symmetry group $G$. For example, the action of T-duality and the $\Odd$ group is a solution generating transformation in ordinary string theory. Via the Buscher rules one can relate \eg the string and the wave. But by going to DFT this duality transformation is lifted to a symmetry transformation of the whole theory. 

Similarly the relation between the membrane and the wave in the $\Slt$ EFT ($D=2$) takes the form of a solution generating transformation since it is not part of the symmetry group. But we have seen that this is indeed the case in the $\Slf$ EFT ($D=4$) where we do have a symmetry group large enough to include the transformation from wave to membrane. The $\Slf$ group and its EFT though are not big enough to include the relation between monopole and fivebrane. To capture this as a symmetry transformation we have to go to the $\Es$ EFT ($D=7$) where all the relations between wave, monopole and the branes are part of a symmetry transformation. One can think of this transformation as an $\Es$ rotation in the exceptionally extended space. But one can also perform an $Sp(56)$ transformation which is a group bigger than $\Es$ (in fact $\Es\subset Sp(56)$). This is then a solution generating transformation but not a symmetry transformation. It is speculated that by going to even larger symmetry groups $G=E_D$ with $D>7$ these transformations might become a symmetry transformation again.

A big challenge that remains is the relaxing of the physical section condition. As long as it is imposed in its strong form, the extended field theories provide very powerful and symmetric reformulations of the known theories but do not contain any extra physics beyond supergravity. Only by demanding a weaker version of the constraint can new insights such as providing a higher dimensional origin for certain gauged supergravities be obtained. It is therefore a highly desirable goal to find formulations of DFT and EFT with less stringent constraints in the hope to discover new physics.

\appendix
\chapter{Solutions satisfying EoMs}
\label{ch:appDFT}

In this appendix we will proof that the DFT wave and DFT monopole solutions are indeed solutions of the theory, \ie satisfy the equations of motion \eqref{eq:DFTeomDilaton} and \eqref{eq:DFTeom} derived from the DFT action \eqref{eq:DFTaction}. Recall the equations of motion were
\begin{equation}
	\RR=0 \qandq {P_{MN}}^{KL}K_{KL} = 0
\end{equation}
where we had the scalar $\RR$
\begin{equation}
\begin{aligned}
\RR 	&= \frac{1}{8}\HH^{MN}\partial_M\HH^{KL}\partial_N\HH_{KL} 
		- \frac{1}{2}\HH^{MN}\partial_M\HH^{KL}\partial_K\HH_{NL} \\
	&\quad+ 4\HH^{MN}\partial_M\partial_N d - \partial_M\partial_N\HH^{MN}
		-4\HH^{MN}\partial_M d \partial_N d + 4\partial_M\HH^{MN}\partial_N d \\
	&\quad+ \frac{1}{2}\eta^{MN}\eta^{KL}\partial_M{\EE^A}_K\partial_N{\EE^B}_L\HH_{AB} \, , 
\end{aligned}
\end{equation}
the tensor $K_{MN}$
\begin{equation}
\begin{aligned}
K_{MN} &= \frac{1}{8}\partial_M\HH^{KL}\partial_N\HH_{KL} + 2\partial_M\partial_N d \\
	&\quad +(\partial_L-2\partial_L d) 		
		\left[\HH^{KL}\left(\partial_{(M}\HH_{N)K}
		- \frac{1}{4}\partial_K\HH_{MN}\right)\right] \\
	&\quad  + \frac{1}{4}\left(\HH^{KL}\HH^{PQ}-2\HH^{KQ}\HH^{LP}\right)
		\partial_K\HH_{MP}\partial_L\HH_{NQ} \\
	&\quad - \eta^{KL}\eta^{PQ}\left(\partial_K d\partial_L{\EE^A}_P 
		- \frac{1}{2}\partial_K\partial_L{\EE^A}_P\right)\HH_{(N|R}{\EE^R}_A\HH_{|M)Q} 
\end{aligned}
\end{equation}
and the projector
\begin{equation}
{P_{MN}}^{KL} = \frac{1}{2}\left[{\delta_M}^{(K}{\delta_N}^{L)} 
	- \HH_{MP}\eta^{P(K}\eta_{NQ}\HH^{L)Q}\right] \, .
\end{equation}

\section{DFT Wave}
\label{sec:wavecheck}
We start by checking the wave solution. For the wave we can actually show that the stronger equation $K_{MN}=0$ instead of the projected equation is satisfied. Both $\RR$ and $K_{MN}$ have three kind of terms: those just containing the generalized metric $\HH_{MN}$, those containing the dilaton $d$ and those with the generalized vielbein ${\EE^A}_M$. The derivatives in the vielbein terms are always contracted with $\eta$ and thus vanish since our solution satisfies the section condition. In our solution $d$ is constant so all the dilaton terms vanish as well as they are always acted on by a derivative operator. This leaves us to check the metric terms. Recall that the harmonic function $H$ is a funtion of the transverse $y^m$ only, so the only derivatives acting on $\HH$ that give a non-vanishing contribution are the $\partial_m$.

We will split this task into several steps. First consider the term that is proportional to $\partial_M\HH^{KL}\partial_N\HH_{KL}$. Using the notation of Section \ref{sec:goldstones} where a winding coordinate is denoted by $\tx^\bmu$, we can expand the indices to get
\begin{equation}
\begin{aligned}
\partial_M\HH^{KL}\partial_M\HH_{KL} 
	&\rightarrow \partial_m\HH^{ab}\partial_n\HH_{ab}
		+ \partial_m\HH^{kl}\partial_n\HH_{kl} \\
	&\qquad	+ \partial_m\HH^{\ba\bb}\partial_n\HH_{\ba\bb} 
		+ \partial_m\HH^{\bk\bl}\partial_n\HH_{\bk\bl}
		+ 2\partial_m\HH^{a\bb}\partial_n\HH_{a\bb}  \\
	&= \II^{ab}\II_{ab}\partial_m H \partial_n(2-H) 
		+ \II^{\ba\bb}\II_{\ba\bb}\partial_m(2-H)\partial_n H  \\
	&\qquad	+ 2\JJ^{a\bb}\JJ_{a\bb}\partial_m(H-1)\partial_n(H-1) \\	
	&= (-2-2+4)\partial_m H \partial_n H \\
	&= 0. 
\end{aligned}
\label{eq:appendixsum}
\end{equation}
Next consider the term in $\RR$ proportional to $\partial_M\HH^{KL}\partial_K\HH_{NL}$. It vanishes as well
\begin{equation}
\partial_M\HH^{KL}\partial_K\HH_{NL} \rightarrow \partial_m\HH^{kl}\partial_k\HH_{nl} = 0.
\end{equation} 
Similarly the terms in $K_{MN}$ and $\RR$ where the derivatives are contracted with the generalized metric (in any combination) vanish since the only derivative we need to consider is $\partial_m$, but upon contraction this forces both indices on $\HH$
to be of $kl$ type and $\HH^{kl}=\delta^{kl}$ so its derivative vanishes.

This leaves us with two more terms in $K_{MN}$ to check
\begin{equation}
\begin{aligned}
\HH^{KL}\partial_K\partial_L\HH_{MN} 
	&\rightarrow \delta^{kl}\partial_k\partial_l\HH_{MN} = 0 \quad\mathrm{since}\quad \delta^{kl}\partial_k\partial_l H=0 \\
\HH^{KL}\HH^{PQ}\partial_K\HH_{MP}\partial_L\HH_{NQ} &\rightarrow \delta^{kl}\HH^{PQ}\partial_k\HH_{MP}\partial_l\HH_{NQ} \, .
\end{aligned}
\end{equation}
The first one vanishes since $\HH_{MN}$ is a linear function of $H$ which is harmonic and thus annihilated by the Laplacian. The second expression has to be expanded for all possible values $MN$ can take. In each case it vanishes either trivially or the terms sum to zero along the lines of \eqref{eq:appendixsum}.

Thus we have shown that all terms in $K_{MN}$and $\RR$ are zero and therefore the equations of motion is satisfied by our solution.

\section{DFT Monopole}
\label{sec:monopolecheck}
Let us now turn to the DFT monopole solution presented in \eqref{eq:DFTmonopole}. Again we will prove that it satisfies the equations of motion \eqref{eq:DFTeomDilaton} and \eqref{eq:DFTeom}. The components of the generalized metric and its inverse (again using the notation $\tx^\bmu$ for winding coordinates) can be written as
\begin{equation}
\begin{aligned}
\HH_{zz} &= H^{-1}						&	\HH^{zz} &= H(1+H^{-2}A^2)			\\
\HH_{\bz\bz} &= H(1+H^{-2}A^2)			&	\HH^{\bz\bz} &= H^{-1}				\\
\HH_{ij} &= H(\delta_{ij}+H^{-2}A_iA_j)	&	\HH^{ij} &= H^{-1}\delta^{ij}		\\
\HH_{\bi\bj} &= 	H^{-1}\delta_{\bi\bj}	&	
							\HH^{\bi\bj} &=H(\delta^{\bi\bj}+H^{-2}A^\bi A^\bj) 		\\
\HH_{zi} &= H^{-1}A_i					&	\HH^{zi} &= -H^{-1}A^i				\\
\HH_{\bz\bi} &= -H^{-1}A_\bi				&	\HH^{\bz\bi} &= H^{-1}A^\bi			\\
\HH_{ab} &= \eta_{ab} 					&	\HH^{ab} &= \eta^{ab} 				\\
\HH_{\ba\bb} &= \eta_{\ba\bb}			&	\HH^{\ba\bb} &= \eta^{\ba\bb} 
\end{aligned}
\end{equation}
and the DFT dilaton is simply
\begin{equation}
d = \phi_0 - \frac{1}{2}\ln H \, .
\end{equation}
The harmonic function $H$ is a function of $y^i$ only, independent of $z$ and any dual coordinate. Therefore the only relevant derivatives will be $\partial_i$. Furthermore, $H$ obeys the section condition and the Laplace equation. The vector $A_i$ (whose index can be freely raised by $\delta^{ij}$) is a function of $H$ and obeys the same constraints. In addition its divergence vanishes. The relation between $H$ and $A$ given in \eqref{eq:AH} will be used frequently.

Since $\HH$ and $d$ obey the section condition, the last line in both $\RR$ and $K_{MN}$ can be dropped as it vanishes under section. With these simplifications in mind, we can proceed to check the equations of motion. 

Start with $\RR$. Inserting the components of $\HH$, the first line reduces to 
\begin{equation}
\frac{1}{8}\HH^{MN}\partial_M\HH^{KL}\partial_N\HH_{KL} 
		- \frac{1}{2}\HH^{MN}\partial_M\HH^{KL}\partial_K\HH_{NL} =
-H^{-3}\delta^{mn}\partial_mH\partial_nH
\end{equation}
while the second line gives 
\begin{equation}
4\HH^{MN}\partial_M\partial_N d - \partial_M\partial_N\HH^{MN}
		-4\HH^{MN}\partial_M d \partial_N d + 4\partial_M\HH^{MN}\partial_N d = H^{-3}\delta^{mn}\partial_mH\partial_nH
\end{equation}
and we thus have $R=0$. 

Next we compute the components of $K_{MN}$. By inspection it can be seen that $K_{aM}$ and $K_{\ba M}$ vanish for any index $M$. Also $K_{z\bz}, K_{m\bn}, K_{z\bm}$ and $K_{\bz m}$ vanish trivially. The non-zero components are
\begin{equation}
\begin{aligned}
K_{mn} &= \frac{1}{4}H^{-2}\delta^{kl}\left[\partial_kA_m\partial_lA_n 
			- \delta_{mn}\partial_kH\partial_lH\right]  
			- H^{-3}\delta^{kl}A_{(m}\partial_{n)}A_k\partial_lH \\
	&\qquad	- \frac{1}{4}H^{-4}A_mA_n\delta^{kl}\delta^{pq}\partial_kA_p\partial_lA_q \\
K_{\bm\bn} &= \frac{1}{4}H^{-4}\delta^{kl}\left[\partial_kA_\bm\partial_lA_\bn 
			- \delta_{\bm\bn}\partial_kH\partial_lH\right]	\\
K_{zz} &= - \frac{1}{4}H^{-4}\delta^{kl}\delta^{pq}\partial_kA_p\partial_lA_q	\\
K_{\bz\bz} &= - \frac{1}{4}H^{-2}\delta^{kl}\delta^{pq}\partial_kA_p\partial_lA_q 
			+ H^{-3}\delta^{kl}\delta^{pq}A_p\partial_kA_q\partial_lH  \\
	&\qquad + \frac{1}{4}H^{-4}\delta^{kl}\left[A^pA^q\partial_kA_p\partial_lA_q 
			- A^2\partial_kH\partial_lH\right] \\
K_{mz} &= -\frac{1}{2}H^{-3}\delta^{kl}\left[2\partial_mA_k-\partial_kA_m\right]\partial_lH
			- \frac{1}{4}H^{-4}\delta^{kl}\delta^{pq}A_m\partial_kA_p\partial_lA_q \\
K_{\bm\bz} &= -\frac{1}{2}H^{-3}\delta^{kl}\partial_kA_\bm\partial_lH
			+ \frac{1}{4}H^{-4}\delta^{kl}\left[A_\bm\partial_kH\partial_lH 
			- \delta^{pq}A_p\partial_kA_\bm\partial_lA_q\right] \, .
\end{aligned}
\end{equation}
Now expand the projected equations of motion component-wise. For example, the $mn$ component of the equation reads
\begin{equation}
2{P_{mn}}^{KL}K_{KL} = K_{mn} 
	- \eta_{m\bm}\left[\HH^{\bm\bk}K_{\bk\bl}\HH^{\bl\bn} 
	+ \HH^{\bm\bz}K_{\bz\bz}\HH^{\bz\bn} 
	+ 2\HH^{\bm\bk}K_{\bk\bz}\HH^{\bz\bn}\right]\eta_{\bn n} \, .
\end{equation}
Inserting the components of $K_{MN}$ computed above into this expression yields zero once all terms are summed up properly. The same holds for all the other components of the equations of motion. They are thus satisfied by our solution.

It is interesting to note the action of the projector here. Whereas the general significance of the projector in the equations of motion was pointed out in \cite{Berkeley:2014nza}, it turned out that its presence was not strictly needed to show that the DFT wave was a solution as all the components of $K_{MN}$ vanished for it independently as seen in the previous section. 

In contrast here for the DFT monopole, not all components of $K_{MN}$ are zero and only once the projector acts are the equations of motion satisfied. This might be due to different properties of the wave and monopole solution, the former being conformally invariant while the latter is not.

\section{$SL(5)$ Wave}
\label{sec:SL5check}
For completeness, we also demonstrate how the wave solution of the $\Slf$ invariant theory presented in \eqref{eq:SL5ppwave} is actually a solution to the relevant equations of motion \eqref{eq:SL5eom} and $R=0$. To do this we will use the five-dimensional coordinate representation $X^{ab}$ with $a,b=1,\dots,5$ (the index pair is antisymmetric) given in \eqref{eq:SL5coords} and split the coordinates into worldvolume and transverse parts. Note that the membrane in four dimensions only has one transverse direction. By introducing $m,n=1,2,3$, the coordinates read
\begin{equation}
X^M = X^{ab} = (X^{m5};X^{45},X^{m4},X^{mn}) = (x^m;x^4,y^{mn},y^{m4}).
\end{equation}
In this notation the non-zero components of the generalized metric for the $SL(5)$ wave given in \eqref{eq:SL5ppwave} can be written as
\begin{equation}
\begin{aligned}
\MM_{m5,n5} &= (2-H)\II_{mn}			&	\MM^{m5,n5} &= H\II^{mn} 		\\
\MM_{m4,n4} &= -H\II_{mn}			&	\MM^{m4,n4} &= -(2-H)\II^{mn} 	\\
\MM_{m4,n5} &= -(H-1)\II_{mn}		&	\MM^{m4,n5} &= -(H-1)\II^{mn} 	\\
\MM_{mn,kl} &= \II_{mn,kl}			&	\MM^{mn,kl} &= \II^{mn,kl}		\\
\MM_{45,45} &= 1						&	\MM^{45,45} &= 1
\end{aligned}
\end{equation}
where the harmonic function $H$ is a function of the transverse coordinate $X^{45}=x^4$ only and for convenience these two matrices are introduced
\begin{align}
\II_{mn} &=  \begin{pmatrix} -1 & 0 & 0 \\ 0 & 1 & 0 \\ 0 & 0 & 1 \end{pmatrix} = \II^{mn} \, ,  &
\II_{mn,kl} &= \begin{pmatrix} 1 & 0 & 0 \\ 0 & 1 & 0 \\ 0 & 0 & -1 \end{pmatrix} = \II^{mn,kl}.
\end{align}

Before we insert the metric into $K_{MN}$ and $R$, we note some simplifications. Since $H$ only depends on $X^{45}$, the only derivative that yields a non-zero result is $\partial_{45}$ which we will simply denote by $\partial$. Thus, just as in the DFT case, terms like
\begin{equation}
\partial_K \MM^{KL}\partial_L \MM_{MN}, \quad
\partial_M \MM^{KL}\partial_L \MM_{KN}, \quad
\MM^{KL} \partial_L\partial_M \MM_{KN},
\end{equation}
that is terms where a derivative acts on a metric which is contracted with a derivative, vanish since $\MM_{45,45}=1$.

The volume factor $\Delta$ is a constant for our solution, so all terms with $\partial \Delta$ also vanish. Furthermore, since $H$ is a harmonic function, it is annihilated by the Laplacian and therefore
\begin{equation}
\MM^{KL} \partial_K \partial_L \MM_{MN} = \partial^2 \MM_{MN} = 0
\end{equation}
since all the components of $\MM_{MN}$ are linear functions of $H$.

With these simplifications in mind, most of the terms in $K_{MN}$ and $R$ vanish trivially. We only need to check two terms explicitly, namely
\begin{equation}
\begin{aligned}
\partial_M \MM^{KL}\partial_N \MM_{KL} \qandq
\MM^{KL} \MM^{PQ} \partial_K \MM_{MP} \partial_L \MM_{NQ}.
\end{aligned}
\label{eq:terms2check}
\end{equation}
We start with the first expression and expand the indices to get
\begin{equation}
\begin{aligned}
\partial_M \MM^{KL}\partial_N \MM_{KL} 
	&\rightarrow \partial\MM^{k5,l5}\partial\MM_{k5,l5}
		+ \partial\MM^{k4,l4}\partial\MM_{k4,l4}
		+ 2\partial\MM^{k5,l4}\partial\MM_{k5,l4}  \\
	&\qquad + \partial\MM^{kl,pq}\partial\MM_{kl,pq}
		+ \partial\MM^{45,45}\partial\MM_{45,45} \\
	&= \II^{mn}\II_{mn}\left[\partial H\partial(2-H) 
		+ \partial(2-H)\partial H
		+ 2\partial(H-1)\partial(H-1) \right]\\
	&= 3\left[-1-1+2\right] \partial H \partial H \\
	&=0.
\end{aligned}
\end{equation}
Similarly we can show that the other expression in \eqref{eq:terms2check} vanishes
\begin{equation}
\begin{aligned}
\MM^{KL}\MM^{PQ}\partial_K\MM_{MP}\partial_L\MM_{NQ} 
	&\rightarrow \MM^{p5,q5}\partial\MM_{M,p5}\partial\MM_{N,q5}
		+ \MM^{p4,q4}\partial\MM_{M,p4}\partial\MM_{N,q4} \\
	&\qquad	+ 2\MM^{p5,q4}\partial\MM_{M,p5}\partial\MM_{N,q4} \\
	&\qquad 	+ \MM^{kl,pq}\partial\MM_{M,kl}\partial\MM_{N,pq} 
		+ \MM^{45,45}\partial\MM_{M,45}\partial\MM_{N,45}\\
	&\rightarrow \left[H-(2-H)-2(H-1)\right]\II^{pq}\II_{mp}\II_{nq}
		\partial H \partial H \\
	&= \left[H-2+H-2H+2\right]\II_{mn}\partial H \partial H \\
	&=0.
\end{aligned}
\end{equation}

We have thus shown that all the terms in $K_{MN}$ and $R$ vanish and the equations of motion are therefore satisfied by our solution.

\chapter{Relating Solutions in $E_7$}
\label{ch:appE7}

In this appendix we fill in the details of how the extended solutions of the $E_7$ duality invariant theory can be rewritten by using a Kaluza-Klein ansatz to obtain solutions in ordinary spacetime.

\section{From Wave to Fivebrane}
\label{sec:appWave}
In Section \ref{sec:M2M5wave} it is explained how the extended wave solution can be rotated to carry momentum along a fivebrane wrapping direction. From a ordinary spacetime point of view, this is then the M5-brane solution of supergravity. Here this calculation is presented in detail. 

After the rotation \eqref{eq:wave2fivebrane}, the wave solution \eqref{eq:E7wave} reads
\begin{equation}
\begin{aligned}
\dd s^2 &= (2-H)\left[-(\dd X^t)^2 + \delta^{mn}\dd Y_{mz} \dd Y_{nz} 
						+ \delta_{mn}\dd X^m \dd X^n - (\dd Y_{tz})^2 \right] 
						- (\dd W_z)^2 \\
		&\qquad + H\left[(\dd Z^{tz})^2 - \delta^{mn}\dd W_m \dd W_n 
						- \delta_{mn}\dd Z^{mz} \dd Z^{nz} + (\dd W_t)^2 \right]	
						+ (\dd X^z)^2 \\	
		&\qquad + 2(H-1)\left[\dd X^t \dd Z^{tz} - \delta^{mn}\dd W_m \dd Y_{nz} 
						+ \delta_{mn}\dd X^m \dd Z^{nz} - \dd W_t\dd Y_{tz} \right]  \\	
		&\qquad + \delta_{mn} \dd Z^{tm} \dd Z^{tn} + \delta^{mn,kl}\dd Y_{mn} \dd Y_{kl} 
				- \delta_{mn,kl}\dd Z^{mn} \dd Z^{kl} - \delta^{mn}\dd Y_{tm} \dd Y_{tn} \, .
\end{aligned}
\label{eq:E7fivebrane}
\end{equation}
The KK-reduction ansatz to reduce the extended dimensions is based on the line element given in \eqref{eq:genmetric1}
\begin{equation}
\begin{aligned}
\dd s^2 &= g^{-1/2}\left\{\left[g_{\mu\nu} + \frac{1}{2}e^{2\gamma_1}
		\left(g_{\mu\nu}U^\rho U_\rho - U_\mu U_\nu\right)\right]\dd X^\mu X^\nu \right.  \\
	&\qquad + \left[e^{2\alpha_1}g^{\rho\sigma,\lambda\tau} 
			- \frac{1}{2}e^{2\gamma_2}U^{[\rho}g^{\sigma][\lambda}U^{\tau]}\right] 
				\dd Y_{\rho\sigma}\dd Y_{\lambda\tau} \\
	&\qquad + e^{2\alpha_2}g^{-1} g_{\rho\sigma,\lambda\tau}
				\dd Z^{\rho\sigma}\dd Z^{\lambda\tau}  
		+ e^{2\alpha_3}g^{-1}g^{\mu\nu}\dd W_\mu \dd W_\nu \\
	&\qquad + \frac{2}{\sqrt{2}}e^{2\beta_1}g^{-1/2}
			g_{\mu[\lambda}U_{\tau]}\dd X^\mu \dd Z^{\lambda\tau}
	+ \frac{2}{\sqrt{2}}e^{2\beta_2}g^{-1/2}U^{[\rho}g^{\sigma]\nu} 
				\dd Y_{\rho\sigma}\dd W_\nu
		\left. \vphantom{\frac{1}{2}}\right\} 
\end{aligned}
\label{eq:KKansatzU}
\end{equation}
where the scale factors $e^{2\alpha}$, $e^{2\beta}$ and $e^{2\gamma}$ are undetermined. They arise naturally in such a reduction ansatz which attempts to reduce 49 dimensions at once and will be determined by consistency.

By comparing \eqref{eq:KKansatzU} to \eqref{eq:E7fivebrane} term by term, one can step by step work out the fields of the reduced solution. The term with $\dd W^2$ gives
\begin{align}
e^{2\alpha_3}g^{-3/2}g^{zz} &= -1 &
e^{2\alpha_3}g^{-3/2}g^{tt} &= H &
e^{2\alpha_3}g^{-3/2}g^{mn} &= -H\delta^{mn}
\label{eq:WreductionGinv}
\end{align}
while the $\dd Z^2$ term gives
\begin{equation}
\begin{aligned}
e^{2\alpha_2}g^{-3/2}g_{tz,tz} &= H \, , & \qquad
e^{2\alpha_2}g^{-3/2}g_{zm,zn} &= -H\delta_{mn} \\
e^{2\alpha_2}g^{-3/2}g_{tm,tn} &= \delta_{mn}\, , & \qquad
e^{2\alpha_2}g^{-3/2}g_{mn,kl} &= -\delta_{mn,kl} \, .
\end{aligned}
\label{eq:WreductionGG}
\end{equation}
Using \eqref{eq:WreductionGinv}, the cross-term $\dd Y\dd W$ gives an expression for $U^\mu$ which encodes the six-form potential
\begin{equation}
\left.
\begin{aligned}
-e^{2\beta_2}g^{-1}U^zg^{tt} &= -(H-1) \\
-e^{2\beta_2}g^{-1}U^zg^{mn} &= (H-1)\delta^{mn}
\end{aligned}
\right\}  \quad \longrightarrow \quad
e^{2\beta_2-2\alpha_3}g^{1/2}U^z = \frac{H-1}{H}  \, .
\label{eq:WreductionU}
\end{equation}

Next consider the $\dd Y^2$ term which gives
\begin{equation}
\begin{aligned}
e^{2\alpha_1}g^{-1/2}g^{mz,nz} + e^{2\gamma_2}g^{-1/2}g^{mn}U^zU^z 
	&= (2-H)\delta^{mn}\, , & \quad
e^{2\alpha_1}g^{-1/2}g^{mn,kl} &= \delta^{mn,kl}  \\ 
e^{2\alpha_1}g^{-1/2}g^{tz,tz} + e^{2\gamma_1}g^{-1/2}g^{tt}U^zU^z 
	&= -(2-H)\, , &  \quad
e^{2\alpha_1}g^{-1/2}g^{tm,tn} &= -\delta^{mn} 
\end{aligned}
\end{equation}
and using \eqref{eq:WreductionGinv} and \eqref{eq:WreductionU} one can extract
\begin{equation}
\begin{aligned}
e^{2\alpha_1}g^{-1/2}g^{zm,zn} &= 
	\left[(2-H) + H\frac{(H-1)^2}{H^2}e^{2\gamma_2+2\alpha_3-4\beta_2}\right]\delta^{mn} = H^{-1}\delta^{mn} \\
e^{2\alpha_1}g^{-1/2}g^{tz,tz} &= 
	-\left[(2-H) + H\frac{(H-1)^2}{H^2}e^{2\gamma_2+2\alpha_3-4\beta_2}\right] = -H^{-1} 
\end{aligned}
\end{equation}
if the factor $e^{2\gamma_2+2\alpha_3-4\beta_2}$ is equal to 1. The penultimate step is to look at the $\dd X\dd Z$ term 
\begin{align}
e^{2\beta_1}g^{-1}g_{tt}U_z &= (H-1) \, , & 
e^{2\beta_1}g^{-1}g_{mn}U_z &= -(H-1)\delta_{mn} 
\end{align}
and the $\dd X^2$ term which gives 
\begin{equation}
\begin{aligned}
g^{-1/2}g_{tt} + e^{2\gamma_1}g^{-1/2}g_{tt}U^zU_z &= -(2-H) \\
g^{-1/2}g_{mn} + e^{2\gamma_1}g^{-1/2}g_{mn}U^zU_z &= (2-H) \\
g^{-1/2}g_{zz} &= 1 \, .
\end{aligned}
\end{equation}
They can all be combined to determine the two remaining components of the metric 
\begin{equation}
\begin{aligned}
g^{-1/2}g_{tt} &= -\left[(2-H) + \frac{(H-1)^2}{H}e^{2\gamma_1+2\alpha_3-2\beta_1-2\beta_2}\right] 
	= -H^{-1} \\
g^{-1/2}g_{mn} &= \left[(2-H) + \frac{(H-1)^2}{H}e^{2\gamma_1+2\alpha_3-2\beta_1-2\beta_2}\right]\delta_{mn} 
	= H^{-1}\delta_{mn} 
\end{aligned}
\end{equation}
provided that $e^{2\gamma_1+2\alpha_3-2\beta_1-2\beta_2}=1$. Collecting all the above results, we have\footnote{The order of the entries in the diagonal matrices have indices $[t,m,z]$ for $g_{\mu\nu}$ and $g^{\mu\nu}$. For $g_{\mu\nu,\rho\sigma}$ and $g^{\mu\nu,\rho\sigma}$ the order is $[tm,tz,mn,mz]$.} 
\begin{equation}
\begin{aligned}
g^{-1/2}g_{\mu\nu} &= H^{-1}\diag[-1,\delta_{mn},H] \\
e^{2\alpha_3}g^{-3/2}g^{\mu\nu} &= -H\diag[-1,\delta^{mn},H^{-1}] \\
e^{2\alpha_2}g^{-3/2}g_{\mu\nu,\rho\sigma} &= -\diag[-\delta_{mn},-H,\delta_{mn,kl},H\delta_{mn}] \\
e^{2\alpha_1}g^{-1/2}g^{\mu\nu,\rho\sigma} &= \diag[-\delta^{mn},-H^{-1},\delta^{mn,kl},H^{-1}\delta^{mn}] \, .
\end{aligned}
\end{equation}
From the first line the determinant of the spacetime metric can be computed as $g=-H^{12/5}$ and thus $g_{\mu\nu}$ is finally determined. The three objects in the other lines, the inverse metric $g^{\mu\nu}$, $g_{\mu\nu,\rho\sigma}$ and $g^{\mu\nu,\rho\sigma}$, are all related to the metric. For this to be consistent and the constraints mentioned above to be satisfied, the factors $e^{2\alpha}$, $e^{2\beta}$ and $e^{2\gamma}$ have to be
\begin{equation}
\begin{aligned}
e^{2\alpha_1} &= H^{8/5} = |g|^{2/3} & \qquad
e^{2\beta_1} &= H^{2} = |g|^{5/6} & \qquad
e^{2\gamma_1} &= H^{4/5} = |g|^{1/3} \\
e^{2\alpha_2} &= H^{16/5} = |g|^{4/3} & \qquad
e^{2\beta_2} &= H^{18/5} = |g|^{3/2} & \qquad
e^{2\gamma_2} &= H^{12/5} = |g| \\
e^{2\alpha_3} &= H^{24/5} = |g|^{2}\, . & 
\end{aligned}
\end{equation} 
With this the factor in front of $U^z$ in \eqref{eq:WreductionU} now also vanishes and the six-form potential can be worked out from \eqref{eq:defU} as
\begin{equation}
U^z = \frac{H-1}{H} \quad \longrightarrow \quad
\tC_{tx^1x^2x^3x^4x^5} = \frac{H-1}{H} = -(H^{-1}-1) \, .
\end{equation}
Thus the result of reducing the full solution \eqref{eq:E7fivebrane} down to seven dimensions is
\begin{equation}
\begin{aligned}
\dd s^2 &= H^{1/5}\left[-\dd t + \dd\vec{x}_{(5)}^{\, 2} + H \dd z^2\right] \\
\tC_{tx^1x^2x^3x^4x^5} &= -(H^{-1}-1)  \\
H &= 1 + \frac{h}{z}  \, .
\end{aligned}
\end{equation}
where the harmonic function has to be smeared over the reduced directions. This is precisely the fivebrane solution in seven dimensions, obtained from reducing \eqref{eq:classicfivebrane} on $x^3,x^4,x^5$ and $x^6$ (and smearing $H$).

\section{From Monopole to Fivebrane}
\label{sec:appMonopole}
In Section \ref{sec:M5monopole} the extended monopole solution with its KK-circle in a membrane wrapping direction was shown to give the fivebrane coupled to its magnetic potential in ordinary spacetime. The details of this calculation are given here.

The monopole solution \eqref{eq:E7monopole} is transformed by \eqref{eq:monopole2fivebrane} to have its KK-circle along $Y_{wz}$. The extended line element then reads
\begin{equation}
\begin{aligned}
\dd s^2 &= (1+H^{-2}A^2)\left[(X^w)^2 + (\dd Y_{uv})^2
		+ (\dd X^z)^2 + H^{-2}(\dd Z^{wz})^2\right] \\
	&\quad + (1+H^{-2}A_1^2)\left[(\dd X^1)^2 + H^{-2}(\dd Z^{1z})^2 
		+ H^{-2}(\dd Y_{23})^2 + H^{-2}(\dd Z^{w1})^2\right] \\
	&\qquad + (1+H^{-2}A_2^2)\left[\dots\right]+ (1+H^{-2}A_3^2)\left[\dots\right] \\
	&\quad + (1+H^{-2}A_1^2+H^{-2}A_2^2)
		\left[H^{-1}(\dd Y_{u3})^2 + H^{-1}(\dd Y_{v3})^2\right] \\
	&\qquad + (1+H^{-2}A_1^2+H^{-2}A_3^2)	\left[\dots\right]
		+ (1+H^{-2}A_2^2+H^{-2}A_3^2)\left[\dots\right] \\
	&\quad + 2H^{-2}A_1A_2\left[\dd X^1 \dd X^2 - H^{-1}\dd Y_{u1}\dd Y_{u2}
		- H^{-1}\dd Y_{v1}\dd Y_{v2} \right. \\
	&\hspace{3cm} \left. + H^{-2}\dd Z^{1z}\dd Z^{2z} - H^{-2}\dd Y_{12}\dd Y_{23}
		+ H^{-2}\dd Z^{w1}\dd Z^{w2}\right] \\
	&\qquad + 2H^{-2}A_1A_3\left[\dots\right] + 2H^{-2}A_2A_3\left[\dots\right] \\
	&\quad + 2H^{-1}A_1\left[H^{-1}(-\dd X^1\dd Y_{wz} + \dd X^w\dd Y_{1z} 
		+ \dd Y_{uv}\dd Z^{23} + \dd X^z\dd Y_{w1})\right. \\
	&\hspace{2.6cm}\left. + H^{-2}(-\dd Z^{v3}\dd Y_{u2} + \dd Z^{v2}\dd Y_{u3}
		- \dd Y_{v3}\dd Z^{u2} + \dd Y_{v2}\dd Z^{u3}) \right. \\
	&\hspace{2.6cm}\left. + H^{-3}(\dd Z^{1z}\dd W_w + \dd Y_{23}\dd Z^{uv}
		+ \dd Z^{w1}\dd W_z - \dd W_1\dd Z^{wz})\right] \\
	&\qquad + 2H^{-1}A_2\left[\dots\right] + 2H^{-1}A_3\left[\dots\right] \\
	&\quad  + H^{-1}\left[\delta_{ab}\dd X^a\dd X^b + (\dd Y_{uz})^2
		+ \delta^{ab} \dd Y_{aw}\dd Y_{bw} + (\dd Y_{vz})^2\right] \\
	&\quad + H^{-2}\left[(\dd Y_{wz})^2  + \delta^{ij}\dd Y_{iz}\dd Y_{jz} 
		+ \delta_{ij,kl}\dd Z^{ij}\dd Z^{kl} + \delta^{ij} \dd Y_{wi}\dd Y_{wj}\right] \\
	&\quad + H^{-3}\left[\delta^{ab}\dd W_a\dd W_b + (\dd Z^{uz})^2  
		+ \delta_{ab} \dd Z^{aw}\dd Z^{bw} + (\dd Z^{vz})^2 \right. \\
	&\hspace{2cm} \left. + \delta_{ij} \dd Z^{ui}\dd Z^{uj} 
		+ \delta_{ij} \dd Z^{vi}\dd Z^{vj}\right] \\
	&\quad + H^{-4}\left[(\dd W_w)^2 + (\dd Z^{uv})^2 + (\dd W_z)^2
		+ \delta^{ij} \dd W_i\dd W_{j}\right] \, .
\end{aligned}
\label{eq:E7fivebrane2}
\end{equation}
A suitable KK-ansatz to extract the spacetime metric and three-form potential is based on \eqref{eq:genmetric2}
\begin{align}
\dd s^2 &= g^{-1/2}\left\{\left[g_{\mu\nu} 
		+ \frac{1}{2}e^{2\gamma_1}C_{\mu\rho\sigma}g^{\rho\sigma,\lambda\tau}
			C_{\lambda\tau\nu} \right]\dd X^\mu \dd X^\nu \right. \notag\\ 
	&\qquad 	
		+ \left[e^{2\alpha_1}g^{\mu_1\mu_2,\nu_1\nu_2} 
		+ \frac{1}{2}e^{2\gamma_2}V^{\mu_1\mu_2\rho\sigma}
			g_{\rho\sigma,\lambda\tau}V^{\lambda\tau\nu_1\nu_2}\right] 
				\dd Y_{\mu_1\mu_2}\dd Y_{\nu_1\nu_2} \notag\\
	&\qquad  + g^{-1} \left[e^{2\alpha_2}g_{\mu_1\mu_2,\nu_1\nu_2} 
		+ \frac{1}{2}e^{2\gamma_3}C_{\mu_1\mu_2\rho}g^{\rho\sigma}C_{\sigma\nu_1\nu_2}\right]
				\dd Z^{\mu_1\mu_2}\dd Z^{\nu_1\nu_2} \notag\\ 
	&\qquad
		+ e^{2\alpha_3}g^{-1}g^{\mu\nu}\dd W_\mu \dd W_\nu \notag\\ 
	&\qquad	
		+ \frac{2}{\sqrt{2}}e^{2\beta_1}C_{\mu\rho\sigma}g^{\rho\sigma,\lambda\tau}
			\dd X^\mu \dd Y_{\lambda\tau} \notag\\ 
	&\qquad	
		+ \frac{2}{\sqrt{2}}e^{2\beta_2}g^{-1/2}
			V^{\mu_1\mu_2\rho\sigma}g_{\rho\sigma,\nu_1\nu_2} 
			\dd Y_{\mu_1\mu_2}\dd Z^{\nu_1\nu_2}  \notag\\ 
	&\qquad \left.
		+ \frac{2}{\sqrt{2}}e^{2\beta_3}g^{-1/2}
			C_{\mu_1\mu_2\rho}g^{\rho\nu}
			\dd Z^{\mu_1\mu_2}\dd W_\nu   \right\}
	\label{eq:KKansatzC}
\end{align}

where again the a priori undetermined scale factors $e^{2\alpha}$, $e^{2\beta}$ and $e^{2\gamma}$ have to be included. We now proceed in the usual way, comparing \eqref{eq:KKansatzC} to \eqref{eq:E7fivebrane2} term by term to determine all the fields. The scale factors are then picked to ensure a consistent solution. Start with the $\dd W^2$ term
\begin{equation}
\begin{aligned}
e^{2\alpha_3}g^{-3/2}g^{ab} &= H^{-3}\delta^{ab}\, , & \quad
e^{2\alpha_3}g^{-3/2}g^{ww} &= H^{-4} \\
e^{2\alpha_3}g^{-3/2}g^{ij} &= H^{-4}\delta^{ij}\, , & \quad
e^{2\alpha_3}g^{-3/2}g^{zz} &= H^{-4}
\end{aligned}
\label{eq:MreductionGinv}
\end{equation}
which can be used in the $\dd Z\dd W$ term to find an expression for the three-form potential 
\begin{equation}
\left.
\begin{aligned}
e^{2\beta_3}g^{-1}C_{wzi}g^{ij} &= -H^{-3}A^j \\
e^{2\beta_3}g^{-1}C_{izw}g^{ww} &= H^{-3}A_i \\
e^{2\beta_3}g^{-1}C_{wiz}g^{zz} &= H^{-3}A_i
\end{aligned}
\right\} \quad  \longrightarrow \quad
e^{2\beta_3-2\alpha_3}g^{1/2}C_{izw} = A_i \, .
\label{eq:MreductionC}
\end{equation}
Once this is established, it can be used in the $\dd Z^2$ terms 
\begin{equation}
\begin{aligned}
e^{2\alpha_2}g^{-3/2}g_{wz,wz} + e^{2\gamma_3}g^{-1/2}C_{wzi}g^{ij}C_{jwz} &= H^{-2}+H^{-4}A^2 \\
e^{2\alpha_2}g^{-3/2}g_{wi,wj} + e^{2\gamma_3}g^{-1/2}C_{wiz}g^{zz}C_{zwj} 
	&= H^{-2}\delta_{ij}+H^{-4}A_iA_j \\
e^{2\alpha_2}g^{-3/2}g_{iz,jz} + e^{2\gamma_3}g^{-1/2}C_{iz}g^{ww}C_{wjz} 
	&= H^{-2}\delta_{ij}+H^{-4}A_iA_j
\end{aligned}
\end{equation}	
together with \eqref{eq:MreductionGinv} to find
\begin{equation}
\begin{aligned}
e^{2\alpha_2}g^{-3/2}g_{wz,wz} &= H^{-2} + H^{-4}A^2 
	- e^{2\gamma_3+2\alpha_3-4\beta_3}H^{-4}A^2 = H^{-2} \\
e^{2\alpha_2}g^{-3/2}g_{wi,wj} &= H^{-2}\delta_{ij}+H^{-4}A_iA_j
	- e^{2\gamma_3+2\alpha_3-4\beta_3}H^{-4}A_iA_j = H^{-2}\delta_{ij} \\
e^{2\alpha_2}g^{-3/2}g_{iz,jz} &= H^{-2}\delta_{ij}+H^{-4}A_iA_j
	- e^{2\gamma_3+2\alpha_3-4\beta_3}H^{-4}A_iA_j = H^{-2}\delta_{ij}
\end{aligned}
\end{equation}
provided that $e^{2\gamma_3+2\alpha_3-4\beta_3}$ is equal to 1. The remaining components of $g_{\mu\nu,\rho\sigma}$ are
\begin{equation}
\begin{aligned}
e^{2\alpha_2}g^{-3/2}g_{ij,kl} &= H^{-2}\delta_{ij,kl} \, , & \quad
e^{2\alpha_2}g^{-3/2}g_{ai,bj} &= H^{-3}\delta_{ab}\delta_{ij} \\
e^{2\alpha_2}g^{-3/2}g_{aw,bw} &= H^{-3}\delta_{ab} \, , & \quad
e^{2\alpha_2}g^{-3/2}g_{az,bz} &= H^{-3}\delta_{ab} \\
e^{2\alpha_2}g^{-3/2}g_{uv,uv} &= H^{-4} \, .
\end{aligned}
\label{eq:MreductionGG}
\end{equation}	
We continue with the $\dd Y\dd Z$ terms containing the object $V^{\mu\nu\rho\sigma}$. They are all of the same form (up to a sign), for example
\begin{equation}
e^{2\beta_2}g^{-1}V^{u2v3}g_{v3,v3} = -H^{-3}A_1 \quad \longrightarrow \quad
e^{2\beta_2-2\alpha_2}g^{1/2}V^{u2v3} = -A_1
\end{equation}
where \eqref{eq:MreductionGG} was used. Looking at all the terms with the relevant sign and taking the order of the $i$-type index into account, the general expression is  
\begin{equation}
e^{2\beta_2-2\alpha_2}g^{1/2}V^{uvij} = \epsilon^{ijk}A_k  \, .
\label{eq:MreductionV}
\end{equation} 
This can in turn be used in the $\dd Y^2$ terms
\begin{equation}
\begin{aligned}
e^{2\alpha_1}g^{-1/2}g^{uv,uv} + e^{2\gamma_2}g^{-1/2}V^{uvij}g_{ij,kl}V^{kluv} &= 		
	1+H^{-2}A^2 \\
e^{2\alpha_1}g^{-1/2}g^{23,23} + e^{2\gamma_2}g^{-1/2}V^{23uv}g_{uv,uv}V^{uv23} &= 
	H^{-2}  + H^{-4}A_1^2 \\
e^{2\alpha_1}g^{-1/2}g^{a3,b3} + e^{2\gamma_2}g^{-1/2}V^{a3ci}g_{ci,dj}V^{djb3} &= 
	H^{-1}\delta^{ab} + H^{-3}\delta^{ab}(A_1^2+A_2^2)
\end{aligned}
\end{equation}
together with \eqref{eq:MreductionGG} to find
\begin{equation}
\begin{aligned}
e^{2\alpha_1}g^{-1/2}g^{uv,uv} &=   		
	1+H^{-2}A^2 - e^{2\gamma_2+2\alpha_2-4\beta_2}H^{-2}A^2 = 1 \\
e^{2\alpha_1}g^{-1/2}g^{23,23} &=
  H^{-2} + H^{-4}A_1^2 - e^{2\gamma_2+2\alpha_2-4\beta_2}H^{-4}A_1^2 = H^{-2} \\
e^{2\alpha_1}g^{-1/2}g^{a3,b3} &= 
	H^{-1}\delta^{ab} + H^{-3}\delta^{ab}(A_1^2+A_2^2) - e^{2\gamma_2+2\alpha_2-4\beta_2}H^{-3}\delta^{ab} (A_1^2+A_2^2) \\
	&= H^{-1}\delta^{ab}
\end{aligned}
\end{equation}
provided that $e^{2\gamma_2+2\alpha_2-4\beta_2}$ is equal to 1. The same holds for other values of the $i$-type index. The remaining components of $g^{\mu\nu,\rho\sigma}$ are
\begin{equation}
\begin{aligned}
e^{2\alpha_1}g^{-1/2}g^{aw,bw} &= H^{-1}\delta^{ab}  \, , & \quad
e^{2\alpha_1}g^{-1/2}g^{az,bz} &= H^{-1}\delta^{ab} \\
e^{2\alpha_1}g^{-1/2}g^{wi,wj} &= H^{-2}\delta^{ij}  \, , & \quad
e^{2\alpha_1}g^{-1/2}g^{iz,jz} &= H^{-2}\delta^{ij} \\
e^{2\alpha_1}g^{-1/2}g^{wz,wz} &= H^{-2}  \, .
\end{aligned}
\label{eq:MreductionGGinv}
\end{equation}

The final cross-term to consider is the $\dd X\dd Y$ term which together with \eqref{eq:MreductionGGinv} yields another expression for the three-form potential
\begin{equation}
\left.
\begin{aligned}
e^{2\beta_1}g^{-1/2}C_{iwz}g^{wz,wz} &= -H^{-2}A_i \\
e^{2\beta_1}g^{-1/2}C_{wiz}g^{iz,jz} &= H^{-3}A^j \\
e^{2\beta_1}g^{-1/2}C_{zwi}g^{wi,wj} &= H^{-3}A^j
\end{aligned}
\right\} \quad  \longrightarrow \quad
e^{2\beta_1-2\alpha_1}g^{1/2}C_{izw} = A_i \, .
\label{eq:MreductionCC}
\end{equation}
In a last step, the $\dd X^2$ terms 
\begin{equation}
\begin{aligned}
g^{-1/2}g_{ww} + e^{2\gamma_1}g^{-1/2}C_{wiz}g^{iz,jz}C_{jzw} &= 1+H^{-2}A^2 \\
g^{-1/2}g_{zz} + e^{2\gamma_1}g^{-1/2}C_{wiz}g^{wi,wj}C_{wjz} &= 1+H^{-2}A^2 \\
g^{-1/2}g_{ij} + e^{2\gamma_1}g^{-1/2}C_{iwz}g^{wz,wz}C_{wzj} &= \delta_{ij}+H^{-2}A_iA_j \\
g^{-1/2}g_{ab} &= H^{-1}
\end{aligned}
\end{equation}
are combined with previous statements to to determine the spacetime metric
\begin{equation}
\begin{aligned}
g^{-1/2}g_{ww} &= 1+H^{-2}A^2 - e^{2\gamma_1+2\alpha_1-4\beta_1}H^{-2}A^2 = 1 \\
g^{-1/2}g_{zz} &= 1+H^{-2}A^2 - e^{2\gamma_1+2\alpha_1-4\beta_1}H^{-2}A^2 = 1 \\
g^{-1/2}g_{ij} &= \delta_{ij}+H^{-2}A_iA_j - e^{2\gamma_1+2\alpha_1-4\beta_1}A_iA_j = \delta_{ij}
\end{aligned}
\end{equation}
provided that $e^{2\gamma_1+2\alpha_1-4\beta_1}$ is equal to 1. Collecting all the above results, we have\footnote{The order of the entries in the diagonal matrices have indices $[a,w,i,z]$ for $g_{\mu\nu}$ and $g^{\mu\nu}$. For $g_{\mu\nu,\rho\sigma}$ and $g^{\mu\nu,\rho\sigma}$ the order is $[ab,aw,ai,az,wi,wz,ij,iz]$.} 
\begin{equation}
\begin{aligned}
g^{-1/2}g_{\mu\nu} &= H^{-1}\diag[\delta_{ab},H,H\delta_{ij},H] \\
e^{2\alpha_3}g^{-3/2}g^{\mu\nu} 
	&= H^{-3}\diag[\delta^{ab},H^{-1},H^{-1}\delta^{ij},H^{-1}] \\
e^{2\alpha_2}g^{-3/2}g_{\mu\nu,\rho\sigma} 
	&= H^{-4}\diag[1,H\delta_{ab},H\delta_{ab}\delta_{ij},H\delta_{ab}, 
		H^2\delta_{ij},H^2,H^2\delta_{ij,kl},H^2\delta_{ij}] \\
e^{2\alpha_1}g^{-1/2}g^{\mu\nu,\rho\sigma} 
	&= \diag[1,H^{-1}\delta^{ab},H^{-1}\delta^{ab}\delta^{ij},H^{-1}\delta^{ab}, \\
	&\hspace{5cm}	H^{-2}\delta^{ij},H^{-2},H^{-2}\delta^{ij,kl},H^{-2}\delta^{ij}] \, .
\end{aligned}
\end{equation}
From the first line the determinant of the spacetime metric can be computed as $g=H^{4/5}$ and thus $g_{\mu\nu}$ is finally determined. The three objects in the other lines, the inverse metric $g^{\mu\nu}$, $g_{\mu\nu,\rho\sigma}$ and $g^{\mu\nu,\rho\sigma}$, are all related to the metric. For this to be consistent and the constraints mentioned above to be satisfied, the factors $e^{2\alpha}$, $e^{2\beta}$ and $e^{2\gamma}$ have to be
\begin{equation}
\begin{aligned}
e^{2\alpha_1} &= H^{-4/5} = g^{-1} & \qquad
e^{2\beta_1} &= H^{-6/5} = g^{-3/2} & \qquad
e^{2\gamma_1} &= H^{-8/5} = g^{-2} \\
e^{2\alpha_2} &= H^{-8/5} = g^{-2} & \qquad
e^{2\beta_2} &= H^{-10/5} = g^{-5/2} & \qquad
e^{2\gamma_2} &= H^{-12/5} = g^{-3} \\
e^{2\alpha_3} &= H^{-12/5} = g^{-3} & \qquad
e^{2\beta_3} &= H^{-14/5} = g^{-7/2} & \qquad
e^{2\gamma_3} &= H^{-16/5} = g^{-4} \, .  
\end{aligned}
\end{equation}
Having set the scale factors, the prefactors in \eqref{eq:MreductionC}, \eqref{eq:MreductionV} and \eqref{eq:MreductionCC} vanish and $V^{\mu\nu\rho\sigma}$ can be converted into $C_{\mu\nu\rho}$ via \eqref{eq:defV} which all boils down to $C_{izw}=A_i$. Thus the result of reducing the full solution \eqref{eq:E7fivebrane2} down to seven dimensions is
\begin{equation}
\begin{aligned}
\dd s^2 &= H^{-3/5}[\dd\vec{x}_{(2)}^{\, 2} 
	+ H(\dd w^2 + \dd\vec{y}_{(3)}^{\, 2} + \dd z^2)]  \\
C_{izw} &= A_i  \\
H &= 1 + \frac{h}{|w^2 + \vec{y}_{(3)}^{\, 2} + z^2|^{3/2}} \, .
\end{aligned}
\end{equation}
where the harmonic function is smeared over the reduced directions. This is precisely the fivebrane solution in seven dimensions, obtained from reducing \eqref{eq:classicfivebrane} on $x^3,x^4,x^5$ and $t$ (and smearing $H$) with its magnetic potential.

\chapter{Embedding the Type II Theories into EFT}
\label{ch:appEFT}

In this appendix we show how the Type II theories can be embedded into EFT. The difference between Type IIA and Type IIB arises from applying different solutions of the section condition to the EFT equations.

\section{The Type IIA Theory}
The ten-dimensional Type IIA theory is a simple reduction of eleven-dimensional supergravity on a circle. It is thus possible to embed it into EFT as well using the same solution to the section condition given in Section \ref{sec:embedding}. Instead of a $4+7$ coordinate split in the KK-decomposition we now have a $4+6$ split. The dictionary for embedding the Type IIA fields into EFT can be obtained from the results of Section \ref{sec:embedding} by simply splitting the internal seven-dimensional sector into $6+1$ by doing another KK-decomposition. 

Under the split $y^m = (y^\bm,y^\theta)$ with $\bm=1,\dots,6$ and $y^\theta=\theta$ the coordinate of the circle, the corresponding generalized coordinates read
\begin{equation}
Y^M = (y^\bm, y^\theta, y_{\bm\bn}, y_{\bm\theta} , 
		y_\bm, y_\btheta, y^{\bm\bn},y^{\bm\btheta}) \, .
\label{eq:gencoordIIA}
\end{equation}
Noting that the internal metric $g_{mn}$ of our solution is diagonal, no KK-vector will arise from this decomposition. We thus simply have
\begin{equation}
g_{mn} = \diag[ e^{-\phi/6}\bg_{\bm\bn}, e^{4\phi/3} ]
\end{equation}
where $e^{2\phi}$ is the string theory dilaton of the Type IIA theory and the precise numerical powers have been chosen to be in the Einstein frame. Inserting this ansatz into the generalized metric for embedding supergravity into EFT \eqref{eq:genmetricSUGRA} gives
\begin{align}
\MM_{MN}(\bg_{\bm\bn},\phi) = \bg^{1/2}\diag [&
	\bg_{\bm\bn}, e^{3\phi/2}, e^{\phi/2} \bg^{\bm\bn,\bk\bl}, e^{-\phi} \bg^{\bm\bn}, \notag\\
&	\bg^{-1} \bg^{\bm\bn}, e^{-3\phi/2}\bg^{-1}, e^{-\phi/2}\bg^{-1}\bg_{\bm\bn,\bk\bl},
		e^{\phi}\bg^{-1}\bg_{\bm\bn}]
\label{eq:genmetricIIA}
\end{align}
where $\bg$, $\bg_{\bm\bn,\bk\bl}$ and $\bg^{\bm\bn,\bk\bl}$ are defined in terms of $\bg_{\bm\bn}$ as before (barred quantities are six-dimensional). As in the eleven-dimensional case, we are not considering any internal components of the RR or NSNS gauge potentials. The only non-zero components are the one-form parts which are in the EFT vector potential ${\AAA_\mu}^M$ as before. The vector is also split under the above decomposition resulting in a component for each of the directions given in \eqref{eq:gencoordIIA}
\begin{equation}
\left\{ {\AAA_\mu}^\bm, {\AAA_\mu}^\theta, \AAA_{\mu\ \bm\bn}, \AAA_{\mu\ \bm\theta},
\AAA_{\mu\ \bm}, \AAA_{\mu\ \theta}, {\AAA_\mu}^{\bm\bn}, {\AAA_\mu}^{\bm\theta} \right\} \, .
\label{eq:vecpotIIA}
\end{equation}
All these parts of ${\AAA_\mu}^M$ encode a component of a field from the Type IIA theory except $\AAA_{\mu\ \bm}$ which relates to the dual graviton. The first one, ${\AAA_\mu}^\bm$ is just the KK-vector of the original $4+6$ decomposition. The RR-fields $C_1$, $C_3$, $C_5$ and $C_7$ are encoded in ${\AAA_\mu}^\theta$, $\AAA_{\mu\ \bm\bn}$, ${\AAA_\mu}^{\bm\bn}$ and $\AAA_{\mu\ \theta}$ respectively where the latter two have to be dualized on the internal six-dimensional space. The remaining two, $\AAA_{\mu\ \bm\theta}$ and ${\AAA_\mu}^{\bm\theta}$ contain the NSNS-fields $B_2$ and $B_6$, where again the second one has to be dualized. It is nice to see how the self-duality of the EFT vector contains all the known dualities between the form fields in the Type IIA theory.

\section{The Type IIB Theory}
Unlike the Type IIA theory, the Type IIB theory does not follow from the solution to the section condition that gives eleven-dimensional supergravity. There is another, inequivalent solution \cite{Hohm:2013uia} which is related to a different decomposition of the fundamental representation of $E_7$. The relevant maximal subgroup is $GL(6)\times SL(2)$ and we have
\begin{equation}
\mathbf{56} \rightarrow (6,1) + (6,2) + (20,1) + (6,2) + (6,1)
\end{equation}  
which translates to the following splitting of the extended internal coordinates
\begin{equation}
Y^M = (y^\bm, y_{\bm\ a}, y_{\bm\bn\bk}, y^{\bm\ a}, y_\bm)
\label{eq:gencoordIIB}
\end{equation}
where again $\bm=1,\dots,6$ and $a=1,2$ is an $SL(2)$ index. The middle component is totally antisymmetric in all three indices. Note that the six-dimensional index is \emph{not} the same as in the $6+1$ Type IIA decomposition above. Here we rather have a $5+2$ split where the $y_{ab}$ (a single component) is reinterpreted as the sixth component of $y^\bm$. Loosely speaking this comes from the fact that Type IIB on a circle is related to M-theory on a torus. This is made precise at the end of this section.

From \cite{Hohm:2013uia}, the generalized metric (again without any contribution from the internal components of the form fields) for this case is given by
\begin{equation}
\MM_{MN}(\bg_{\bm\bn},\gamma_{ab}) = \bg^{1/2}\diag [ \bg_{\bm\bn}, \bg^{\bm\bn}\gamma^{ab}, \bg^{-1}\bg_{\bm\bk\bp,\bn\bl\bq},
	\bg^{-1}\bg_{\bm\bn}\gamma_{ab}, \bg^{-1}\bg^{\bm\bn}]
\label{eq:genmetricIIBapp}
\end{equation}
where $\bg_{mkp,nlq}=\bg_{m[n|}\bg_{k|l|}\bg_{p|q]}$ (in analogy to $g_{mn,kl}$ above) and $\gamma_{ab}$ is the metric on the torus
\begin{equation}
\gamma_{ab} = \frac{1}{\mathrm{Im\, } \tau}
\begin{pmatrix}
|\tau|^2 & \mathrm{Re\, } \tau \\ \mathrm{Re\, } \tau & 1
\end{pmatrix},
\qquad
\tau = C_{0} + i e^{-\phi}
\label{eq:gamma}
\end{equation}
with the complex torus parameter $\tau$ (the ``axio-dilaton'') given in terms of the RR-scalar $C_0$ and the string theory dilaton $e^{2\phi}$. We will come back to this setup at the end of this section.
 
The EFT vector is also decomposed and has a component for each direction in \eqref{eq:gencoordIIB}
\begin{equation}
\left\{ {\AAA_\mu}^\bm, \AAA_{\mu\ \bm \ a}, \AAA_{\mu\ \bm\bn\bk},
	{\AAA_\mu}^{\bm\ a}, \AAA_{\mu\ \bm} \right\} \, .
\label{eq:vecpotIIB}
\end{equation}
As before, these parts each encode a component of a field from the Type IIB theory except $\AAA_{\mu\ \bm}$ which relates to the dual graviton. As always, ${\AAA_\mu}^\bm$ is the KK-vector of the original $4+6$ decomposition. The components $\AAA_{\mu\ \bm \ a}$ and ${\AAA_\mu}^{\bm\ a}$ contain the $SL(2)$ doublets $B_2/C_2$ and $B_6/C_6$ where the latter one needs to be dualized on the internal space. Here $B$ denotes a NSNS-field and $C$ the dual RR-field. The component $\AAA_{\mu\ \bm\bn\bk}$ corresponds to the self-dual four-form $C_4$. Again it can be seen that the self-duality of the EFT vector gives the duality relations between the form fields in the Type IIB theory.

Let us conclude by checking how the Type IIB theory on a circle is related to the eleven-dimensional theory on a torus. Since both theories have the external four-dimensional spacetime in common, we will only look at the internal sector. The seven dimensions are split into $5+2$ such that the coordinates are $y^m=(y^\dm,y^a)$ where $\dm=1,\dots,5$ and $a=1,2$. Starting from \eqref{eq:gencoordM}, the generalized coordinates then decompose as
\begin{equation}
Y^M = (y^\dm, y^a, y_{\dm\dn}, y_{\dm a}, y_{ab}, 
	y_\dm, y_a, y^{\dm\dn}, y^{\dm a}, y^{ab})\, .
\end{equation}
By noting that $y_{ab}$ has only a single component (by antisymmetry), $y_{12}$, these coordinates can be repackaged into $y^\bm=(y^\dm,y_{12})$ and similar for the dual coordinates to make contact with \eqref{eq:gencoordIIB}. We thus have
\begin{equation}
\begin{array}{rrl}
(6,1): 		& 	y^\bm &= (y^\dm,y_{12}) \\
(6,2): 		&	y_{\bm a} &= (y_{\dm a}, y^a)\\
(20,1):		&	y_{\bm\bn\bk} &= (y_{\dm\dn},y^{\dm\dn})\\
(6,2):		&	y^{\bm a} &= (y^{\dm a}, y_a) \\
(6,1):		&	y_\bm &= (y_\dm,y^{12})
\end{array}
\end{equation}
justifying the presence of the six-dimensional index $\bm$ above. Now turn to the seven-dimensional metric $g_{mn}$. Again omitting a KK-vector for cross-terms, the ansatz for the decomposition is (a dot denotes a five-dimensional quantity)
\begin{equation}
g_{mn} = \diag[ \dg_{\dm\dn}, e^\Delta \gamma_{ab} ]
\end{equation}
with $\gamma_{ab}$ as given in \eqref{eq:gamma}. There the torus metric is conformal and has unit determinant. For completeness, we include a volume factor for the torus in the discussion here, such that the determinant of the $2\times 2$ sector $g_{ab}$ is $\det|e^\Delta\gamma_{ab}| =e^{2\Delta}$. This ansatz can be inserted into the generalized metric for embedding supergravity into EFT \eqref{eq:genmetricSUGRA} to give
\begin{align}
\MM_{MN} = \dg^{1/2} \diag[& e^\Delta\dg_{\dm\dn},e^{2\Delta}\gamma_{ab}, 
		e^\Delta\dg^{\dm\dn,\dpp\dq},
		\dg^{\dm\dn}\gamma^{ab}, e^{-\Delta}\gamma^{ab,cd}, \notag\\
&		\dg^{-1}e^{-\Delta}\dg^{\dm\dn}, \dg^{-1}e^{-2\Delta}\gamma^{ab},
		\dg^{-1}e^{-\Delta}\dg_{\dm\dn,\dpp\dq}, \\
&		\dg^{-1}\dg_{\dm\dn}\gamma_{ab},
		\dg^{-1}e^\Delta\gamma_{ab,cd}] \, . \notag
\end{align}
It is easy to check that the object $\gamma_{ab,cd}=\gamma_{a[c}\gamma_{d]b}$ has only one component $\gamma_{12,12}$ which evaluates to $1$ (and similarly for the inverse $\gamma^{ab,cd}$). With this in mind, the components of the generalized metric can be repackaged in terms of a six-dimensional metric $g_{\bm\bn}=e^{\Delta/2}\diag[\dg_{\dm\dn},e^{-2\Delta}]$ and determinant $\bg=e^\Delta\dg$ just like the coordinates. The five parts of the generalized metric thus read
\begin{equation}
\begin{array}{rrll}
(6,1): 		& 	\bg_{\bm\bn} &= e^{\Delta/2}&\diag[\dg_{\dm\dn},e^{-2\Delta}] \\
(6,2): 		&	\bg^{\bm\bn}\gamma^{ab} &= e^{-\Delta/2}&\diag[\dg^{\dm\dn}\gamma^{ab}, 		
												e^{2\Delta}\gamma_{ab}] \\
(20,1):		&	\bg^{-1}\bg_{\bm\bk\bp,\bn\bl\bq} &= e^{\Delta/2}&\diag[\dg^{\dm\dn,\dpp\dq}, 		
												e^{-2\Delta}\dg^{-1}\dg_{\dm\dn,\dpp\dq}] \\
(6,2):		&	\bg^{-1}\bg_{\bm\bn}\gamma_{ab} &= e^{-\Delta/2} 
												&\diag[\dg^{-1}\dg_{\dm\dn}\gamma_{ab}, 		
												e^{-2\Delta}\dg^{-1}\gamma^{ab}] \\
(6,1):		&	\bg^{-1}\bg^{\bm\bn} &= e^{-3\Delta/2}&\diag[\dg^{-1}\dg_{\dm\dn},
												\dg^{-1}e^{2\Delta}]
\end{array}
\end{equation}
which is in agreement with \eqref{eq:gencoordIIB}. These identifications here are not obvious, but can be checked by an explicit calculation of individual components.

\chapter{Harmonic Function on $\mathbb{R}^3\times T^2$}
\label{ch:apptorus}

In this appendix we give the derivation of the harmonic function localized on $\mathbb{R}^3\times T^2$ that is presented at the end of Section \ref{sec:localization}. Since the precise form of the function is -- as far as we know -- a novel result, we want to give some details of how it has been obtained. 

Solutions $H$ to the Laplace equation $\nabla^2 H = 0$ are called harmonic functions. In a general space with metric $g_{\mu\nu}$ and arbitrary coordinates $x^\mu$ the Laplacian $\Delta=\nabla^2$ (or Laplace-Beltrami operator in general) is given by
\begin{equation}
\nabla^2 H(x) = \frac{1}{\sqrt{g}}\partial_\mu\Big(\sqrt{g}g^{\mu\nu}\partial_\nu H(x)\Big)
\end{equation}
where $g$ is its determinant of the metric. The harmonic functions of supergravity solutions are usually of the form
\begin{equation}
H(r) = 1 + h f(r)
\end{equation}
where $f(r)$ solves the Laplacian up additive and multiplicative constants such as $h$ and is a function of the radial coordinate $r=\sqrt{x_\mu x^\mu}$ only. We will keep working with $f$ without worrying about the constants. 

The fivebrane in eleven dimensions has five transverse directions and is therefore described by a harmonic function on $\mathbb{R}^5$ of the form $f(r) = 1/r^3 = |x_\mu x^\mu|^{-3/2}$. The monopole on the other hand only depends on three of those coordinates, essentially it is given by a harmonic function on $\mathbb{R}^3$ (which in polars reads $\mathbb{R}^+\times S^2$ where the $S^2$ is part of the Hopf fibration of $S^3$) and takes the form $f(r) = 1/r = |x_\mu x^\mu|^{-1/2}$. Since the self-dual EFT solution of Section \ref{sec:selfdual} is smeared, it is also given in terms of the harmonic function on $\mathbb{R}^3$. We now want to extend the coordinate dependence to the remaining two directions to fully localize the solution. For consistency these two dimensions have to be compact, their topology is taken to be that of a torus $T^2=S^1_{R_1} \times S^1_{R_2}$ where $R_1$ and $R_2$ are the radii of the two circles. Hence we are looking for a solution to the Laplace operator on the space $\mathbb{R}^3\times T^2$.

If the two-torus is treated as a one-dimensional complex manifold, it is a Riemann surface of genus one. It can be defined as the quotient of the complex plane $\mathbb{C}$ with a lattice $\Lambda$, \ie $T^2=\mathbb{C}/\Lambda$. The lattice is spanned by two complex periods $\omega_1$ and $\omega_2$
\begin{equation}
\Lambda = \{m\omega_1 + n\omega_2 | m,n\in \mathbb{Z\}} \, .
\end{equation}
Instead of the complex periods the information of the shape and size of a cell in the lattice can be encoded in 
\begin{equation}
\tau = \frac{\omega_2}{\omega_1} \qandq V =  \im \omega_2 \bar{\omega}_1
\end{equation}
where $V$ is the area of the cell which gives the volume of the torus. It is important that $\im \tau >0$ so that the two complex periods are not parallel. Two lattices $\Lambda$ and $\Lambda'$ give rise to the same torus if $\Lambda' = \lambda\Lambda$ with $\lambda\in\mathbb{C}/\{0\}$. Therefore any lattice with $m\omega_1 + n\omega_2$ is equivalent to a lattice with $m + \tau n$ since 
\begin{equation}
\omega_1(m + \frac{\omega_2}{\omega_1}n)=\lambda(m + \tau n) 
\end{equation}
is of the form $\Lambda' = \lambda\Lambda$. This simplification of the lattice in terms of $\tau$ can be seen as the freedom to pick a coordinate system such that $\omega_1$ is aligned with the real axis of the complex plane. To incorporate the proper periodicity of the lattice, we set the lengths of the complex periods equal to the radii of the torus. Therefore 
\begin{equation}
	\omega_1 = R_1  \qandq  \omega_2 =  R_2 e^{i\theta}
\end{equation}
where $\theta$ is the angle between the two circles. In this setup we have
\begin{equation}
\tau = \tau_1 + i\tau_2 = \frac{R_2}{R_1}[\cos\theta + i\sin\theta] \qandq V =  R_1R_2\sin\theta \, .
\end{equation}
The complex coordinate $z$ on the torus is doubly periodic which can be described by $z\sim z + \sqrt{\frac{V}{\tau_2}}w$ for $w\in\Lambda$. The real circle coordinates $z^a$ with $a=1,2$ and whose periods of course are $R_1$ and $R_2$ respectively, are related to the complex coordinate via
\begin{equation}
	z=z^1 + \tau z^2 \, .
\end{equation}
For the real coordinates $z^a$ on the complex surface, the metric on the torus can be written as
\begin{equation}
\gamma_{ab} = \frac{V}{\tau_2}
\begin{pmatrix}
1 & \tau_1 \\ \tau_1 & |\tau|^2
\end{pmatrix}
\end{equation}
which has determinant $\gamma=V^2$. This metric contains all the information of the shape of the torus via $\tau$ and its size since the volume is given by $\sqrt{\gamma}$. In terms of the complex coordinate $z$ the only non-zero component of the metric is $\gamma_{z\bz} = V/2\tau_2$, \ie
\begin{equation}
\dd s^2 = 2\gamma_{z\bz}\dd z \dd\bz = \frac{V}{\tau_2}\dd z \dd\bz 
\end{equation}
and the determinant is $-V^2/4\tau_2$. This metric for the complex coordinate can be rescaled by the square root of its determinant which is just a conformal rescaling to retain a unimodular metric. It then reads $\dd s^2 = -\dd z\dd \bz$ with $\gamma_{z\bz} = -1/2$ (see \cite{Nakahara:2003nw} for this form of the torus metric).

With this setup in mind, the Laplace equation on $\mathbb{R}^3 \times T^2$ can be written as
\begin{equation}
\nabla^2f(r,z,\bz;\tau)	= \frac{1}{r^2}\partial_r\Big(r^2\partial_r f\Big)
					- 2 \partial_z\partial_\bz f = 0 \, . 
\end{equation}
where for the $\mathbb{R}^3$ part the flat metric in polar coordinates is used (and $f$ does not depend on the angular coordinates). This is the equation we want to solve and find the corresponding harmonic function. We will separate the variables by introducing the functions $\rho(r)$ and $\zeta(z,\bz)$. Set $f(r,z,\bz) = \rho(r)\zeta(z,\bz)/r$ and divide the equation by $f$ to get
\begin{equation}
\frac{1}{\rho(r)}\partial_r^2\rho(r) 
	- \frac{2}{\zeta(z,\bz)}\partial_z\partial_\bz\zeta(z,\bz) = 0 \, . 
\end{equation}
Each term is equal to a constant $\lambda^2$ giving the two separate equations
\begin{equation}
\partial_r^2 \rho(r) = -\lambda^2\rho(r) \qandq 
-2\partial_z\partial_\bz\zeta(z,\bz) = \lambda^2\zeta(z,\bz)
\end{equation}
which can be solved. For $\lambda^2\neq 0$ and $\lambda^2=0$ respectively, the radial solution is
\begin{equation}
\rho(r) = Ae^{i\lambda r} + Be^{-i\lambda r} \qandq
\rho(r) = Cr + D
\end{equation}
where as usual $A,B,C$ and $D$ are complex constants.

The second equation is the eigenvalue problem for the Laplacian on the torus. The eigenfunctions and eigenvalues are labeled by two integers $m,n$ and depend on the complex modulus $\tau$ and volume $V$ of the torus. This then gives the correct double periodicity of the harmonic function on the torus. It can be checked that 
\begin{equation}
-2\partial_z\partial_\bz\zeta_{m,n} = \lambda_{m,n}^2\zeta_{m,n}
\end{equation}
is satisfied by
\begin{align}
\zeta_{m,n}(z,\bz) &= \zeta_0\exp\left\{\frac{\pi}{\sqrt{\tau_2V}}\Big[z(m-\btau n) 
										-\bz(m-\tau n)\Big]\right\} \\
\lambda_{m,n}^2 &= \frac{2\pi^2}{\tau_2V}(m-\tau n)(m-\btau n) \,. 								
\end{align}
This includes the case $\lambda^2=0$ by setting $m=n=0$ which just gives a constant term in the solution. There cannot be a linear term in $z$ or $\bz$ or actually not even a separated term of the form $F_1[z]+F_2[\bz]$ for any functions $F_1,F_2$ since these do not respect the double periodicity.

The $\lambda$ in the radial function $\rho(r)$ is of course given by the eigenvalue $\lambda_{m,n}$ of the torus function $\zeta(z,\bz)$. Putting everything back together and summing over all possible solutions (principle of superposition), we obtain via $f = \rho\zeta/r$ 
\begin{equation}
\begin{aligned}
f(r,z,\bz;\tau,V) &= C_0  
			+ \frac{1}{r}\sum_{m,n}\sum_\pm C^\pm_{m,n} \exp\left\{\frac{\pi}{\sqrt{\tau_2V}}
	\Big[z(m+\btau n) - \bz(m+\tau n) \right.\\ 
	& \left. \hspace{6cm}\pm i r \sqrt{2(m+n\tau)(m+\btau n)}\Big]\vphantom{\frac{\pi}{\sqrt{\tau_2V}}}\right\}
\end{aligned}	
\end{equation}
where $C_0$ and $C^\pm_{m,n}$ are constants of integration. This is the desired harmonic function on the space $\mathbb{R}^3 \times T^2$ which is doubly periodic in $z$ and $\bz$ on the torus ($z\sim z + \sqrt{\frac{V}{\tau_2}}w$ and $\bz\sim\bz+\sqrt{\frac{V}{\tau_2}}\bw$ where $w=m+\tau n\in\Lambda$).

Note that the factor of $V$ here is like the $R$ in the harmonic function on the circle \eqref{eq:Hlocalized}. It is the volume of the space and if we take it to infinity (but at the same time keeping the shape of the torus, \ie $\tau$, fixed) the compact space is decompactified.

\chapter{$SL(2)\times\mathbb{R}^+$ EFT Appendices}
\label{ch:appSL2}

The field content of exceptional field theories include in addition to the external metric and generalized metric a sequence of forms transforming under various representations of $\G$. These constitute the tensor hierarchy of EFT (similar to that of gauged supergravity \cite{deWit:2005hv,deWit:2008ta}). As well as being forms with respect to the ``external'' directions $x^\mu$, one may think of these fields as providing an analogue of forms from the point of view of the extended space. In this appendix we discuss the definitions and properties of these fields.

\section{Cartan Calculus}
\label{sec:cartancalc}

Here we summarize the ``Cartan calculus'' relevant for the $\G$ EFT, discussed in \cite{Hohm:2015xna,Wang:2015hca}, in order to introduce our conventions. The form fields that we consider are valued in various representations, $\TAw$, $\TBw$, $\dots$ of $\G$. We list these in Table \ref{t:Forms}. The value $w$ in brackets is the weight under generalized Lie derivatives.

\begin{table}[h]
\centering
\begin{tabular}{|c|c|c|c|}
  \hline
   Module ($w$) & Representation & Gauge field & Field strength \\ \hline
   $\TAw$ & $\mathbf{2}_{1}\oplus\mathbf{1}_{-1}$ & $\Aa^{\alpha} \oplus \Aa^{s}$ & $\Fa^{\alpha} \oplus \Fa^{s}$ \\
   $\TBw$ & $\mathbf{2}_{0}$ & $\Ab^{\alpha,s}$ & $\Fb^{\alpha,s}$ \\
   $\TCw$ & $\mathbf{1}_{1}$ & $\Ac^{\alpha\beta,s}$ & $\Fc^{\alpha\beta,s}$ \\
   $\TDw$ & $\mathbf{1}_{0}$ & $\Ad^{\alpha\beta,ss}$ & $\Fd^{\alpha\beta,ss}$ \\
   $\TEw$ & $\mathbf{2}_{1}$ & $\Ae^{\gamma,\alpha\beta,ss}$ & $\Fe^{\gamma,\alpha\beta,ss}$ \\
   $\TFw$ & $\mathbf{2}_{0}\oplus\mathbf{1}_{2}$ & $\Af_{\alpha} \oplus \Af_{s}$ & -- \\
   \hline
\end{tabular}
\caption{Modules, gauge fields and field strengths relevant for the tensor hierarchy and their representations under $\G$. The subscript denotes the weight under the $\mathbb{R}^+$, $w$ denotes the weights of the elements of the module under the generalized Lie derivative.}
\label{t:Forms}
\end{table}

Given these representations, the key ingredients of the Cartan calculus are then a nilpotent derivative $\hpartial$ and an ``outer product'' $\p$ which act to map between the various modules listed in Table \ref{t:Forms}. As first discussed in \cite{Cederwall:2013naa}, the chain complex
\begin{equation}
 \TAw \xleftarrow{~\hpartial~} \TBw \xleftarrow{~\hpartial~} \TCw \xleftarrow{~\hpartial~} \TDw \xleftarrow{~\hpartial~} \TEw \xleftarrow{~\hpartial~} \TFw \,,
\end{equation}
formed from these modules together with the nilpotent derivative can be seen as a generalisation of the deRham-complex and thus of differential forms. The bilinear product $\p$ is defined between certain modules such that it maps as follows:

\begin{equation}
\begin{array}{|c|cccccc|}
\hline
\bullet & \TAw & \TBw & \TCw & \TDw & \TEw & \TFw \\
\hline
\TAw & \TBw & \TCw & \TDw & \TEw & \TFw & \TSw \\
\TBw & \TCw & \TDw & \TEw & \TFw & \TSw & \\
\TCw & \TDw & \TEw & \TFw & \TSw & & \\
\TDw & \TEw & \TFw & \TSw & & & \\
\TEw & \TFw & \TSw & & & & \\
\TFw & \TSw & & & & &  \\
\hline 
\end{array}
\label{t:modules}
\end{equation}

%\begin{table}[h]
%\begin{center}
%\begin{tabular}

%\end{tabular}
%\end{center}
%\caption{}

%\end{table}

\noindent The nilpotent derivative $\hpartial$ and the product $\p$ satisfy the following ``magic identity'' \cite{Hohm:2015xna,Wang:2015hca}
\begin{equation}
 \gL_\Lambda X = \Lambda \p \hpartial X + \hpartial \left( \Lambda \p X \right) \,,
\end{equation}
for all $X \in \TBw$, $\TCw$, $\TDw$, $\TEw$ and $\Lambda \in \TAw$. Here $\gL$ denotes the generalized Lie derivative \eqref{eq:gld}. Explicitly, the product is defined as
\begin{align}
  \left(\Aa_1 \p \Aa_2\right)^{\alpha,s} &= \Aa^\alpha_1 \Aa^s_2 + \Aa^s_1 \Aa^\alpha_2 \,, &  \left(\Aa \p \Ab\right)^{[\alpha\beta],s} &= 2 \Aa^{[\alpha} \Ab^{\beta],s} \,, \nonumber \\
  \left(\Aa \p \Ac\right)^{[\alpha\beta],ss} &= \Aa^s \Ac^{[\alpha\beta],s} \,, &  \left( \Aa \p \Ad \right)^{\gamma,[\alpha\beta],ss} &= \Aa^{\gamma} \Ad^{[\alpha\beta],ss} \,, \nonumber \\
  \left( \Aa \p \Ae \right)_{\gamma} &= \epsilon_{\gamma\delta} \Aa^s \Ae^{\gamma,[\alpha\beta],ss} \,, & \left( \Aa \p \Ae \right)_s &= \frac{1}{2}\epsilon_{\alpha\beta}\epsilon_{\gamma\delta} \Aa^{\gamma} E^{\delta,[\alpha\beta],ss} \,, \nonumber \\
  \left( \Aa \p \Af \right) &= \Aa^{\alpha} \Af_\alpha + \Aa^s \Af_s \,, & \left( \Ab_1 \p \Ab_2 \right)^{[\alpha\beta],ss} &= 2\Ab^{[\alpha|,s}_1 \Ab_2^{\beta],s} \,, \nonumber \\
  \left( \Ab \p \Ac \right)^{\gamma,[\alpha\beta],ss} &= \Ab^{\gamma,s} \Ac^{[\alpha\beta],s} \,, & \left( \Ab \p \Ad \right)_{\gamma} &= \frac{1}{2} \epsilon_{\alpha\beta} \epsilon_{\gamma\delta} \Ab^{\delta,s} \Ad^{[\alpha\beta],ss} \,, \nonumber \\
  \left( \Ab \p \Ad \right)_{s} &= 0 \,, & \left( \Ab \p \Ae \right) &= \frac{1}{2} \epsilon_{\alpha\beta} \epsilon_{\gamma\delta} \Ab^{\gamma,s} \Ae^{\delta,\alpha\beta,ss} \,, \nonumber \\
  \left( \Ac_1 \p \Ac_2 \right)_\gamma &= 0 \,, & \left( \Ac_1 \p \Ac_2 \right)_s &= \Ac_1^{[\alpha\beta],s} \Ac_2^{[\gamma\delta],s} \,, \nonumber \\
  \left( \Ac \p \Ad \right) &= \frac{1}{4} \epsilon_{\alpha\beta} \epsilon_{\gamma\delta} \Ac^{[\alpha\beta],s} \Ad^{[\gamma\delta],ss} \,, &  &
\end{align}
and is symmetric when acting on two different modules. The nilpotent derivative is defined as
\begin{align}
 \left( \hpartial \Ab \right)^{\alpha} &= \partial_s \Ab^{\alpha,s} \,, & \left( \hpartial \Ab \right)^{s} &= \partial_\alpha \Ab^{\alpha,s} \,, & \left( \hpartial \Ac \right)^{\alpha,s} &= \partial_\beta \Ac^{[\beta\alpha],s} \,, \nonumber \\
 \left( \hpartial \Ad \right)^{[\alpha\beta],s} &= \partial_s \Ad^{[\alpha\beta],ss} \,, & \left( \hpartial \Ae \right)^{[\alpha\beta],ss} &= \partial_\gamma \Ae^{\gamma,[\alpha\beta],ss} \,, & \left( \hpartial \Af \right)^{\gamma,[\alpha\beta],ss} &= \epsilon^{\gamma\delta} \partial_s F_\delta \,.
\end{align}
Nilpotency follows from the section condition \eqref{eq:constraint}.

Let us finally discuss some properties of the generalized Lie derivative which will be important in the construction of the tensor hierarchy in the next section. From now onwards we will often omit the $\G$ indices on elements of the modules. First, note that for any $\Aa_1$, $\Aa_2 \in \TAw$, the symmetric part of the Lie derivative is given by
\begin{equation}
 2 \left(\Aa_1 \,, \Aa_2 \right) \equiv \left( \gL_{\Aa_1} \Aa_2 + \gL_{\Aa_2} \Aa_1 \right) = \hpartial \left( \Aa_1 \p \Aa_2 \right) \,. \label{eq:SymBracket}
\end{equation}
Using the explicit formulas, one can see that this generates a vanishing generalized Lie derivative, \ie
\begin{equation}
 \gL_{\left(\Aa_1 \,, \Aa_2\right)} = 0 \,.
\end{equation}
It will also be useful to write the generalized Lie derivative in terms of its symmetric and antisymmetric parts
\begin{equation}
 \gL_{\Aa_1} \Aa_2 = \left[ \Aa_1 \,, \Aa_2 \right]_E + \left( \Aa_1 \,, \Aa_2 \right) \,, \label{eq:gLsa}
\end{equation}
where the $E$-bracket is the antisymmetric part of the generalized Lie derivative
\begin{equation}
 \left[ \Aa_1 \,, \Aa_2 \right]_E = \frac12 \left( \gL_{\Aa_1} \Aa_2 - \gL_{\Aa_2} \Aa_1 \right) \,. \label{eq:EBracket}
\end{equation}
Finally, the Jacobiator of the $E$-bracket is proportional to terms that generate vanishing generalized Lie derivatives
\begin{equation}
 \left[ \left[ \Aa_1 \,, \Aa_2 \right]_E \,, \Aa_3 \right] + \textrm{cycl.} = \frac{1}{3} \left( \left[ \Aa_1 \,, \Aa_2 \right]_E \,, \Aa_3 \right) + \textrm{cycl.} \,, \label{eq:Jacobiator}
\end{equation}
so that the Jacobiator of generalized Lie derivatives does not vanish but lies in the kernel of the generalized Lie derivative when viewed as a map from generalized vectors to generalized tensors.

\section{Tensor Hierarchy}
\label{sec:tensorhier}
The EFT is invariant under generalized diffeomorphisms, generated by a generalized vector field $\Lambda(x,Y) \in \TAw$. From the perspective of the ``extended space'' it induces gauge transformations and diffeomorphisms, while from the nine-dimensional perspective, it induces non-abelian gauge transformations of the scalar sector. Correspondingly, one introduces a gauge field $\Aa \in \TAw$ such that
\begin{equation}
 \delta_\Lambda \Aa = \D_\mu \Lambda \,,
\end{equation}
where $\D_\mu = \partial_\mu - \gL_{\Aa_\mu}$ is the nine-dimensional covariant derivative. The naive form of the field strength would resemble the Yang-Mills field strength
\begin{equation}
 F_{\mu\nu} = 2 \partial_{[\mu} \Aa_{\nu]} - \left[ \Aa_\mu \,, \Aa_\nu \right]_E \,,
\end{equation}
which involves the $E$-bracket in order to be a two-form. However, this fails to be gauge invariant
\begin{equation}
 \delta F_{\mu\nu} = 2 \D_{[\mu} \delta \Aa_{\nu]} + \hpartial \left( \Aa_{[\mu} \p \delta \Aa_{\nu]} \right) \,,
\end{equation}
using \eqref{eq:gLsa}. In order to define a gauge-invariant field strength we are led to modify the usual Yang-Mills definition as follows
\begin{equation}
 \Fa_{\mu\nu} = 2 \partial_{[\mu} \Aa_{\nu]} - \left[ \Aa_{\mu}, \Aa_{\nu} \right]_E + \hpartial \Ab_{\mu\nu} \,,
\end{equation}
where $\Ab_{\mu\nu} \in \TBw$ is a two-form. The modified field strength is now gauge-invariant if we define the variation of $\Ab_{\mu\nu}$ to be
\begin{equation}
 \Delta_\Lambda \Ab_{\mu\nu} = \Lambda \p \Fa_{\mu\nu} \,,
\end{equation}
where
\begin{equation}
 \Delta \Ab_{\mu\nu} \equiv \delta \Ab_{\mu\nu} + \Aa_{[\mu} \p \delta \Aa_{\nu]} \,.
\end{equation}
The definition of both the naive field strength $F_{\mu\nu}$ and the covariant field strength $\Fa_{\mu\nu}$ is compatible with the commutator of covariant derivatives
\begin{equation}
 \left[ \D_{\mu} ,\, \D_{\nu} \right] = - \gL_{F_{\mu\nu}} = - \gL_{\Fa_{\mu\nu}} \,,
\end{equation}
since their difference is of the form \eqref{eq:SymBracket} and thus generates vanishing generalized Lie derivative. Mirroring the tensor hierarchy of gauged supergravities \cite{deWit:2005hv,deWit:2008ta}, one can introduce a gauge transformation and field strength for $\Ab_{\mu\nu}$, which in turn requires a new three-form potential. This way one obtains a hierarchy of $p$-form fields up to a five-form gauge potential and its six-form field strength. The six-form potential does not appear in the action and so we do not define its field strength. Here we give this construction explicitly for the $\G$ EFT for the first time. In the following, the expressions for the five-form potential and six-form field strengths are new while the lower form potentials are also given in \cite{Hohm:2015xna}. Let us begin with the definition of the field strengths
\begin{align}
\Fa_{\1\2} &= 2\partial_{[\1} \Aa_{\2]} - [\Aa_\1,\Aa_\2]_E + \hd\Ab_{\1\2} \,, \notag\\
\Fb_{\1\2\3} &= 3\D_{[\1}\Ab_{\2\3]} - 3\partial_{[\1}\Aa_{\2}\pl\Aa_{\3]} + \Aa_{[\1}\pl[\Aa_\2,\Aa_{\3]}]_E + \hd\Ac_{\1\2\3} \,, \notag\\
\Fc_{\1\ldots\4} &= 4\D_{[\1}\Ac_{\2\3\4]} + 3\hd \Ab_{[\1\2}\pl\Ab_{\3\4]} - 6\Fa_{[\1\2}\pl\Ab_{\3\4]} + 4\Aa_{[\1}\pl(\Aa_\2\pl\partial_\3\Aa_{\4]}) \notag\\
 & \quad - \Aa_{[\1}\pl(\Aa_\2\pl[\Aa_\3,\Aa_{\4]}]_E) + \hd\Ad_{\1\2\3\4} \,, \notag\\
\Fd_{\1\ldots\5} &= 5\D_{[\1}\Ad_{\2\ldots\5]} + 15\Ab_{[\1\2}\pl\D_3\Ab_{\4\5]} - 10\Fa_{[\1\2}\pl\Ac_{\3\4\5]} \notag\\ 
& \quad - 30\Ab_{[\1\2}\pl(\Aa_\3\pl\partial_\4\Aa_{\5]}) + 10\Ab_{[\1\2}\pl(\Aa_\3\pl[\Aa_\4,\Aa_{\5]}]_E) \notag\\
 & \quad - 5\Aa_{[\1}\pl(\Aa_\2\pl(\Aa_\3\pl\partial_\4\Aa_{\5]})) + \Aa_{[\1}\pl(\Aa_\2\pl(\Aa_\3\pl[\Aa_\4,\Aa_{\5]}]_E)) + \hd\Ae_{\1\ldots\5} \,, \notag\\
\Fe_{\1\ldots\6} &= 6\D_{[\1}\Ae_{\2\ldots\6]} - 15\Fa_{[\1\2}\pl\Ad_{\3\ldots\6]} - 10\Ac_{[\1\2\3}\pl\hd\Ac_{\4\5\6]} \notag \\
 & \quad - 20\Fb_{[\1\2\3}\pl\Ac_{\4\5\6]} - 45\Ab_{[\1\2}\pl(\hd\Ab_{\3\4}\pl\Ab_{\5\6}]) \notag\\
    &\quad  - 90\Ab_{[\1\2}\pl(\partial\Aa_{\3\4}\pl\Ab_{\5\6]}) + 45\Ab_{[\1\2}\pl([\Aa_\3,\Aa_\4]_E\,\pl\Ab_{\5\6}])   \notag\\
    & \quad + 60\Ab_{[\1\2}\pl(\Aa_\3\pl(\Aa_\4\pl\partial_\5\Aa_{\6]}))- 15\Ab_{[\1\2}\pl(\Aa_\3\pl(\Aa_\4\pl[\Aa_\5,\Aa_{\6]}]_E)) \notag\\
    & \quad 	+ 6\Aa_{[\1}\pl(\Aa_\2\pl(\Aa_\3\pl(\Aa_\4\pl\partial_\5\Aa_{\6]})))-  \Aa_{[\1}\pl(\Aa_\2\pl(\Aa_\3\pl(\Aa_\4\pl[\Aa_\5,\Aa_{\6]}]_E)))\notag\\
    & \quad  + \hd\Af_{\1\ldots\6} 
\label{eq:fieldstrengths}
\end{align}
Their variations are given by
\begin{align}
 \delta\Fa_{\1\2} &= 2\D_{[\1}\delta\Aa_{\2]} + \hd\Delta\Ab_{\1\2} \,, \notag\\
 \delta\Fb_{\1\2\3} &= 3\D_{[\1}\Delta\Ab_{\2\3]} - 3\delta\Aa_{[\1}\pl\Fa_{\2\3]} + \hd\Delta\Ac_{\1\2\3} \,, \notag\\
 \delta\Fc_{\1\ldots\4} &= 4\D_{[\1}\Delta\Ac_{\2\3\4]} - 4\delta\Aa_{[\1}\pl\Fb_{\2\3\4]} - 6\Fa_{[\1}\pl\Delta\Ab_{\2\3\4]} + \hd\Delta\Ad_{\1\ldots\4} \,, \notag\\
 \delta\Fd_{\1\ldots\5} &= 5\D_{[\1}\Delta\Ad_{\2\ldots\5]} - 5\delta\Aa_{[\1}\pl\Fc_{\2\ldots\5]} - 10\Fa_{[\1\2}\pl\Delta\Ac_{\3\4\5]} \notag\\
 & \quad - 10\Fb_{[\1\2\3}\pl\Delta\Ab_{\4\5]}  + \hd\Delta\Ae_{\1\ldots\5} \,, \notag\\
 \delta\Fe_{\1\ldots\6} &= 6\D_{[\1}\Delta\Ae_{\2\ldots\6]} - 6\delta\Aa_{[\1}\pl\Fd_{\2\ldots\6]} - 15\Fa_{[\1\2}\pl\Delta\Ad_{\3\ldots\6]} \notag\\
 & \quad - 20\Fb_{[\1\2\3}\pl\Delta\Ac_{\4\5\6]}  
 + 15\Fc_{[\1\ldots\4}\pl\Delta\Ab_{\5\6]} + \hd\Delta\Af_{\1\ldots\6} \,,
\label{eq:fieldstrengthvariation}
\end{align}
where have we defined the ``covariant'' gauge field variations as
\begin{align}
 \Delta\Ab_{\1\2} &= \delta\Ab_{\1\2} + \Aa_{[\1}\pl\delta\Aa_{\2]} \,, \notag\\
 \Delta\Ac_{\1\2\3} &= \delta\Ac_{\1\2\3} - 3\delta\Aa_{[\1}\pl\Ab_{\2\3]} + \Aa_{[\1}\pl(\Aa_\2\pl\delta\Aa_{\3]}) \,, \notag\\
 \Delta\Ad_{\1\ldots\4} &= \delta\Ad_{\1\ldots\4} - 4\delta\Aa_{[\1}\pl\Ac_{\2\ldots\4]} + 3\Ab_{[\1\2}\pl\left(\delta\Ab_{\3\4]}+2\Aa_{\3}\pl\delta\Aa_{\4]}\right)  \notag\\
 & \quad  + \Aa_{[\1}\!\pl(\Aa_\2\pl(\Aa_\3\pl\delta\Aa_{\4]})) \,, \notag\\
 \Delta\Ae_{\1\ldots\5} &= \delta\Ae_{\1\ldots\5} - 5\delta\Aa_{[\1}\pl\Ad_{\2\ldots\5]} - 10\delta\Ab_{[\1\2}\pl\Ac_{\3\ldots\5]}  \notag\\
 & \quad  - 15\Ab_{[\1\2}\pl\left(\delta\Aa_\3\pl\Ab_{\4\5]}\right) -10\left(\Aa_{[\1}\pl\delta\Aa_\2\right)\pl\Ac_{\3\4\5]} \notag\\
 & \quad + 10\Ab_{[\1\2}\pl\left(\Aa_\3\pl\left(\Aa_\4\pl\delta\Aa_\5\right)\right)   + \Aa_{[\1}\!\pl\left(\Aa_\2\pl\left(\Aa_\3\pl\left(\Aa_\4\pl\delta\Aa_{\5]}\right)\right)\right) \,, \notag\\
 \Delta\Af_{\1\ldots\6} &= \delta\Af_{\1\ldots\6} - 6\delta\Aa_{[\1}\pl\Ae_{\2\ldots\6]} - 15\delta\Ab_{[\1\2}\pl\Ad_{\3\ldots\6]}  \notag\\
 & \quad - 15\left(\Aa_{[\1}\pl\delta\Aa_\2\right)\,\pl\Ad_{\3\ldots\6]} - 10\delta\Ac_{[\1\2\3}\pl\Ac_{\4\5\6]} \notag\\
 &\quad + 60(\delta\Aa_{[\1}\pl\Ab_{\2\3}\pl\Ac_{\4\5\6]} - 20\left(\Aa_{[\1}\pl\left(\Aa_\2\pl\delta\Aa_\3\right)\right)\,\pl\Ac_{\4\5\6]} \notag\\
 &\quad	- 45 \Ab_{[\1\2}\pl\left(\delta\Ab_{\3\4}\pl\Ab_{\5\6]}\right) + 45 \Ab_{[\1\2}\pl\left(\Ab_{\3\4}\pl(\Aa_\5\pl\delta\Aa_{\6]}\right) \notag \\
 &\quad	+ 15 \Ab_{[\1\2}\pl\left(\Aa_\3\pl\left(\Aa_\4\pl\left(\Aa_\5\pl\delta\Aa_{\6]}\right)\right)\right)  \notag\\
 & \quad  + \Aa_{[\1}\pl\left(\Aa_\2\pl\left(\Aa_\3\pl\left(\Aa_\4\pl\left(\Aa_\5\pl\delta\Aa_{\6]}\right)\right)\right)\right) \,.
\label{eq:gaugefieldvariation}
\end{align}
It is now easy to check that the field strengths \eqref{eq:fieldstrengths} are invariant under the following gauge transformations
\begin{align}
  \delta \Aa_{\1} &= D_{\1} \Lambda - \hat{\partial}\, \Xi_\1 \,, \notag\\
  \Delta \Ab_{\1\2} &= \Lambda\, \pl \Fa_{\1\2} + 2 D_{[\1} \Xi_{\2]} - \hat{\partial} \Theta_{\1\2} \,, \notag\\
  \Delta \Ac_{\1\ldots\3} &= \Lambda\, \pl \Fb_{\1\2\3} + 3 \Fa_{[\1\2} \pl \Xi_{\3]} + 3 D_{[\1} \Theta_{\2\3]} - \hat{\partial} \Omega _{\1\ldots\3} \,, \notag\\
  \Delta \Ad_{\1\ldots\4} &= \Lambda\, \pl \Fc_{\1\ldots \4} - 4 \Fb_{[\1\ldots\3} \pl \Xi_{\4]} + 6 \Fa_{[\1\2} \pl \Theta_{\3\4]} + 4 D_{[\1} \Omega_{\2\ldots\4]}  \notag\\
 & \quad - \hat{\partial} \Upsilon_{\1\ldots\4} \,, \notag\\
  \Delta \Ae_{\1\ldots\5} &= \Lambda\, \pl \Fd_{\1\ldots\5} - 5 \Fc_{[\1\ldots\4} \pl \Xi_{\5]} - 10 \Fb_{[\1\ldots\3} \pl \Theta_{\4\5]} + 10 \Fa_{[\1\2} \pl \Omega_{\3\ldots \5]} \notag\\
  & \quad + 5 D_{[\1} \Upsilon_{\2 \ldots \5]} - \hat{\partial} \Phi_{\1\ldots\5} \,, \notag\\
  \Delta \Af_{\1 \ldots \6} &= \Lambda\, \pl {\cal L}_{\1\ldots \6} + 6 \Fd_{[\1\ldots\5} \pl \Xi_{\6]} + 15 \Fc_{\1\ldots\4} \pl \Theta_{\5\6]} - 20 \Fb_{[\1\ldots\3} \pl \Omega_{\4\ldots\6]} \notag\\
  & \quad + \Fa_{[\1\2} \pl \Upsilon_{\3\ldots\6]} + 6 D_{[\1} \Phi_{\2 \ldots \6]} \,.
\end{align}
Finally, the field strengths \eqref{eq:fieldstrengths} satisfy the following Bianchi identities, as can be seen from their definitions.
\begin{equation}
\begin{split}
3\D_{[\1}\Fa_{\2\3]} &= \hd\Fb_{\1\ldots\3} \,, \\
4\D_{[\1}\Fb_{\2\ldots\4]} + 3\Fa_{[\1\2}\pl\Fa_{\3\4]} &= \hd\Fc_{\1\ldots\4} \,, \\
5\D_{[\1}\Fc_{\2\ldots\5]} + 10\Fa_{[\1\2}\pl\Fb_{\3\4\5]} &= \hd\Fd_{\1\ldots\5} \,, \\
6\D_{[\1}\Fd_{\2\ldots\6]} + 15\Fa_{[\1\2}\pl\Fc_{\3\ldots\6]} -10\Fb_{[\1\2\3}\pl\Fb_{\4\5\6]} &= \hd\Fe_{\1\ldots\6} \,.
\end{split}
\label{eq:Bianchi}
\end{equation}
While the first three equations have appeared before, the final identity is new.

\section{Topological Term}
\label{sec:topterm}
Maximal supergravity theories contain a topological term, which is mirrored in the corresponding EFT. Armed with the Cartan calculus and the tensor hierarchy we can now construct a topological term for the action. It is given by
\begin{equation}
\begin{split}
  S_{top} &= \kappa \int \dd^{10}x\, \dd^3Y\, \varepsilon^{\1\ldots\mt} \\
  & \hspace{1.5cm}
  \left[ \frac{1}{5} \hpartial \Fd_{\1\ldots\5} \p \Fd_{\6\ldots\mt} 
    - \frac{5}{2} \left( \Fa_{\1\2} \p \Fc_{\3\ldots\6} \right) \p \Fc_{\7\ldots\mt} \right. \\
  & \hspace{2cm} \left.
    + \frac{10}{3} \left( \Fb_{\1\ldots\3} \p \Fb_{\4\ldots\6} \right) \p \Fc_{\7 \ldots \mt} \right] \,,
\end{split} 
\label{eq:ToptTermApp}
\end{equation}
where we have abused the notation to also denote the ten-dimensional indices by $\1,\ldots,\mt=1,\ldots,10$, and $\varepsilon^{\1\ldots\mt} = \pm 1$ is the ten-dimensional alternating symbol. This term is a manifestly gauge-invariant boundary term in ten dimensions and has weight one under generalized diffeomorphisms, as required. Instead of explicitly showing that it is a boundary term itself, we will just show that its variation is a boundary term. Using the variations of the field strengths \eqref{eq:fieldstrengthvariation} and the Bianchi identities \eqref{eq:Bianchi}, one finds
\begin{equation}
\begin{split}
  \delta S_{top} &= \kappa \int\!\! \dd^{10}x\, \dd^3Y\, \varepsilon^{\1\ldots\mt} \\
  & \hspace{1.5cm}  D_{\1} \bigg[
   	- 5 \left( \delta \Aa_{\2} \bullet {\cal J}_{\3 \ldots \6} \right) \bullet {\cal J}_{\7 \ldots \mt}  \\
  & \hspace{2.4cm}  
    + 20 \left( \Delta \Ab_{\2\3} \bullet {\cal H}_{\4\ldots\6} \right) \bullet {\cal J}_{\7\ldots \mt}  \\
  & \hspace{2.4cm}  
  	- 20 \left( {\cal F}_{\2\3} \bullet {\cal J}_{\4 \ldots \7} \right) \bullet \Delta \Ac_{\8 \ldots \mt} \\
  & \hspace{2.4cm} 	
    + \frac{40}{3} \left( {\cal H}_{\2\ldots\4} \bullet {\cal H}_{\5\ldots\7} \right) \bullet \Delta \Ac_{\8\ldots\mt}  \\
  & \hspace{2.4cm} 
  	+ 2 \hat{\partial} \Delta \Ad_{\2\ldots\5} \bullet {\cal K}_{\6 \ldots \mt} \bigg] \,. 
\end{split}
\label{eq:TopVariation}
\end{equation}
Throughout we assume that the ``extended space'' parametrized by the $Y$'s does not have a boundary. One can easily check using \eqref{eq:gaugefieldvariation} that the term is invariant under gauge transformations. One can now use \eqref{eq:TopVariation} to fix the overall coefficient relative to the other terms in the action by requiring invariance under external diffeomorphisms \cite{Berman:2015rcc}.

\chapter{Glossary of Supergravity Solutions}
\label{ch:glossary}

The purpose of this appendix is not only to collect all the fundamental, solitonic and Dirichlet solutions of ten- and eleven-dimensional supergravities as they can be found in any standard text book (for us Ortin's book \cite{Ortin:2004ms} was an invaluable source), but also to present them with their fields rearranged according to a $4+7$ Kaluza-Klein coordinate split relevant for the EFT solutions in Chapter \ref{ch:EFTsol} since it is the decomposed fields that are extracted from the EFT solutions in the main text. It also highlights some interesting similarities between these solutions, such as that they \emph{all} have the same four-dimensional external spacetime under the decomposition.

\section{Classic Supergravity Solutions}
\label{sec:classicsol}
In eleven-dimensional supergravity there are four classic solutions: the wave, the membrane, the fivebrane and the monopole. They are all related by T- and S-duality and upon reduction on a circle they give rise to the spectrum of string theory solutions in ten dimensions.

Here we will briefly present these four solutions in terms of the bosonic fields $C_3,C_6$ and $g$ which in turn are given terms of an harmonic function $H$. To begin with we will give them in a standardized coordinate system where the time direction is denoted by $t$, the spatial worldvolme directions are $\vec{x}$ and the transverse directions are $\vec{y}$. If there is a ``special'' coordinate to be singled out such as the direction of propagation of the wave or the isometry of the monopole, it will be labeled as $z$. Later we will break these coordinates down for the various KK-splits needed for the EFT solutions. 

Let's start with the ``pure gravity'' solutions, the pp-wave and the KK-monopole. They do not come with a gauge potential and are given just in terms of the metric. The pp-wave consists of parallel rays carrying momentum in the $z$ direction with transverse plane wavefronts spanned by $\vec{y}_{(9)}$
\begin{equation}
\begin{aligned}
\dd s^2 &= -H^{-1}\dd t^2 + H\left[\dd z - (H^{-1}-1)\dd t\right]^2 
				+ \dd\vec{y}_{(9)}^{\, 2}   \\
		&= (H-2)\dd t^2 + 2(H-1)\dd t \dd z + H\dd z^2 + \dd\vec{y}_{(9)}^{\, 2} \\
H &= 1 + \frac{h}{|\vec{y}_{(9)}|^7}  
\end{aligned}
\label{eq:classicwave}
\end{equation}
where $h$ is some constant proportional to the momentum carried.

The KK-monopole solution has a non-trivial topology which is given by the Hopf fibration of $S^3$ (in a local patch this becomes $S^2\times S^1$). Whereas the momentum of the wave solution can be seen as \emph{gravito-static} charge, the monopole carries topological or \emph{gravito-magnetic} charge which is given by the first Chern class, hence the name ``monopole''. This solution is expressed in terms of a vector potential $A_i$ whose curl is the gradient of the harmonic function. The $S^1$ is an isometry of the solution and we will refer to it as the ``monopole circle'' or ``KK-circle''. It is labeled by the coordinate $z$. The three transverse coordinates $y^i$ are supplemented by six worldvolume coordinates $x^a$ to form a KK-brane. The full monopole solution is thus given by
\begin{align}
\dd s^2 &= -\dd t^2 + \dd\vec{x}_{(6)}^{\, 2} 
				+ H^{-1}\left[\dd z + A_i\dd y^i\right]^2 + H\dd\vec{y}_{(3)}^{\, 2}  \notag\\
		&= -\dd t^2 + \delta_{ab}\dd x^a\dd x^b + H^{-1}\dd z^2 + 2H^{-1}A_i\dd y^i\dd z
				+ H\left(\delta_{ij}+H^{-2}A_iA_j\right)\dd y^i\dd y^j \notag\\
H &= 1 + \frac{h}{|\vec{y}_{(3)}|} \, , \qquad
\partial_{[i}A_{j]} = \frac{1}{2}{\epsilon_{ij}}^k\partial_k H \, .
\label{eq:classicmonopole}
\end{align}
Again $h$ is a constant, here it is proportional to the magnetic charge.

Now turn to the extended solutions, the M2-brane and the M5-brane. These branes naturally couple to the $C_3$ and $C_6$ gauge potentials respectively. This can be seen as the natural \emph{electric} coupling. The harmonic function $H$ in each case is a function of the transverse directions. The membrane solution is given by
\begin{equation}
\begin{aligned}
\dd s^2 &= H^{-2/3}[-\dd t^2 + \dd\vec{x}_{(2)}^{\, 2}] + H^{1/3}\dd\vec{y}_{(8)}^{\, 2}   \\
C_{tx^1x^2} &= -(H^{-1}-1), \qquad C_{iy^4y^5y^6y^7y^8} = A_i\\
H &= 1 + \frac{h}{|\vec{y}_{(8)}|^6} 
\end{aligned}
\label{eq:classicmembrane}
\end{equation}
and the fivebrane solution reads 
\begin{equation}
\begin{aligned}
\dd s^2 &= H^{-1/3}[-\dd t^2 + \dd\vec{x}_{(5)}^{\, 2}] + H^{2/3}\dd\vec{y}_{(5)}^{\, 2}   \\
C_{tx^1x^2x^3x^4x^5} &= -(H^{-1}-1),  \qquad C_{iy^4y^5} = A_i  \\
H &= 1 + \frac{h}{|\vec{y}_{(5)}^{\, 2}|^3}  \, .
\end{aligned}
\label{eq:classicfivebrane}
\end{equation}
In both cases both the electric and magnetic potentials are shown. The latter ones can be found by dualizing the corresponding field strengths. The field strength of the electric potential is proportional to $F\sim \partial H^{-1}\sim \partial H$ which is dualized into  $\tF\sim\epsilon\partial H\sim \partial A$ where we use the expression in \eqref{eq:classicmonopole} to relate $H$ and $A$. Therefore the vector potential $A_i$ appears in the components of the magnetic potentials ($i=1,2,3$ as before).

The four solutions recapped above are all related to each other by M-theory dualities. The wave and the membrane are T-dual to each other, in the same way the wave and the fundamental string are related by T-duality in string theory. Similarly the monopole and the fivebrane are T-duals, again as for the monopole and NS5-brane in string theory (\cf Chapter \ref{ch:DFTsol}). Furthermore, the membrane and fivebrane are related by S-duality, they are electromagnetic duals of each other. To complete the picture, there is a S-duality relation between the wave and the monopole. This can only be made manifest in the full non-truncated exceptional field theory.

\bigskip

In Section \ref{sec:E7monopole} solutions to the truncated $\Es$ EFT are constructed. If the classic solutions are carried over from eleven-dimensional supergravity to EFT, the underlying spacetime has to be reduced from eleven to seven dimensions in order to build the 56-dimensional extended space. There are various ways of picking the seven and four out of the eleven. The notation used in that particular section is as follows. For all four solutions mentioned above, we now have the time coordinate $t$ and the ``special'' coordinate $z$. There are three transverse coordinates $\vec{y}_{(3)}=y^i$ and the remaining six coordinates $\vec{x}_{(6)}=x^a$ are partly worldvolume, partly transverse, depending on the solution. In Table \ref{tab:classicsol} the character of each of the eleven dimensions for each of the four solutions is illustrated.
 
\begin{table}[h]
\begin{center}
\begin{tabular}{|l|lllllllllll|}
\hline
solution    & $t$ & $x^1$ & $x^2$ & $x^3$ & $x^4$ & $x^5$ & $x^6$ & $y^1$   & $y^2$   & $y^3$   & $z$     \\ \hline
pp-wave     & -   &       &       &       &       &       &       &         &         &         & -       \\
KK-monopole & -   & -     & -     & -     & -     & -     & -     & $\bullet$ & $\bullet$ & $\bullet$ & $\bullet$ \\
M2-brane    & -   & -     & -     &       &       &       &       & $\circ$   & $\circ$   & $\circ$   & $\circ$   \\
M5-brane    & -   & -     & -     & -     & -     & -     &       & $\circ$   & $\circ$   & $\circ$   & $\circ$ \\ \hline
\end{tabular}
\end{center}
\caption{In this table a dash denotes that the solution is extended in that direction while a blank denotes a transverse direction. For the monopole, the four transverse directions (denoted by a dot) are special in the sense that the magnetic potential $A_i$ and the KK-circle $z$ encapsulate all the non-trivial features of the monopole. These four directions are of interest for the M2 and M5 because they are the directions (denoted by a circle) through which the electric or magnetic fluxes will flow.}
\label{tab:classicsol}
\end{table}

Note that in order to keep the notation simple the following conventions are used. If the directions $x^3,x^4$ and $x^5$ are reduced, we retain $x^a$ with $a=1,2$ for the first two $x$'s or alternatively label them as $x^1=u$ and $x^2=v$. Similarly $x^6=w$ is used where necessary. The order of these coordinates is important for the extended coordinates with an antisymmetric pair of indices since for example $Y_{tz}=-Y_{zt}$. It is fixed by defining the permutation symbol $\epsilon_{tx^1x^2x^3x^4x^5x^6y^1y^2y^3z}=+1$. This order is kept also after reductions when some of the coordinates drop out.

\section{Kaluza-Klein Decomposition of Solutions}
\label{sec:KKdecomp}

Now we turn to the KK-decomposition of the supergravity solution which is relevant for the self-dual EFT solution of Section \ref{sec:selfdual}. Let us briefly outline what exactly we mean by Kaluza-Klein decomposition.

The coordinates $\hx^\hmu=(x^\mu,x^m)$ are either split into $11\rightarrow 4+7$ or $10\rightarrow 4+6$ and the corresponding KK-decomposition takes the form 
\begin{equation}
\hg_{\hmu\hnu} = 
			\begin{pmatrix}
				g_{\mu\nu} + {A_\mu}^m{A_\nu}^ng_{mn} & {A_\mu}^mg_{mn} \\
				g_{mn}{A_\nu}^n & g_{mn}
			\end{pmatrix}
\end{equation}
where hatted quantities are ten- or eleven-dimensional and the internal sector is six- or seven-dimensional. The off-diagonal or cross-term ${A_\mu}^m$ is called the KK-vector and will mostly be zero except for the wave and the monopole. The four-dimensional external metric $g_{\mu\nu}$ has to be rescaled by the determinant of the internal metric $g_{mn}$ to remain in the Einstein frame. This is crucial for comparing solutions and takes the form
\begin{equation}
g_{\mu\nu} \rightarrow |\det g_{mn}|^{1/2} g_{\mu\nu} \, .
\end{equation}
The power of the determinant in the rescaling depends on the number of external dimensions and is $1/2$ in our case.

The eleven-dimensional supergravity solutions are specified in terms of the metric $\hg_{\hmu\hnu}$ and the three-form and the six-form potentials $C_3$ and $C_6$ which are duals of each other. In the NSNS-sector, the fields of the ten-dimensional Type II solutions are the metric $\hg_{\hmu\hnu}$, the string theory dilaton\footnote{The constant part of the dilaton is denoted by $e^{2\phi_0}$ where $\phi_0$ is a constant which can be set to zero if convenient.} $e^{2\phi}$ and the two-form and six-form Kalb-Ramond potentials $B_2$ and $B_6$ which again are duals. In the RR-sector we have the $C_p$ potentials with $p=1,\dots,7$ in this thesis. The odd ones belong to the Type IIA theory and the even ones to the Type IIB theory.

From an EFT point of view, the external metric is simply the rescaled $g_{\mu\nu}$. The form fields and the KK-vector ${A_\mu}^m$ constitute the components of the EFT vector ${\AAA_\mu}^M$. The generalized metric $\MM_{MN}$ is constructed from the internal metric $g_{mn}$ according to \eqref{eq:genmetricSUGRA}. The dilaton $\phi$ in Type IIA or the axio-dilaton $\tau$ in Type IIB also enter the generalized metric as in \eqref{eq:genmetricIIA} and \eqref{eq:genmetricIIB} respectively.

Each solution is presented with its full field content in terms of the harmonic function $H$ which has a functional dependence on the transverse directions of each solution. Then we perform the explained KK-decomposition by picking time and three of the transverse directions to be in the four-dimensional external sector and the world volume directions together with the remaining transverse ones to be in the six- or seven-dimensional internal sector. As part of the decomposition the solution is \emph{smeared} over those transverse directions in the internal sector so that it is only localized in the three transverse directions in the external sector, \ie $H=1+h/|r|$ with $r^2=\delta_{ij}w^iw^j$ and the $w$'s denote these three directions.

A final note on the notation: $t$ is the time coordinate, $z$ is the ``special'' direction of the wave and the monopole, $\vec{x}_{(p)}$ denotes the $p$ world volume directions of a p-brane and $\vec{y}_{(D-1-p)}$ the remaining $D-1-p$ transverse directions, the first three of which are usually taken to be in the external sector as explained above, \ie $w^i=y^i$ for $i,=1,2,3$.

All the ten- and eleven-dimensional solutions listed in the following originate from the single self-dual solution of Section \ref{sec:selfdual} and are listed in Table \ref{tab:solutions}.

\subsection{Wave, Membrane, Fivebrane and Monopole in $D=11$}
\label{sec:SUGRAsolutions}

\subsubsection*{The Wave - WM}
\begin{equation}
\begin{aligned}
\dd s^2 &= -H^{-1}\dd t^2 + H\left[\dd z - (H^{-1}-1)\dd t\right]^2 
				+ \dd\vec{y}_{(9)}^{\, 2}   \\
H &= 1 + \frac{h}{|\vec{y}_{(9)}|^7}  
\end{aligned}
\label{eq:WM}
\end{equation}
KK-decomposition: $x^\mu = (t, \vec{y}_{(3)})$ and $x^m = (z, \vec{y}_{(6)})$
\begin{equation}
\begin{aligned}
g_{mn} &= \diag[H,\delta_6] \, , & \det g_{mn}&=H  \\
g_{\mu\nu} &= \diag[-H^{-1/2},H^{1/2}\delta_{ij}] \, , &  {A_t}^z &= -(H^{-1}-1) 
\end{aligned}
\end{equation}

\subsubsection*{The Membrane - M2}
\begin{equation}
\begin{aligned}
\dd s^2 &= H^{-2/3}[-\dd t^2 + \dd\vec{x}_{(2)}^{\, 2}] + H^{1/3}\dd\vec{y}_{(8)}^{\, 2}   \\
C_{tx^1x^2} &= -(H^{-1}-1), \qquad C_{iy^4y^5y^6y^7y^8} = A_i\\
H &= 1 + \frac{h}{|\vec{y}_{(8)}|^6} 
\end{aligned}
\label{eq:M2}
\end{equation}
KK-decomposition: $x^\mu = (t, \vec{y}_{(3)})$ and $x^m = (\vec{x}_{(2)}, \vec{y}_{(5)})$
\begin{equation}
\begin{aligned}
g_{mn} &= H^{1/3}\diag[H^{-1}\delta_2,\delta_5] \, , & \det g_{mn}&=H^{1/3}  \\
g_{\mu\nu} &= \diag[-H^{-1/2},H^{1/2}\delta_{ij}] 
\end{aligned}
\end{equation}

\subsubsection*{The Fivebrane - M5}
\begin{equation}
\begin{aligned}
\dd s^2 &= H^{-1/3}[-\dd t^2 + \dd\vec{x}_{(5)}^{\, 2}] + H^{2/3}\dd\vec{y}_{(5)}^{\, 2}   \\
C_{tx^1x^2x^3x^4x^5} &= -(H^{-1}-1),  \qquad C_{iy^4y^5} = A_i  \\
H &= 1 + \frac{h}{|\vec{y}_{(5)}|^3}  \, .
\end{aligned}
\label{eq:M5}
\end{equation}
KK-decomposition: $x^\mu = (t, \vec{y}_{(3)})$ and $x^m = (\vec{x}_{(5)}, \vec{y}_{(2)})$
\begin{equation}
\begin{aligned}
g_{mn} &= H^{2/3}\diag[H^{-1}\delta_5,\delta_2] \, , & \det g_{mn}&=H^{-1/3}  \\
g_{\mu\nu} &= \diag[-H^{-1/2},H^{1/2}\delta_{ij}] 
\end{aligned}
\end{equation}

\subsubsection*{The Monopole - KK7}
\begin{equation}
\begin{aligned}
\dd s^2 &= -\dd t^2 + \dd\vec{x}_{(6)}^{\, 2} 
				+ H^{-1}\left[\dd z + A_i\dd y^i\right]^2 + H\dd\vec{y}_{(3)}^{\, 2}  \\
H &= 1 + \frac{h}{|\vec{y}_{(3)}|} \, , \qquad
\partial_{[i}A_{j]} = \frac{1}{2}{\epsilon_{ij}}^k\partial_k H \, .
\end{aligned}
\label{eq:KK7}
\end{equation}
KK-decomposition: $x^\mu = (t, \vec{y}_{(3)})$ and $x^m = (z, \vec{x}_{(6)})$
\begin{equation}
\begin{aligned}
g_{mn} &= \diag[H^{-1},\delta_6] \, , & \det g_{mn}&=H^{-1}  \\
g_{\mu\nu} &= \diag[-H^{-1/2},H^{1/2}\delta_{ij}] \, , &  {A_i}^z &= A_i 
\end{aligned}
\end{equation}

\subsubsection*{The M2/M5 Bound State}
\begin{equation}
\begin{aligned}
\dd s^2 &= H^{-2/3}\Xi^{1/3}[\dd t^2 + \dd\vec{x}_{(2)}^{\, 2}] 
			+ H^{1/3}\Xi^{1/3}\dd\vec{y}_{(5)}^{\, 2}
			+ H^{1/3}\Xi^{-2/3}\dd\vec{z}_{(3)}^{\, 2}   \\
C_{tx^1x^2} &= -(H^{-1}-1)\sin\xi,  \qquad C_{iy^4y^5z^1x^2z^3} = A_i\sin\xi  \\
C_{iy^4y^5} &= A_i\cos\xi,  \qquad C_{tx^1z^2z^1x^2z^3} = -(H^{-1}-1)\cos\xi \\ 
C_{z^1z^2z^3} &= -(H-1)\Xi^{-1}\sin\xi\cos\xi \\
H &= 1 + \frac{h}{|\vec{y}_{(3)}|},  \qquad\Xi = \sin^2\xi + H \cos^2\xi \, .
\end{aligned}
\label{eq:M2M5}
\end{equation}
KK-decomposition: $x^\mu = (t, \vec{y}_{(3)})$ and $x^m = (\vec{x}_{(2)}, \vec{y}_{(2)}, \vec{z}_{(3)})$
\begin{equation}
\begin{aligned}
g_{mn} &= H^{1/3}\Xi^{1/3}\diag[H^{-1}\delta_2,\delta_2,\Xi^{-1}\delta_3] \, , & 
\det g_{mn}&=H^{1/3}\Xi^{-2/3}  \\
g_{\mu\nu} &= \diag[-H^{-1/2},H^{1/2}\delta_{ij}] 
\end{aligned}
\end{equation}

\subsection{Wave, String, Fivebrane and Monopole in $D=10$}
\label{sec:stringsolutions}

\subsubsection*{The Wave - WA/B}
\begin{equation}
\begin{aligned}
\dd s^2 &= -H^{-1}\dd t^2 + H\left[\dd z - (H^{-1}-1)\dd t\right]^2 
				+ \dd\vec{y}_{(8)}^{\, 2}   \\
H &= 1 + \frac{h}{|\vec{y}_{(8)}|^6} \, ,  \qquad e^{2\phi}=e^{2\phi_0}
\end{aligned}
\label{eq:WA}
\end{equation}
KK-decomposition: $x^\mu = (t, \vec{y}_{(3)})$ and $x^\bm = (z, \vec{y}_{(5)})$
\begin{equation}
\begin{aligned}
\bg_{\bm\bn} &= \diag[H,\delta_5] \, , & \det \bg_{\bm\bn}& = H \\
g_{\mu\nu} &= \diag[-H^{-1/2},H^{1/2}\delta_{ij}] \, , &  {A_t}^z &= -(H^{-1}-1) 
\end{aligned}
\end{equation}

\subsubsection*{The Fundamental String - F1}
\begin{equation}
\begin{aligned}
\dd s^2 &= H^{-3/4}[-\dd t^2 + \dd x^2] + H^{1/4}\dd\vec{y}_{(8)}^{\, 2}   \\
B_{tx} &= -(H^{-1}-1), \qquad B_{iy^4y^5y^6y^7y^8} = A_i\\
H &= 1 + \frac{h}{|\vec{y}_{(8)}|^6} , \qquad e^{2\phi} = H^{-1}e^{2\phi_0}
\end{aligned}
\label{eq:F1}
\end{equation}
KK-decomposition: $x^\mu = (t, \vec{y}_{(3)})$ and $x^\bm = (x, \vec{y}_{(5)})$
\begin{equation}
\begin{aligned}
\bg_{\bm\bn} &= H^{1/4}\diag[H^{-1},\delta_5] \, , & \det \bg_{\bm\bn}&=H^{1/2}  \\
g_{\mu\nu} &= \diag[-H^{-1/2},H^{1/2}\delta_{ij}] 
\end{aligned}
\end{equation}

\subsubsection*{The Solitonic Fivebrane - NS5}
\begin{equation}
\begin{aligned}
\dd s^2 &= H^{-1/4}[-\dd t^2 + \dd\vec{x}_{(5)}^{\, 2}] + H^{3/4}\dd\vec{y}_{(4)}^{\, 2}   \\
B_{tx^1x^2x^3x^4x^5} &= -(H^{-1}-1), \qquad B_{iy^4} = A_i\\
H &= 1 + \frac{h}{|\vec{y}_{(4)}|^2} , \qquad e^{2\phi} = He^{2\phi_0}
\end{aligned}
\label{eq:NS5}
\end{equation}
KK-decomposition: $x^\mu = (t, \vec{y}_{(3)})$ and $x^\bm = (\vec{x}_{(5)}, y^4)$
\begin{equation}
\begin{aligned}
\bg_{\bm\bn} &= H^{3/4}\diag[H^{-1}\delta_5, 1] \, , & \det \bg_{\bm\bn}&=H^{-1/2}  \\
g_{\mu\nu} &= \diag[-H^{-1/2},H^{1/2}\delta_{ij}] 
\end{aligned}
\end{equation}

\subsubsection*{The Monopole - KK6A/B}
\begin{equation}
\begin{aligned}
\dd s^2 &= -\dd t^2 + \dd\vec{x}_{(5)}^{\, 2} 
				+ H^{-1}\left[\dd z + A_i\dd y^i\right]^2 + H\dd\vec{y}_{(3)}^{\, 2}  \\
H &= 1 + \frac{h}{|\vec{y}_{(3)}|} \, , \qquad e^{2\phi} = e^{2\phi_0}
\end{aligned}
\label{eq:KK6A}
\end{equation}
KK-decomposition: $x^\mu = (t, \vec{y}_{(3)})$ and $x^\bm = (z, \vec{x}_{(5)})$
\begin{equation}
\begin{aligned}
\bg_{\bm\bn} &= \diag[H^{-1},\delta_5] \, , & \det \bg_{\bm\bn}&=H^{-1}  \\
g_{\mu\nu} &= \diag[-H^{-1/2},H^{1/2}\delta_{ij}] \, , &  {A_i}^z &= A_i 
\end{aligned}
\end{equation}

\subsection{D-Branes in $D=10$}
\label{sec:Dbranes}

\subsubsection*{The Dp-Brane for $p=0,\dots,6$}
\begin{equation}
\begin{aligned}
\dd s^2 &= H^{\frac{p-7}{8}}[-\dd t^2 + \dd\vec{x}_{(p)}^{\, 2}] 
			+ H^{\frac{p+1}{8}}\dd\vec{y}_{(9-p)}^{\, 2}   \\
C_{tx^1\dots x^p} &= -(H^{-1}-1), \qquad C_{iy^4\dots y^{9-p}} = A_i\\
H &= 1 + \frac{h}{|\vec{y}_{(9-p)}|^{7-p}} , \qquad e^{2\phi} = H^{\frac{3-p}{2}} e^{2\phi_0}
\end{aligned}
\label{eq:Dbrane}
\end{equation}
KK-decomposition: $x^\mu = (t, \vec{y}_{(3)})$ and $x^\bm = (\vec{x}_{(p)}, \vec{y}_{(6-p)})$
\begin{equation}
\begin{aligned}
\bg_{\bm\bn} &= H^{\frac{p+1}{8}}\diag[H^{-1}\delta_p,\delta_{6-p}] \, , & 
	\det \bg_{\bm\bn}&=H^{\frac{3-p}{4}}  \\
g_{\mu\nu} &= \diag[-H^{-1/2},H^{1/2}\delta_{ij}] 
\end{aligned}
\end{equation}
Note: In Type IIB the D1-brane forms an S-duality doublet with the F1-string. This means they are identical solutions up to an $SL(2)$ transformation and their dilatons are inverses of each other. The same applies for the D5-brane and the NS5-brane.

\newpage
\addcontentsline{toc}{chapter}{Bibliography}
\bibliography{mybib}

\end{document}